\documentclass[aip,jmp,reprint,amsmath,amssymb,onecolumn,nofootinbib,longbibliography]{revtex4-2}
\pdfoutput=1
\usepackage[utf8]{inputenc}
\usepackage{graphicx,bbm}
\usepackage[colorlinks=true,urlcolor=blue,anchorcolor=blue,citecolor=blue,filecolor=blue,linkcolor=blue,menucolor=blue,linktocpage=true,unicode=true,bookmarksopen=true,pdfa=true]{hyperref}
\usepackage{mathtools}
\mathtoolsset{mathic} 
\usepackage{lmodern}
\RequirePackage[T1]{fontenc}
\DeclareFontShape{OMX}{cmex}{m}{n}{
 <-7.5> cmex7
 <7.5-8.5> cmex8
 <8.5-9.5> cmex9
 <9.5-> cmex10
}{}
\SetSymbolFont{largesymbols}{normal}{OMX}{cmex}{m}{n}
\SetSymbolFont{largesymbols}{bold} {OMX}{cmex}{m}{n}

\DeclareSymbolFontAlphabet{\mathbb}{AMSb}
\DeclareMathAlphabet{\mathsfi}{OT1}{cmss}{m}{sl}
\DeclareMathAlphabet{\mathbfi}{OML}{cmm}{b}{it}
\DeclareSymbolFont{rsfs}{U}{rsfs}{m}{n}
\DeclareSymbolFontAlphabet{\mathscr}{rsfs}

\usepackage{graphicx}
\usepackage{subcaption}

\let\originalleft\left
\let\originalright\right
\renewcommand{\left}{\mathopen{}\mathclose\bgroup\originalleft}
\renewcommand{\right}{\aftergroup\egroup\originalright}

\makeatletter
\newenvironment{equations}[1][]{\subequations\ifx\relax#1\relax\else\label{#1}\fi\align\ignorespaces}{\endalign\ignorespacesafterend\endsubequations}
\def\@spliteq#1{\begin{equation}\begin{split}#1\end{split}\end{equation}}
\def\@spliteqstar#1{\begin{equation*}\begin{split}#1\end{split}\end{equation*}}
\def\splitequation{\collect@body\@spliteq}

\expandafter\def\csname splitequation*\endcsname{\collect@body\@spliteqstar}
\expandafter\def\csname endsplitequation*\endcsname{\ignorespacesafterend}
\makeatother

\usepackage{accents}
\newlength{\dhatheight}
\newcommand{\dhat}[1]{%
 \settoheight{\dhatheight}{\ensuremath{\hat{#1}}}%
 \addtolength{\dhatheight}{-0.25ex}%
 \hat{\vphantom{\rule{1pt}{\dhatheight}}%
 \smash{\hat{#1}}}}

\renewcommand{\vec}[1]{{\ifnum9<1#1\mathbf{#1}\else\ifcat\noexpand#1\relax\boldsymbol{#1}\else\mathbfi{#1}\fi\fi}}
\newcommand{\mathe}{\mathrm{e}}
\newcommand{\mathi}{\mathrm{i}}
\let\oldre\Re
\let\oldim\Im
\renewcommand{\Re}{\oldre\mathfrak{e}\,}
\renewcommand{\Im}{\oldim\mathfrak{m}\,}
\newcommand{\total}{\mathop{}\!\mathrm{d}}

\newcommand{\1}{\mathbbm{1}}

\newcommand{\eqend}[1]{\,#1}
\newcommand{\bigo}[1]{\mathcal{O}\left({#1}\right)}

\newcommand{\brst}{\mathop{}\!\mathsf{s}\hskip 0.05em\relax}
\newcommand{\expect}[1]{\left\langle{#1}\right\rangle}
\newcommand{\bessel}[3]{\mathop{}\!\mathrm{#1}_{#2}\left(#3\right)}

\allowdisplaybreaks

\begin{document}

\title{Heat kernel coefficients for massive gravity}

\author{Renata Ferrero}
\email{renata.ferrero@fau.de}
\affiliation{Institut f{\"u}r Physik (THEP), Johannes Gutenberg-Universit{\"a}t Mainz,\\ Staudingerweg 7, 55128 Mainz, Germany}
\affiliation{Institute for Quantum Gravity, Friedrich-Alexander-Universit{\"a}t Erlangen-N{\"u}rnberg, Staudtstr. 7, 91058 Erlangen, Germany}

\author{Markus B. Fr\"ob}
\email{mfroeb@itp.uni-leipzig.de}
\affiliation{Institut f{\"u}r Theoretische Physik, Universit{\"a}t Leipzig,\\ Br{\"u}derstra{\ss}e 16, 04103 Leipzig, Germany}

\author{William C. C. Lima}
\email{williamcclima@gmail.com}
\affiliation{Institut f{\"u}r Theoretische Physik, Universit{\"a}t Leipzig,\\ Br{\"u}derstra{\ss}e 16, 04103 Leipzig, Germany}

\begin{abstract}
We compute the heat kernel coefficients that are needed for the regularization and renormalization of massive gravity. Starting from the Stueckelberg action for massive gravity, we determine the propagators of the different fields (massive tensor, vector and scalar) in a general linear covariant gauge depending on four free gauge parameters. We then compute the non-minimal heat kernel coefficients for all the components of the scalar, vector and tensor sector, and employ these coefficients to regularize the propagators of all the different fields of massive gravity. We also study the massless limit and discuss the appearance of the van Dam--Veltman--Zakharov discontinuity. In the course of the computation, we derive new identities relating the heat kernel coefficients of different field sectors, both massive and massless.
\end{abstract}

\date{08. January 2024}
\revised{22. May 2024}

\maketitle

\section{Introduction}
\label{sec:intro}

In both mathematics and physics the heat kernel technique is a well-established method for the computation of traces of differential operators~\cite{Avramidi:2000bm,Barvinsky:1985an,Barvinsky:1990up,Vilkovisky:1992za,Vassilevich:2003xt,Codello:2012kq}. Its area of application ranges from fluctuations of quantum fields on curved spacetimes, the determination of ultraviolet divergences, effective actions and quantum anomalies to quantum gravity and further. Concretely, it allows to define the regularized propagator and using this the one-loop effective action, and to compute counterterms and anomalies in a way that is naturally extended to field theories on curved space. Furthermore, it constitutes an essential ingredient in solving the gravitational functional renormalization group equation~\cite{Reuter:2019byg,Percacci:2017fkn}.

The interest in the heat kernel can be traced back to the beginning of the theory of quantum fields in curved spacetimes. In the fifties, Morette and DeWitt established the path integral approach to propagators in curved spacetime, and in the sixties Schwinger and DeWitt developed the proper time formalism and the associated heat kernel expansion. The main idea is to express a given Green’s function as an integral of the heat kernel (which satisfies the heat equation with time $\tau$, the proper time) over the proper time. The heat kernel depends on the background geometry, especially the background metric, and possibly other background fields such as gauge fields. Except for some special manifolds, such as maximally symmetric spacetimes, it is not possible to compute it exactly in general. However, it is possible to compute it approximately, and up to now two main expansion schemes have been developed: the small-time or local expansion (also known as the Seeley--DeWitt technique), and a non-local expansion for small curvature. The first scheme is employed in quantum field theory to compute ultraviolet divergences and anomalies, since these are directly determined by the lowest-order heat kernel coefficients in the small-time expansion. The non-local expansion is used instead to compute the finite part of the effective action in a covariant manner. In the framework of quantum gravity, non-local heat kernels are crucial in renormalization group studies, such as tests of the Asymptotic Safety conjecture, and heat kernels on maximally symmetric spaces are used to find $\beta$ functions for gravitational couplings~\cite{reuter_prd_1998,lauscher_reuter_prd_2002,Codello:2008vh,Knorr:2021slg}. A recent development in this framework is the consideration of Lorentzian renormalization group (RG) flows~\cite{Manrique:2011jc,Fehre:2021eob,Banerjee:2022xvi,Saueressig:2023tfy,DAngelo:2023ssw,DAngelo:2023wje}, which is important since a generic metric does not admit a Wick rotation. In general, including couplings with increasing scaling dimension such as higher curvature terms, an increasing number of heat kernel coefficients is needed, whose computation becomes increasingly difficult.

For gauge theories of Yang--Mills type it is straightforward to determine classical observables using the BRST formalism~\cite{becchietal1975,weinberg_v2,barnichetal2000}. Since these generally are not linear in the elementary fields, in the quantum theory they become composite operators which need additional renormalization, beyond the usual renormalization of couplings~\cite{collinsbook}. This can also be done using the heat kernel technique, point-splitting the classical expression and subtracting as many terms of the small-time expansion of the heat kernel as are needed to obtain a finite coincidence limit (depending on the scaling dimension of the operator). When using an effective action, one has to introduce an additional term coupling the composite operator to an external source, renormalize the extended effective action, and then take functional derivatives with respect to the source to obtain correlation functions including insertions of this operator. In the exact renormalization group framework, this has been recently used to study flows of volume and length operators in gravity~\cite{Pagani:2016dof,Houthoff:2020zqy,Becker:2019fhi}, see also Refs.~\onlinecite{Ambjorn:1997di,Hamber:2009zz}. For example, when studying the correlator of metric fluctuations in the vielbein formalism, the metric itself is a composite operator which needs to be regularized and renormalized in a suitable way, such as the point-splitting method~\cite{DeWitt:1964mxt,DeWitt:1975ys,Christensen:1976vb} or the heat kernel renormalization method explained above.

However, when defining observables in gravity another complication arises. Namely, in contrast to Yang--Mills gauge theories where the gauge symmetry is an internal symmetry and acts locally, i.e., it transforms fields at the same point, the diffeomorphism symmetry of gravity moves points around. It follows immediately that a local field cannot be gauge-invariant, and that observables must be necessarily nonlocal. Various approaches to construct such nonlocal observables have been considered, for example defining correlation functions involving the geodesic distance between points~\cite{mandelstam1962,mandelstam1968,tsamiswoodard1992,teitelboim1993,hamber1994,modanese1994,ambjorn1997,khavkine2012,bongakhavkine2014,froeb2018,klitgaardloll2018,beckerpagani2019} or relational observables. Nevertheless, once one has suitable observables at hand, also for gravity the usual framework of regularization and renormalization applies, and a natural question that arises is the gauge dependence of those regularized and renormalized quantities in the quantum theory. While the gauge independence of the observables is clear in the classical theory, in the quantum theory anomalies may arise which then might render the final result gauge-dependent. To check the gauge independence of the renormalized observables, it is useful to work in a general family of gauges that depend on a certain number of gauge parameters, and then verify that the end result is independent of these parameters. For this, it is necessary to know the Green's functions and the heat kernel coefficients in such a general family of gauges. In Ref.~\onlinecite{froebtaslimitehrani2018}, the authors constructed Green's functions in a globally hyperbolic spacetime in general linear covariant gauges, both for vector gauge bosons and linearized Einstein gravity in the presence of a cosmological constant. In this work, we will generalize their result to the massive gravity case, relating the propagators of the massive gravity field sectors in a general linearized gauge to their expressions in the corresponding Feynman gauge.

The choice of the massive gravity theory is motivated from both a phenomelogical and cosmological point of view, and for theoretical interest. Modifications of Einstein gravity, of which massive gravity theories are a subset, give rise to an effective cosmological fluid which can explain dark matter and/or dark energy~\cite{damicoetal2011,derham2014,koyama2015}. On the other hand, while there are stringent upper bounds on the possible mass of the graviton~\cite{gravitonmass1,gravitonmass2}, a tiny mass is not ruled out completely, even in light of the van Dam--Veltman--Zakharov (vDVZ) discontinuity~\cite{vandamveltman1970,zakharov1970}. On the theoretical side, to the best of our knowledge the propagators for a massive spin-2 field in a general gauge have not been determined, and consequently also the corresponding heat kernel expansion is unknown. To avoid breaking diffeomorphism invariance, which a simple Fierz--Pauli mass term~\cite{fierz1939,paulifierz1939,fierzpauli1939} would do, we use the Stueckelberg formalism~\cite{stueckelberg1938a,stueckelberg1938b} and add additional spin-1 and spin-0 fields that transform in such a way to keep diffeomorphism invariance of the complete theory. That is, we consider in this work only the free massive gravity theory, which determines the propagators. Moreover, we work with an on-shell background satisfying the background Einstein equations, which means that the propagators that we derive could, for example, later on be used to determine an on-shell effective action for gravity~\cite{Parker:2009uva}. While this is not strictly necessary, since one could also work with the off-shell (Vilkovisky--deWitt) invariant effective action~\cite{VILKOVISKY1984125}, the resulting expressions are already complicated enough for an on-shell background, and would become unmanageable off-shell.

This article is structured as follows: In Section~\ref{sec:massivegravity} we give some more background on massive gravity, determine its linearized action including the Stueckelberg fields that are necessary to preserve diffeomorphism invariance, and using the BRST formalism add the gauge-fixing and ghost terms that determine a quite general linear covariant gauge for the fields. Section~\ref{sec:propagators} is devoted to the determination of the propagators in this gauge, which comprises three free gauge parameters, as well as their massless limit, and the verification of the Ward identities in the free theory. In Section~\ref{sec:heatkernel}, we give an overview of the small-time expansion of the heat kernel, compute the leading three heat kernel coefficients, and determine relations between the heat kernel coefficients of different spins. Section~\ref{sec:propagatorheatkernel} comprises our main results, the heat kernel expansion of the propagators for all fields in massive gravity, including the massless limit. Finally, in Section~\ref{sec:discussion} we discuss our result and present an outlook for future work.

\paragraph*{Conventions:} Our conventions are a mostly plus metric, the Riemann tensor defined by $\nabla_\mu \nabla_\nu v_\rho - \nabla_\nu \nabla_\mu v_\rho = R_{\mu\nu\rho\sigma} v^\sigma$, and the Ricci tensor $R_{\mu\nu} = R^\rho{}_{\mu\rho\nu}$. We work in $n$ dimensions, use geometrical units $\hbar = c = 1$, and set $\kappa \equiv \sqrt{ 16 \pi G_\text{N} } \propto \ell_\text{Pl}$ with Newton's constant $G_\text{N}$ and the Planck length $\ell_\text{Pl}$.

\section{Massive gravity}
\label{sec:massivegravity}

Even though General Relativity (GR) has been very well tested and successfully describes gravity both a small and large scales~\cite{will_lrr_2014}, there are also some tensions. Of those, let us mention the observed galaxy rotation curves~\cite{corbelli_salucci_mnras_1999,deblok_aa_2010} and primodial inflation~\cite{guth_prd_1981,linde_plb_1982,planck_vi_2020,planck_x_2020}, which cannot be explained using GR alone. Various models have been proposed to explain these effects and alleviate the tensions, of which dark matter -- matter that only interacts gravitationally, or only very weakly with Standard Model particles --- can explain the galaxy rotation curves and dark energy --- an effective fluid with negative pressure --- can explain inflation.\footnote{We note that the introduction of a cosmological constant is not enough to explain both the primordial accelerated expansion and the current one, since they occur at widely different scales; a dynamical model is thus necessary.} While dark matter can be easily added as a new fundamental particle (or several) such as axions, and the current expansion can be explained by a cosmological constant, primordial dark energy is usually modelled by a single fundamental scalar field (the inflaton) with a suitable potential, see Refs.~\onlinecite{castelloilickunz2021,escamillaetal_arxiv_2023} for recent reviews. Other fundamental explanations of dark matter and dark energy come in the form of modified gravity models~\cite{shankaranarayanan_johnson_grg_2022}, of which massive gravity in its various incarnations~\cite{Hinterbichler:2011tt,derham2014,Tolley:2015oxa,deRham:2014naa,deRham:2023ngf} is a popular one.

In massive gravity, the mediator of gravitational interactions, the spin-2 field usually called graviton, is given a small mass. From a particle physics perspective, this is a natural modification since GR can be thought of as the unique theory of a massless spin-2 particle~\cite{deser_grg_1970,wald_prd_1986,deser_cqg_1987}; it also seems natural since we know that the carrier particles of the electroweak forces acquire a mass through the Higgs mechanism. Since the force mediated by massive particles falls off exponentially at large distances, a graviton with a small mass of the order of the Hubble rate $m \sim H$ could also mimic the effects of the current accelerated expansion of our universe without invoking a cosmological constant or other forms of dark energy. Another motivation for infrared modifications of gravity comes from brane world models~\cite{randall_sundrum_prl_1999,gregory_rubakov_sibiryakov_prl_2000}, where extra dimensions can be of large or infinite size and the effective four-dimensional graviton propagator on our brane can behave like a massive one. We note that giving a mass to a free spin-2 field is easy, and was already done in 1939 by Fierz and Pauli~\cite{fierz1939,paulifierz1939,fierzpauli1939}; the construction of interacting theories of massive gravity is much harder. One possibility for a non-linear generalization of the Fierz--Pauli mass term was explored in~\cite{Hassan:2011zd}, where an additional fully dynamical ``reference'' metric was introduced, rendering the model invariant under general coordinate transformations. Such theories are called \emph{bi-metric theories of gravity}, and have been of particular interest due to the presence of accelerating cosmological solutions. Observational data from both the early evolution of the universe and solar system tests of gravity can be used to constrain the parameters of these theories~\cite{Caravano:2021aum,Hogas:2021fmr,Hogas:2021lns,Hogas:2021saw}. Another possibility is topologically massive gravity~\cite{deser_jackiw_templeton_prl_1982,deser_jackiw_templeton_annals_1982} and its generalizations~\cite{bergshoeff_hohm_townsend_prl_2009}, which was the first interacting model of massive gravity in three dimensions. Its $\beta$ functions can be computed in the asymptotic safety approach, and it turns out that as Einstein gravity in four dimensions~\cite{reuter_prd_1998,lauscher_reuter_prd_2002}, topologically massive gravity is asymptotically safe~\cite{ohta_percacci_cqg_2014}. Since Lorentz-invariant models of massive gravity often suffer from the presence of ghosts (fields with a wrong sign of the kinetic term), a possibility which recently also attracted attention is to allow for a violation of Lorentz invariance~\cite{Blas:2009my,Rubakov:2004eb,Rubakov:2008nh,Dubovsky:2004ud,Arkani-Hamed:2003pdi}. From the cosmological point of view, this is particularly interesting because one can construct a consistent model of gravity where the tensor graviton mode is massive, while the linearized equations for scalar and vector metric perturbations are not modified. The Friedmann equation then acquires an effective dark-energy component, leading to an accelerated expansion, while gravity in the solar system is not modified.

However, in this work we are interested in the linearized Fierz--Pauli theory that preserves Lorentz invariance, and which can be obtained as the first approximation of a general massive gravity theory around a given fixed background. It turns out that even in the massless limit, the extra polarizations of a massive spin-2 field do not completely decouple in general, unlike what happens for an Abelian gauge theory. Instead, a scalar mode remains coupled to the tensor modes, and the resulting theory differs from the completely massless one, a effect that is known as the van Dam--Veltman--Zakharov (vDVZ) discontinuity~\cite{vandamveltman1970,zakharov1970}.\footnote{A similar discontinuity also appears in Yang--Mills theory, see Ref.~\onlinecite{hell2022} and references therein.} Nevertheless, whether decoupling occurs depends strongly on the concrete theory, which in our case means the background around which we study the theory. Let us mention here that the discontinuity is absent when one considers GR with a non-vanishing cosmological constant $\Lambda$. While for an anti-de Sitter background with $\Lambda < 0$ the resulting theory is ghost-free for all masses~\cite{Kogan:2000uy,Porrati:2000cp}, in $n$-dimensional de Sitter spacetime with $\Lambda = (n-1) (n-2)/2 H^2 > 0$ the theory has a scalar ghost if $0 < m^2 < (n-2) H^2$; the upper bound $(n-2) H^2$ is known as the Higuchi bound~\cite{Higuchi:1986py}. For a modern account of these issues we refer the reader to Refs.~\onlinecite{Arkani-Hamed:2002bjr,Dubovsky:2002jm,Buoninfante:2023ryt} and references therein.

Even though Fierz--Pauli theory preserves Lorentz invariance, the mass term breaks linearized diffeomorphisms, which are a symmetry of the massless theory that descends from the full Einstein--Hilbert action for gravity. To remedy this, we employ the Stueckelberg trick~\cite{stueckelberg1938a,stueckelberg1938b} and add auxiliary fields that transform in such a way as to keep the full action invariant under linearized diffeomorphisms. The original massive Fierz--Pauli theory can then be recovered in a special gauge, the so-called unitary gauge. While both perturbative and non-perturbative computations in gravity have used heat kernel expansions (see for example Ref.~\onlinecite{Kolar:2023mkw} for a recent study), and also massive vector fields have been studied in great detail~\cite{Belokogne:2015etf,Belokogne:2016dvd,Buchbinder:2017zaa}, to our knowledge the present work represents the first computation of the heat kernel coefficients for a massive theory of gravity in a general gauge.


\subsection{Action}

We consider the Einstein--Hilbert action for gravity including a cosmological constant, and add Stueckelberg fields to make the theory massive while keeping gauge invariance~\cite{stueckelberg1938a,stueckelberg1938b}. Working in a perturbative approach, we separate the full metric $\tilde{g}_{\mu\nu}$ into a background metric $g_{\mu\nu}$ and perturbations $h_{\mu\nu}$ according to
\begin{equation}
\label{eq:metric_split}
\tilde{g}_{\mu\nu} = g_{\mu\nu} + \kappa h_{\mu\nu} \eqend{,}
\end{equation}
where $\kappa \equiv \sqrt{16 \pi G_\text{N}}$ is our perturbation parameter. Expanding the Einstein--Hilbert action to second order in $h_{\mu\nu}$, we obtain
\begin{splitequation}
\label{eq:action_eh_offshell}
S_\text{EH} &= \frac{1}{\kappa^2} \int \left( R - 2 \Lambda \right) \sqrt{-g} \total^n x - \frac{1}{\kappa} \int \Big( R_{\mu\nu} - \frac{1}{2} R g_{\mu\nu} + \Lambda g_{\mu\nu} \Big) h^{\mu\nu} \sqrt{-g} \total^n x \\
&\quad+ \frac{1}{4} \int \Big[ h^{\mu\nu} \left( \nabla^2 h_{\mu\nu} - 2 \nabla^\rho \nabla_\mu h_{\nu\rho} + 2 \nabla_\mu \nabla_\nu h \right) - h \nabla^2 h \\
&\hspace{4em}+ \Big( \Lambda - \frac{1}{2} R \Big) \left( 2 h^{\mu\nu} h_{\mu\nu} - h^2 \right) + 2 R_{\mu\nu} \left( 2 h^{\mu\rho} h^\nu{}_\rho - h h^{\mu\nu} \right) \Big] \sqrt{-g} \total^n x + \bigo{\kappa} \eqend{,} \raisetag{4.8em}
\end{splitequation}
where $\Lambda$ is the cosmological constant. The background equations of motion are obtained by varying $S_\text{EH}$ with respect to $h_{\mu\nu}$ and setting $h$ to zero, which results in the well-known Einstein equations
\begin{equation}
\label{eq:einstein_eom}
R_{\mu\nu} - \frac{1}{2} R g_{\mu\nu} + \Lambda g_{\mu\nu} = 0 \eqend{,} \quad R_{\mu\nu} = \frac{2}{n-2} \Lambda g_{\mu\nu} \eqend{.}
\end{equation}
When these are satisfied, we say that the theory has an on-shell background or simply is on-shell; the terms quadratic in the perturbation $h_{\mu\nu}$ in the action~\eqref{eq:action_eh_offshell} then simplify to
\begin{splitequation}
\label{eq:action_eh_onshell}
S_\text{EH}^{(2)} &= \frac{1}{4} \int \Big[ h^{\mu\nu} \left( \nabla^2 h_{\mu\nu} - 2 \nabla^\rho \nabla_\mu h_{\nu\rho} + 2 \nabla_\mu \nabla_\nu h \right) - h \nabla^2 h \\
&\hspace{6em}+ \frac{2}{n-2} \left( 2 h^{\mu\nu} h_{\mu\nu} - h^2 \right) \Lambda \Big] \sqrt{-g} \total^n x \eqend{,}
\end{splitequation}
A straightforward computation using the background Einstein equations~\eqref{eq:einstein_eom} shows that this action is invariant under the gauge transformation
\begin{equation}
\label{eq:hmunu_gauge}
\delta_\xi h_{\mu\nu} = \nabla_\mu \xi_\nu + \nabla_\nu \xi_\mu \eqend{,}
\end{equation}
which arises from the linearization of the full diffeomorphism symmetry $\delta_\xi \tilde{g}_{\mu\nu} = \mathscr{L}_\xi \tilde{g}_{\mu\nu}$ around the background~\eqref{eq:metric_split} after the rescaling of the gauge parameter $\xi^\mu \to \kappa \xi^\mu$.

To obtain the (free) massive gravity theory, we add the Fierz--Pauli mass term~\cite{fierz1939,paulifierz1939,fierzpauli1939}
\begin{equation}
\label{eq:action_fp}
S_\text{FP} = - \frac{1}{4} \int m^2 \left( h^{\mu\nu} h_{\mu\nu} - h^2 \right) \sqrt{-g} \total^n x \eqend{,}
\end{equation}
which is however not invariant under the gauge symmetry~\eqref{eq:hmunu_gauge}. To preserve the symmetry, we add a compensating Stueckelberg field $A_\mu$ that transforms as $\delta_\xi A_\mu = \xi_\mu$, and replace $h_{\mu\nu} \to h_{\mu\nu} - \nabla_\mu A_\nu - \nabla_\nu A_\mu$ in the Fierz--Pauli action~\eqref{eq:action_fp}. To obtain a smooth limit $m \to 0$, we further introduce another (secondary) Stueckelberg field $\phi$ and a gauge symmetry
\begin{equation}
\label{eq:amuphi_gauge}
\delta_f A_\mu = \partial_\mu f \eqend{,} \quad \delta_f \phi = f \eqend{,}
\end{equation}
and use the combination $A_\mu - \partial_\mu \phi$ instead of $A_\mu$. Note that $\phi$ is inert under linearized diffeomorphisms. Taking all together, we obtain the extended Fierz--Pauli--Stueckelberg action
\begin{splitequation}
\label{eq:action_fps}
S_\text{FPS} &= - \frac{1}{4} m^2 \int \Big[ \left( h^{\mu\nu} - \nabla^\mu A^\nu - \nabla^\nu A^\mu + 2 \nabla^\mu \nabla^\nu \phi \right) \left( h_{\mu\nu} - \nabla_\mu A_\nu - \nabla_\nu A_\mu + 2 \nabla_\mu \nabla_\nu \phi \right) \\
&\qquad\qquad- \left( h - 2 \nabla_\mu A^\mu + 2 \nabla^2 \phi \right)^2 \Big] \sqrt{-g} \total^n x \eqend{,} \raisetag{1.5em}
\end{splitequation}
which is invariant under both the gravitational gauge symmetry~\eqref{eq:hmunu_gauge} and the new gauge symmetry~\eqref{eq:amuphi_gauge}. The unitary gauge, in which we recover the original theory, is achieved by making a gauge transformation with parameters $\xi_\mu = - A_\mu - \partial_\mu \phi$ and $f = - \phi$, which sets both the new $A_\mu$ and $\phi$ to zero.

For vanishing cosmological constant $\Lambda = 0$ and in the limit $m \to 0$, the different sectors (tensor $h_{\mu\nu}$, vector $A_\mu$ and scalar $\phi$) actually decouple. To see this, we follow Ref.~\onlinecite{Hinterbichler:2011tt} and rescale $A_\mu \to m^{-1} A_\mu$, $\phi \to m^{-2} \phi$. This results in
\begin{splitequation}
\label{eq:action_fps_rescaled}
S_\text{FPS} &= - \frac{1}{4} m^2 \int \left( h^{\mu\nu} h_{\mu\nu} - h^2 \right) \sqrt{-g} \total^n x - m \int A^\mu \left( \nabla^\nu h_{\mu\nu} - \nabla_\mu h \right) \sqrt{-g} \total^n x \\
&\quad- \int \phi \left( \nabla_\mu \nabla_\nu h^{\mu\nu} - \nabla^2 h \right) \sqrt{-g} \total^n x - \frac{1}{4} \int \left( F^{\mu\nu} F_{\mu\nu} - 4 R^{\mu\nu} A_\mu A_\nu \right) \sqrt{-g} \total^n x \\
&\quad+ m^{-2} \int R^{\mu\nu} \nabla_\mu \phi \nabla_\nu \phi \sqrt{-g} \total^n x - 2 m^{-1} \int A_\mu R^{\mu\nu} \nabla_\nu \phi \sqrt{-g} \total^n x
\end{splitequation}
and the rescaled gauge transformations
\begin{equations}
\label{eq:gauge_trafo}
\delta_\xi h_{\mu\nu} &= \nabla_\mu \xi_\nu + \nabla_\nu \xi_\mu \eqend{,} \quad \delta_\xi A_\mu = m \xi_\mu \eqend{,} \\
\delta_f A_\mu &= \partial_\mu f \eqend{,} \hspace{5.8em} \delta_f \phi = m f \eqend{.}
\end{equations}
Using now the on-shell equations of motion (EOM)~\eqref{eq:einstein_eom} and adding the second-order Einstein--Hilbert action~\eqref{eq:action_eh_onshell}, we obtain the full action that we consider in this work:
\begin{splitequation}
\label{eq:full_action_onshell}
S &= S_\text{EH}^{(2)} + S_\text{FPS} \\
&= \frac{1}{4} \int \Big[ h^{\mu\nu} \left( \nabla^2 h_{\mu\nu} - 2 \nabla^\rho \nabla_\mu h_{\nu\rho} + 2 \nabla_\mu \nabla_\nu h \right) - h \nabla^2 h \\
&\hspace{6em}+ \frac{2}{n-2} \left( 2 h^{\mu\nu} h_{\mu\nu} - h^2 \right) \Lambda \Big] \sqrt{-g} \total^n x \\
&\quad- \frac{1}{4} m^2 \int \left( h^{\mu\nu} h_{\mu\nu} - h^2 \right) \sqrt{-g} \total^n x - m \int A^\mu \left( \nabla^\nu h_{\mu\nu} - \nabla_\mu h \right) \sqrt{-g} \total^n x \\
&\quad- \int \phi \left( \nabla_\mu \nabla_\nu h^{\mu\nu} - \nabla^2 h \right) \sqrt{-g} \total^n x - \frac{1}{4} \int \Big( F^{\mu\nu} F_{\mu\nu} - \frac{8 \Lambda}{n-2} A^\mu A_\mu \Big) \sqrt{-g} \total^n x \\
&\quad+ \frac{2}{n-2} \frac{\Lambda}{m^2} \int \nabla^\mu \phi \nabla_\mu \phi \sqrt{-g} \total^n x - \frac{4}{n-2} \frac{\Lambda}{m} \int A^\mu \nabla_\mu \phi \sqrt{-g} \total^n x \eqend{,}
\end{splitequation}
and for $\Lambda = 0$ the last two terms (which would be problematic as $m \to 0$) vanish, as well as the mass term for the vector $A_\mu$. If we then take $m \to 0$, the last two gauge transformations in~\eqref{eq:gauge_trafo} vanish and we are only left with the original gravitational transformation~\eqref{eq:hmunu_gauge} and the usual electromagnetic $\mathrm{U}(1)$ gauge transformation. Moreover, the full action~\eqref{eq:full_action_onshell} can be rewritten as
\begin{splitequation}
\label{eq:full_action_onshell_zerolambdamass}
\lim_{m \to 0} \lim_{\Lambda \to 0} S &= \frac{1}{4} \int \Big[ \tilde{h}^{\mu\nu} \left( \nabla^2 \tilde{h}_{\mu\nu} - 2 \nabla^\rho \nabla_\mu \tilde{h}_{\nu\rho} + 2 \nabla_\mu \nabla_\nu \tilde{h} \right) - \tilde{h} \nabla^2 \tilde{h} \Big] \sqrt{-g} \total^n x \\
&\quad+ \frac{n-1}{n-2} \int \phi \nabla^2 \phi \sqrt{-g} \total^n x - \frac{1}{4} \int F^{\mu\nu} F_{\mu\nu} \sqrt{-g} \total^n x
\end{splitequation}
with the new field $\tilde{h}_{\mu\nu} \equiv h_{\mu\nu} - 2/(n-2) g_{\mu\nu} \phi$, which is just a sum of the free actions and thus the tensor, vector and scalar sectors decouple in this limit. However, we will keep both $\Lambda$ and $m^2$ non-zero throughout the remainder of this work, and in fact define
\begin{equation}
\label{eq:lambda_def}
\lambda \equiv \frac{4}{n-2} \frac{\Lambda}{m^2}
\end{equation}
to shorten the expressions. Note that $\lambda$ is a dimensionless parameter, such that the mass $m$ is the only dimensionful parameter that is left.

\subsection{BRST, gauge fixing and ghosts}

To quantize the Stueckelberg gauge theory of (free) massive gravity with action~\eqref{eq:full_action_onshell}, we use the standard BRST formalism~\cite{becchietal1975,weinberg_v2,barnichetal2000}, where we introduce ghost, antighost and auxiliary fields for every symmetry. The gauge transformations~\eqref{eq:gauge_trafo} are replaced by the action of the fermionic BRST operator $\brst$ whose action on fields is obtained by replacing the gauge parameter by the corresponding ghost field, and the action on the ghost, antighost and auxiliary fields must be determined such that the action is nilpotent: $\brst^2 = 0$.

We therefore need a diffeomorphism ghost $c_\mu$ and a Stueckelberg ghost $c$. We also define their antighosts $\bar{c}_\mu$ and $\bar{c}$, and introduce for each an auxiliary field $B_\mu$ and $B$. Then the BRST transformations on fields and ghosts are obtained from Eq.~\eqref{eq:gauge_trafo} and read
\begin{equations}[eq:brst_fields]
\brst h_{\mu\nu} &= \nabla_\mu c_\nu + \nabla_\nu c_\mu \eqend{,} \\
\brst A_\mu &= m c_\mu + \partial_\mu c \eqend{,} \\
\brst \phi &= m c \eqend{,} \\
\brst c_\mu &= 0 = \brst c \eqend{.}
\end{equations}
In our case, the ghosts have trivial BRST transformations since we consider a free theory (but on a non-trivial background) such that the gauge transformations~\eqref{eq:gauge_trafo} are linear. The antighosts and auxiliary fields correspondingly form trivial pairs:
\begin{equations}[eq:brst_ghosts]
\brst \bar{c}_\mu &= B_\mu \eqend{,} \quad \brst B_\mu = 0 \eqend{,} \\
\brst \bar{c} &= B \eqend{,} \hspace{2em} \brst B = 0 \eqend{.}
\end{equations}

To obtain the gauge-fixing and ghost terms, we add a BRST-exact term to the action. We use the gauge conditions
\begin{equation}
\label{eq:action_gaugecondition}
F_\mu \equiv \nabla^\nu h_{\mu\nu} - \frac{1}{2 \zeta} \nabla_\mu h - \theta m A_\mu + \sigma \nabla_\mu \phi \eqend{,} \quad G \equiv \nabla_\mu A^\mu + \beta m \phi - \frac{\gamma}{2} m h
\end{equation}
for the graviton and Stueckelberg fields, which are the most general ones compatible with the scaling dimension of the fields and a regular limit $m \to 0$. We thus set
\begin{equation}
S_\text{gf} + S_\text{gh} = \brst \int \bigg[ \bar{c}^\mu \left( \frac{\xi}{2} B_\mu + F_\mu \right) + \bar{c} \left( \frac{\alpha}{2} B + G \right) \bigg] \sqrt{-g} \total^n x \eqend{,}
\end{equation}
which after performing the BRST transformation reads
\begin{equation}
\label{eq:action_gf}
S_\text{gf} = \int \bigg[ \frac{\xi}{2} \left( B^\mu + \frac{1}{\xi} F^\mu \right) \left( B_\mu + \frac{1}{\xi} F_\mu \right) + \frac{\alpha}{2} \left( B + \frac{1}{\alpha} G \right)^2 - \frac{1}{2 \alpha} G^2 - \frac{1}{2 \xi} F^\mu F_\mu \bigg] \sqrt{-g} \total^n x
\end{equation}
and
\begin{splitequation}
\label{eq:action_gh}
S_\text{gh} &= - \int \bigg[ \bar{c}^\mu \left( \nabla^2 c_\mu + R_{\mu\nu} c^\nu + \left( 1 - \frac{1}{\zeta} \right) \nabla_\mu \nabla^\nu c_\nu - \theta m^2 c_\mu - (\theta-\sigma) m \nabla_\mu c \right) \\
&\qquad\qquad+ \bar{c} \Big( (1-\gamma) m \nabla_\mu c^\mu + \nabla^2 c + \beta m^2 c \Big) \bigg] \sqrt{-g} \total^n x \eqend{.}
\end{splitequation}
We see that the shifted auxiliary fields $B_\mu + \xi^{-1} F_\mu$ and $B + \alpha^{-1} G$ decouple, and that the remainder of the gauge-fixing action~\eqref{eq:action_gf} gives the required quadratic gauge-fixing terms. Since the gauge conditions~\eqref{eq:action_gaugecondition} couple the different sectors, also the ghost action~\eqref{eq:action_gh} couples the diffeomorphism and Stueckelberg ghosts. To determine the propagators of the fields and ghosts with reasonable effort, we thus make two simplifications: we work with the on-shell background~\eqref{eq:einstein_eom} and the on-shell action~\eqref{eq:full_action_onshell}, and make a choice for some of the gauge parameters, namely
\begin{equation}
\label{eq:gauge_fixing}
\gamma = 1 \eqend{,} \quad \theta = \xi \eqend{,} \quad \beta = 0 \eqend{,} \quad \sigma = \lambda \eqend{.}
\end{equation}
We recall that $\lambda = 4/(n-2) \Lambda/m^2$~\eqref{eq:lambda_def}, such that the on-shell background has $R_{\mu\nu} = \lambda/2 m^2 g_{\mu\nu}$. The full action for the fields including gauge fixing now reads
\begin{splitequation}
\label{eq:finalaction}
S + S_\text{gf} &= \frac{1}{4} \int \bigg[ h^{\mu\nu} \nabla^2 h_{\mu\nu} - 2 \left( 1 - \frac{1}{\xi} \right) h^{\mu\nu} \nabla_\mu \nabla^\rho h_{\nu\rho} + 2 \left( 1 - \frac{1}{\zeta \xi} \right) h \nabla_\mu \nabla_\nu h^{\mu\nu} \\
&\qquad\quad- \left( 1 - \frac{1}{2 \zeta^2 \xi} \right) h \nabla^2 h + 2 h^{\mu\nu} R_{\mu\rho\nu\sigma} h^{\rho\sigma} - m^2 h^{\mu\nu} h_{\mu\nu} \\
&\qquad\quad+ \left( 1 - \frac{1}{2 \alpha} - \frac{\lambda}{2} \right) m^2 h^2 \bigg] \sqrt{-g} \total^n x - m \left( 1 - \frac{1}{2 \zeta} - \frac{1}{2 \alpha} \right) \int h \nabla_\mu A^\mu \sqrt{-g} \total^n x \\
&\quad- \int \phi \left[ \left( 1 - \frac{\lambda}{\xi} \right) \nabla_\mu \nabla_\nu h^{\mu\nu} - \left( 1 - \frac{\lambda}{2 \xi \zeta} \right) \nabla^2 h \right] \sqrt{-g} \total^n x \\
&\quad+ \frac{1}{2} \int \left[ A^\mu \nabla^2 A_\mu - \left( 1 - \frac{1}{\alpha} \right) A^\mu \nabla_\mu \nabla^\nu A_\nu - \left( \xi - \frac{\lambda}{2} \right) m^2 A^\mu A_\mu \right] \sqrt{-g} \total^n x \\
&\quad+ \frac{\lambda (\lambda-\xi)}{2 \xi} \int \phi \nabla^2 \phi \sqrt{-g} \total^n x + \int \left( \frac{\xi}{2} \tilde{B}^\mu \tilde{B}_\mu + \frac{\alpha}{2} \tilde{B}^2 \right) \sqrt{-g} \total^n x \eqend{,} \raisetag{2em}
\end{splitequation}
where we defined
\begin{equations}[eq:tildeB]
\tilde{B}_\mu &= B_\mu + \frac{1}{\xi} \left( \nabla^\nu h_{\mu\nu} - \frac{1}{2 \zeta} \nabla_\mu h - \xi m A_\mu + \lambda \nabla_\mu \phi \right) \eqend{,} \\
\tilde{B} &= B + \frac{1}{\alpha} \left( \nabla_\mu A^\mu - \frac{1}{2} m h \right) \eqend{,}
\end{equations}
and the ghost action is
\begin{splitequation}
\label{eq:ghostaction}
S_\text{gh} &= - \int \bigg[ \bar{c}^\mu \left( \nabla^2 c_\mu + \left( \frac{\lambda}{2} - \xi \right) m^2 c_\mu + \left( 1 - \frac{1}{\zeta} \right) \nabla_\mu \nabla^\nu c_\nu - (\xi-\lambda) m \nabla_\mu c \right) \\
&\qquad\quad+ \bar{c} \nabla^2 c \bigg] \sqrt{-g} \total^n x \eqend{.}
\end{splitequation}
We see that the various sectors are still coupled. For a full decoupling, we would have to make a different gauge choice (the analogue of Feynman gauge), for which
\begin{equation}
\label{eq:gauge_feynman}
\zeta = 1 \eqend{,} \quad \xi = 1 \eqend{,} \quad \sigma = 1 \eqend{,} \quad \theta = 1 \eqend{,} \quad \alpha = 1 \eqend{,} \quad \gamma = 1 \eqend{,} \quad \beta = \lambda - 1 \eqend{,}
\end{equation}
and then use the new field $\tilde{h}_{\mu\nu}$ as for the limit~\eqref{eq:full_action_onshell_zerolambdamass}.\footnote{For $\lambda = 1$, this choice is contained in the class of gauge fixings~\eqref{eq:gauge_fixing}.} However, to verify the gauge dependence of the on-shell heat kernel coefficients, we have to keep at least some of the gauge parameters arbitrary, and the choice we made above is the one that is actually feasible.

\section{Propagators}
\label{sec:propagators}

To determine the propagators of the different fields in a general gauge, we have to relate them to the propagators in Feynman-type gauges, where the corresponding equation of motion is normally hyperbolic. In normally hyperbolic equations of motion, second-order derivatives only appear via the d'Alembertian $\nabla^2$, and their (retarded and advanced) propagators or Green's functions can be constructed in an arbitrary globally hyperbolic spacetime using successive approximations~\cite{baerginouxpfaeffle2007,baerginoux2012,baer2015}. As we will see below, in a general gauge where second-order derivatives appear also in a different form the propagators can be obtained as a linear combination of the Feynman-gauge propagator and derivatives of propagators for fields of lower spin. For this, we will first derive various identities for the Feynman-gauge propagators, then treat the ghost sector where a spin-1 propagator in a general gauge is needed, and afterwards consider the fields sector where in addition also the spin-2 propagator in a general gauge appears. To derive these identities, we will make repeated use of the fact that the solution of a normally hyperbolic equation with retarded or advanced boundary conditions is unique, in particular that such an equation with vanishing source only has a vanishing solution. Lastly, we verify that our propagators satisfy the relevant Ward(--Takahashi--Slavnov--Taylor) identities.

\subsection{Feynman-type gauges}

We recall that the scalar propagator of (squared) mass $M^2$ fulfills the EOM
\begin{equation}
\label{eq:scalar_eom}
\left( \nabla^2 - M^2 \right) G_{M^2}(x,x') = \frac{\delta^n(x-x')}{\sqrt{-g}} \eqend{.}
\end{equation}
For the massive vector propagator, the Feynman gauge action reads
\begin{splitequation}
S_\text{V,FG} &= - \frac{1}{4} \int F^{\mu\nu} F_{\mu\nu} \sqrt{-g} \total^n x - \frac{1}{2} \int M^2 A^\mu A_\mu \sqrt{-g} \total^n x - \frac{1}{2} \int \left( \nabla_\mu A^\mu \right)^2 \sqrt{-g} \total^n x \\
&= \frac{1}{2} \int A_\mu \left[ g^{\mu\nu} \left( \nabla^2 - M^2 \right) - R^{\mu\nu} \right] A_\nu \sqrt{-g} \total^n x \eqend{,}
\end{splitequation}
and the Feynman gauge massive vector propagator $G^{\text{FG},M^2}$ thus fulfills the equation
\begin{equation}
\label{eq:vector_fg_eom_pre}
\left[ g^{\mu\nu} \left( \nabla^2 - M^2 \right) - R^{\mu\nu} \right] G^{\text{FG},M^2}_{\nu\rho'}(x,x') = g^\mu_{\rho'} \frac{\delta^n(x-x')}{\sqrt{-g}} \eqend{.}
\end{equation}
Here and later on, primed indices refer to the point $x'$ and unprimed ones to $x$, and we have introduced the bitensor of parallel transport $g^\mu_{\nu'}$~\cite{synge1960,dewittbrehme1960}, which is defined as the unique solution to the equation
\begin{equation}
\label{eq:bitensor_parallel}
\nabla^\rho \sigma \nabla_\rho g_{\mu\beta'} = 0 \eqend{,} \qquad \lim_{x' \to x} g_{\mu\beta'} = g_{\mu\beta} \eqend{.}
\end{equation}
For the on-shell background with $R_{\mu\nu} = \lambda/2 m^2 g_{\mu\nu}$ that we are using, Eq.~\eqref{eq:vector_fg_eom_pre} reduces to
\begin{equation}
\label{eq:vector_fg_eom}
g^{\mu\nu} \left( \nabla^2 - M^2 - \frac{\lambda}{2} m^2 \right) G^{\text{FG},M^2}_{\nu\rho'}(x,x') = g^\mu_{\rho'} \frac{\delta^n(x-x')}{\sqrt{-g}} \eqend{,}
\end{equation}
and we see that there is a mismatch between the explicit mass $M^2$ of the propagator and the effective one $M^2 + \lambda/2 m^2$ appearing in the equation.\footnote{Alternatively, this equation could have been obtained directly from the vector part of the total massive gravity action~\eqref{eq:finalaction} with gauge parameter $\alpha = 1$.}

Taking the divergence of the equation of motion~\eqref{eq:vector_fg_eom} for the Feynman gauge vector propagator and commuting covariant derivatives, we obtain
\begin{splitequation}
\nabla^\nu \left( \nabla^2 - M^2 - \frac{\lambda}{2} m^2 \right) G^{\text{FG},M^2}_{\nu\rho'}(x,x') &= \left( \nabla^2 - M^2 \right) \nabla^\nu G^{\text{FG},M^2}_{\nu\rho'}(x,x') \\
&= \nabla_\mu \left[ g^\mu_{\rho'} \frac{\delta^n(x-x')}{\sqrt{-g}} \right] = - \nabla_{\rho'} \frac{\delta^n(x-x')}{\sqrt{-g}} \eqend{,}
\end{splitequation}
where the last equality follows from properties of the parallel transport bitensor (see Eq.~(13.3) in Ref.~\onlinecite{poissonpoundvega2011}), and combining with the scalar equation of motion~\eqref{eq:scalar_eom} we get
\begin{equation}
\left( \nabla^2 - M^2 \right) \left[ \nabla^\nu G^{\text{FG},M^2}_{\nu\rho'}(x,x') + \nabla_{\rho'} G_{M^2}(x,x') \right] = 0 \eqend{.}
\end{equation}
Since $\nabla^2 - M^2$ is a normally hyperbolic differential operator, which has unique (retarded and advanced) solutions, we obtain the identity~\cite{dewittbrehme1960}
\begin{equation}
\label{eq:vector_fg_div}
\nabla^\nu G^{\text{FG},M^2}_{\nu\rho'}(x,x') = - \nabla_{\rho'} G_{M^2}(x,x') \eqend{,}
\end{equation}
and upon exchanging $x$ and $x'$ also
\begin{equation}
\label{eq:vector_fg_divs}
\nabla^{\rho'} G^{\text{FG},M^2}_{\nu\rho'}(x,x') = - \nabla_\nu G_{M^2}(x,x') \eqend{.}
\end{equation}

For the massive tensor propagator, the Feynman gauge action is the tensor part of the total massive gravity action~\eqref{eq:finalaction} with gauge parameters $\xi = \zeta = \alpha = 1$ and reads
\begin{equation}
S = \frac{1}{2} \int h_{\mu\nu} P^{\mu\nu\rho\sigma} h_{\rho\sigma} \sqrt{-g} \total^n x
\end{equation}
with
\begin{splitequation}
P^{\mu\nu\rho\sigma} &= \frac{1}{2} g^{\mu(\rho} g^{\sigma)\nu} \nabla^2 - \frac{1}{4} g^{\mu\nu} g^{\rho\sigma} \nabla^2 - R^{\mu(\rho\sigma)\nu} - \frac{1}{2} m^2 g^{\mu(\rho} g^{\sigma)\nu} + \frac{1-\lambda}{4} m^2 g^{\mu\nu} g^{\rho\sigma} \eqend{.}
\end{splitequation}
However, this EOM operator is not normally hyperbolic, since the coefficient of $\nabla^2$ is not (a multiple of) the identity on symmetric rank-2 tensors. Instead, we consider the operator
\begin{splitequation}
\label{}
Q^{\mu\nu\rho\sigma} &= 2 P^{\mu\nu\rho\sigma} - \frac{2}{n-2} g^{\rho\sigma} g_{\alpha\beta} P^{\mu\nu\alpha\beta} \\
&= g^{\mu(\rho} g^{\sigma)\nu} \nabla^2 - 2 R^{\mu(\rho\sigma)\nu} - m^2 g^{\mu(\rho} g^{\sigma)\nu} \eqend{,}
\end{splitequation}
which is normally hyperbolic. Therefore, it has unique Green's functions which can be constructed using standard methods~\cite{baerginouxpfaeffle2007,baerginoux2012,baer2015}, and we can assert that there exists a (rescaled) trace-reversed Feynman-type gauge propagator $G^{\text{TG},m^2}_{\mu\nu\rho'\sigma'}$ satisfying
\begin{equation}
\label{eq:tensor_tg_eom}
Q^{\mu\nu\rho\sigma} G^{\text{TG},m^2}_{\rho\sigma\alpha'\beta'}(x,x') = g_{\alpha'}^{(\mu} g_{\beta'}^{\nu)} \frac{\delta^n(x-x')}{\sqrt{-g}} \eqend{.}
\end{equation}
The Feynman-type gauge tensor propagator is then obtained by reversing this operation:
\begin{equation}
\label{eq:tensor_fg_tg_relation}
G^{\text{FG},m^2}_{\rho\sigma\alpha'\beta'}(x,x') = 2 G^{\text{TG},m^2}_{\rho\sigma\alpha'\beta'}(x,x') - \frac{2}{n-2} g_{\rho\sigma} g^{\mu\nu} G^{\text{TG},m^2}_{\mu\nu\alpha'\beta'}(x,x') \eqend{,}
\end{equation}
and satisfies
\begin{equation}
\label{eq:tensor_fg_eom}
P^{\mu\nu\rho\sigma} G^{\text{FG},m^2}_{\rho\sigma\alpha'\beta'}(x,x') = g_{\alpha'}^{(\mu} g_{\beta'}^{\nu)} \frac{\delta^n(x-x')}{\sqrt{-g}} \eqend{.}
\end{equation}

We now need to derive the trace and divergence identities for this tensor propagator. The easier one is the trace identity, for which we contract the propagator equation~\eqref{eq:tensor_fg_eom} with $g_{\mu\nu}$ to obtain
\begin{equation}
- \frac{n-2}{4} g^{\rho\sigma} \Big[ \nabla^2 - (1-\lambda) m^2 \Big] G^{\text{FG},m^2}_{\rho\sigma\alpha'\beta'}(x,x') = g_{\alpha'\beta'} \frac{\delta^n(x-x')}{\sqrt{-g}} \eqend{.}
\end{equation}
and from this
\begin{equation}
\Big[ \nabla^2 - (1-\lambda) m^2 \Big] \left[ \frac{n-2}{4} g^{\rho\sigma} G^{\text{FG},m^2}_{\rho\sigma\alpha'\beta'}(x,x') + g_{\alpha'\beta'} G_{(1-\lambda) m^2}(x,x') \right] = 0 \eqend{.}
\end{equation}
The differential operator is again normally hyperbolic, such that we have the unique solution
\begin{equation}
\label{eq:tensor_fg_tr}
g^{\rho\sigma} G^{\text{FG},m^2}_{\rho\sigma\alpha'\beta'}(x,x') = - \frac{4}{n-2} g_{\alpha'\beta'} G_{(1-\lambda) m^2}(x,x') \eqend{,}
\end{equation}
and upon exchanging $x$ and $x'$ also
\begin{equation}
\label{eq:tensor_fg_trs}
g^{\alpha'\beta'} G^{\text{FG},m^2}_{\rho\sigma\alpha'\beta'}(x,x') = - \frac{4}{n-2} g_{\rho\sigma} G_{(1-\lambda) m^2}(x,x') \eqend{.}
\end{equation}
These trace identities generalize the known ones for the massless spin-2 field~\cite{lichnerowicz1961}.

Using this relation, we can rewrite the propagator equation~\eqref{eq:tensor_fg_eom} as
\begin{splitequation}
&\left[ \frac{1}{2} g^{\mu(\rho} g^{\sigma)\nu} \nabla^2 - R^{\mu(\rho\sigma)\nu} - \frac{1}{2} m^2 g^{\mu(\rho} g^{\sigma)\nu} \right] G^{\text{FG},m^2}_{\rho\sigma\alpha'\beta'}(x,x') \\
&\quad= \left( g_{\alpha'}^{(\mu} g_{\beta'}^{\nu)} - \frac{1}{n-2} g^{\mu\nu} g_{\alpha'\beta'} \right) \frac{\delta^n(x-x')}{\sqrt{-g}} \eqend{.}
\end{splitequation}
Contracting with $\nabla_\mu$ and commuting covariant derivatives, from this formula we obtain
\begin{splitequation}
&\left[ \nabla^2 - \left( 1 - \frac{\lambda}{2} \right) m^2 \right] \nabla^\rho G^{\text{FG},m^2}_{\rho\sigma\alpha'\beta'}(x,x') = \nabla_\mu \left[ \left( 2 g_{\sigma(\alpha'} g_{\beta')}^\mu - \frac{2}{n-2} \delta^\mu_\sigma g_{\alpha'\beta'} \right) \frac{\delta^n(x-x')}{\sqrt{-g}} \right] \\
&\hspace{8em}= - 2 \nabla_{(\alpha'} \left[ g_{|\sigma|\beta')} \frac{\delta^n(x-x')}{\sqrt{-g}} \right] - \frac{2}{n-2} g_{\alpha'\beta'} \nabla_\sigma \frac{\delta^n(x-x')}{\sqrt{-g}} \eqend{,}
\end{splitequation}
where we also used the contracted second Bianchi identity
\begin{equation}
\nabla_\mu R^{\mu\rho\nu\sigma} = \nabla^\nu R^{\sigma\rho} - \nabla^\sigma R^{\nu\rho} \eqend{,}
\end{equation}
whose right-hand side vanishes for our on-shell background. On the other hand, using the equation of motion~\eqref{eq:vector_fg_eom} for the Feynman gauge vector propagator of mass $M^2$, we compute
\begin{equation}
\left[ \nabla^2 - \left( 1 - \frac{\lambda}{2} \right) m^2 \right] \nabla_{(\alpha'} G^{\text{FG},(1-\lambda)m^2}_{|\rho|\beta')}(x,x') = \nabla_{(\alpha'} \left[ g_{|\rho|\beta')} \frac{\delta^n(x-x')}{\sqrt{-g}} \right] \eqend{,}
\end{equation}
such that we infer
\begin{splitequation}
&\left[ \nabla^2 - \left( 1 - \frac{\lambda}{2} \right) m^2 \right] \left[ \nabla^\rho G^{\text{FG},m^2}_{\rho\sigma\alpha'\beta'}(x,x') + 2 \nabla_{(\alpha'} G^{\text{FG},(1-\lambda)m^2}_{|\sigma|\beta')}(x,x') \right] \\
&\quad= - \frac{2}{n-2} g_{\alpha'\beta'} \nabla_\sigma \frac{\delta^n(x-x')}{\sqrt{-g}} = - \frac{2}{n-2} g_{\alpha'\beta'} \nabla_\sigma \left( \nabla^2 - M^2 \right) G_{M^2}(x,x') \\
&\quad= - \frac{2}{n-2} g_{\alpha'\beta'} \left[ \nabla^2 - \left( M^2 + \frac{\lambda}{2} m^2 \right) \right] \nabla_\sigma G_{M^2}(x,x') \eqend{.}
\end{splitequation}
Finally choosing $M^2 = (1-\lambda) m^2$, it follows that
\begin{equation}
\label{eq:tensor_fg_div}
\nabla^\rho G^{\text{FG},m^2}_{\rho\sigma\alpha'\beta'}(x,x') = - 2 \nabla_{(\alpha'} G^{\text{FG},(1-\lambda)m^2}_{|\sigma|\beta')}(x,x') - \frac{2}{n-2} g_{\alpha'\beta'} \nabla_\sigma G_{(1-\lambda) m^2}(x,x') \eqend{,}
\end{equation}
which is the massive generalization of the known identity in the massless case~\cite{lichnerowicz1961}. Exchanging $x$ and $x'$, we also obtain
\begin{equation}
\label{eq:tensor_fg_divs}
\nabla^{\alpha'} G^{\text{FG},m^2}_{\rho\sigma\alpha'\beta'}(x,x') = - 2 \nabla_{(\rho} G^{\text{FG},(1-\lambda)m^2}_{\sigma)\beta'}(x,x') - \frac{2}{n-2} g_{\rho\sigma} \nabla_{\beta'} G_{(1-\lambda) m^2}(x,x') \eqend{.}
\end{equation}

For the trace-reversed Feynman-type gauge propagator $G^{\text{TG},m^2}_{\mu\nu\rho'\sigma'}$~\eqref{eq:tensor_fg_tg_relation}, the trace and divergence identities~\eqref{eq:tensor_fg_tr}, \eqref{eq:tensor_fg_trs}, \eqref{eq:tensor_fg_div} and~\eqref{eq:tensor_fg_divs} simplify to
\begin{equations}[eq:tensor_tg_trdiv]
g^{\rho\sigma} G^{\text{TG},m^2}_{\rho\sigma\alpha'\beta'}(x,x') &= g_{\alpha'\beta'} G_{(1-\lambda) m^2}(x,x') \eqend{,} \\
g^{\alpha'\beta'} G^{\text{TG},m^2}_{\rho\sigma\alpha'\beta'}(x,x') &= g_{\rho\sigma} G_{(1-\lambda) m^2}(x,x') \eqend{,} \\
\nabla^\rho G^{\text{TG},m^2}_{\rho\sigma\alpha'\beta'}(x,x') &= - \nabla_{(\alpha'} G^{\text{FG},(1-\lambda)m^2}_{|\sigma|\beta')}(x,x') \eqend{,} \\
\nabla^{\alpha'} G^{\text{TG},m^2}_{\rho\sigma\alpha'\beta'}(x,x') &= - \nabla_{(\rho} G^{\text{FG},(1-\lambda)m^2}_{\sigma)\beta'}(x,x') \eqend{.}
\end{equations}

\subsection{The ghost sector}

To determine the propagators of all fields in massive gravity, we start with the simpler ghost sector. If we introduce the composite index $A$, we can define $c_A = (c_\mu,c)$ and $\bar{c}_A = (\bar{c}_\mu,\bar{c})$ and write the ghost action~\eqref{eq:ghostaction} as
\begin{equation}
S_\text{gh} = \int \bar{c}_A P^{AB} c_B \sqrt{-g} \total^n x
\end{equation}
with the components of the equation-of-motion operator $P^{AB}$ given by
\begin{equations}
P^{\mu\nu} &= - g^{\mu\nu} \left( \nabla^2 + \frac{1}{2} \lambda m^2 - \xi m^2 \right) - \left( 1 - \frac{1}{\zeta} \right) \nabla^\mu \nabla^\nu \eqend{,} \\
P^{\mu\circ} &= (\xi-\lambda) m \nabla^\mu \eqend{,} \\
P^{\circ\mu} &= 0 \eqend{,} \\
P^{\circ\circ} &= - \nabla^2 \eqend{.}
\end{equations}
The propagator
\begin{equation}
G_{AB'}(x,x') \equiv - \mathi \expect{ c_A(x) \bar{c}_{B'}(x') }
\end{equation}
satisfies
\begin{equation}
P^{AB} G_{BC'}(x,x') = \delta^A_{C'} \frac{\delta^n(x-x')}{\sqrt{-g}} \eqend{,}
\end{equation}
where a sum over the composite index $B$ is implicit, or in components
\begin{equations}[eq:ghost_eom]
P^{\mu\rho} G_{\rho\nu'}(x,x') + P^{\mu\circ} G_{\circ\nu'}(x,x') &= g^\mu_{\nu'} \frac{\delta^n(x-x')}{\sqrt{-g}} \eqend{,} \\
P^{\mu\rho} G_{\rho\circ}(x,x') + P^{\mu\circ} G_{\circ\circ}(x,x') &= 0 \eqend{,} \\
P^{\circ\rho} G_{\rho\nu'}(x,x') + P^{\circ\circ} G_{\circ\nu'}(x,x') &= 0 \eqend{,} \\
P^{\circ\rho} G_{\rho\circ}(x,x') + P^{\circ\circ} G_{\circ\circ}(x,x') &= \frac{\delta^n(x-x')}{\sqrt{-g}} \eqend{.}
\end{equations}

We make the general ansatz
\begin{equations}[eq:ghost_propagator_ansatz]
G_{\mu\nu'}(x,x') &= - G^{\text{FG},M_A^2}_{\mu\nu'}(x,x') + \nabla_\mu \nabla_{\nu'} G_D(x,x') \eqend{,} \\
G_{\mu\circ}(x,x') &= \nabla_\mu G_B(x,x') \eqend{,} \\
G_{\circ\nu'}(x,x') &= \nabla_{\nu'} G_C(x,x') \eqend{,} \\
G_{\circ\circ}(x,x') &= G_E(x,x') \eqend{,}
\end{equations}
where $G_{A/B/C/D}$ are scalar propagators of various masses, or linear combinations thereof, and $M_A$ is an unspecified mass. Plugging the ansatz~\eqref{eq:ghost_propagator_ansatz} into the equations of motion~\eqref{eq:ghost_eom} and commuting covariant derivatives (using that on-shell we have $R_{\mu\nu} = \lambda/2 m^2 g_{\mu\nu}$), we obtain the system of equations
\begin{equations}
\begin{split}
&g^{\mu\rho} \left[ \nabla^2 - \left( \xi - \frac{\lambda}{2} \right) m^2 \right] G^{\text{FG},M_A^2}_{\rho\nu'}(x,x') + \left( 1 - \frac{1}{\zeta} \right) \nabla^\mu \nabla^\rho G^{\text{FG},M_A^2}_{\rho\nu'}(x,x') \\
&\quad- \nabla^\mu \left[ \left( 2 - \frac{1}{\zeta} \right) \nabla^2 - (\xi-\lambda) m^2 \right] \nabla_{\nu'} G_D(x,x') \\
&\quad+ (\xi-\lambda) m \nabla^\mu \nabla_{\nu'} G_C(x,x') = g^\mu_{\nu'} \frac{\delta^n(x-x')}{\sqrt{-g}} \eqend{,}
\end{split} \\
&\nabla^\mu \bigg[ \left( 2 - \frac{1}{\zeta} \right) \nabla^2 G_B(x,x') - (\xi-\lambda) m^2 G_B(x,x') - (\xi-\lambda) m G_E(x,x') \bigg] = 0 \eqend{,} \\
&\nabla^2 \nabla_{\nu'} G_C(x,x') = 0 \eqend{,} \\
&\nabla^2 G_E(x,x') = - \frac{\delta^n(x-x')}{\sqrt{-g}} \eqend{.}
\end{equations}
The third equation is solved by $G_C(x,x') = 0$, the last equation tells us that $G_E = - G_0(x,x')$, and the second equation is solved by
\begin{equation}
G_B(x,x') = \frac{1}{m} \left[ G_0(x,x') - G_{M_B^2}(x,x') \right]
\end{equation}
with $M_B^2 = \zeta (2\zeta-1)^{-1} (\xi-\lambda) m^2$. Using the divergence identity~\eqref{eq:vector_fg_div} and the equation of motion~\eqref{eq:vector_fg_eom} for the Feynman gauge vector propagator, we obtain $M_A^2 = (\xi-\lambda) m^2$ and the equation
\begin{equation}
\left[ \left( 2 - \frac{1}{\zeta} \right) \nabla^2 - M_A^2 \right] G_D(x,x') = - \left( 1 - \frac{1}{\zeta} \right) G_{M_A^2}(x,x') \eqend{,}
\end{equation}
which is solved by
\begin{equation}
G_D(x,x') = \frac{1}{M_A^2} \left[ G_{M_B^2}(x,x') - G_{M_A^2}(x,x') \right] \eqend{.}
\end{equation}
So we obtain the ghost propagators
\begin{equations}[eq:ghost_propagator]
G_{\mu\nu'}(x,x') &= - \mathi \expect{ \bar{c}_\mu(x) c_{\nu'}(x') } = - G^{\text{FG},M_A^2}_{\mu\nu'}(x,x') + \frac{1}{M_A^2} \nabla_\mu \nabla_{\nu'} \left[ G_{M_B^2}(x,x') - G_{M_A^2}(x,x') \right] \eqend{,} \\
G_{\mu\circ}(x,x') &= - \mathi \expect{ \bar{c}_\mu(x) c(x') } = \frac{1}{m} \nabla_\mu \left[ G_0(x,x') - G_{M_B^2}(x,x') \right] \eqend{,} \\
G_{\circ\nu'}(x,x') &= - \mathi \expect{ \bar{c}(x) c_{\nu'}(x') } = 0 \eqend{,} \\
G_{\circ\circ}(x,x') &= - \mathi \expect{ \bar{c}(x) c(x') } = - G_0(x,x') \label{eq:ghost_propagator_00}
\end{equations}
with the masses
\begin{equation}
\label{eq:ghost_masses}
M_A^2 = (\xi-\lambda) m^2 \eqend{,} \quad M_B^2 = \frac{\zeta}{2\zeta-1} (\xi-\lambda) m^2 = \frac{\zeta}{2\zeta-1} M_A^2 \eqend{.}
\end{equation}

We can also easily compute the massless limit. For this, let us define the mass derivatives
\begin{equation}
\label{eq:scalar_massderivative}
\hat{G}_{M^2}(x,x') \equiv \left. \frac{\partial G_{m^2}(x,x')}{\partial m^2} \right\rvert_{m^2 = M^2}
\end{equation}
of the scalar propagator, which fulfill the equation of motion
\begin{equation}
\label{eq:scalar_massderivative_eom}
\left( \nabla^2 - M^2 \right) \hat{G}_{M^2}(x,x') = G_{M^2}(x,x') \eqend{.}
\end{equation}
We then have
\begin{equation}
\label{eq:scalar_massexpansion}
G_{M^2}(x,x') = G_0(x,x') + M^2 \hat{G}_0(x,x') + \bigo{M^4} \eqend{,}
\end{equation}
and, keeping $\lambda$ fixed, obtain the massless ghost propagators
\begin{equations}[eq:ghost_propagator_massless]
\lim_{m \to 0} G_{\mu\nu'}(x,x') \big\rvert_{\lambda = \text{cte}} &= - G^{\text{FG},0}_{\mu\nu'}(x,x') + \frac{1-\zeta}{2\zeta-1} \nabla_\mu \nabla_{\nu'} \hat{G}_0(x,x') \eqend{,} \\
\lim_{m \to 0} G_{\mu\circ}(x,x') \big\rvert_{\lambda = \text{cte}} &= 0 \eqend{,} \label{eq:ghost_propagator_massless_mu0} \\
\lim_{m \to 0} G_{\circ\nu'}(x,x') \big\rvert_{\lambda = \text{cte}} &= 0 \eqend{,} \\
\lim_{m \to 0} G_{\circ\circ}(x,x') \big\rvert_{\lambda = \text{cte}} &= - G_0(x,x') \eqend{.}
\end{equations}
Note that $\lambda$ fixed means that both $\Lambda$ and $m^2$ tend to zero, with their ratio fixed, and we see that the massless limit of the ghost sector does not depend on this ratio. However, we can also keep $\Lambda$ fixed and send $m^2 \to 0$. For this, we simply replace again $\lambda = 4/(n-2) \Lambda m^{-2}$, whence the limit $m^2 \to 0$ results in
\begin{equations}[eq:ghost_propagator_massless_lambda]
\begin{split}
\lim_{m \to 0} G_{\mu\nu'}(x,x') \big\rvert_{\Lambda = \text{cte}} &= - G^{\text{FG}, - \frac{4}{n-2} \Lambda}_{\mu\nu'}(x,x') \\
&\quad- \frac{n-2}{4 \Lambda} \nabla_\mu \nabla_{\nu'} \left[ G_{- \frac{\zeta}{2\zeta-1} \frac{4}{n-2} \Lambda}(x,x') - G_{- \frac{4}{n-2} \Lambda}(x,x') \right] \eqend{,}
\end{split} \\
\lim_{m \to 0} G_{\mu\circ}(x,x') \big\rvert_{\Lambda = \text{cte}} &\sim \frac{1}{m} \nabla_\mu \left[ G_0(x,x') - G_{- \frac{\zeta}{2\zeta-1} \frac{4}{n-2} \Lambda}(x,x') \right] \eqend{,} \\
\lim_{m \to 0} G_{\circ\nu'}(x,x') \big\rvert_{\Lambda = \text{cte}} &= 0 \eqend{,} \\
\lim_{m \to 0} G_{\circ\circ}(x,x') \big\rvert_{\Lambda = \text{cte}} &= - G_0(x,x') \eqend{,}
\end{equations}
and we see that the propagator $G_{\mu\circ}(x,x') = - \mathi \expect{ \bar{c}_\mu(x) c(x') }$ is actually divergent. This is related to the rescaling that we did at the beginning, which only has a sensible limit for vanishing cosmological constant~\cite{Hinterbichler:2011tt}.

On the other hand, if we first take the limit $\Lambda \to 0$ and then $m \to 0$, using the mass expansion~\eqref{eq:scalar_massexpansion} we obtain the finite result
\begin{equations}[eq:ghost_propagator_lambda0]
\lim_{m \to 0} \lim_{\Lambda \to 0} G_{\mu\nu'}(x,x') &= - G^{\text{FG},0}_{\mu\nu'}(x,x') + \frac{1-\zeta}{2\zeta-1} \nabla_\mu \nabla_{\nu'} \hat{G}_0(x,x') \eqend{,} \\
\lim_{m \to 0} \lim_{\Lambda \to 0} G_{\mu\circ}(x,x') &= 0 \eqend{,} \\
\lim_{m \to 0} \lim_{\Lambda \to 0} G_{\circ\nu'}(x,x') &= 0 \eqend{,} \\
\lim_{m \to 0} \lim_{\Lambda \to 0} G_{\circ\circ}(x,x') &= - G_0(x,x') \eqend{.}
\end{equations}
We thus see that the limits $\Lambda \to 0$ and $m \to 0$ do not commute, which is an example of the van Dam--Veltman--Zakharov (vDVZ) discontinuity~\cite{vandamveltman1970,zakharov1970}.

\subsection{The field sector}

Analogously to the ghost sector, we write the full gauge-fixed action~\eqref{eq:finalaction} as
\begin{equation}
S = \frac{1}{2} \int H_A P^{AB} H_B \sqrt{-g} \total^n x
\end{equation}
with the composite field $H_A = (h_{\mu\nu}, A_\mu, \phi)$ and the Hermitean equation-of-motion operator $P_{AB}$. Since the shifted auxiliary fields $\tilde{B}_\mu$ and $\tilde{B}$ decouple, we can treat them separately at the end. The propagator
\begin{equation}
G_{AB'}(x,x') \equiv - \mathi \expect{ H_A(x) H_{B'}(x') }
\end{equation}
satisfies
\begin{equation}
P^{AB} G_{BC'}(x,x') = \delta^A_{C'} \frac{\delta^n(x-x')}{\sqrt{-g}} \eqend{,}
\end{equation}
where a sum over the composite index $B$ is implicit, or in components
\begin{equations}[eq:fields_eom]
&P^{\mu\nu\rho\sigma} G_{\rho\sigma\alpha'\beta'}(x,x') + P^{\mu\nu\rho\circ} G_{\rho\circ\alpha'\beta'}(x,x') + P^{\mu\nu\circ\circ} G_{\circ\circ\alpha'\beta'}(x,x') = g_{(\alpha'}^\mu g_{\beta')}^\nu \frac{\delta^n(x-x')}{\sqrt{-g}} \eqend{,} \\
&P^{\mu\nu\rho\sigma} G_{\rho\sigma\alpha'\circ}(x,x') + P^{\mu\nu\rho\circ} G_{\rho\circ\alpha'\circ}(x,x') + P^{\mu\nu\circ\circ} G_{\circ\circ\alpha'\circ}(x,x') = 0 \eqend{,} \\
&P^{\mu\nu\rho\sigma} G_{\rho\sigma\circ\circ}(x,x') + P^{\mu\nu\rho\circ} G_{\rho\circ\circ\circ}(x,x') + P^{\mu\nu\circ\circ} G_{\circ\circ\circ\circ}(x,x') = 0 \eqend{,} \\
&P^{\mu\circ\rho\sigma} G_{\rho\sigma\alpha'\beta'}(x,x') + P^{\mu\circ\rho\circ} G_{\rho\circ\alpha'\beta'}(x,x') + P^{\mu\circ\circ\circ} G_{\circ\circ\alpha'\beta'}(x,x') = 0 \eqend{,} \\
&P^{\mu\circ\rho\sigma} G_{\rho\sigma\alpha'\circ}(x,x') + P^{\mu\circ\rho\circ} G_{\rho\circ\alpha'\circ}(x,x') + P^{\mu\circ\circ\circ} G_{\circ\circ\alpha'\circ}(x,x') = g_{\alpha'}^\mu \frac{\delta^n(x-x')}{\sqrt{-g}} \eqend{,} \\
&P^{\mu\circ\rho\sigma} G_{\rho\sigma\circ\circ}(x,x') + P^{\mu\circ\rho\circ} G_{\rho\circ\circ\circ}(x,x') + P^{\mu\circ\circ\circ} G_{\circ\circ\circ\circ}(x,x') = 0 \eqend{,} \\
&P^{\circ\circ\rho\sigma} G_{\rho\sigma\alpha'\beta'}(x,x') + P^{\circ\circ\rho\circ} G_{\rho\circ\alpha'\beta'}(x,x') + P^{\circ\circ\circ\circ} G_{\circ\circ\alpha'\beta'}(x,x') = 0 \eqend{,} \\
&P^{\circ\circ\rho\sigma} G_{\rho\sigma\alpha'\circ}(x,x') + P^{\circ\circ\rho\circ} G_{\rho\circ\alpha'\circ}(x,x') + P^{\circ\circ\circ\circ} G_{\circ\circ\alpha'\circ}(x,x') = 0 \eqend{,} \\
&P^{\circ\circ\rho\sigma} G_{\rho\sigma\circ\circ}(x,x') + P^{\circ\circ\rho\circ} G_{\rho\circ\circ\circ}(x,x') + P^{\circ\circ\circ\circ} G_{\circ\circ\circ\circ}(x,x') = \frac{\delta^n(x-x')}{\sqrt{-g}} \eqend{.}
\end{equations}
The equation-of-motion operators $P^{AB}$ can be read off from the full action~\eqref{eq:finalaction}, and are given by
\begin{equations}
\begin{split}
P^{\mu\nu\rho\sigma} &= \frac{1}{2} g^{\mu(\rho} g^{\sigma)\nu} \nabla^2 - \left( 1 - \frac{1}{\xi} \right) \nabla^{(\mu} g^{\nu)(\rho} \nabla^{\sigma)} + \frac{1}{2} \left( 1 - \frac{1}{\zeta \xi} \right) g^{\mu\nu} \nabla^\rho \nabla^\sigma \\
&\quad+ \frac{1}{2} \left( 1 - \frac{1}{\zeta \xi} \right) g^{\rho\sigma} \nabla^\mu \nabla^\nu - \frac{1}{2} \left( 1 - \frac{1}{2 \zeta^2 \xi} \right) g^{\mu\nu} g^{\rho\sigma} \nabla^2 - R^{\mu(\rho\sigma)\nu} \\
&\quad- \frac{1}{2} m^2 g^{\mu(\rho} g^{\sigma)\nu} + \frac{1}{2} \left( 1 - \frac{1}{2 \alpha} - \frac{\lambda}{2} \right) m^2 g^{\mu\nu} g^{\rho\sigma} \eqend{,}
\end{split} \\
P^{\mu\nu\rho\circ} &= - P^{\rho\circ\mu\nu} = - m \left( 1 - \frac{1}{2 \zeta} - \frac{1}{2 \alpha} \right) g^{\mu\nu} \nabla^\rho \eqend{,} \\
P^{\mu\nu\circ\circ} &= P^{\circ\circ\mu\nu} = - \left( 1 - \frac{\lambda}{\xi} \right) \nabla^\mu \nabla^\nu + \left( 1 - \frac{\lambda}{2 \xi \zeta} \right) g^{\mu\nu} \nabla^2 \eqend{,} \\
P^{\mu\circ\nu\circ} &= g^{\mu\nu} \nabla^2 - \left( 1 - \frac{1}{\alpha} \right) \nabla^\mu \nabla^\nu - \left( \xi - \frac{\lambda}{2} \right) m^2 g^{\mu\nu} \eqend{,} \\
P^{\mu\circ\circ\circ} &= - P^{\circ\circ\mu\circ} = 0 \eqend{,} \\
P^{\circ\circ\circ\circ} &= \frac{\lambda (\lambda-\xi)}{\xi} \nabla^2 \eqend{.}
\end{equations}

As for the ghost fields, we make a general ansatz
\begin{equations}
\begin{split}
G_{\mu\nu\rho'\sigma'}(x,x') &= - \mathi \expect{ h_{\mu\nu}(x) h_{\rho'\sigma'}(x') } = G^{\text{FG},m^2}_{\mu\nu\rho'\sigma'}(x,x') + 4 \nabla_{(\mu} \nabla_{(\rho'} G^A_{\nu)\sigma')}(x,x') \\
&\hspace{11em}+ \nabla_\mu \nabla_\nu \nabla_{\rho'} \nabla_{\sigma'} G_C(x,x') + g_{\mu\nu} g_{\rho'\sigma'} G_D(x,x') \\
&\hspace{11em}+ \left( g_{\mu\nu} \nabla_{\rho'} \nabla_{\sigma'} + g_{\rho'\sigma'} \nabla_\mu \nabla_\nu \right) G_B(x,x') \eqend{,}
\end{split} \\
\begin{split}
G_{\mu\nu\rho'\circ}(x,x') &= - \mathi \expect{ h_{\mu\nu}(x) A_{\rho'}(x') } = 2 \nabla_{(\mu} G^E_{\nu)\rho'}(x,x') + \nabla_\mu \nabla_\nu \nabla_{\rho'} G_F(x,x') \\
&\hspace{11em}+ g_{\mu\nu} \nabla_{\rho'} G_G(x,x') \eqend{,}
\end{split} \\
\begin{split}
G_{\mu\circ\rho'\sigma'}(x,x') &= - \mathi \expect{ A_\mu(x) h_{\rho'\sigma'}(x') } = 2 \nabla_{(\rho'} G^E_{|\mu|\sigma')}(x,x') + \nabla_\mu \nabla_{\rho'} \nabla_{\sigma'} G_F(x,x') \\
&\hspace{11em}+ g_{\rho'\sigma'} \nabla_\mu G_G(x,x') \eqend{,}
\end{split} \\
G_{\mu\nu\circ\circ}(x,x') &= - \mathi \expect{ h_{\mu\nu}(x) \phi(x') } = \nabla_\mu \nabla_\nu G_H(x,x') + g_{\mu\nu} G_J(x,x') \eqend{,} \\
G_{\circ\circ\rho'\sigma'}(x,x') &= - \mathi \expect{ \phi(x) h_{\rho'\sigma'}(x') } = \nabla_{\rho'} \nabla_{\sigma'} G_H(x,x') + g_{\rho'\sigma'} G_J(x,x') \eqend{,} \\
G_{\mu\circ\rho'\circ}(x,x') &= - \mathi \expect{ A_\mu(x) A_{\rho'}(x') } = G^K_{\mu\rho'}(x,x') + \nabla_\mu \nabla_{\rho'} G_L(x,x') \eqend{,} \\
G_{\mu\circ\circ\circ}(x,x') &= - \mathi \expect{ A_\mu(x) \phi(x') } = \nabla_\mu G_M(x,x') \eqend{,} \\
G_{\circ\circ\rho'\circ}(x,x') &= - \mathi \expect{ \phi(x) A_{\rho'}(x') } = \nabla_{\rho'} G_M(x,x') \eqend{,} \\
G_{\circ\circ\circ\circ}(x,x') &= - \mathi \expect{ \phi(x) \phi(x') } = G_N(x,x') \eqend{,}
\end{equations}
where the propagators with two indices are (linear combinations of) Feynman-gauge vector propagators of different masses, and the ones without indices (linear combinations of) scalar propagators of different masses. This system is solved as for the ghost sector, using the divergence and trace identities for the Feynman gauge propagators. We do not give any details of the lengthy computation (which we performed using the \textsc{xAct} suite of tensor algebra packages~\cite{xact,martingarcia2008,brizuelaetal2009,nutma2014}), but only the result:
\begin{equations}[eq:field_propagator]
\begin{split}
G_{\mu\nu\rho'\sigma'}(x,x') &= G^{\text{FG},m^2}_{\mu\nu\rho'\sigma'}(x,x') + \frac{4}{m^2} \nabla_{(\mu} \nabla_{(\rho'} \left[ G^{\text{FG},M_C^2}_{\nu)\sigma')}(x,x') - G^{\text{FG},M_A^2}_{\nu)\sigma')}(x,x') \right] \\
&\quad+ \frac{4 (\rho+\alpha)}{n m^2} \left( g_{\mu\nu} \nabla_{\rho'} \nabla_{\sigma'} + g_{\rho'\sigma'} \nabla_\mu \nabla_\nu \right) \left[ G_{M_C^2}(x,x') - G_0(x,x') \right] \\
&\quad+ \frac{4}{m^2} \nabla_\mu \nabla_\nu \nabla_{\rho'} \nabla_{\sigma'} \bigg[ \frac{2 \mu}{m^2} \left[ G_0(x,x') - G_{M_B^2}(x,x') \right] + M_B^2 \frac{\mu}{m^2} \hat{G}_{M_B^2}(x,x') \\
&\hspace{10em}- \frac{1}{M_A^2} \left[ G_{M_A^2}(x,x') - G_{M_B^2}(x,x') \right] + \rho \hat{G}_0(x,x') \\
&\hspace{10em}+ \frac{n-2}{n m^2} (\rho+\alpha) \left[ G_{M_C^2}(x,x') - G_0(x,x') \right] \bigg] \\
&\quad+ \frac{4 (\rho+\alpha)}{n (n-2)} g_{\mu\nu} g_{\rho'\sigma'} G_{M_C^2}(x,x') \eqend{,}
\end{split} \\
G_{\mu\nu\rho'\circ}(x,x') &= - \frac{2 \mu}{m^3} \nabla_\mu \nabla_\nu \nabla_{\rho'} \left[ G_{M_B^2}(x,x') - G_0(x,x') - M_B^2 \hat{G}_{M_B^2}(x,x') \right] \eqend{,} \\
G_{\mu\circ\rho'\sigma'}(x,x') &= - \frac{2 \mu}{m^3} \nabla_\mu \nabla_{\rho'} \nabla_{\sigma'} \left[ G_{M_B^2}(x,x') - G_0(x,x') - M_B^2 \hat{G}_{M_B^2}(x,x') \right] \eqend{,} \\
\begin{split}
G_{\mu\nu\circ\circ}(x,x') &= 2 \nabla_\mu \nabla_\nu \left[ \frac{\mu}{m^2} \left( G_{M_B^2}(x,x') - G_0(x,x') \right) - \rho \hat{G}_0(x,x') \right] \\
&\quad+ \frac{2}{n} (\rho+\alpha) g_{\mu\nu} G_0(x,x') \eqend{,}
\end{split} \\
\begin{split}
G_{\circ\circ\rho'\sigma'}(x,x') &= 2 \nabla_{\rho'} \nabla_{\sigma'} \left[ \frac{\mu}{m^2} \left( G_{M_B^2}(x,x') - G_0(x,x') \right) - \rho \hat{G}_0(x,x') \right] \\
&\quad+ \frac{2}{n} (\rho+\alpha) g_{\rho'\sigma'} G_0(x,x') \eqend{,}
\end{split} \\
\begin{split}
G_{\mu\circ\rho'\circ}(x,x') &= G^{\text{FG},M_A^2}_{\mu\rho'}(x,x') \\
&\quad+ \nabla_\mu \nabla_{\rho'} \left[ \frac{1}{M_A^2} \left( G_{M_A^2}(x,x') - G_{M_B^2}(x,x') \right) + M_B^2 \frac{\mu}{m^2} \hat{G}_{M_B^2}(x,x') \right] \eqend{,}
\end{split} \\
G_{\mu\circ\circ\circ}(x,x') &= \frac{\mu}{m} \nabla_\mu \left[ G_{M_B^2}(x,x') - G_0(x,x') \right] \eqend{,} \\
G_{\circ\circ\rho'\circ}(x,x') &= \frac{\mu}{m} \nabla_{\rho'} \left[ G_{M_B^2}(x,x') - G_0(x,x') \right] \eqend{,} \\
G_{\circ\circ\circ\circ}(x,x') &= \frac{n-2}{n} (\rho+\alpha) G_0(x,x') + \rho m^2 \hat{G}_0(x,x') \eqend{,}
\end{equations}
where we recall the masses~\eqref{eq:ghost_masses}
\begin{equation}
\label{eq:field_masses}
M_A^2 = (\xi-\lambda) m^2 \eqend{,} \quad M_B^2 = \frac{\zeta}{2\zeta-1} (\xi-\lambda) m^2 = \frac{\zeta}{2\zeta-1} M_A^2 \eqend{,} \quad M_C^2 = (1-\lambda) m^2 \eqend{,}
\end{equation}
and defined the abbreviations
\begin{equation}
\label{eq:field_abbreviations}
\rho \equiv \frac{n}{2 (n-1) - (n-2) \lambda} - \alpha \eqend{,} \quad \mu \equiv \frac{m^2}{M_A^2} - \alpha \frac{m^2}{M_B^2} = \frac{\zeta - \alpha (2\zeta-1)}{\zeta (\xi-\lambda)} \eqend{.}
\end{equation}

Also for the field propagators, we can take the massless limit, and again we have the choice of keeping $\lambda$ fixed (and thus sending $\Lambda \to 0$ together with $m$), or holding $\Lambda$ fixed. In the first case, we obtain
\begin{equations}[eq:field_propagator_massless]
\begin{split}
\lim_{m \to 0} G_{\mu\nu\rho'\sigma'}(x,x') \big\rvert_{\lambda = \text{cte}} &= G^{\text{FG},0}_{\mu\nu\rho'\sigma'}(x,x') + 4 (1-\xi) \nabla_{(\mu} \nabla_{(\rho'} \hat{G}^{\text{FG},0}_{\nu)\sigma')}(x,x') \\
&\quad+ \frac{4 (1-\lambda)}{n} (\rho+\alpha) \left( g_{\mu\nu} \nabla_{\rho'} \nabla_{\sigma'} + g_{\rho'\sigma'} \nabla_\mu \nabla_\nu \right) \hat{G}_0(x,x') \\
&\quad+ 2 \left[ \frac{n-2}{n} (\rho+\alpha) (1-\lambda)^2 - \frac{(1-3\zeta) (1-\zeta)}{(2\zeta-1)^2} (\xi-\lambda) \right] \\
&\quad\qquad\times \nabla_\mu \nabla_\nu \nabla_{\rho'} \nabla_{\sigma'} \dhat{G}_0(x,x') \\
&\quad+ \frac{4}{n (n-2)} (\rho+\alpha) g_{\mu\nu} g_{\rho'\sigma'} G_0(x,x') \eqend{,}
\end{split} \\
\lim_{m \to 0} G_{\mu\nu\rho'\circ}(x,x') \big\rvert_{\lambda = \text{cte}} &= 0 \eqend{,} \label{eq:field_propagator_massless_munurho0} \\
\lim_{m \to 0} G_{\mu\circ\rho'\sigma'}(x,x') \big\rvert_{\lambda = \text{cte}} &= 0 \eqend{,} \label{eq:field_propagator_massless_mu0rho0} \\
\lim_{m \to 0} G_{\mu\nu\circ\circ}(x,x') \big\rvert_{\lambda = \text{cte}} &= 2 \left[ \frac{\zeta}{2\zeta-1} - (\rho+\alpha) \right] \nabla_\mu \nabla_\nu \hat{G}_0(x,x') + \frac{2}{n} (\rho+\alpha) g_{\mu\nu} G_0(x,x') \eqend{,} \\
\lim_{m \to 0} G_{\circ\circ\rho'\sigma'}(x,x') \big\rvert_{\lambda = \text{cte}} &= 2 \left[ \frac{\zeta}{2\zeta-1} - (\rho+\alpha) \right] \nabla_{\rho'} \nabla_{\sigma'} \hat{G}_0(x,x') + \frac{2}{n} (\rho+\alpha) g_{\rho'\sigma'} G_0(x,x') \eqend{,} \\
\lim_{m \to 0} G_{\mu\circ\rho'\circ}(x,x') \big\rvert_{\lambda = \text{cte}} &= G^{\text{FG},0}_{\mu\rho'}(x,x') + (1-\alpha) \nabla_\mu \nabla_{\rho'} \hat{G}_0(x,x') \eqend{,} \\
\lim_{m \to 0} G_{\mu\circ\circ\circ}(x,x') \big\rvert_{\lambda = \text{cte}} &= 0 \eqend{,} \label{eq:field_propagator_massless_mu000} \\
\lim_{m \to 0} G_{\circ\circ\rho'\circ}(x,x') \big\rvert_{\lambda = \text{cte}} &= 0 \eqend{,} \\
\lim_{m \to 0} G_{\circ\circ\circ\circ}(x,x') \big\rvert_{\lambda = \text{cte}} &= \frac{n-2}{n} (\rho+\alpha) G_0(x,x') \eqend{,}
\end{equations}
where we also defined the mass derivative of the Feynman-gauge vector propagator
\begin{equation}
\label{eq:vector_massderivative}
\hat{G}^{\text{FG},M^2}_{\mu\rho'}(x,x') \equiv \left. \frac{\partial}{\partial m^2} G^{\text{FG},m^2}_{\mu\rho'}(x,x') \right\rvert_{m^2 = M^2} \eqend{,}
\end{equation}
which fulfills the equation of motion
\begin{equation}
\label{eq:vector_massderivative_eom}
\left( \nabla^2 - M^2 - \frac{\lambda}{2} m^2 \right) \hat{G}^{\text{FG},M^2}_{\mu\rho'}(x,x') = G^{\text{FG},M^2}_{\mu\rho'}(x,x') \eqend{,}
\end{equation}
and the second mass derivative of the scalar propagator
\begin{equation}
\label{eq:scalar_massderivative2}
\dhat{G}_{M^2} \equiv \left. \frac{\partial \hat{G}_{m^2}}{\partial m^2} \right\rvert_{m^2 = M^2} \eqend{,}
\end{equation}
which fulfills
\begin{equation}
\label{eq:scalar_massderivative_eom2}
\left( \nabla^2 - M^2 \right) \dhat{G}_{M^2} = \hat{G}_{M^2} \eqend{.}
\end{equation}
We see that while the vector $A_\mu$ decouples in this limit, such that the mixed propagators of $A_\mu$ and other fields vanish, the scalar $\phi$ is still coupled to the graviton $h_{\mu\nu}$. This is again related to the vDVZ discontinuity~\cite{Hinterbichler:2011tt,vandamveltman1970,zakharov1970}: the scalar mode of the graviton does not decouple in the massless limit, and instead still interacts with other scalar fields.

On the other hand, we can also take the massless limit while holding $\Lambda$ fixed. Defining $\tilde{\Lambda} \equiv 4/(n-2) \Lambda$, this results in
\begin{equations}[eq:field_propagator_massless_lambda]
\begin{split}
\lim_{m \to 0} G_{\mu\nu\rho'\sigma'}(x,x') \big\rvert_{\Lambda = \text{cte}} &\sim G^{\text{FG},0}_{\mu\nu\rho'\sigma'}(x,x') + 4 (\xi-1) \nabla_{(\mu} \nabla_{(\rho'} \hat{G}^{\text{FG},- \tilde{\Lambda}}_{\nu)\sigma')}(x,x') \\
&\quad- \frac{4}{(n-2) \tilde{\Lambda}} \left( g_{\mu\nu} \nabla_{\rho'} \nabla_{\sigma'} + g_{\rho'\sigma'} \nabla_\mu \nabla_\nu \right) \left[ G_{- \tilde{\Lambda}}(x,x') - G_0(x,x') \right] \\
&\quad+ \nabla_\mu \nabla_\nu \nabla_{\rho'} \nabla_{\sigma'} \bigg[ \frac{4}{m^2} \frac{\zeta - \alpha (2\zeta-1)}{2\zeta-1} \hat{G}_{- \frac{\zeta}{2\zeta-1} \tilde{\Lambda}}(x,x') \\
&\hspace{2.5em}+ \frac{4}{m^2 \tilde{\Lambda}} \frac{2 \alpha (2\zeta-1) - \zeta}{\zeta} \left[ G_0(x,x') - G_{- \frac{\zeta}{2\zeta-1} \tilde{\Lambda}}(x,x') \right] \\
&\hspace{2.5em}- \frac{4}{m^2} \alpha \hat{G}_0(x,x') + \frac{4 \xi}{\tilde{\Lambda}^2} \frac{\zeta - 2 \alpha (2\zeta-1)}{\zeta} G_{- \frac{\zeta}{2\zeta-1} \tilde{\Lambda}}(x,x') \\
&\hspace{2.5em}+ \frac{8}{(n-2) \tilde{\Lambda}^2} \left[ (n-1) - (n-2) \frac{\zeta - \alpha (2\zeta-1)}{\zeta} \xi \right] G_0(x,x') \\
&\hspace{2.5em}+ \frac{4}{(n-2) \tilde{\Lambda}^2} \left[ (n-2) \xi - 2 (n-1) \right] G_{- \tilde{\Lambda}}(x,x') \\
&\hspace{2.5em}+ \frac{4 \xi}{\tilde{\Lambda}} \frac{\zeta - 2 \alpha (2\zeta-1)}{2\zeta-1} \hat{G}_{- \frac{\zeta}{2\zeta-1} \tilde{\Lambda}}(x,x') + (\xi-1) \frac{4}{\tilde{\Lambda}} \hat{G}_{- \tilde{\Lambda}}(x,x') \\
&\hspace{2.5em}- \frac{4 n}{(n-2) \tilde{\Lambda}} \hat{G}_0(x,x') + 4 \xi \zeta \frac{\zeta - \alpha (2\zeta-1)}{(2\zeta-1)^2} \dhat{G}_{- \frac{\zeta}{2\zeta-1} \tilde{\Lambda}}(x,x') \bigg] \eqend{,}
\end{split} \\
\begin{split}
\lim_{m \to 0} G_{\mu\nu\rho'\circ}(x,x') \big\rvert_{\Lambda = \text{cte}} &\sim \frac{2}{m} \frac{\zeta - \alpha (2\zeta-1)}{\zeta \tilde{\Lambda}} \nabla_\mu \nabla_\nu \nabla_{\rho'} \bigg[ G_{- \frac{\zeta}{2\zeta-1} \tilde{\Lambda}}(x,x') - G_0(x,x') \\
&\hspace{13em}+ \frac{\zeta}{2\zeta-1} \tilde{\Lambda} \hat{G}_{- \frac{\zeta}{2\zeta-1} \tilde{\Lambda}}(x,x') \bigg] \eqend{,}
\end{split} \\
\begin{split}
\lim_{m \to 0} G_{\mu\circ\rho'\sigma'}(x,x') \big\rvert_{\Lambda = \text{cte}} &\sim \frac{2}{m} \frac{\zeta - \alpha (2\zeta-1)}{\zeta \tilde{\Lambda}} \nabla_\mu \nabla_{\rho'} \nabla_{\sigma'} \bigg[ G_{- \frac{\zeta}{2\zeta-1} \tilde{\Lambda}}(x,x') - G_0(x,x') \\
&\hspace{13em}+ \frac{\zeta}{2\zeta-1} \tilde{\Lambda} \hat{G}_{- \frac{\zeta}{2\zeta-1} \tilde{\Lambda}}(x,x') \bigg] \eqend{,}
\end{split} \\
\begin{split}
\lim_{m \to 0} G_{\mu\nu\circ\circ}(x,x') \big\rvert_{\Lambda = \text{cte}} &= 2 \nabla_\mu \nabla_\nu \bigg[ \frac{\alpha (2\zeta-1) - \zeta}{\zeta \tilde{\Lambda}} \left( G_{- \frac{\zeta}{2\zeta-1} \tilde{\Lambda}}(x,x') - G_0(x,x') \right) \\
&\hspace{5em}+ \alpha \hat{G}_0(x,x') \bigg] \eqend{,}
\end{split} \\
\begin{split}
\lim_{m \to 0} G_{\circ\circ\rho'\sigma'}(x,x') \big\rvert_{\Lambda = \text{cte}} &= 2 \nabla_{\rho'} \nabla_{\sigma'} \bigg[ \frac{\alpha (2\zeta-1) - \zeta}{\zeta \tilde{\Lambda}} \left( G_{- \frac{\zeta}{2\zeta-1} \tilde{\Lambda}}(x,x') - G_0(x,x') \right) \\
&\hspace{5em}+ \alpha \hat{G}_0(x,x') \bigg] \eqend{,}
\end{split} \\
\lim_{m \to 0} G_{\mu\circ\rho'\circ}(x,x') \big\rvert_{\Lambda = \text{cte}} &= G^{\text{FG},0}_{\mu\rho'}(x,x') - \frac{1}{\tilde{\Lambda}} \nabla_\mu \nabla_{\rho'} \left[ G_{- \tilde{\Lambda}}(x,x') - G_{- \frac{\zeta}{2\zeta-1} \tilde{\Lambda}}(x,x') \right] \eqend{,} \\
\lim_{m \to 0} G_{\mu\circ\circ\circ}(x,x') \big\rvert_{\Lambda = \text{cte}} &= 0 \eqend{,} \\
\lim_{m \to 0} G_{\circ\circ\rho'\circ}(x,x') \big\rvert_{\Lambda = \text{cte}} &= 0 \eqend{,} \\
\lim_{m \to 0} G_{\circ\circ\circ\circ}(x,x') \big\rvert_{\Lambda = \text{cte}} &= 0 \eqend{.}
\end{equations}
As in the ghost sector, we have a divergence for small masses, which comes from the fact that the limit $m \to 0$ is not continuous for $\Lambda \neq 0$, i.e., ultimately from the fact that we have rescaled some fields. However, since the divergent terms are pure gauge, namely symmetrized derivatives, they will not contribute to correlators of invariant observables.

On the other hand, if we first take the limit $\Lambda \to 0$ and then $m \to 0$, we again obtain a finite result
\begin{equations}[eq:field_propagator_lambda0]
\begin{split}
\lim_{m \to 0} \lim_{\Lambda \to 0} G_{\mu\nu\rho'\sigma'}(x,x') &= G^{\text{FG},0}_{\mu\nu\rho'\sigma'}(x,x') + 4 (1-\xi) \nabla_{(\mu} \nabla_{(\rho'} \hat{G}^{\text{FG},0}_{\nu)\sigma')}(x,x') \\
&\quad+ \frac{2}{n-1} \left( g_{\mu\nu} \nabla_{\rho'} \nabla_{\sigma'} + g_{\rho'\sigma'} \nabla_\mu \nabla_\nu \right) \hat{G}_0(x,x') \\
&\quad+ 2 \left[ \frac{n-2}{2 (n-1)} - \frac{(1-3\zeta) (1-\zeta)}{(2\zeta-1)^2} \xi \right] \nabla_\mu \nabla_\nu \nabla_{\rho'} \nabla_{\sigma'} \dhat{G}_0(x,x') \\
&\quad+ \frac{2}{(n-1) (n-2)} g_{\mu\nu} g_{\rho'\sigma'} G_0(x,x') \eqend{,}
\end{split} \\
\lim_{m \to 0} \lim_{\Lambda \to 0} G_{\mu\nu\rho'\circ}(x,x') &= 0 \eqend{,} \\
\lim_{m \to 0} \lim_{\Lambda \to 0} G_{\mu\circ\rho'\sigma'}(x,x') &= 0 \eqend{,} \\
\lim_{m \to 0} \lim_{\Lambda \to 0} G_{\mu\nu\circ\circ}(x,x') &= \left[ \frac{2 \zeta}{2\zeta-1} - \frac{n}{n-1} \right] \nabla_\mu \nabla_\nu \hat{G}_0(x,x') + \frac{1}{n-1} g_{\mu\nu} G_0(x,x') \eqend{,} \\
\lim_{m \to 0} \lim_{\Lambda \to 0} G_{\circ\circ\rho'\sigma'}(x,x') &= \left[ \frac{2 \zeta}{2\zeta-1} - \frac{n}{n-1} \right] \nabla_{\rho'} \nabla_{\sigma'} \hat{G}_0(x,x') + \frac{1}{n-1} g_{\rho'\sigma'} G_0(x,x') \eqend{,} \\
\lim_{m \to 0} \lim_{\Lambda \to 0} G_{\mu\circ\rho'\circ}(x,x') &= G^{\text{FG},0}_{\mu\rho'}(x,x') + (1-\alpha) \nabla_\mu \nabla_{\rho'} \hat{G}_0(x,x') \eqend{,} \\
\lim_{m \to 0} \lim_{\Lambda \to 0} G_{\mu\circ\circ\circ}(x,x') &= 0 \eqend{,} \\
\lim_{m \to 0} \lim_{\Lambda \to 0} G_{\circ\circ\rho'\circ}(x,x') &= 0 \eqend{,} \\
\lim_{m \to 0} \lim_{\Lambda \to 0} G_{\circ\circ\circ\circ}(x,x') &= \frac{n-2}{2 (n-1)} G_0(x,x') \eqend{,}
\end{equations}
and also in this limit the scalar $\phi$ is still coupled to the graviton $h_{\mu\nu}$, a manifestation of the vDZV discontinuity.

\subsection{Auxiliary fields}

Finally, we consider the shifted auxiliary fields $\tilde{B}_\mu$ and $\tilde{B}$ defined by~\eqref{eq:tildeB}, which decoupled in the action~\eqref{eq:finalaction}. Since they have no kinetic term, their propagator simply reads
\begin{equation}
\label{eq:tildeB_propagator}
\expect{ \tilde{B}_\mu(x) \tilde{B}_{\nu'}(x') } = \frac{\mathi}{\xi} g_{\mu\nu'} \frac{\delta^n(x-x')}{\sqrt{-g}} \eqend{,} \quad \expect{ \tilde{B}(x) \tilde{B}(x') } = \frac{\mathi}{\alpha} \frac{\delta^n(x-x')}{\sqrt{-g}} \eqend{.}
\end{equation}
Using the propagators~\eqref{eq:field_propagator}, we can then determine the propagators involving the original auxiliary fields $B_\mu$ and $B$. First we determine propagators involving a single $B_\mu$ or $B$, which follow from the fact that propagators involving $\tilde{B}_\mu$ or $\tilde{B}$ and a dynamical field $\phi$, $A_\mu$ or $h_{\mu\nu}$ vanish. This results in
\begin{equations}[eq:B_propagator_linear]
\begin{split}
\expect{ B(x) h_{\rho'\sigma'}(x') } &= - \frac{1}{\alpha} \nabla_\mu \expect{ A^\mu(x) h_{\rho'\sigma'}(x') } + \frac{m}{2 \alpha} \expect{ h(x) h_{\rho'\sigma'}(x') } \\
&= \frac{2 \mathi}{m} \nabla_{\rho'} \nabla_{\sigma'} \left[ G_{M_B^2}(x,x') - G_0(x,x') \right] \eqend{,}
\end{split} \\
\expect{ B(x) A_{\rho'}(x') } &= - \frac{1}{\alpha} \nabla_\mu \expect{ A^\mu(x) A_{\rho'}(x') } + \frac{m}{2 \alpha} \expect{ h(x) A_{\rho'}(x') } = \mathi \nabla_{\rho'} G_{M_B^2}(x,x') \eqend{,} \\
\expect{ B(x) \phi(x') } &= - \frac{1}{\alpha} \nabla_\mu \expect{ A^\mu(x) \phi(x') } + \frac{m}{2 \alpha} \expect{ h(x) \phi(x') } = \mathi m G_0(x,x') \eqend{,} \\
\begin{split}
\expect{ B_\mu(x) h_{\rho'\sigma'}(x') } &= - \frac{1}{\xi} \nabla^\nu \expect{ h_{\mu\nu}(x) h_{\rho'\sigma'}(x') } + \frac{1}{2 \xi \zeta} \nabla_\mu \expect{ h(x) h_{\rho'\sigma'}(x') } \\
&\quad+ m \expect{ A_\mu(x) h_{\rho'\sigma'}(x') } - \frac{\lambda}{\xi} \nabla_\mu \expect{ \phi(x) h_{\rho'\sigma'}(x') } \\
&= 2 \mathi \nabla_{(\rho'} G^{\text{FG},M_A^2}_{|\mu|\sigma')}(x,x') + \frac{2 \mathi}{M_A^2} \nabla_\mu \nabla_{\rho'} \nabla_{\sigma'} \left[ G_{M_A^2}(x,x') - G_{M_B^2}(x,x') \right] \eqend{,}
\end{split} \raisetag{4.6em} \\
\begin{split}
\expect{ B_\mu(x) A_{\rho'}(x') } &= - \frac{1}{\xi} \nabla^\nu \expect{ h_{\mu\nu}(x) A_{\rho'}(x') } + \frac{1}{2 \xi \zeta} \nabla_\mu \expect{ h(x) A_{\rho'}(x') } \\
&\quad+ m \expect{ A_\mu(x) A_{\rho'}(x') } - \frac{\lambda}{\xi} \nabla_\mu \expect{ \phi(x) A_{\rho'}(x') } \\
&= \mathi m G^{\text{FG},M_A^2}_{\mu\rho'}(x,x') + \frac{\mathi m}{M_A^2} \nabla_\mu \nabla_{\rho'} \left[ G_{M_A^2}(x,x') - G_{M_B^2}(x,x') \right] \eqend{,}
\end{split} \\
\begin{split}
\expect{ B_\mu(x) \phi(x') } &= - \frac{1}{\xi} \nabla^\nu \expect{ h_{\mu\nu}(x) \phi(x') } + \frac{1}{2 \xi \zeta} \nabla_\mu \expect{ h(x) \phi(x') } \\
&\quad+ m \expect{ A_\mu(x) \phi(x') } - \frac{\lambda}{\xi} \nabla_\mu \expect{ \phi(x) \phi(x') } = 0 \eqend{.}
\end{split}
\end{equations}
These results were again computed with $\textsc{xAct}$, using the trace~\eqref{eq:tensor_fg_tr}, \eqref{eq:tensor_fg_trs} and divergence identities~\eqref{eq:tensor_fg_div}, \eqref{eq:tensor_fg_divs} for the Feynman-type gauge tensor propagator, as well as the divergence identities~\eqref{eq:vector_fg_div}, \eqref{eq:vector_fg_divs} for the Feynman gauge vector propagator and its equation of motion~\eqref{eq:vector_fg_eom}. We now take the massless limit, holding either $\lambda$ or $\Lambda$ fixed. In the first case, we obtain
\begin{equations}[eq:B_propagator_linear_massless]
\lim_{m \to 0} \expect{ B(x) h_{\rho'\sigma'}(x') } \big\rvert_{\lambda = \text{cte}} &= 0 \eqend{,} \\
\lim_{m \to 0} \expect{ B(x) A_{\rho'}(x') } \big\rvert_{\lambda = \text{cte}} &= \mathi \nabla_{\rho'} G_0(x,x') \eqend{,} \\
\lim_{m \to 0} \expect{ B(x) \phi(x') } \big\rvert_{\lambda = \text{cte}} &= 0 \eqend{,} \\
\lim_{m \to 0} \expect{ B_\mu(x) h_{\rho'\sigma'}(x') } \big\rvert_{\lambda = \text{cte}} &= 2 \mathi \nabla_{(\rho'} G^{\text{FG},0}_{|\mu|\sigma')}(x,x') + 2 \mathi \frac{\zeta-1}{2\zeta-1} \nabla_\mu \nabla_{\rho'} \nabla_{\sigma'} \hat{G}_0(x,x') \eqend{,} \\
\lim_{m \to 0} \expect{ B_\mu(x) A_{\rho'}(x') } \big\rvert_{\lambda = \text{cte}} &= 0 \eqend{,} \\
\lim_{m \to 0} \expect{ B_\mu(x) \phi(x') } \big\rvert_{\lambda = \text{cte}} &= 0 \eqend{,}
\end{equations}
and we see that most propagators vanish, except the one between the diffeomorphism auxiliary field $B_\mu$ and the graviton $h_{\mu\nu}$ and the one between the Stueckelberg auxiliary field $B$ and the Stueckelberg vector $A_\mu$. That is, in the massless limit the auxiliary field sector decouples into separate tensor and vector sectors, which is of course very reasonable. In the second case, we obtain instead
\begin{equations}[eq:B_propagator_linear_massless_lambda]
\lim_{m \to 0} \expect{ B(x) h_{\rho'\sigma'}(x') } \big\rvert_{\Lambda = \text{cte}} &\sim \frac{2 \mathi}{m} \nabla_{\rho'} \nabla_{\sigma'} \left[ G_{- \frac{\zeta}{2\zeta-1} \tilde{\Lambda}}(x,x') - G_0(x,x') \right] \eqend{,} \\
\lim_{m \to 0} \expect{ B(x) A_{\rho'}(x') } \big\rvert_{\Lambda = \text{cte}} &= \mathi \nabla_{\rho'} G_{- \frac{\zeta}{2\zeta-1} \tilde{\Lambda}}(x,x') \eqend{,} \\
\lim_{m \to 0} \expect{ B(x) \phi(x') } \big\rvert_{\Lambda = \text{cte}} &= 0 \eqend{,} \\
\begin{split}
\lim_{m \to 0} \expect{ B_\mu(x) h_{\rho'\sigma'}(x') } \big\rvert_{\Lambda = \text{cte}} &= 2 \mathi \nabla_{(\rho'} G^{\text{FG},-\tilde{\Lambda}}_{|\mu|\sigma')}(x,x') \\
&\quad- \frac{2 \mathi}{\tilde{\Lambda}} \nabla_\mu \nabla_{\rho'} \nabla_{\sigma'} \left[ G_{- \tilde{\Lambda}}(x,x') - G_{- \frac{\zeta}{2\zeta-1} \tilde{\Lambda}}(x,x') \right] \eqend{,}
\end{split} \\
\lim_{m \to 0} \expect{ B_\mu(x) A_{\rho'}(x') } \big\rvert_{\Lambda = \text{cte}} &= 0 \eqend{,} \\
\lim_{m \to 0} \expect{ B_\mu(x) \phi(x') } \big\rvert_{\Lambda = \text{cte}} &= 0 \eqend{,}
\end{equations}
where we set as before $\tilde{\Lambda} \equiv 4/(n-2) \Lambda = \lambda m^2$. We see once more that this is not a sensible limit, since some propagators diverge. Taking thus first the limit $\Lambda \to 0$ and afterwards the massless one, we obtain the same limit as in Eqs.~\eqref{eq:B_propagator_linear_massless}.

Using the propagators~\eqref{eq:B_propagator_linear} and~\eqref{eq:tildeB_propagator}, we can then also determine the propagators involving two auxiliary fields $B_\mu$ or $B$. We obtain
\begin{equations}[eq:B_propagator_quadratic]
\begin{split}
\expect{ B(x) B(x') } &= \frac{\mathi}{\alpha} \frac{\delta^n(x-x')}{\sqrt{-g}} - \frac{1}{\alpha} \nabla_{\rho'} \expect{ B(x) A^{\rho'}(x') } - \frac{1}{\alpha} \nabla_\mu \expect{ A^\mu(x) B(x') } \\
&\quad+ \frac{m}{2 \alpha} \expect{ B(x) h(x') } + \frac{m}{2 \alpha} \expect{ h(x) B(x') } - \frac{1}{\alpha^2} \nabla_\mu \nabla_{\rho'} \expect{ A^\mu(x) A^{\rho'}(x') } \\
&\quad+ \frac{m}{2 \alpha^2} \nabla_\mu \expect{ A^\mu(x) h(x') } + \frac{m}{2 \alpha^2} \nabla_{\rho'} \expect{ h(x) A^{\rho'}(x') } - \frac{m^2}{4 \alpha^2} \expect{ h(x) h(x') } \\
&= 0 \eqend{,}
\end{split} \raisetag{9.2em} \\
\begin{split}
\expect{ B_\mu(x) B(x') } &= - \frac{1}{\alpha} \nabla_{\rho'} \expect{ B_\mu(x) A^{\rho'}(x') } + \frac{m}{2 \alpha} \expect{ B_\mu(x) h(x') } - \frac{1}{\xi} \nabla^\nu \expect{ h_{\mu\nu}(x) B(x') } \\
&\quad+ \frac{1}{2 \xi \zeta} \nabla_\mu \expect{ h(x) B(x') } + m \expect{ A_\mu(x) B(x') } - \frac{\lambda}{\xi} \nabla_\mu \expect{ \phi(x) B(x') } \\
&\quad- \frac{1}{\alpha \xi} \nabla^\nu \nabla_{\rho'} \expect{ h_{\mu\nu}(x) A^{\rho'}(x') } + \frac{1}{2 \alpha \xi \zeta} \nabla_\mu \nabla_{\rho'} \expect{ h(x) A^{\rho'}(x') } \\
&\quad+ \frac{m}{\alpha} \nabla_{\rho'} \expect{ A_\mu(x) A^{\rho'}(x') } - \frac{\lambda}{\alpha \xi} \nabla_\mu \nabla_{\rho'} \expect{ \phi(x) A^{\rho'}(x') } - \frac{m^2}{2 \alpha} \expect{ A_\mu(x) h(x') } \\
&\quad+ \frac{m}{2 \alpha \xi} \nabla^\nu \expect{ h_{\mu\nu}(x) h(x') } - \frac{m}{4 \alpha \xi \zeta} \nabla_\mu \expect{ h(x) h(x') } + \frac{m \lambda }{2 \alpha \xi} \nabla_\mu \expect{ \phi(x) h(x') } \\
&= 0 \eqend{,}
\end{split} \raisetag{1.4em} \\
\begin{split}
\expect{ B_\mu(x) B_{\rho'}(x') } &= \frac{\mathi}{\xi} g_{\mu\rho'} \frac{\delta^n(x-x')}{\sqrt{-g}} - \frac{1}{\xi} \nabla^{\sigma'} \expect{ B_\mu(x) h_{\rho'\sigma'}(x') } + \frac{1}{2 \xi \zeta} \nabla_{\rho'} \expect{ B_\mu(x) h(x') } \\
&\quad+ m \expect{ B_\mu(x) A_{\rho'}(x') } - \frac{\lambda}{\xi} \nabla_{\rho'} \expect{ B_\mu(x) \phi(x') } - \frac{1}{\xi} \nabla^\nu \expect{ h_{\mu\nu}(x) B_{\rho'}(x') } \\
&\quad+ \frac{1}{2 \zeta \xi} \nabla_\mu \expect{ h(x) B_{\rho'}(x') } + m \expect{ A_\mu(x) B_{\rho'}(x') } - \frac{\lambda}{\xi} \nabla_\mu \expect{ \phi(x) B_{\rho'}(x') } \\
&\quad- \frac{1}{\xi^2} \nabla^\nu \nabla^{\sigma'} \expect{ h_{\mu\nu}(x) h_{\rho'\sigma'}(x') } + \frac{1}{2 \zeta \xi^2} \nabla^\nu \nabla_{\rho'} \expect{ h_{\mu\nu}(x) h(x') } \\
&\quad+ \frac{1}{2 \zeta \xi^2} \nabla_\mu \nabla^{\sigma'} \expect{ h(x) h_{\rho'\sigma'}(x') } - \frac{1}{4 \zeta^2 \xi^2} \nabla_\mu \nabla_{\rho'} \expect{ h(x) h(x') } \\
&\quad+ \frac{m}{\xi} \nabla^\nu \expect{ h_{\mu\nu}(x) A_{\rho'}(x') } - \frac{m}{2 \zeta \xi} \nabla_\mu \expect{ h(x) A_{\rho'}(x') } + \frac{m}{\xi} \nabla^{\sigma'} \expect{ A_\mu(x) h_{\rho'\sigma'}(x') } \\
&\quad- \frac{m}{2 \zeta \xi} \nabla_{\rho'} \expect{ A_\mu(x) h(x') } - \frac{\lambda}{\xi^2} \nabla^\nu \nabla_{\rho'} \expect{ h_{\mu\nu}(x) \phi(x') } + \frac{\lambda}{2 \zeta \xi^2} \nabla_\mu \nabla_{\rho'} \expect{ h(x) \phi(x') } \\
&\quad- \frac{\lambda}{\xi^2} \nabla_\mu \nabla^{\sigma'} \expect{ \phi(x) h_{\rho'\sigma'}(x') } + \frac{\lambda}{2 \zeta \xi^2} \nabla_\mu \nabla_{\rho'} \expect{ \phi(x) h(x') } - m^2 \expect{ A_\mu(x) A_{\rho'}(x') } \\
&\quad+ \frac{m \lambda}{\xi} \nabla_{\rho'} \expect{ A_\mu(x) \phi(x') } + \frac{m \lambda}{\xi} \nabla_\mu \expect{ \phi(x) A_{\rho'}(x') } - \frac{\lambda^2}{\xi^2} \nabla_\mu \nabla_{\rho'} \expect{ \phi(x) \phi(x') } \\
&= 0 \eqend{.}
\end{split} \raisetag{1.4em}
\end{equations}
We see that the auxiliary fields do not propagate, such that the mixed propagators~\eqref{eq:B_propagator_linear} only become relevant when the auxiliary fields appear in external states, i.e., explicitly in the correlation function. This is relevant for the verification of Ward identities, which we do in the next subsection.

\subsection{Ward identities}

Ward(--Takahashi--Slavnov--Taylor) identities~\cite{ward1950,takahashi1957,taylor1971,slavnov1972} are relations between correlation functions that stem from the gauge invariance of the underlying classical theory. In the BRST formalism, they arise from the fact that any BRST-exact term has vanishing expectation value in a physical state, i.e., $\expect{ \brst (\text{anything}) } = 0$. In our case, they relate the different propagators of the fields, ghosts and auxiliary fields, and we will display explicitly a subset of all possible identities.

A non-trivial group of identities is obtained by considering the expectation value of the BRST transformation of a field and an antighost. Using the explicit BRST transformations~\eqref{eq:brst_fields} and~\eqref{eq:brst_ghosts}, we obtain
\begin{equations}[eq:ward_ghost_b]
0 &= \expect{ \brst \left[ h_{\mu\nu}(x) \bar{c}_{\rho'}(x') \right] } = \nabla_\mu \expect{ c_\nu(x) \bar{c}_{\rho'}(x') } + \nabla_\nu \expect{ c_\mu(x) \bar{c}_{\rho'}(x') } + \expect{ h_{\mu\nu}(x) B_{\rho'}(x') } \eqend{,} \\
0 &= \expect{ \brst \left[ A_\mu(x) \bar{c}_{\rho'}(x') \right] } = m \expect{ c_\mu(x) \bar{c}_{\rho'}(x') } + \nabla_\mu \expect{ c(x) \bar{c}_{\rho'}(x') } + \expect{ A_\mu(x) B_{\rho'}(x') } \eqend{,} \\
0 &= \expect{ \brst \left[ \phi(x) \bar{c}_{\rho'}(x') \right] } = m \expect{ c(x) \bar{c}_{\rho'}(x') } + \expect{ \phi(x) B_{\rho'}(x') } \eqend{,} \\
0 &= \expect{ \brst \left[ h_{\mu\nu}(x) \bar{c}(x') \right] } = \nabla_\mu \expect{ c_\nu(x) \bar{c}(x') } + \nabla_\nu \expect{ c_\mu(x) \bar{c}(x') } + \expect{ h_{\mu\nu}(x) B(x') } \eqend{,} \\
0 &= \expect{ \brst \left[ A_\mu(x) \bar{c}(x') \right] } = m \expect{ c_\mu(x) \bar{c}(x') } + \nabla_\mu \expect{ c(x) \bar{c}(x') } + \expect{ A_\mu(x) B(x') } \eqend{,} \\
0 &= \expect{ \brst \left[ \phi(x) \bar{c}(x') \right] } = m \expect{ c(x) \bar{c}(x') } + \expect{ \phi(x) B(x') } \eqend{,}
\end{equations}
which connects the ghost propagators~\eqref{eq:ghost_propagator} with the ones containing an auxiliary field~\eqref{eq:B_propagator_linear}. Inserting the explicit expressions for the propagators, one checks easily that the identities~\eqref{eq:ward_ghost_b} are indeed fulfilled.

Also the non-propagation of the auxiliary fields, i.e., that fact that their propagators vanish~\eqref{eq:B_propagator_quadratic}, is the consequence of a Ward identity. Namely, we consider the expectation value of the BRST transformation of an auxiliary field and an antighost. Using the explicit BRST transformations for antighosts and auxiliary fields~\eqref{eq:brst_ghosts}, it follows that
\begin{equations}[eq:ward_auxiliary]
0 &= \expect{ \brst \left[ B_\mu(x) \bar{c}_{\rho'}(x') \right] } = \expect{ B_\mu(x) B_{\rho'}(x') } \eqend{,} \\
0 &= \expect{ \brst \left[ B_\mu(x) \bar{c}(x') \right] } = \expect{ B_\mu(x) B(x') } \eqend{,} \\
0 &= \expect{ \brst \left[ B(x) \bar{c}_{\rho'}(x') \right] } = \expect{ B(x) B_{\rho'}(x') } \eqend{,} \\
0 &= \expect{ \brst \left[ B(x) \bar{c}(x') \right] } = \expect{ B(x) B(x') } \eqend{.}
\end{equations}
Various other identities can also be derived as consequences of a Ward identity, for example the divergence identities~\eqref{eq:vector_fg_div} and~\eqref{eq:vector_fg_divs} for the Feynman gauge vector propagator and the trace~\eqref{eq:tensor_fg_tr}, \eqref{eq:tensor_fg_trs} and divergence identities~\eqref{eq:tensor_fg_div}, \eqref{eq:tensor_fg_divs} for the Feynman-type gauge tensor propagator. For details of the derivation and further identities, we refer the reader to Ref.~\onlinecite{froebtaslimitehrani2018}.

\section{Heat kernel}
\label{sec:heatkernel}

The heat kernel technique was introduced in quantum field theory as a way to treat functional traces and determinants of local differential operators of Laplace type, i.e., second-order elliptic differential operators whose principal symbol is the metric (also called minimal operators). Later on, it was generalized to also include non-minimal operators, where a more general principal symbol is admissible, and which arise for example in gauge theories in a gauge different from Feynman gauge, as well as hyperbolic operators. In the following we will give a short overview of the technique, and refer to the reviews and books~\onlinecite{Barvinsky:1985an,choquet1982analysis,Fulling:1989nb,camporesi1990harmonic,Avramidi:2000bm,Vassilevich:2003xt} for more in-depth results.

Formally, the heat kernel associated to a Laplace-type operator $P$ is given by
\begin{equation}
\label{eq:HK}
K(\tau) = \mathe^{\tau P} \quad\text{for}\quad \tau \geq 0 \eqend{,}
\end{equation}
i.e., it is the kernel of the heat semigroup whose generator is the operator $P$, which we take to be negative definite (or more generally, $-P$ should be bounded from below). More concretely, we consider a Laplace-type operator $P^{AB}$ acting on sections of a vector bundle whose elements are denoted by $v_A$, which is a differential operator of the form
\begin{equation}
\label{eq:diffop}
P^{AB} = \1^{AB} \nabla^2 + E^{AB}
\end{equation}
with $\1^{AB}$ the identity on the bundle under consideration, and where $E^{AB}$ contains at most first-order derivatives. For example, for the standard scalar Laplacian acting on (spin-0) functions, we have $\1 = 1$ and $E = - \xi R$ with the scalar curvature $R$ and a parameter $\xi \in \mathbb{R}$, for spin-1 vector fields we have $\1^{\mu\nu} = g^{\mu\nu}$ and for a spin-2 symmetric tensor we have $\1^{(\mu\nu)(\rho\sigma)} = g^{\mu(\rho} g^{\sigma)\nu}$; in general, $\1^{AB}$ is a properly symmetrized product of metrics for a spin-$s$ tensor field. The heat kernel $K_{AB'}(x,x',\tau)$ is then the solution to the heat equation
\begin{equation}
\label{eq:heatkernel_diffeq}
\left( P^{AB} - \1^{AB} \partial_\tau \right) K_{BC'}(x,x',\tau) = 0
\end{equation}
with boundary condition
\begin{equation}
\lim_{\tau \to 0^+} K^A{}_{B'}(x,x',\tau) = \1^A{}_{B'} \frac{\delta^n(x-x')}{\sqrt{-g}} \eqend{,}
\end{equation}
which is unique assuming some mild conditions on the growth at infinity~\cite{berline2003heat,kawakami2010global}. Furthermore, $K$ is bounded for large $x$ or $x'$ for any $\tau > 0$~\cite{berline2003heat}.

For $n$-dimensional flat space with $P^{AB} = \1^{AB} \partial^2$, the heat kernel is known explicitly~\cite{choquet1982analysis,camporesi1990harmonic}:
\begin{equation}
K_\text{flat}{}^A{}_{B'}(x,x',\tau) = \1^A{}_{B'} \frac{1}{(4 \pi \tau)^\frac{n}{2}} \mathe^{- \frac{(x-x')^2}{4 \tau}} = \1^A{}_{B'} \int \mathe^{- \tau p^2} \mathe^{\mathi p (x-x')} \frac{\total^n p}{(2\pi)^n} \eqend{.}
\end{equation}
In Fourier space, it is clear that for $\tau = 0$ we obtain a $\delta$ distribution and the correct normalisation. However, for a general Riemannian manifold, no explicit closed-form expression can be given. Instead, it is possible to determine successive approximations to the heat kernel, and two main schemes are known: the first one is the asymptotic local or early-time expansion in terms of powers of $\tau$ (see Ref.~\onlinecite[Thm.~2.30]{berline2003heat})
\begin{equation}
\label{eq:smalltime}
K^A{}_{B'}(x,x',\tau) \sim \frac{\sqrt{\Delta(x,x')}}{(4 \pi \tau)^\frac{n}{2}} \mathe^{- \frac{\sigma(x,x')}{2 \tau}} \sum_{k=0}^\infty \tau^k A_k{}^A{}_{B'}(x,x') \eqend{,}
\end{equation}
where $\sigma$ is the Synge world function~\cite{Synge:1931zz}, equal to half of the square of the geodesic distance between $x$ and $x'$, and $\Delta$ is the van Vleck--Morette determinant~\cite{VanVleck:1928zz}. The quantities $A_k{}^A{}_{B'}$ are called \emph{heat kernel coefficients}, and their coincidence limits $A_k{}^A{}_{B'}(x,x)$ are local functions of the curvature tensors and their covariant derivatives. The second scheme is based on a non-local expansion of the heat kernel around the flat-space limit in powers of the curvature tensors~\cite{Barvinsky:1987uw,Barvinsky:1990up,Groh:2011dw}, and resums certain powers of $\tau$ in the expansion. Since we will only use the first scheme in this work, we do not enter into further details.

For non-minimal operators, there is no direct heat kernel expansion. Unfortunately, many of the operators that are of physical interest are not of Laplace-type. A well-known example is Yang-Mills theory in a covariant gauge with parameter $\xi$ different from Feynman gauge $\xi = 1$, where the bundle is composed from the tangent bundle and an internal Lie algebra bundle (depending on the gauge group). While for a semi-simple Lie algebra the internal part of $P^{AB}$ (the Cartan--Killing form) is proportional to the identity $\delta^{ab}$, the covariant spacetime derivatives are uncontracted, and $P^{AB} = \delta^{ab} \left[ g^{\mu\nu} \nabla^2 - (1-\xi) \nabla^\mu \nabla^\nu \right]$ is a non-minimal differential operator. One possibility to treat such operators is the use of covariant projectors which lead to Laplace-type differential operators acting on subspaces of the original field space, i.e., on a different vector bundle. Typical examples for such projections are the transverse decomposition of a vector field, or the transverse traceless decomposition of a symmetric tensor field, for which one can in particular use the second approximation scheme explained above~\cite{Decanini:2005gt,Anselmi:2007eq,Benedetti:2010nr,Groh:2011vn}. Alternatively, one can write the heat kernel for the non-minimal operator as the sum of the one for the minimal operator and covariant derivatives of the heat kernel of lower spin fields as in Ref.~\onlinecite[Sec.~2]{Barvinsky:1985an}. This is akin to the determination of the corresponding Green's functions in terms of the Feynman-type gauge ones, and is based on the Ward identities that hold in the free theory analogously to the trace and divergence identities that we computed in Sec.~\ref{sec:propagators}. In fact, we will derive below trace and divergence identities for the (minimal, Feynman-type gauge) heat kernels (and their coefficients), using which the corresponding non-minimal heat kernel could be computed. Unfortunately, the relation between the two approaches to non-minimal operators and their heat kernel expansions is not straightforward, mainly because the known expansions are in general only asymptotic. One can obtain a convergent expansion by multiplying the coefficients $A_k{}^A{}_{B'}(x,x')$ in the local expansion~\eqref{eq:smalltime} by cutoff functions whose support decreases quickly enough with $k$ (see Ref.~\onlinecite[Sec.~5.2]{Hollands:2001nf}). In this way, the same coincidence limits (as $x' \to x$) as for the unmodified expansion hold, but in any finite neighborhood the sum only has a finite number of terms; only for analytic spacetimes one can hope to have a convergent local expansion without cutoffs. Fortunately, for most purposes it is enough to know the coincidence limits of a finite number of coefficients and their derivatives, which are always well-defined.

Given the heat kernel, the unique solution of the differential equation
\begin{equation}
\left( P^{AB} - \1^{AB} m^2 \right) G_{BC'}(x,x') = \1^A{}_{C'} \frac{\delta^n(x-x')}{\sqrt{-g}}
\end{equation}
for the Green's function of $P - m^2$ is then obtained by integrating over the proper time $\tau$:
\begin{equation}
\label{eq:propagator_from_heatkernel}
G_{AB'}(x,x') = - \int_0^\infty K_{AB'}(x,x',\tau) \, \mathe^{- m^2 \tau} \total \tau \eqend{.}
\end{equation}
Using integration by parts and the fact that $K$ is bounded, one sees that this indeed fulfills the required differential equation:
\begin{splitequation}
\left( P^{AB} - \1^{AB} m^2 \right) G_{BC'}(x,x') &= - \int_0^\infty \left( \partial_\tau - m^2 \right) K^A{}_{C'}(x,x',\tau) \, \mathe^{- m^2 \tau} \total \tau \\
&= K^A{}_{C'}(x,x',0) - \lim_{\tau \to \infty} \left[ K^A{}_{C'}(x,x',\tau) \, \mathe^{- m^2 \tau} \right] \\
&= \1^A{}_{C'} \frac{\delta^n(x-x')}{\sqrt{-g}} \eqend{.}
\end{splitequation}
For later convenience, we have separated the mass so that $P$ is the massless differential operator. As we see, a positive mass ensures convergence for large $\tau$.

We now assume that the differential operator has the form~\eqref{eq:diffop} where $E$ does not involve derivatives, which is all that we need. Inserting the asymptotic expansion~\eqref{eq:smalltime} into the equation~\eqref{eq:heatkernel_diffeq} satisfied by the heat kernel and comparing powers of $\tau$, we obtain the transport equations
\begin{equation}
\label{eq:transport}
\Big( \nabla^\nu \sigma \nabla_\nu + k \Big) A_k^A{}_{C'} = \frac{1}{\sqrt{\Delta}} P^A{}_B \Big( \sqrt{\Delta} \, A_{k-1}^B{}_{C'} \Big)
\end{equation}
with the initial condition
\begin{equation}
\nabla^\nu \sigma \nabla_\nu A_0^A{}_{B'} = 0 \eqend{,}
\end{equation}
where we also used the following identities for the Synge world function~\cite{poissonpoundvega2011}:
\begin{equation}
\label{eq:synge_identities}
\nabla^\mu \sigma \nabla_\mu \sigma = 2 \sigma \eqend{,} \quad \nabla^2 \sigma = n - \nabla^\mu \sigma \nabla_\mu \ln \Delta \eqend{.}
\end{equation}
Taking into account the boundary condition for the heat kernel, the initial condition is solved by
\begin{equation}
\label{eq:transport_initial}
A_0^A{}_{B'}(x,x') = \mathcal{G}^A{}_{B'}(x,x')
\end{equation}
with the spin-$s$ parallel propagator $\mathcal{G}$. For integer spin, it is given by the symmetrized product of spin-1 parallel propagators $g_{\mu\rho'}$:
\begin{equation}
\mathcal{G}^A{}_{B'}(x,x') = g^{(\mu_1}{}_{\rho_1'}(x,x') \cdots g^{\mu_s)}{}_{\rho_s'}(x,x') \eqend{,}
\end{equation}
where $A = \mu_1 \cdots \mu_s$, $B' = \rho_1' \cdots \rho_s'$, and where $g_{\mu\rho'}$ satisfies~\cite{poissonpoundvega2011}
\begin{equation}
\label{eq:parallel_propagator}
\nabla^\nu \sigma \nabla_\nu g_{\mu\rho'} = 0 \eqend{,} \quad \lim_{x' \to x} g^\mu{}_{\rho'} = \delta^\mu_\rho \eqend{.}
\end{equation}

The solution for the higher-order coefficients $A_k^A{}_{B'}$ can be given exactly in Riemann normal coordinates~\cite{Sakai,Berger1971LeSD,chavel1984eigenvalues}. In these coordinates, the geodesics from $x'$ to $x$ are straight lines $y^\mu(\lambda) = (x')^\mu + \lambda [ x^\mu - (x')^\mu ]$, such that $\sigma(x,x') = \frac{1}{2} g_{\mu\nu}(x) [ x^\mu - (x')^\mu ] [ x^\nu - (x')^\nu ]$ and $\nabla^\mu \sigma(x,x') = x^\mu - (x')^\mu$. For $k > 0$, we thus compute
\begin{splitequation}
&\Big[ \nabla^\nu \sigma(x,x') \nabla^x_\nu + k \Big] \int_0^1 J(y,x') \lambda^{k-1} \total \lambda \\
&\quad= \int_0^1 \Big[ ( x^\nu - (x')^\nu ) \lambda ( \nabla_\nu J )(y,x') + k J(y,x') \Big] \lambda^{k-1} \total \lambda \\
&\quad= \int_0^1 \Big( \lambda \partial_\lambda + k \Big) J(y,x') \lambda^{k-1} \total \lambda = J(x,x') \eqend{,}
\end{splitequation}
and it follows that
\begin{equation}
\label{eq:normal}
A_k^A{}_{C'}(x,x') = \int_0^1 \left[ \frac{1}{\sqrt{\Delta}} P^A{}_B \Big( \sqrt{\Delta} \, A_{k-1}^B{}_{C'} \Big) \right](y,x') \lambda^{k-1} \total \lambda \eqend{.}
\end{equation}

The main application of the heat kernel in quantum field theory is in the renormalisation of the effective action or of composite operators, for which only the coincidence limit of a finite number of coefficients and their derivatives are needed. In particular, for the renormalisation of the stress tensor in $n$ dimensions one needs the coincidence limit of the coefficients $A_k$ with $k = 0,\ldots,n$. To check that the renormalized stress tensor is covariantly conserved, one needs in addition the coincidence limit of their derivatives $\nabla_{\mu_1} \cdots \nabla_{\mu_\ell} A_k$ with $k + \ell \leq n+1$. To determine these coincidence limits, one can either use the explicit formula~\eqref{eq:normal} in normal coordinates, or one can take covariant derivatives of the transport equations~\eqref{eq:transport} and their coincidence limit. In turn, these can be obtained recursively using the coincidence limits of derivatives of the Synge world function $\sigma$ and the van Vleck--Morette determinant $\Delta$.

Several methods have been developed to find explicit expressions for the heat kernel coefficients~\cite{Avramidi:1995qe}. DeWitt~\cite{DeWitt:1964mxt} determined the first two coefficients with a covariant recursive method. Sakai~\cite{Sakai} relied on the Riemann normal coordinates to find the third coefficient in the scalar case (i.e., for a single scalar field on a curved space). For the general case, this coefficient was found by Gilkey~\cite{Gilkey1979} using a noncovariant pseudo-differential-operator technique (see also Ref.~\onlinecite{Kirsten:1999qjn}). The integrated and traced fourth and fifth coefficients for an arbitrary field theory in flat space were found in~\cite{vandeven1985} through the evaluation of a noncovariant Feynman graph. Avramidi~\cite{Avramidi:1986mj,Avramidi:1989ik} presented a new covariant non-recursive procedure and found the fourth coefficient for the general case; the coefficient $a_5$ in flat space was computed in~\cite{vandeven1985,vandeVen:1997pf}. Various resummation methods have also been developed, see Ref.~\onlinecite{FranchinoVinas2023} and references therein. Finally, the connection between the Riemannian heat kernel and the Lorentzian analogue, which we treat in the following, was elucidated for stationary spacetimes by Strohmaier and Zelditch~\cite{Strohmaier:2023wsv}.

In the Lorentzian case, the heat kernel expansion is only formal even for positive mass, since $\nabla^2$ is now an unbounded operator whose spectrum (in general) is the full real line. We thus consider instead the Wick-rotated version of the heat kernel, which fulfills~\cite{Fulling:1989nb,Estrada:1997qt,Moretti:1999ez}
\begin{equation}
\label{eq:lorentzian_heateq}
\left( \mathi P^{AB} - \1^{AB} \partial_\tau \right) K_{BC'}(x,x',\tau) = 0
\end{equation}
with the boundary condition
\begin{equation}
K^A{}_{B'}(x,x',0) = \1^A{}_{B'} \frac{\delta^n(x-x')}{\sqrt{-g}} \eqend{.}
\end{equation}
In this case the heat kernel satisfies a Schrödinger-like equation with $P^{AB}$ playing the role of the Hamiltonian, since the problem has been reformulated into an equivalent ``quantum mechanical'' problem, solving the ``evolution'' equation~\eqref{eq:lorentzian_heateq} in the Schwinger proper time \cite{camporesi1990harmonic}. We assume that there exists a unique solution of this equation which does not grow faster than exponentially. The Feynman propagator corresponding to $P - m^2$ can then be defined as
\begin{equation}
\label{eq:feynman_from_heatkernel}
G^\text{F}_{AB'}(x,x') \equiv - \mathi \lim_{\epsilon \to 0^+} \int_0^\infty K_{AB'}(x,x',\tau) \, \mathe^{- \mathi ( m^2 - \mathi \epsilon ) \tau} \total \tau \eqend{,}
\end{equation}
where the limit $\epsilon \to 0$ is to be taken in a distributional sense, i.e., after integrating over $\tau$ and smearing with test functions. Since we assume that $K$ does not grow too fast, the $\tau$ integral is convergent for $\epsilon > 0$, and analogously to the Riemannian case we can easily check that the Feynman propagator fulfills the right equation:
\begin{splitequation}
\left( P^{AB} - \1^{AB} m^2 \right) G^\text{F}_{BC'}(x,x') &= \lim_{\epsilon \to 0^+} \int_0^\infty \left( - \partial_\tau + \mathi m^2 \right) K^A{}_{C'}(x,x',\tau) \mathe^{- \mathi ( m^2 - \mathi \epsilon ) \tau} \total \tau \\
&= K^A{}_{C'}(x,x',0) - \lim_{\tau \to \infty} \left[ K^A{}_{C'}(x,x',\tau) \, \mathe^{- \mathi ( m^2 - \mathi \epsilon ) \tau} \right] \\
&\quad- \lim_{\epsilon \to 0^+} \left[ \epsilon \int_0^\infty K^A{}_{C'}(x,x',\tau) \mathe^{- \mathi ( m^2 - \mathi \epsilon ) \tau} \total \tau \right] \\
&= \1^A{}_{C'} \frac{\delta^n(x-x')}{\sqrt{-g}} \eqend{.}
\end{splitequation}

For $n$-dimensional flat space with $P^{AB} = \1^{AB} \partial^2$, we again have an explicit expression for the heat kernel~\cite{choquet1982analysis,camporesi1990harmonic}
\begin{splitequation}
K_\text{flat}^A{}_{B'}(x,x',\tau) &= \1^A{}_{B'} \frac{1}{(4 \pi \tau)^\frac{n}{2}} \mathe^{\frac{\mathi (x-x')^2}{4 \tau}} \mathe^{- \mathi \frac{n-2}{4} \pi} \\
&= \1^A{}_{B'} \lim_{\epsilon \to 0^+} \int \mathe^{- \mathi \tau p^2} \mathe^{- \epsilon [ (p^0)^2 + \vec{p}^2 ]} \mathe^{\mathi p (x-x')} \frac{\total^n p}{(2\pi)^n} \eqend{.}
\end{splitequation}
In Fourier space, it is again clear that for $\tau = 0$ we obtain a $\delta$ distribution and the correct normalisation. For a general pseudo-Riemannian manifold, we make an ansatz analogous to the Riemannian case~\eqref{eq:smalltime} for the small-$\tau$ asymptotic expansion~\cite{Fulling:1989nb,camporesi1990harmonic,Avramidi:2000bm}:
\begin{equation}
\label{eq:smalltimeL}
K^A{}_{B'}(x,x',\tau) \sim \frac{\sqrt{\Delta(x,x')}}{(4 \pi \tau)^\frac{n}{2}} \mathe^{\frac{\mathi \sigma(x,x')}{2 \tau}} \mathe^{- \mathi \frac{n-2}{4} \pi} \sum_{k=0}^\infty (\mathi \tau)^k A_k^A{}_{B'}(x,x') \eqend{,}
\end{equation}
and assume the same form~\eqref{eq:diffop} for the differential operator. Let us remark that this expansion, as the one in the Riemannian case, is only valid for minimal operators, such as a vector field in Feynman gauge. For non-minimal operators, such as the Proca operator for a transverse vector field, one has to use either covariant projectors or add the appropriate multiple of derivatives of the heat kernel of lower spin fields as explained above.

Inserting the expansion~\eqref{eq:smalltimeL} into the equation~\eqref{eq:lorentzian_heateq} satisfied by the heat kernel and comparing powers of $\tau$, we obtain the transport equations
\begin{equation}
\Big( \nabla^\mu \sigma \nabla_\mu + k \Big) A_k^A{}_{C'}(x,x') = \frac{1}{\sqrt{\Delta}} P^A{}_B \Big( \sqrt{\Delta} \, A_{k-1}^B{}_{C'}(x,x') \Big)
\end{equation}
and the initial condition
\begin{equation}
\nabla^\nu \sigma \nabla_\nu A_0^A{}_{B'}(x,x') = 0 \eqend{.}
\end{equation}
We see that we obtain the same transport equations as in the Riemannian case, which is due to our inclusion of an explicit factor of $\mathi^k$ in the expansion~\eqref{eq:smalltimeL}.

\subsection{Small time regularisation}

For $x \neq x'$, the factor $\mathe^{\frac{\mathi \sigma(x,x')}{2 \tau}}$ in the heat kernel expansion~\eqref{eq:smalltimeL}, or $\mathe^{- \frac{\sigma(x,x')}{2 \tau}}$ in the expansion~\eqref{eq:smalltime}, ensures that the proper time integrals~\eqref{eq:propagator_from_heatkernel} and~\eqref{eq:feynman_from_heatkernel} converge at $\tau = 0$. However, in the coincidence limit $x' = x$ which is needed for applications, this factor becomes 1 and these integrals are divergent for small $\tau$. This is of course the well-known divergence of the propagators in the coincidence limit.\footnote{In the Lorentzian case, the factor $\mathe^{\frac{\mathi \sigma(x,x')}{2 \tau}}$ also vanishes for light-like separations, and we recover the well-known divergence of the propagator as one approaches the light cone.} In particular, it is clear from the expansions~\eqref{eq:smalltime} and~\eqref{eq:smalltimeL} that the coefficients $A_0$, ..., $A_{n/2-1}$ have divergent prefactors if one computes the coincidence limit of the propagator, and more if one is interested in the coincidence limit of derivatives acting on it.

To regulate this divergence and extract the finite part of the result, analytic regularisation schemes are well suited for the heat kernel expansion. A well-known such scheme is dimensional regularisation, where one continues the dimension $n$ of spacetime into the complex plane, computes results in a region where all integrals are convergent, and then analytically continues the result back to the physical dimension. In particular, for $\Re n$ small enough (depending on the number of covariant derivatives acting on the heat kernel), the proper time integrals are convergent for small $\tau$ even in the coincidence limit. The original divergence of the propagator then manifests itself in poles as $n$ approaches the physical dimension. While dimensional regularisation is versatile and usually the simplest one, it also has problems, namely when the objects under study cannot naturally be defined in arbitrary dimension $n$. In particular, this concerns chiral, conformal and supersymmetric theories, where the symmetries are dimension-dependent. These symmetries can then be broken in the quantum theory and one obtains chiral, conformal and supersymmetry anomalies~\cite{Piguet:1980fa,Itoyama:1985qi}. Of course anomalies are physical and cannot depend on the choice of a regularisation scheme, but only on renormalisation conditions (as can be seen in the dichotomy between covariant and consistent anomalies~\cite{Bardeen:1984pm}). Therefore, for dimension-dependent symmetries where the advantages of dimensional regularisation are absent, it can be easier to employ a different regularisation scheme. For the heat kernel, such a scheme is obtained by inserting a factor $\tau^\delta$ in the integrals~\eqref{eq:smalltime} and~\eqref{eq:smalltimeL} and performing the analytic continuation in $\delta$.

\subsection{Generic bundle}

We would like to compute the coincidence limit of up to three covariant derivatives of the propagator in $n = 4$ dimensions. For this, we need the coefficients $A_k$ with $k \leq 4$ and certain derivatives of them, whose coincidence limit we compute in arbitrary dimensions. As ingredients, we need the coincidence limit of derivatives of the Synge world function $\sigma$ and the van Vleck--Morette determinant $\Delta$, which enter the transport equations~\eqref{eq:transport} (we recall that those are the same in the Riemannian and Lorentzian case). These can be computed recursively using equations~\eqref{eq:synge_identities}, taking the required number of covariant derivatives and then the coincidence limit. The results are dimension-independent and have been determined many times, see for example~\cite{VanVleck:1928zz,Morette:1951zz,Avramidi_2016}; for completeness, we here reproduce the required ones:
\begin{equations}[eq:synge_identities_long]
\lim_{x' \to x} \sigma &= 0 = \lim_{x' \to x} \nabla_\mu \sigma \eqend{,} \\
\lim_{x' \to x} \nabla_\mu \nabla_\nu \sigma &= g_{\mu\nu} \eqend{,} \\
\lim_{x' \to x} \nabla_{(\mu_1} \cdots \nabla_{\mu_k)} \sigma &= 0 \qquad (k \geq 3) \eqend{,} \\
\lim_{x' \to x} \ln \Delta &= 0 = \lim_{x' \to x} \nabla_\mu \ln \Delta \eqend{,} \\
\lim_{x' \to x} \nabla_\mu \nabla_\nu \ln \Delta &= \frac{1}{3} R_{\mu\nu} \eqend{,} \\
\lim_{x' \to x} \nabla_\mu \nabla_\nu \nabla_\rho \ln \Delta &= \frac{1}{2} \nabla_{(\mu} R_{\nu\rho)} \eqend{,} \\
\lim_{x' \to x} \nabla_{(\mu} \nabla_\nu \nabla_\rho \nabla_{\sigma)} \ln \Delta &= \frac{3}{5} \nabla_{(\mu} \nabla_\nu R_{\rho\sigma)} + \frac{2}{15} R_{\alpha(\mu\nu|\beta|} R^\alpha{}_{\rho\sigma)}{}^\beta \eqend{,} \\
\lim_{x' \to x} \nabla_{(\mu} \nabla_\nu \nabla_\rho \nabla_\sigma \nabla_{\tau)} \ln \Delta &= \frac{2}{3} \nabla_{(\mu} \nabla_\nu \nabla_\rho R_{\sigma\tau)} + \frac{2}{3} R_{\alpha(\mu\nu|\beta|} \nabla_\rho R^\alpha{}_{\sigma\tau)}{}^\beta \eqend{.}
\end{equations}
Non-symmetrized derivatives can be obtained from this by expanding the symmetrized expressions and commuting the covariant derivatives into the required order. Such computations are best done with the help of a computer algebra package such as xAct~\cite{xact,martingarcia2008,brizuelaetal2009,nutma2014}, see also~\cite{Ottewill:2009uj} for a specialist package. In particular, we have
\begin{equations}
\begin{split}
\lim_{x' \to x} \nabla_\mu \nabla_\nu \nabla^2 \ln \Delta &= \frac{4}{45} R_\mu{}^{\alpha\beta\rho} R_{\nu(\alpha\beta)\rho} + \frac{1}{15} R^{\alpha\beta} R_{\mu\alpha\nu\beta} - \frac{11}{45} R_\mu{}^\rho R_{\rho\nu} \\
&\quad+ \frac{1}{10} \nabla^2 R_{\mu\nu} + \frac{3}{10} \nabla_\mu \nabla_\nu R \eqend{,}
\end{split} \\
\begin{split}
\lim_{x' \to x} \nabla_\mu \nabla^2 \nabla^2 \ln \Delta &= \frac{8}{45} R^{\alpha\beta\gamma\delta} \nabla_\mu R_{\alpha(\beta\gamma)\delta} - \frac{2}{9} R^{\alpha\beta} \nabla_\beta R_{\mu\alpha} - \frac{11}{45} R^{\alpha\beta} \nabla_\mu R_{\alpha\beta} \\
&\quad- \frac{1}{9} R_{\mu\alpha} \nabla^\alpha R + \frac{2}{5} \nabla_\mu \nabla^2 R \eqend{,}
\end{split}
\end{equations}
where we used the second Bianchi identity to simplify the results.

The coefficients $A_k$ can then be recursively computed using the transport equations~\eqref{eq:transport} with the initial condition~\eqref{eq:transport_initial}, taking into account the explicit form of the differential operator~\eqref{eq:diffop}. We first rewrite the transport equations~\eqref{eq:transport} as
\begin{splitequation}
\label{eq:transport_explicit}
&\Big( \nabla^\nu \sigma \nabla_\nu + k \Big) A_k{}^A{}_{B'}(x,x') = E^A{}_C A_{k-1}{}^C{}_{B'}(x,x') \\
&\qquad+ \left( \nabla^2 + \nabla^\nu \ln \Delta \nabla_\nu + \frac{1}{4} \nabla^\nu \ln \Delta \nabla_\nu \ln \Delta + \frac{1}{2} \nabla^2 \ln \Delta \right) A_{k-1}{}^A{}_{B'}(x,x') \eqend{,}
\end{splitequation}
and using the results~\eqref{eq:synge_identities_long}, take the coincidence limit of this equation, which gives
\begin{equation}
\label{eq:transport_coincidence}
k A_k{}^A{}_{B'}(x,x) = E^A{}_C A_{k-1}{}^C{}_{B'}(x,x) + \lim_{x' \to x} \left( \nabla^2 + \frac{1}{6} R \right) A_{k-1}{}^A{}_{B'}(x,x') \eqend{.}
\end{equation}
Taking a covariant derivative of equation~\eqref{eq:transport_explicit} and then the coincidence limit, we obtain
\begin{splitequation}
\label{eq:transport_der_coincidence}
&(k+1) \lim_{x' \to x} \nabla_\mu A_k{}^A{}_{B'}(x,x') \\
&\quad= \lim_{x' \to x} \bigg[ \nabla_\mu \nabla^2 A_{k-1}{}^A{}_{B'}(x,x') + \nabla_\mu \left( E^A{}_C + \frac{1}{6} \1^A{}_C R \right) A_{k-1}{}^C{}_{B'}(x,x') \\
&\qquad\qquad+ \frac{1}{3} \left[ \1^A{}_C \left( R_{\mu\nu} + \frac{1}{2} R g_{\mu\nu} \right) + 3 E^A{}_C g_{\mu\nu} \right] \nabla^\nu A_{k-1}{}^C{}_{B'}(x,x') \bigg] \eqend{,}
\end{splitequation}
where we also used the contracted second Bianchi identity $\nabla^\nu R_{\mu\nu} = \frac{1}{2} \nabla_\mu R$. The pattern continues: to determine the coincidence limit of $\ell$ covariant derivatives acting on the coefficient $A_k$, we need to know the coincidence limit of up to $\ell+2$ covariant derivatives acting on the coefficient $A_{k-1}$. To determine the coincidence limit of $A_2$ and its first derivative, which are needed to check conservation of the stress tensor in four dimensions, we thus need to know the coincidence limit of up to three derivatives acting on $A_1$ and up to five derivatives acting on $A_0$.

Generalizing the coincidence limit~\eqref{eq:parallel_propagator} for the spin-1 parallel propagator, the spin-$s$ parallel propagator satisfies~\cite{Allen:1986qj,poissonpoundvega2011}
\begin{equation}
\mathcal{G}^A{}_{B'}(x,x) = \1^A{}_B \eqend{,} \quad \lim_{x' \to x} \nabla_{(\mu_1} \cdots \nabla_{\mu_k)} \mathcal{G}^A{}_{B'}(x,x') = 0 \eqend{.}
\end{equation}
To obtain the coincidence limit of unsymmetrized derivatives, we have to commute them into the right order, using that
\begin{equation}
\left( \nabla_\mu \nabla_\nu - \nabla_\nu \nabla_\mu \right) T^A = R_{\mu\nu}{}^{AB} T_B \eqend{,}
\end{equation}
where $R_{\mu\nu}{}^{AB}$ is the curvature tensor of the inner bundle described by the composite indices $A$. In particular, for scalars (spin 0) we have $R_{\mu\nu}{}^{\cdot\cdot} = 0$, while for vectors (spin 1) $R_{\mu\nu}{}^{\rho\sigma}$ is the usual Riemann curvature tensor, and for symmetric tensors (spin 2) with $A = \rho\sigma$ and $B = \alpha\beta$ we have
\begin{equation}
R_{\mu\nu}{}^{AB} = R_{\mu\nu}{}^{\rho(\alpha} g^{\beta)\sigma} + R_{\mu\nu}{}^{\sigma(\alpha} g^{\beta)\rho} \eqend{.}
\end{equation}
A straightforward computation~\cite{kazdan1981,schlicht1995} shows that the second Bianchi identity is fulfilled:
\begin{equation}
\label{eq:bianchi_identity}
\nabla_{[\rho} R_{\mu\nu]}{}^{AB} T_B = 2 \nabla_{[\rho} \nabla_\mu \nabla_{\nu]} T^A - R_{[\mu\nu}{}^{AB} \nabla_{\rho]} T_B = R_{[\mu\nu\rho]\sigma} \nabla^\sigma T^A = 0 \eqend{.}
\end{equation}
From the above equations, we thus obtain the coincidence limits of $A_1$ and its first derivative:
\begin{equations}[eq:coincidence_a1_der]
A_1{}^A{}_{B'}(x,x) &= E^A{}_B + \frac{1}{6} \1^A{}_B R \eqend{,} \\
\begin{split}
\lim_{x' \to x} \nabla_\mu A_1{}^A{}_{B'}(x,x') &= - \frac{1}{6} \nabla^\rho R_{\rho\mu}{}^A{}_B + \frac{1}{2} \nabla_\mu \left( E^A{}_B + \frac{1}{6} \1^A{}_B R \right) \\
&= \frac{1}{2} \nabla_\mu \lim_{x' \to x} A_1{}^A{}_{B'}(x,x') - \frac{1}{6} \nabla^\rho R_{\rho\mu}{}^A{}_B \eqend{.}
\end{split}
\end{equations}

To obtain the same for $A_2$, we need in addition the coincidence limit of two and three derivatives acting on $A_1$. Acting with two covariant derivatives on the transport equation~\eqref{eq:transport_explicit} for $k = 1$, symmetrising them when they act on $\sigma$ or $\Delta$, using the explicit result~\eqref{eq:transport_initial} for $A_0$ and taking the coincidence limit we obtain
\begin{splitequation}
&\lim_{x' \to x} \nabla_\mu \nabla_\nu A_1{}^A{}_{B'}(x,x') = \frac{1}{3} \nabla_\mu \nabla_\nu E^A{}_B + \frac{1}{3} R_{\mu\nu}{}^A{}_C E^C{}_B + \frac{1}{6} E^A{}_C R_{\mu\nu}{}^C{}_B \\
&\qquad+ \frac{1}{6} \nabla_{(\mu} \nabla^\rho R_{\nu)\rho}{}^A{}_B + \frac{1}{6} R_{(\mu}{}^{\rho A}{}_C R_{\nu)\rho}{}^C{}_B + \frac{1}{12} R R_{\mu\nu}{}^A{}_B \\
&\qquad+ \left( \frac{1}{90} R_\mu{}^{\alpha\beta\gamma} R_{\nu\alpha\beta\gamma} + \frac{1}{90} R^{\alpha\beta} R_{\mu\alpha\nu\beta} - \frac{1}{45} R_\mu{}^\alpha R_{\nu\alpha} + \frac{1}{60} \nabla^2 R_{\mu\nu} + \frac{1}{20} \nabla_\mu \nabla_\nu R \right) \1^A{}_B \eqend{.}
\end{splitequation}
To derive this result, we used the \textsc{xAct} tensor package~\cite{xact,martingarcia2008,brizuelaetal2009,nutma2014} and the \textsc{FieldsX} extension~\cite{fieldsx}, the contracted second Bianchi identity and the identity
\begin{equation}
R_\mu{}^{\alpha\beta\gamma} R_{\nu\beta\alpha\gamma} = \frac{1}{2} R_\mu{}^{\alpha\beta\gamma} R_{\nu\alpha\beta\gamma} \eqend{,}
\end{equation}
which follows from the first Bianchi identity. The same procedure for three derivatives gives
\begin{splitequation}
\lim_{x' \to x} \nabla_\mu \nabla^2 A_1{}^A{}_{B'}(x,x') &= \frac{3}{4} \nabla_\mu \lim_{x' \to x} \nabla^2 A_1{}^A{}_{B'}(x,x') + \frac{1}{4} \nabla^\rho \left( R_{\mu\rho}{}^A{}_C E^C{}_B + E^A{}_C R_{\mu\rho}{}^C{}_B \right) \\
&\quad- \frac{1}{6} E^A{}_C \nabla^\rho R_{\mu\rho}{}^C{}_B - \frac{1}{6} R_{\mu\rho} \nabla^\rho E^A{}_B - \frac{1}{10} R^{\rho\sigma A}{}_B \nabla_\rho \left( R_{\mu\sigma} + g_{\mu\sigma} R \right) \\
&\quad+ \frac{1}{20} \nabla^2 \nabla^\rho R_{\mu\rho}{}^A{}_B + \frac{1}{18} R \nabla^\rho R_{\mu\rho}{}^A{}_B - \frac{13}{180} R_{\mu\rho} \nabla_\sigma R^{\rho\sigma A}{}_B \\
&\quad+ \frac{1}{30} R^{\rho\sigma} \nabla_\rho R_{\mu\sigma}{}^A{}_B + \frac{1}{30} R_{\mu\alpha\beta\gamma} \nabla^\alpha R^{\beta\gamma A}{}_B - \frac{1}{36} R_{\mu\rho} \nabla^\rho R \, \1^A{}_B \\
&\quad+ \frac{1}{20} \left( R_{\mu\rho}{}^A{}_C \nabla_\sigma R^{\rho\sigma C}{}_B - \nabla_\sigma R^{\rho\sigma A}{}_C R_{\mu\rho}{}^C{}_B \right) \\
&\quad- \frac{1}{40} \left( R^{\alpha\beta A}{}_C \nabla_\mu R_{\alpha\beta}{}^C{}_B - \nabla_\mu R_{\alpha\beta}{}^A{}_C R^{\alpha\beta C}{}_B \right) \eqend{.}
\end{splitequation}
We now can insert these results into the equations~\eqref{eq:transport_coincidence} and~\eqref{eq:transport_der_coincidence} for $k = 2$, which together with the previous results~\eqref{eq:coincidence_a1_der} gives us
\begin{splitequation}
\label{eq:coincidence_a2}
A_2{}^A{}_{B'}(x,x) &= \frac{1}{6} \nabla^2 E^A{}_B + \frac{1}{2} \left( E^A{}_C + \frac{1}{6} \1^A{}_C R \right) \left( E^C{}_B + \frac{1}{6} \1^C{}_B R \right) + \frac{1}{12} R^{\alpha\beta A}{}_C R_{\alpha\beta}{}^C{}_B \\
&\quad+ \left( \frac{1}{180} R^{\alpha\beta\gamma\delta} R_{\alpha\beta\gamma\delta} - \frac{1}{180} R^{\alpha\beta} R_{\alpha\beta} + \frac{1}{30} \nabla^2 R \right) \1^A{}_B \raisetag{2em}
\end{splitequation}
and
\begin{splitequation}
\label{eq:coincidence_a2_der}
&\lim_{x' \to x} \nabla_\mu A_2{}^A{}_{B'}(x,x') = \frac{1}{2} \nabla_\mu \lim_{x' \to x} A_2{}^A{}_{B'}(x,x') + \frac{1}{12} \nabla^\rho \left( R_{\mu\rho}{}^A{}_C E^C{}_B + E^A{}_C R_{\mu\rho}{}^C{}_B \right) \\
&\qquad- \frac{1}{12} \left( E^A{}_C + \frac{1}{6} \1^A{}_C R \right) \overleftrightarrow{\nabla}\!_\mu \left( E^C{}_B + \frac{1}{6} \1^C{}_B R \right) + \frac{1}{60} \nabla^2 \nabla^\rho R_{\mu\rho}{}^A{}_B + \frac{1}{108} R \nabla^\rho R_{\mu\rho}{}^A{}_B \\
&\qquad+ \frac{1}{30} \nabla^\rho R R_{\mu\rho}{}^A{}_B - \frac{1}{180} R_{\mu\rho} \nabla_\sigma R^{\rho\sigma A}{}_B + \frac{1}{90} R^{\rho\sigma} \nabla_\rho R_{\mu\sigma}{}^A{}_B - \frac{1}{30} \nabla_\rho R_{\mu\sigma} R^{\rho\sigma A}{}_B \\
&\qquad+ \frac{1}{90} R_{\mu\alpha\beta\gamma} \nabla^\alpha R^{\beta\gamma A}{}_B - \frac{1}{120} R^{\alpha\beta A}{}_C \overleftrightarrow{\nabla}\!_\mu R_{\alpha\beta}{}^C{}_B \\
&\qquad+ \frac{1}{60} \left( R_{\mu\rho}{}^A{}_C \nabla_\sigma R^{\rho\sigma C}{}_B - \nabla_\sigma R^{\rho\sigma A}{}_C R_{\mu\rho}{}^C{}_B \right) \eqend{,} \raisetag{1.6em}
\end{splitequation}
where we defined
\begin{equation}
A \overleftrightarrow{\nabla}\!_\mu B \equiv A \nabla_\mu B - ( \nabla_\mu A ) B \eqend{.}
\end{equation}

In the following sections, we specialize these results to scalar, vector and tensor coefficients.

\subsection{Scalar coefficients}

For scalars, the inner bundle is one-dimensional and trivial, and consequently its curvature tensor vanishes, $R_{\mu\nu}{}^{\cdot \cdot} = 0$. However, we consider a non-minimal coupling to curvature such that the differential operator~\eqref{eq:diffop} reads $\nabla^2 - \xi R$, and thus $E = - \xi R$. Our results~\eqref{eq:coincidence_a2} and~\eqref{eq:coincidence_a2_der} then specialize to
\begin{equations}
&A_2(x,x) = \frac{1}{180} R^{\alpha\beta\gamma\delta} R_{\alpha\beta\gamma\delta} - \frac{1}{180} R^{\alpha\beta} R_{\alpha\beta} + \frac{1}{2} \left( \xi - \frac{1}{6} \right)^2 R^2 + \frac{1-5\xi}{30} \nabla^2 R \eqend{,} \\
&\lim_{x' \to x} \nabla_\mu A_2(x,x') = \frac{1}{2} \nabla_\mu \lim_{x' \to x} A_2(x,x') \eqend{,}
\end{equations}
and we recall that this result is valid in any dimension.

In four dimensions, we can express the Riemann tensor in terms of the Weyl tensor to write $A_2$ in terms of the square of the Weyl tensor
\begin{equation}
\label{eq:weyl_squared}
C^{\alpha\beta\gamma\delta} C_{\alpha\beta\gamma\delta} = R^{\alpha\beta\gamma\delta} R_{\alpha\beta\gamma\delta} - 2 R^{\alpha\beta} R_{\alpha\beta} + \frac{1}{3} R^2
\end{equation}
and the Euler density
\begin{equation}
\label{eq:euler_density}
\mathcal{E} \equiv R^{\alpha\beta\gamma\delta} R_{\alpha\beta\gamma\delta} - 4 R^{\alpha\beta} R_{\alpha\beta} + R^2 \eqend{.}
\end{equation}
This results in
\begin{equation}
A_2(x,x) = \frac{1}{120} C^{\alpha\beta\gamma\delta} C_{\alpha\beta\gamma\delta} - \frac{1}{360} \mathcal{E} + \frac{1}{2} \left( \xi - \frac{1}{6} \right)^2 R^2 + \frac{1-5\xi}{30} \nabla^2 R \eqend{,}
\end{equation}
and agrees with known results in the literature (see for example Ref.~\onlinecite{Groh:2011dw}).

\subsection{Vector coefficients}

For vectors, the inner bundle is the tangent bundle of the manifold, and thus the bundle curvature tensor is the usual Riemann curvature tensor. We also consider a non-minimal coupling to curvature of the form $E^{\mu\nu} = - \xi R^{\mu\nu} - \zeta g^{\mu\nu} R$; the vector propagator in Feynman gauge~\eqref{eq:vector_fg_eom} has $\xi = 1$ and $\zeta = 0$. Our results~\eqref{eq:coincidence_a2} and~\eqref{eq:coincidence_a2_der} then specialize to
\begin{splitequation}
A_2{}^\mu{}_{\nu'}(x,x) &= - \frac{1}{6} \nabla^2 \left( \xi R^\mu{}_\nu + \zeta \delta^\mu_\nu R \right) + \frac{1}{2} \left[ \xi R^{\mu\rho} + \left( \zeta - \frac{1}{6} \right) g^{\mu\rho} R \right] \left[ \xi R_{\rho\nu} + \left( \zeta - \frac{1}{6} \right) g_{\rho\nu} R \right] \\
&\quad- \frac{1}{12} R^{\mu\alpha\beta\gamma} R_{\nu\alpha\beta\gamma} + \left( \frac{1}{180} R^{\alpha\beta\gamma\delta} R_{\alpha\beta\gamma\delta} - \frac{1}{180} R^{\alpha\beta} R_{\alpha\beta} + \frac{1}{30} \nabla^2 R \right) \delta^\mu_\nu
\end{splitequation}
and
\begin{splitequation}
&\lim_{x' \to x} \nabla_\rho A_2{}^\mu{}_{\nu'}(x,x') = \frac{1}{2} \nabla_\rho \lim_{x' \to x} A_2{}^\mu{}_{\nu'}(x,x') + \frac{1-5\zeta}{30} R_{\rho\alpha}{}^\mu{}_\nu \nabla^\alpha R \\
&\qquad+ \frac{1}{60} R^{\mu\alpha}{}_{\rho\beta} \left( \nabla_\nu R_\alpha{}^\beta - \nabla_\alpha R_\nu{}^\beta - 5 \xi \nabla^\beta R_{\nu\alpha} \right) + \frac{1}{90} R^{\alpha\beta} \nabla_\alpha R_{\rho\beta}{}^\mu{}_\nu - \frac{1}{30} R^{\alpha\beta\mu}{}_\nu \nabla_\alpha R_{\rho\beta} \\
&\qquad- \frac{1}{60} R_{\nu\alpha\rho\beta} \left( \nabla^\mu R^{\alpha\beta} - \nabla^\alpha R^{\mu\beta} - 5 \xi \nabla^\beta R^{\mu\alpha} \right) - \frac{1}{180} R_{\rho\alpha} \left( \nabla_\nu R^{\mu\alpha} - \nabla^\mu R_\nu{}^\alpha \right) \\
&\qquad+ \frac{1}{90} R_{\rho\alpha\beta\gamma} \nabla^\alpha R^{\beta\gamma\mu}{}_\nu + \frac{1}{120} R^{\mu\alpha\beta\gamma} \overleftrightarrow{\nabla}\!_\rho R_{\nu\alpha\beta\gamma} - \frac{\xi}{12} R_{\nu\alpha} \left(\nabla^\alpha R^\mu{}_\rho - \nabla^\mu R^\alpha{}_\rho \right) \\
&\qquad+ \left( \frac{1}{60} \nabla^2 + \frac{1}{108} R - \frac{\zeta}{6} R \right) \left( \nabla_\nu R^\mu{}_\rho - \nabla^\mu R_{\nu\rho} \right) + \frac{\xi}{12} R^{\mu\alpha} \left( \nabla_\alpha R_{\nu\rho} - \nabla_\nu R_{\alpha\rho} \right) \\
&\qquad- \frac{1}{12} \left[ \xi R^{\mu\alpha} + \left( \zeta - \frac{1}{6} \right) g^{\mu\alpha} R \right] \overleftrightarrow{\nabla}\!_\rho \left[ \xi R_{\alpha\nu} + \left( \zeta - \frac{1}{6} \right) g_{\alpha\nu} R \right] \eqend{,} \raisetag{2em}
\end{splitequation}
where we used the contracted second Bianchi identity to simplify the result. We note that the coincidence limit of $A_2{}^\mu{}_{\nu'}(x,x')$ is symmetric in $\mu$ and $\nu$, which follows from the fact that the heat kernel coefficients themselves are symmetric~\cite{Moretti:1999ez,Moretti:1999fb,Kaminski:2019adk}: $A_2{}^\mu{}_{\nu'}(x,x') = A_2{}_{\nu'}{}^\mu(x',x)$. However, the difference $\lim_{x' \to x} \nabla_\rho A_2{}^\mu{}_{\nu'}(x,x') - \frac{1}{2} \nabla_\rho A_2{}^\mu{}_{\nu'}(x,x)$ is antisymmetric in $\mu$ and $\nu$, which follows from this symmetry together with the Synge rule~\cite{poissonpoundvega2011}
\begin{equation}
\label{eq:synge_rule}
\nabla_\mu \lim_{x' \to x} F(x,x') = \lim_{x' \to x} \left[ \nabla_\mu F(x,x') + g_\mu{}^{\mu'} \nabla_{\mu'} F(x,x') \right] \eqend{,}
\end{equation}
where the derivative $\nabla_{\mu'}$ acts on $x'$. This provides a useful check on the computation.

In four dimensions we can again simplify these expressions somewhat by using dimension-dependent identities~\cite{lovelock1970,edgarhoeglund2002}. These follow from the fact that antisymmetrising over five or more indices gives a vanishing result in four dimensions, and in general antisymmetrising over more than $n$ indices in $n$ dimensions. In particular, the Weyl tensor satisfies
\begin{equation}
\label{eq:weyl_4d_identity}
C_\mu{}^{\alpha\beta\gamma} C_{\nu\alpha\beta\gamma} = \frac{1}{4} g_{\mu\nu} C^{\alpha\beta\gamma\delta} C_{\alpha\beta\gamma\delta} \eqend{,}
\end{equation}
such that
\begin{splitequation}
A_2{}^\mu{}_{\nu'}(x,x) &= \frac{3 \xi^2 - 1}{6} \left( R^{\mu\alpha} R_{\nu\alpha} - \frac{1}{4} \delta^\mu_\nu R^{\alpha\beta} R_{\alpha\beta} \right) - \frac{1}{12} \left( 2 \xi - 1 - 12 \xi \zeta \right) \left( R R^\mu{}_\nu - \frac{1}{4} \delta^\mu_\nu R^2 \right) \\
&\quad- \frac{1}{6} \left[ \xi \nabla^2 \left( R^\mu{}_\nu - \frac{1}{4} \delta^\mu_\nu R \right) + R^{\alpha\beta} R^\mu{}_{\alpha\nu\beta} - \frac{1}{4} \delta^\mu_\nu R^{\alpha\beta} R_{\alpha\beta} \right] \\
&\quad+ \bigg[ \frac{15 \xi^2 - 8}{240} C^{\alpha\beta\gamma\delta} C_{\alpha\beta\gamma\delta} - \frac{45 \xi^2 - 13}{720} \mathcal{E} - \frac{5 \xi - 4 + 20 \zeta}{120} \nabla^2 R \\
&\qquad\qquad+ \frac{1 - 6 \xi (1-\xi) + 12 \zeta (3\xi-2) + 72 \zeta^2}{144} R^2 \bigg] \delta^\mu_\nu \raisetag{2em}
\end{splitequation}
in terms of the Weyl tensor and the Euler density~\eqref{eq:euler_density}. We have also separated the full trace, such that the first two lines are traceless.

The coefficients again agree with results in the literature (see for example Ref.~\onlinecite{Groh:2011dw}).

\subsection{Tensor coefficients}

For symmetric tensors with $A = \rho\sigma$ and $B = \alpha\beta$, the bundle curvature tensor reads
\begin{equation}
R_{\mu\nu}{}^{AB} = R_{\mu\nu}{}^{\rho(\alpha} g^{\beta)\sigma} + R_{\mu\nu}{}^{\sigma(\alpha} g^{\beta)\rho} \eqend{,}
\end{equation}
and we also have $\1^A{}_B = \delta^{(\rho}_\alpha \delta^{\sigma)}_\beta$. Again we consider a non-minimal coupling to curvature of the general form
\begin{splitequation}
E^{\mu\nu\rho\sigma} &= - 2 \xi R^{\mu(\rho\sigma)\nu} - \zeta \left( R^{\mu\nu} g^{\rho\sigma} + R^{\rho\sigma} g^{\mu\nu} \right) - 2 \chi \left( R^{\mu(\rho} g^{\sigma)\nu} + R^{\nu(\rho} g^{\sigma)\mu} \right) \\
&\quad- \left( \eta g^{\mu\nu} g^{\rho\sigma} + 2 \theta g^{\mu(\rho} g^{\sigma)\nu} \right) R \eqend{,}
\end{splitequation}
which is the most general form compatible with the symmetries $E^{\mu\nu\rho\sigma} = E^{(\mu\nu)(\rho\sigma)} = E^{(\rho\sigma)(\mu\nu)}$ that make the differential operator hermitean, and we recall that the trace-reversed and rescaled tensor propagator in Feynman gauge~\eqref{eq:tensor_tg_eom} has $\xi = 1$ and $\zeta = \chi = \eta = \theta = 0$. Our results~\eqref{eq:coincidence_a2} and~\eqref{eq:coincidence_a2_der} then specialize to
\begin{splitequation}
A_2{}^{\mu\nu}{}_{\rho'\sigma'}(x,x) &= \frac{1}{6} \nabla^2 E^{\mu\nu}{}_{\rho\sigma} + \frac{1}{2} \tilde{E}^{\mu\nu}{}_{\alpha\beta} \tilde{E}^{\alpha\beta}{}_{\rho\sigma} + \frac{1}{6} R^{\alpha\beta(\mu}{}_{(\rho} R^{\nu)}{}_{\sigma)\alpha\beta} - \frac{1}{6} \delta^{(\mu}_{(\rho} R^{\nu)\alpha\beta\gamma} R_{\sigma)\alpha\beta\gamma} \\
&\quad+ \left( \frac{1}{180} R^{\alpha\beta\gamma\delta} R_{\alpha\beta\gamma\delta} - \frac{1}{180} R^{\alpha\beta} R_{\alpha\beta} + \frac{1}{30} \nabla^2 R \right) \delta^{(\mu}_\rho \delta^{\nu)}_\sigma
\end{splitequation}
with
\begin{splitequation}
\tilde{E}^{\mu\nu}{}_{\rho\sigma} &\equiv E^{\mu\nu}{}_{\rho\sigma} + \frac{1}{6} \delta^\mu_{(\rho} \delta_{\sigma)}^\nu R \\
&= - 2 \xi R^\mu{}_{(\rho\sigma)}{}^\nu - \zeta \left( R^{\mu\nu} g_{\rho\sigma} + R_{\rho\sigma} g^{\mu\nu} \right) - 4 \chi R^{(\mu}{}_{(\rho} \delta_{\sigma)}^{\nu)} \\
&\quad- \left[ \eta g^{\mu\nu} g_{\rho\sigma} + \left( 2 \theta - \frac{1}{6} \right) \delta^\mu_{(\rho} \delta_{\sigma)}^\nu \right] R \eqend{,}
\end{splitequation}
and
\begin{splitequation}
\lim_{x' \to x} \nabla_\tau A_2{}^{\mu\nu}{}_{\rho'\sigma'}(x,x') &= \frac{1}{2} \nabla_\tau \lim_{x' \to x} A_2{}^{\mu\nu}{}_{\rho'\sigma'}(x,x') - \frac{1}{6} \nabla^\gamma \left[ R_{\tau\gamma\alpha}{}^{(\mu} E^{\nu)\alpha}{}_{\rho\sigma} - R_{\tau\gamma}{}^\alpha{}_{(\rho} E^{\mu\nu}{}_{\sigma)\alpha} \right] \\
&\quad- \frac{1}{12} \tilde{E}^{\mu\nu}{}_{\alpha\beta} \overleftrightarrow{\nabla}\!_\tau \tilde{E}^{\alpha\beta}{}_{\rho\sigma} + \frac{1}{15} R_{\tau\alpha}{}^{(\mu}{}_{(\rho} \delta_{\sigma)}^{\nu)} \nabla^\alpha R \\
&\quad+ \frac{1}{30} \nabla^2 \left[ \nabla_{(\rho} R_\tau{}^{(\mu} - \nabla^{(\mu} R_{\tau(\rho} \right] \delta_{\sigma)}^{\nu)} + \frac{1}{54} R \left[ \nabla_{(\rho} R_\tau{}^{(\mu} - \nabla^{(\mu} R_{\tau(\rho} \right] \delta_{\sigma)}^{\nu)} \\
&\quad- \frac{1}{90} R_{\tau\alpha} \left[ \nabla_{(\rho} R_\alpha{}^{(\mu} - \nabla^{(\mu} R_{\alpha(\rho} \right] \delta_{\sigma)}^{\nu)} - \frac{1}{15} R^{\alpha\beta(\mu}{}_{(\rho} \delta_{\sigma)}^{\nu)} \nabla_\alpha R_{\tau\beta} \\
&\quad+ \frac{1}{45} R^{\alpha\beta} \nabla_\alpha R_{\tau\beta}{}^{(\mu}{}_{(\rho} \delta_{\sigma)}^{\nu)} + \frac{1}{30} R_{\tau\alpha}{}^{(\mu}{}_\beta \delta_{(\rho}^{\nu)} \left( \nabla_{\sigma)} R_\alpha{}^\beta - \nabla^\beta R_{\sigma)\alpha} \right) \\
&\quad+ \frac{1}{45} R_{\tau\alpha\beta\gamma} \nabla^\alpha R^{\beta\gamma(\mu}{}_{(\rho} \delta_{\sigma)}^{\nu)} + \frac{1}{60} R^{\alpha\beta(\mu}{}_\gamma \delta^{\nu)}_{(\rho} \overleftrightarrow{\nabla}\!_{|\tau} R_{\alpha\beta|\sigma)}{}^\gamma \\
&\quad- \frac{1}{30} R_{\tau\alpha(\rho}{}^\beta \delta_{\sigma)}^{(\mu} \left( \nabla^{\nu)} R_{\alpha\beta} - \nabla_\beta R^{\nu)}{}_\alpha \right) \eqend{,} \raisetag{2em}
\end{splitequation}
where we used the contracted second Bianchi identity to simplify the result.

In four dimensions, we can again use dimension-dependent identities to simplify the result, in particular~\eqref{eq:weyl_4d_identity}. This gives
\begin{splitequation}
A_2{}^{\mu\nu}{}_{\rho'\sigma'}(x,x) &= \frac{1}{6} \nabla^2 \left[ E^{\mu\nu}{}_{\rho\sigma} - \frac{2}{3} \delta^{(\mu}_{(\rho} E^{\nu)\alpha}{}_{\sigma)\alpha} + \frac{1}{15} \delta^{(\mu}_\rho \delta^{\nu)}_\sigma E^{\alpha\beta}{}_{\alpha\beta} \right] \\
&\quad- \frac{\xi+2\zeta+6\chi}{9} \delta^{(\mu}_{(\rho} \nabla^2 \left[ R^{\nu)}{}_{\sigma)} - \frac{1}{4} \delta^{\nu)}_{\sigma)} R \right] \\
&\quad+ \frac{1}{2} \left[ \tilde{E}^{\mu\nu}{}_{\alpha\beta} \tilde{E}^{\alpha\beta}{}_{\rho\sigma} - \frac{2}{3} \delta^{(\mu}_{(\rho} \tilde{E}^{\nu)\alpha\beta\gamma} \tilde{E}_{\sigma)\alpha\beta\gamma} + \frac{1}{15} \delta^{(\mu}_\rho \delta^{\nu)}_\sigma \tilde{E}^{\alpha\beta}{}_{\gamma\delta} \tilde{E}^{\gamma\delta}{}_{\alpha\beta} \right] \\
&\quad+ \frac{1}{6} \left[ R^{\alpha\beta(\mu}{}_{(\rho} R^{\nu)}{}_{\sigma)\alpha\beta} + \frac{1}{3} \delta^{(\mu}_{(\rho} R^{\nu)\alpha\beta\gamma} R_{\sigma)\alpha\beta\gamma} - \frac{1}{30} \delta^{(\mu}_\rho \delta^{\nu)}_\sigma R^{\alpha\beta\gamma\delta} R_{\alpha\beta\gamma\delta} \right] \\
&\quad+ \frac{1}{3} \delta^{(\mu}_{(\rho} \left[ \tilde{E}^{\nu)\alpha\beta\gamma} \tilde{E}_{\sigma)\alpha\beta\gamma} - \frac{1}{4} \delta^{\nu)}_{\sigma)} \tilde{E}^{\alpha\beta\gamma\delta} \tilde{E}_{\alpha\beta\gamma\delta} \right] \\
&\quad- \frac{4}{9} \delta^{(\mu}_{(\rho} \left[ C^{\nu)}{}_{|\alpha|\sigma)\beta} R^{\alpha\beta} + \frac{1}{6} \left( R^{\nu)}{}_{\sigma)} - \frac{1}{4} \delta^{\nu)}_{\sigma)} R \right) R \right] \\
&\quad+ \Bigg[ \frac{36 \xi^2 - 11 + 24 \zeta (\zeta - \xi + 2 \chi) + 24 \chi (\xi + 3 \chi)}{120} C^{\alpha\beta\gamma\delta} C_{\alpha\beta\gamma\delta} \\
&\qquad\quad- \frac{54 \xi^2 - 17 + 72 \zeta (\zeta - \xi + 2 \chi) + 72 \chi (\xi + 3 \chi)}{360} \mathcal{E} + \frac{\alpha}{360} R^2 \\
&\qquad\quad+ \frac{2 - \xi - 2 \zeta - 4 \eta - 20 \theta - 10 \chi}{60} \nabla^2 R \Bigg] \delta^{(\mu}_\rho \delta^{\nu)}_\sigma \raisetag{2em}
\end{splitequation}
with the constant
\begin{splitequation}
\alpha &\equiv 18 \xi^2 - 6 \xi - 1 + 12 \zeta (-1 + 7 \zeta + 24 \eta + 12 \theta - 4 \xi + 8 \chi) + 24 \theta (- 5 + 30 \theta + 3 \xi + 30 \chi) \\
&\quad+ 12 \chi (- 5 + 4 \xi + 18 \chi) + 24 \eta (-1 + 12 \eta + 12 \theta - 3 \xi + 6 \chi) \eqend{,}
\end{splitequation}
where we also separated the (full) trace. For $\xi = 1$, the trace-reversed and rescaled tensor propagator in Feynman gauge, this further simplifies to
\begin{splitequation}
A_2{}^{\mu\nu}{}_{\rho'\sigma'}(x,x) &= - \frac{1}{3} \left( \nabla^2 + R \right) \left[ R^\mu{}_{(\rho\sigma)}{}^\nu - \frac{1}{3} \delta^{(\mu}_{(\rho} R^{\nu)}{}_{\sigma)} + \frac{1}{30} \delta^{(\mu}_\rho \delta^{\nu)}_\sigma R \right] \\
&\quad- \frac{1}{9} \delta^{(\mu}_{(\rho} \left( \nabla^2 - \frac{4}{3} R \right) \left[ R^{\nu)}{}_{\sigma)} - \frac{1}{4} \delta^{\nu)}_{\sigma)} R \right] + \frac{14}{9} \delta^{(\mu}_{(\rho} C^{\nu)}{}_{|\alpha|\sigma)\beta} R^{\alpha\beta} \\
&\quad+ 2 \left[ R^{\mu(\alpha\beta)\nu} R_{\rho\alpha\beta\sigma} - \frac{1}{2} \delta^{(\mu}_{(\rho} R^{\nu)\alpha\beta\gamma} R_{\sigma)\alpha\beta\gamma} + \frac{1}{20} \delta^{(\mu}_{(\rho} \delta^{\nu)}_{\sigma)} R^{\alpha\beta\gamma\delta} R_{\alpha\beta\gamma\delta} \right] \\
&\quad+ \frac{1}{6} \left[ R^{\alpha\beta(\mu}{}_{(\rho} R^{\nu)}{}_{\sigma)\alpha\beta} + \frac{1}{3} \delta^{(\mu}_{(\rho} R^{\nu)\alpha\beta\gamma} R_{\sigma)\alpha\beta\gamma} - \frac{1}{30} \delta^{(\mu}_\rho \delta^{\nu)}_\sigma R^{\alpha\beta\gamma\delta} R_{\alpha\beta\gamma\delta} \right] \\
&\quad+ \left( \frac{5}{24} C^{\alpha\beta\gamma\delta} C_{\alpha\beta\gamma\delta} - \frac{37}{360} \mathcal{E} + \frac{11}{360} R^2 + \frac{1}{60} \nabla^2 R \right) \delta^{(\mu}_\rho \delta^{\nu)}_\sigma \eqend{.} \raisetag{2em}
\end{splitequation}
Again this agrees with known results in the literature (see for example Ref.~\onlinecite{Groh:2011dw}). Recently, heat kernel coefficients for integer spin were computed for anti-de~Sitter spacetime~\cite{Gopakumar:2011qs,Bobev:2023dwx} and the sphere~\cite{Kluth:2019vkg} (as the Euclidean analogue of de~Sitter space), which also agree with our results if we specialize them to the appropriate background spacetime.

\subsection{Relations between coefficients}

From the trace and divergence identities for the vector and tensor propagators in Feynman-like gauges, we obtain relations between their corresponding heat kernel coefficients.

Inserting the heat kernel expansion~\eqref{eq:smalltimeL} in the divergence identity~\eqref{eq:vector_fg_div} for the vector propagator, we obtain
\begin{splitequation}
\label{eq:heatkernel_expansion_vector_feynman}
0 &= \nabla^\nu G^{\text{FG},M^2}_{\nu\rho'}(x,x') + \nabla_{\rho'} G_{M^2}(x,x') \\
&= - \mathi \lim_{\epsilon \to 0^+} \int_0^\infty \left[ \nabla^\nu K_{\nu\rho'}(x,x',\tau) + \nabla_{\rho'} K(x,x',\tau) \right] \mathe^{- \mathi ( M^2 - \mathi \epsilon ) \tau} \total \tau \\
&\sim \frac{\mathi}{2} \mathe^{- \mathi \frac{n}{4} \pi} \frac{\sqrt{\Delta}}{(4\pi)^\frac{n}{2}} \Big[ A_{0\nu\rho'}(x,x') \nabla^\nu \sigma + A_0(x,x') \nabla_{\rho'} \sigma 
\Big] \lim_{\epsilon \to 0^+} \int_0^\infty \tau^{-\frac{n}{2}-1} \mathe^{\frac{\mathi \sigma}{2 \tau}} \, \mathe^{- \mathi ( M^2 - \mathi \epsilon ) \tau} \total \tau \\
&\quad+ \mathe^{- \mathi \frac{n}{4} \pi} \frac{\sqrt{\Delta}}{(4\pi)^\frac{n}{2}} \lim_{\epsilon \to 0^+} \sum_{k=0}^\infty \mathi^k \bigg[ \nabla^\nu A_{k\nu\rho'}(x,x') + \nabla_{\rho'} A_k(x,x') + \frac{1}{2} A_{k\nu\rho'}(x,x') \nabla^\nu \ln \Delta \\
&\hspace{8em}+ \frac{1}{2} A_k(x,x') \nabla_{\rho'} \ln \Delta - \frac{1}{2} A_{(k+1)\nu\rho'}(x,x') \nabla^\nu \sigma - \frac{1}{2} A_{k+1}(x,x') \nabla_{\rho'} \sigma \bigg] \\
&\hspace{9em}\times \int_0^\infty \tau^{k-\frac{n}{2}} \mathe^{\frac{\mathi \sigma}{2 \tau}} \, \mathe^{- \mathi ( M^2 - \mathi \epsilon ) \tau} \total \tau \eqend{.}
\end{splitequation}
We now employ the small time regularisation, inserting a factor $\tau^\delta$ and then computing the integrals over $\tau$ using the integral representation
\begin{equation}
\bessel{K}{\nu}{z} = \frac{1}{2} \left( \frac{z}{2} \right)^\nu \int_0^\infty \exp\left( - t - \frac{z^2}{4 t} \right) t^{-\nu-1} \total t
\end{equation}
of the modified Bessel function given in Ref.~\onlinecite[Eq.~(10.32.10)]{dlmf}. This results in
\begin{equation}
\int_0^\infty \tau^{-\frac{n}{2}+k+\delta} \mathe^{\frac{\mathi \sigma}{2 \tau}} \, \mathe^{- \mathi ( M^2 - \mathi \epsilon ) \tau} \total \tau = 2 (2 M^2)^{\frac{n-2}{2}-k-\delta} ( - \mathi \Sigma )^{k-\frac{n-2}{2}+\delta} \bessel{K}{\frac{n-2}{2}-k-\delta}{\Sigma}
\end{equation}
with $\Sigma \equiv \sqrt{ 2 M^2 (\sigma+\mathi \epsilon) }$, and we see that the small-time asymptotic expansion of the heat kernel is a large-mass expansion if we hold $\Sigma$ fixed. It follows that each power of $\tau$ in the asymptotic expansion~\eqref{eq:heatkernel_expansion_vector_feynman} must vanish separately, such that we have
\begin{equation}
A_{0\nu\rho'}(x,x') \nabla^\nu \sigma = - A_0(x,x') \nabla_{\rho'} \sigma
\end{equation}
and
\begin{splitequation}
\label{eq:coefficient_vector_div}
\nabla^\nu A_{k\nu\rho'}(x,x') &= - \nabla_{\rho'} A_k(x,x') - \frac{1}{2} A_{k\nu\rho'}(x,x') \nabla^\nu \ln \Delta - \frac{1}{2} A_k(x,x') \nabla_{\rho'} \ln \Delta \\
&\quad+ \frac{1}{2} A_{(k+1)\nu\rho'}(x,x') \nabla^\nu \sigma + \frac{1}{2} A_{k+1}(x,x') \nabla_{\rho'} \sigma
\end{splitequation}
for $k \geq 0$.

Repeating this derivation with the identity~\eqref{eq:vector_fg_divs}, we also obtain
\begin{equation}
A_{0\nu\rho'}(x,x') \nabla^{\rho'} \sigma = - A_0(x,x') \nabla_\nu \sigma
\end{equation}
and
\begin{splitequation}
\label{eq:coefficient_vector_divs}
\nabla^{\rho'} A_{k\nu\rho'}(x,x') &= - \nabla_\nu A_k(x,x') - \frac{1}{2} A_{k\nu\rho'}(x,x') \nabla^{\rho'} \ln \Delta - \frac{1}{2} A_k(x,x') \nabla_\nu \ln \Delta \\
&\quad+ \frac{1}{2} A_{(k+1)\nu\rho'}(x,x') \nabla^{\rho'}\sigma + \frac{1}{2} A_{k+1}(x,x') \nabla_\nu \sigma
\end{splitequation}
for $k \geq 0$. Moreover, taking covariant derivatives of these relations and the coincidence limit, we obtain relations between the coincidence limits of the coefficients. In particular, without covariant derivatives we have
\begin{equations}
\lim_{x' \to x} \nabla^\nu A_{k\nu\rho'}(x,x') &= - \lim_{x' \to x} \nabla_{\rho'} A_k(x,x') \eqend{,} \\
\lim_{x' \to x} \nabla^{\rho'} A_{k\nu\rho'}(x,x') &= - \lim_{x' \to x} \nabla_\nu A_k(x,x') \eqend{,}
\end{equations}
and using one covariant derivative we obtain
\begin{equations}
A_{0\mu\rho'}(x,x) &= g_{\mu\rho} A_0(x,x) = g_{\mu\rho} \eqend{,} \\
\begin{split}
\lim_{x' \to x} \nabla_\mu \nabla^\nu A_{k\nu\rho'}(x,x') &= - \lim_{x' \to x} \nabla_\mu \nabla_{\rho'} A_k(x,x') - \frac{1}{6} R_\mu{}^\nu A_{k\nu\rho'}(x,x) + \frac{1}{6} R_{\rho\mu} A_k(x,x) \\
&\quad+ \frac{1}{2} A_{(k+1)\mu\rho'}(x,x) - \frac{1}{2} g_{\mu\rho} A_{k+1}(x,x) \eqend{,}
\end{split} \\
\begin{split}
\lim_{x' \to x} \nabla_\mu \nabla^{\rho'} A_{k\nu\rho'}(x,x') &= - \lim_{x' \to x} \nabla_\mu \nabla_\nu A_k(x,x') + \frac{1}{6} R_\mu{}^\rho A_{k\nu\rho'}(x,x) - \frac{1}{6} R_{\mu\nu} A_k(x,x) \\
&\quad- \frac{1}{2} A_{(k+1)\nu\mu'}(x,x) + \frac{1}{2} g_{\mu\nu} A_{k+1}(x,x) \eqend{.}
\end{split}
\end{equations}

In a completely analogous way, we derive from the trace and divergence identities~\eqref{eq:tensor_tg_trdiv} for the trace-reversed Feynman gauge tensor propagator relations for the tensor heat kernel coefficients. Inserting the heat kernel expansion~\eqref{eq:smalltimeL}, we obtain
\begin{equation}
\label{eq:heatkernel_expansion_tensor_trace_feynman}
\sum_{k=0}^\infty \mathi^k \int_0^\infty \left[ g^{\rho\sigma} A_{k\rho\sigma\alpha'\beta'}(x,x') - g_{\alpha'\beta'} A_k(x,x') \, \mathe^{\mathi \frac{4}{n-2} \Lambda \tau} \right] \mathe^{\frac{\mathi \sigma}{2 \tau}} \tau^{k-\frac{n}{2}} \mathe^{- \mathi ( m^2 - \mathi \epsilon ) \tau} \total \tau = 0
\end{equation}
and
\begin{splitequation}
\label{eq:heatkernel_expansion_tensor_div_feynman}
&\frac{\mathi}{2} \int_0^\infty \left[ A_{0\rho\sigma\alpha'\beta'}(x,x') \nabla^\rho \sigma + A_{0\sigma(\alpha'}(x,x') \nabla_{\beta')} \sigma \, \mathe^{\mathi \frac{4}{n-2} \Lambda \tau} \right] \mathe^{\frac{\mathi \sigma}{2 \tau}} \tau^{-1-\frac{n}{2}} \mathe^{- \mathi ( m^2 - \mathi \epsilon ) \tau} \total \tau \\
&\quad+ \sum_{k=0}^\infty \mathi^k \int_0^\infty \bigg[ \nabla^\rho A_{k\rho\sigma\alpha'\beta'}(x,x') + \nabla_{(\alpha'} A_{k|\sigma|\beta')}(x,x') \, \mathe^{\mathi \frac{4}{n-2} \Lambda \tau} + \frac{1}{2} A_{k\rho\sigma\alpha'\beta'}(x,x') \nabla^\rho \ln \Delta \\
&\qquad\quad+ \frac{1}{2} A_{k\sigma(\alpha'}(x,x') \nabla_{\beta')} \ln \Delta \, \mathe^{\mathi \frac{4}{n-2} \Lambda \tau} - \frac{1}{2} A_{(k+1)\rho\sigma\alpha'\beta'}(x,x') \nabla^\rho \sigma \\
&\qquad\quad- \frac{1}{2} A_{(k+1)\sigma(\alpha'}(x,x') \nabla_{\beta')} \sigma \, \mathe^{\mathi \frac{4}{n-2} \Lambda \tau} \bigg] \mathe^{\frac{\mathi \sigma}{2 \tau}} \tau^{k-\frac{n}{2}} \mathe^{- \mathi ( m^2 - \mathi \epsilon ) \tau} \total \tau = 0 \eqend{,}
\end{splitequation}
where we have reexpressed $\lambda$ in terms of $\Lambda = \frac{n-2}{4} m^2 \lambda$. Since the relations between coefficients are obtained from a large-mass expansion, it is important that we keep the cosmological constant $\Lambda$ and the mass $m$ separate to derive them, since otherwise we would obtain an inconsistent expansion and wrong relations. Afterwards, we can of course express $\Lambda$ again by the rescaled $\lambda$. As in the vector case, we then obtain the relations for the coefficients by comparing powers of $\tau$, for which we now have to expand the exponential in $\Lambda$ as well. For the trace~\eqref{eq:heatkernel_expansion_tensor_trace_feynman}, that means that the sought relations can be obtained from
\begin{splitequation}
0 &= \frac{\mathi^{-k}}{k!} \frac{\partial^k}{\partial \tau^k} \sum_{j=0}^\infty \mathi^j \tau^j \left[ g^{\rho\sigma} A_{j\rho\sigma\alpha'\beta'}(x,x') - g_{\alpha'\beta'} A_j(x,x') \, \mathe^{\mathi \frac{4}{n-2} \Lambda \tau} \right]_{\tau = 0} \\
&= g^{\rho\sigma} A_{k\rho\sigma\alpha'\beta'}(x,x') - g_{\alpha'\beta'} \sum_{j=0}^k \frac{1}{j!} \left( \frac{4}{n-2} \Lambda \right)^j A_{k-j}(x,x') \eqend{,}
\end{splitequation}
where we used the general Leibniz rule
\begin{equation}
\frac{\partial^n}{\partial x^n} [ f(x) g(x) ] = \sum_{k=0}^n \frac{n!}{k! (n-k)!} \frac{\partial^k}{\partial x^k} f(x) \frac{\partial^{n-k}}{\partial x^{n-k}} g(x) \eqend{.}
\end{equation}
We thus have the relations
\begin{equations}
g^{\rho\sigma} A_{k\rho\sigma\alpha'\beta'}(x,x') &= g_{\alpha'\beta'} \sum_{j=0}^k \frac{1}{j!} m^{2j} \lambda^j A_{k-j}(x,x') \eqend{,} \\
g^{\alpha'\beta'} A_{k\rho\sigma\alpha'\beta'}(x,x') &= g_{\rho\sigma} \sum_{j=0}^k \frac{1}{j!} m^{2j} \lambda^j A_{k-j}(x,x')
\end{equations}
following from the trace identity, and we see that the trace of each tensor coefficient is a polynomial in the cosmological constant. In the same way, from~\eqref{eq:heatkernel_expansion_tensor_div_feynman} (and the analogous equation with $x$ and $x'$ exchanged) we obtain
\begin{equations}
A_{0\rho\sigma\alpha'\beta'}(x,x') \nabla^\rho \sigma &= - A_{0\sigma(\alpha'}(x,x') \nabla_{\beta')} \sigma \eqend{,} \\
A_{0\rho\sigma\alpha'\beta'}(x,x') \nabla^{\alpha'} \sigma &= - A_{0(\rho|\beta'|}(x,x') \nabla_{\sigma)} \sigma \eqend{,} \\
\begin{split}
\nabla^\rho A_{k\rho\sigma\alpha'\beta'}(x,x') &= - \frac{1}{2} A_{k\rho\sigma\alpha'\beta'}(x,x') \nabla^\rho \ln \Delta + \frac{1}{2} A_{(k+1)\rho\sigma\alpha'\beta'}(x,x') \nabla^\rho \sigma \\
&\quad- \sum_{j=0}^k \frac{1}{j!} m^{2j} \lambda^j \bigg[ \nabla_{(\alpha'} A_{(k-j)|\sigma|\beta')}(x,x') + \frac{1}{2} A_{(k-j)\sigma(\alpha'}(x,x') \nabla_{\beta')} \ln \Delta \\
&\qquad- \frac{1}{2} A_{(k+1-j)\sigma(\alpha'}(x,x') \nabla_{\beta')} \sigma \bigg] + \frac{m^{2k+2} \lambda^{k+1}}{2 (k+1)!} A_{0\sigma(\alpha'}(x,x') \nabla_{\beta')} \sigma \eqend{,}
\end{split} \\
\begin{split}
\nabla^{\alpha'} A_{k\rho\sigma\alpha'\beta'}(x,x') &= - \frac{1}{2} A_{k\rho\sigma\alpha'\beta'}(x,x') \nabla^{\alpha'} \ln \Delta + \frac{1}{2} A_{(k+1)\rho\sigma\alpha'\beta'}(x,x') \nabla^{\alpha'} \sigma \\
&\quad- \sum_{j=0}^k \frac{1}{j!} m^{2j} \lambda^j \bigg[ \nabla_{(\rho} A_{(k-j)\sigma)\beta'}(x,x') + \frac{1}{2} A_{(k-j)(\rho|\beta'|}(x,x') \nabla_{\sigma)} \ln \Delta \\
&\qquad- \frac{1}{2} A_{(k+1-j)(\rho|\beta'|}(x,x') \nabla_{\sigma)} \sigma \bigg] + \frac{m^{2k+2} \lambda^{k+1}}{2 (k+1)!} A_{0(\rho|\beta'|}(x,x') \nabla_{\sigma)} \sigma \eqend{.}
\end{split}
\end{equations}

As in the vector case, we obtain relations for coincidence limits of the coefficients by taking covariant derivatives and then the coincidence limit of these relations. In particular, without covariant derivatives we have
\begin{equations}
\lim_{x' \to x} \nabla^\rho A_{k\rho\sigma\alpha'\beta'}(x,x') &= - \sum_{j=0}^{k+1} \frac{1}{j!} m^{2j} \lambda^j \lim_{x' \to x} \nabla_{(\alpha'} A_{(k-j)|\sigma|\beta')}(x,x') \eqend{,} \\
\lim_{x' \to x} \nabla^{\alpha'} A_{k\rho\sigma\alpha'\beta'}(x,x') &= - \sum_{j=0}^{k+1} \frac{1}{j!} m^{2j} \lambda^j \lim_{x' \to x} \nabla_{(\rho} A_{(k-j)\sigma)\beta'}(x,x') \eqend{,}
\end{equations}
and using one covariant derivative we obtain
\begin{equations}
A_{0\rho\sigma\alpha'\beta'}(x,x) &= g_{\rho(\alpha} g_{\beta)\sigma} \eqend{,} \\
\begin{split}
\lim_{x' \to x} \nabla_\mu \nabla^\rho A_{k\rho\sigma\alpha'\beta'}(x,x') &= \frac{1}{2} A_{(k+1)\mu\sigma\alpha'\beta'}(x,x) - \frac{1}{12} \lambda m^2 A_{k\mu\sigma\alpha'\beta'}(x,x) \\
&\quad- \sum_{j=0}^{k+1} \frac{1}{j!} m^{2j} \lambda^j \lim_{x' \to x} \nabla_\mu \nabla_{(\alpha'} A_{(k-j)|\sigma|\beta')}(x,x') \\
&\quad+ \frac{1}{12} \sum_{j=0}^{k+1} \frac{j-6}{j!} m^{2j} \lambda^j A_{(k+1-j)\sigma(\alpha'}(x,x) g_{\beta')\mu} \eqend{,}
\end{split} \\
\begin{split}
\lim_{x' \to x} \nabla_\mu \nabla^{\alpha'} A_{k\rho\sigma\alpha'\beta'}(x,x') &= - \frac{1}{2} A_{(k+1)\rho\sigma\mu\beta'}(x,x) + \frac{1}{12} \lambda m^2 A_{k\rho\sigma\mu\beta'}(x,x) \\
&\quad- \sum_{j=0}^{k+1} \frac{1}{j!} m^{2j} \lambda^j \nabla_\mu \nabla_{(\rho} A_{(k-j)\sigma)\beta'}(x,x') \\
&\quad- \frac{1}{12} \sum_{j=0}^{j=k+1} \frac{j-6}{j!} m^{2j} \lambda^j A_{(k+1-j)(\rho|\beta'|}(x,x') g_{\sigma)\mu} \eqend{,}
\end{split}
\end{equations}
where we also used the previous relations~\eqref{eq:coefficient_vector_div} and~\eqref{eq:coefficient_vector_divs} for the vector coefficients, and took into account that the background is on-shell, $R_{\mu\nu} = \frac{\lambda}{2} m^2 g_{\mu\nu}$.

\section{Heat kernel expansion of propagators}
\label{sec:propagatorheatkernel}

Having obtained all the ingredients, the heat kernel expansion can be used to derive the regularized propagators of a massive theory of gravity. In Sec.~\ref{sec:propagators} we obtained the propagators in the various fields sectors, which we now express using the heat kernel.

\subsection{The ghost sector}

The propagators of the ghost sector are given in Eq.~\eqref{eq:ghost_propagator}. Expressing them in terms of the heat kernel~\eqref{eq:propagator_from_heatkernel} and inserting the heat kernel expansion~\eqref{eq:smalltimeL}, we obtain for the time-ordered propagators the expansions
\begin{splitequation}
\label{eq:ghost_propagator_munu_heatkernel}
&G^\text{F}_{\mu\nu'}(x,x') = \mathi \lim_{\epsilon \to 0^+} \int_0^\infty \left[ K_{\mu\nu'}(x,x',\tau) + \frac{1 - \mathe^{\mathi (M_A^2-M_B^2) \tau}}{M_A^2} \nabla_\mu \nabla_{\nu'} K(x,x',\tau) \right] \, \mathe^{- \mathi ( M_A^2 - \mathi \epsilon ) \tau} \total \tau \\
&\quad\sim \frac{1}{4} \mathe^{- \mathi \frac{n}{4} \pi} \frac{\sqrt{\Delta}}{(4 \pi)^\frac{n}{2}} \nabla_\mu \sigma \nabla_{\nu'} \sigma A_0(x,x') \lim_{\epsilon \to 0^+} \int_0^\infty \frac{1 - \mathe^{\mathi (M_A^2-M_B^2) \tau}}{M_A^2} \mathe^{\frac{\mathi \sigma}{2 \tau}} \tau^{-2-\frac{n}{2}} \mathe^{- \mathi ( M_A^2 - \mathi \epsilon ) \tau} \total \tau \\
&\qquad+ \frac{1}{4} \mathe^{- \mathi \frac{n+2}{4} \pi} \frac{\sqrt{\Delta}}{(4 \pi)^\frac{n}{2}} \Big[ 4 \nabla_{(\mu} \sigma \nabla_{\nu')} A_0(x,x') + \left( 2 \nabla_\mu \nabla_{\nu'} \sigma + 2 \nabla_{(\mu} \ln \Delta \nabla_{\nu')} \sigma \right) A_0(x,x') \\
&\qquad\qquad\quad- \nabla_\mu \sigma \nabla_{\nu'} \sigma A_1(x,x') \Big] \lim_{\epsilon \to 0^+} \int_0^\infty \frac{1 - \mathe^{\mathi (M_A^2-M_B^2) \tau}}{M_A^2} \mathe^{\frac{\mathi \sigma}{2 \tau}} \tau^{-1-\frac{n}{2}} \mathe^{- \mathi ( M_A^2 - \mathi \epsilon ) \tau} \total \tau \\
&\qquad+ \mathe^{- \mathi \frac{n-4}{4} \pi} \frac{\sqrt{\Delta}}{(4 \pi)^\frac{n}{2}} \lim_{\epsilon \to 0^+} \sum_{k=0}^\infty \mathi^k \int_0^\infty \bigg[ A_{k\mu\nu'}(x,x') + \frac{1 - \mathe^{\mathi (M_A^2-M_B^2) \tau}}{M_A^2} \bigg[ \nabla_\mu \nabla_{\nu'} A_k(x,x') \\
&\qquad\qquad\quad+ \frac{1}{2} \left( 2 \nabla_{(\mu} \ln \Delta \nabla_{\nu')} + \nabla_\mu \nabla_{\nu'} \ln \Delta + \frac{1}{2} \nabla_\mu \ln \Delta \nabla_{\nu'} \ln \Delta \right) A_k(x,x') \\
&\qquad\qquad\quad- \frac{1}{2} \left( 2 \nabla_{(\mu} \sigma \nabla_{\nu')} + \nabla_\mu \nabla_{\nu'} \sigma + \nabla_{(\mu} \ln \Delta \nabla_{\nu')} \sigma \right) A_{k+1}(x,x') \\
&\qquad\qquad\quad+ \frac{1}{4} \nabla_\mu \sigma \nabla_{\nu'} \sigma A_{k+2}(x,x') \bigg] \bigg] \mathe^{\frac{\mathi \sigma}{2 \tau}} \tau^{k-\frac{n}{2}} \mathe^{- \mathi ( M_A^2 - \mathi \epsilon ) \tau} \total \tau \eqend{,}
\end{splitequation}
\begin{splitequation}
\label{eq:ghost_propagator_mu0_heatkernel}
&G^\text{F}_{\mu\circ}(x,x') = - \mathi \lim_{\epsilon \to 0^+} \int_0^\infty \frac{1 - \mathe^{- \mathi M_B^2 \tau}}{m} \nabla_\mu K(x,x',\tau) \, \mathe^{- \epsilon \tau} \total \tau \\
&\quad\sim \frac{1}{2} \mathe^{- \mathi \frac{n-2}{4} \pi} \frac{\sqrt{\Delta}}{(4 \pi)^\frac{n}{2}} \nabla_\mu \sigma A_0(x,x') \lim_{\epsilon \to 0^+} \int_0^\infty \frac{1 - \mathe^{- \mathi M_B^2 \tau}}{m} \mathe^{\frac{\mathi \sigma}{2 \tau}} \tau^{-1-\frac{n}{2}} \mathe^{- \epsilon \tau} \total \tau \\
&\qquad+ \mathe^{- \mathi \frac{n}{4} \pi} \frac{\sqrt{\Delta}}{(4 \pi)^\frac{n}{2}} \lim_{\epsilon \to 0^+} \sum_{k=0}^\infty \mathi^k \left[ \nabla_\mu A_k(x,x') + \frac{1}{2} \nabla_\mu \ln \Delta A_k(x,x') - \frac{1}{2} \nabla_\mu \sigma A_{k+1}(x,x') \right] \\
&\qquad\quad\times \int_0^\infty \frac{1 - \mathe^{- \mathi M_B^2 \tau}}{m} \mathe^{\frac{\mathi \sigma}{2 \tau}} \tau^{k-\frac{n}{2}} \mathe^{- \epsilon \tau} \total \tau \eqend{,}
\end{splitequation}
and
\begin{splitequation}
\label{eq:ghost_propagator_00_heatkernel}
G^\text{F}_{\circ\circ}(x,x') &= \mathi \lim_{\epsilon \to 0^+} \int_0^\infty K(x,x',\tau) \, \mathe^{- \epsilon \tau} \total \tau \\
&\sim \mathe^{- \mathi \frac{n-4}{4} \pi} \frac{\sqrt{\Delta}}{(4 \pi)^\frac{n}{2}} \lim_{\epsilon \to 0^+} \sum_{k=0}^\infty \mathi^k A_k(x,x') \, \int_0^\infty \mathe^{\frac{\mathi \sigma}{2 \tau}} \tau^{k-\frac{n}{2}} \mathe^{- \epsilon \tau} \total \tau \eqend{,}
\end{splitequation}
where we recall the masses~\eqref{eq:field_masses}
\begin{equation}
M_A^2 = (\xi-\lambda) m^2 \eqend{,} \quad M_B^2 = \frac{\zeta}{2\zeta-1} (\xi-\lambda) m^2 = \frac{\zeta}{2\zeta-1} M_A^2 \eqend{,} \quad M_C^2 = (1-\lambda) m^2 \eqend{.}
\end{equation}

It is now straightforward to take the massless limit. Since we have seen that the limit $m^2 \to 0$ with $\Lambda$ fixed is divergent, while we obtain a sensible limit keeping $\lambda = \frac{4}{n-2} \frac{\Lambda}{m^2}$ fixed, we only give the expressions for the latter one. Now $G^\text{F}_{\circ\circ}(x,x')$~\eqref{eq:ghost_propagator_00} is already massless, and the limit $m \to 0$ of $G^\text{F}_{\mu\circ}(x,x')$ vanishes~\eqref{eq:ghost_propagator_massless_mu0}, while for the diffeomorphism ghost propagator we obtain
\begin{splitequation}
\label{eq:ghost_propagator_massless_munu_heatkernel}
\lim_{m \to 0} G^\text{F}_{\mu\nu'}(x,x') &\sim \frac{1}{4} \mathe^{- \mathi \frac{n+2}{4} \pi} \frac{\zeta-1}{2\zeta-1} \frac{\sqrt{\Delta}}{(4 \pi)^\frac{n}{2}} \nabla_\mu \sigma \nabla_{\nu'} \sigma A_0(x,x') \lim_{\epsilon \to 0^+} \int_0^\infty \mathe^{\frac{\mathi \sigma}{2 \tau}} \tau^{-1-\frac{n}{2}} \mathe^{- \epsilon \tau} \total \tau \\
&\quad+ \frac{1}{4} \mathe^{- \mathi \frac{n-4}{4} \pi} \frac{\zeta-1}{2\zeta-1} \frac{\sqrt{\Delta}}{(4 \pi)^\frac{n}{2}} \Big[ 4 \nabla_{(\mu} \sigma \nabla_{\nu')} A_0(x,x') - \nabla_\mu \sigma \nabla_{\nu'} \sigma A_1(x,x') \\
&\qquad\qquad+ \left( 2 \nabla_\mu \nabla_{\nu'} \sigma + 2 \nabla_{(\mu} \ln \Delta \nabla_{\nu')} \sigma \right) A_0(x,x') \Big] \lim_{\epsilon \to 0^+} \int_0^\infty \mathe^{\frac{\mathi \sigma}{2 \tau}} \tau^{-\frac{n}{2}} \mathe^{- \epsilon \tau} \total \tau \\
&\quad+ \mathe^{- \mathi \frac{n-4}{4} \pi} \frac{\sqrt{\Delta}}{(4 \pi)^\frac{n}{2}} A_{0\mu\nu'}(x,x') \lim_{\epsilon \to 0^+} \int_0^\infty \mathe^{\frac{\mathi \sigma}{2 \tau}} \tau^{-\frac{n}{2}} \mathe^{- \epsilon \tau} \total \tau \\
&\quad+ \mathe^{- \mathi \frac{n-4}{4} \pi} \frac{\sqrt{\Delta}}{(4 \pi)^\frac{n}{2}} \lim_{\epsilon \to 0^+} \sum_{k=1}^\infty \mathi^k \int_0^\infty \bigg[ A_{k\mu\nu'}(x,x') - \frac{\zeta-1}{2\zeta-1} \bigg[ \nabla_\mu \nabla_{\nu'} A_{k-1}(x,x') \\
&\qquad\qquad+ \frac{1}{2} \left( 2 \nabla_{(\mu} \ln \Delta \nabla_{\nu')} + \nabla_\mu \nabla_{\nu'} \ln \Delta + \frac{1}{2} \nabla_\mu \ln \Delta \nabla_{\nu'} \ln \Delta \right) A_{k-1}(x,x') \\
&\qquad\qquad- \frac{1}{2} \left( 2 \nabla_{(\mu} \sigma \nabla_{\nu')} + \nabla_\mu \nabla_{\nu'} \sigma + \nabla_{(\mu} \ln \Delta \nabla_{\nu')} \sigma \right) A_k(x,x') \\
&\qquad\qquad+ \frac{1}{4} \nabla_\mu \sigma \nabla_{\nu'} \sigma A_{k+1}(x,x') \bigg] \bigg] \mathe^{\frac{\mathi \sigma}{2 \tau}} \tau^{k-\frac{n}{2}} \mathe^{- \epsilon \tau} \total \tau \eqend{,}
\end{splitequation}
where we used the limit
\begin{equation}
\lim_{m \to 0} \frac{1 - \mathe^{\mathi (M_A^2-M_B^2) \tau}}{M_A^2} = - \mathi \tau \frac{\zeta-1}{2\zeta-1} \eqend{.}
\end{equation}
Compared to~\eqref{eq:ghost_propagator_munu_heatkernel}, we have shifted the summation index in the last term to absorb the extra factor of $\tau$.

\subsection{The field sector}

The propagators of the field sector are given in Eq.~\eqref{eq:field_propagator}. As for the ghost sector, we express them in terms of the heat kernel~\eqref{eq:propagator_from_heatkernel} and insert the heat kernel expansion~\eqref{eq:smalltimeL}. However, there are two differences: first, that the propagators in the field sector also contain mass derivatives which are defined in Eqs.~\eqref{eq:scalar_massderivative} and \eqref{eq:vector_massderivative}. For them, we thus obtain instead of Eq.~\eqref{eq:feynman_from_heatkernel} the heat kernel representation
\begin{equation}
\hat{G}^\text{F}_{\nu_1 \cdots \nu_s \rho_1' \cdots \rho_s'}(x,x') \equiv - \lim_{\epsilon \to 0^+} \int_0^\infty \tau K_{\nu_1 \cdots \nu_s \rho_1' \cdots \rho_s'}(x,x',\tau) \, \mathe^{- \mathi ( m^2 - \mathi \epsilon ) \tau} \total \tau \eqend{,}
\end{equation}
which is less singular for small $\tau$, consistent with the fact that the most singular term of the propagator is mass independent. Second, the tensor propagator in Feynman gauge $G^{\text{FG},m^2}$ does not have a straightforward heat kernel expansion, but the trace-reversed and rescaled tensor propagator $G^{\text{TG},m^2}$~\eqref{eq:tensor_fg_tg_relation} has. For the Feynman gauge tensor propagator, we obtain from this
\begin{equation}
G^{\text{FG},m^2}_{\mu\nu\rho'\sigma'}(x,x') = - \mathi \lim_{\epsilon \to 0^+} \int_0^\infty \left[ 2 K_{\mu\nu\rho'\sigma'} - \frac{2}{n-2} g_{\mu\nu} g^{\alpha\beta} K_{\alpha\beta\rho'\sigma'} \right](x,x',\tau) \, \mathe^{- \mathi ( m^2 - \mathi \epsilon ) \tau} \total \tau \eqend{.}
\end{equation}

Taking these differences into account, we obtain for the time-ordered propagators the expansions
\begin{splitequation*}
&G_{\mu\nu\rho'\sigma'}(x,x') = - \mathi \lim_{\epsilon \to 0^+} \int_0^\infty \left[ 2 K_{\mu\nu\rho'\sigma'} - \frac{2}{n-2} g_{\mu\nu} g^{\alpha\beta} K_{\alpha\beta\rho'\sigma'} \right](x,x',\tau) \, \mathe^{- \mathi ( m^2 - \mathi \epsilon ) \tau} \total \tau \\
&\qquad- \frac{4 \mathi}{m^2} \lim_{\epsilon \to 0^+} \int_0^\infty \nabla_{(\mu} \nabla_{(\rho'} K_{\nu)\sigma'}(x,x',\tau) \left( \mathe^{- \mathi M_C^2 \tau} - \mathe^{- \mathi M_A^2 \tau} \right) \, \mathe^{- \epsilon \tau} \total \tau \\
&\qquad+ \frac{4 \mathi}{2 (n-1) - (n-2) \lambda} \lim_{\epsilon \to 0^+} \int_0^\infty \left( g_{\mu\nu} \nabla_{\rho'} \nabla_{\sigma'} + g_{\rho'\sigma'} \nabla_\mu \nabla_\nu \right) K(x,x',\tau) \frac{1 - \mathe^{- \mathi M_C^2 \tau}}{m^2} \mathe^{- \epsilon \tau} \total \tau \\
&\qquad- \frac{4 \mathi}{m^2} \lim_{\epsilon \to 0^+} \int_0^\infty \nabla_\mu \nabla_\nu \nabla_{\rho'} \nabla_{\sigma'} K(x,x',\tau) \bigg[ \frac{2 \mu}{m^2} \left[ 1 - \mathe^{- \mathi M_B^2 \tau} \right] - \frac{1}{M_A^2} \left[ \mathe^{- \mathi M_A^2 \tau} - \mathe^{- \mathi M_B^2 \tau} \right] \\
&\qquad\qquad+ \frac{n-2}{2 (n-1) - (n-2) \lambda} \frac{1}{m^2} \left[ \mathe^{- \mathi M_C^2 \tau} - 1 \right] - \mathi \tau \left( \rho + M_B^2 \frac{\mu}{m^2} \mathe^{- \mathi M_B^2 \tau} \right) \bigg] \mathe^{- \epsilon \tau} \total \tau \\
&\qquad- \frac{4 \mathi}{(n-2) [ 2 (n-1) - (n-2) \lambda ]} g_{\mu\nu} g_{\rho'\sigma'} \lim_{\epsilon \to 0^+} \int_0^\infty K(x,x',\tau) \, \mathe^{- \mathi ( M_C^2 - \mathi \epsilon ) \tau} \total \tau \\
&\quad\sim \mathe^{- \mathi \frac{n}{4} \pi} \frac{\sqrt{\Delta}}{(4 \pi)^\frac{n}{2}} \sum_{k=0}^\infty \mathi^k \left[ 2 A_{k\mu\nu\rho'\sigma'}(x,x') - \frac{2}{n-2} g_{\mu\nu} g_{\rho'\sigma'} \sum_{j=0}^k \frac{1}{j!} m^{2j} \lambda^j A_{k-j}(x,x') \right] \\
&\qquad\quad\times \lim_{\epsilon \to 0^+} \int_0^\infty \mathe^{\frac{\mathi \sigma}{2 \tau}} \mathe^{- \mathi m^2 \tau} \tau^{k - \frac{n}{2}} \mathe^{- \epsilon \tau} \total \tau \\
&\qquad+ \frac{\sqrt{\Delta}}{(4 \pi)^\frac{n}{2}} \mathe^{- \mathi \frac{n-4}{4} \pi} \nabla_{(\rho'} \sigma \nabla_{|(\mu} \sigma A_{0\nu)|\sigma')}(x,x') \lim_{\epsilon \to 0^+} \int_0^\infty \mathe^{\frac{\mathi \sigma}{2 \tau}} \frac{\mathe^{- \mathi M_C^2 \tau} - \mathe^{- \mathi M_A^2 \tau}}{m^2} \tau^{-2 - \frac{n}{2}} \mathe^{- \epsilon \tau} \total \tau \\
&\qquad+ \frac{\sqrt{\Delta}}{(4 \pi)^\frac{n}{2}} \mathe^{- \mathi \frac{n-2}{4} \pi} \bigg[ \bigg( \nabla_{(\rho'} \ln \Delta \nabla_{|(\mu} \sigma + \nabla_{(\rho'} \sigma \nabla_{|(\mu} \ln \Delta + 2 \nabla_{(\rho'} \nabla_{|(\mu} \sigma + 2 \nabla_{(\rho'} \sigma \nabla_{|(\mu} + \nabla_{(\mu} \sigma \nabla_{(\rho'} \bigg) \\
&\qquad\qquad\times A_{0\nu)|\sigma')}(x,x') - \nabla_{(\rho'} \sigma \nabla_{|(\mu} \sigma A_{1\nu)|\sigma')}(x,x') \bigg] \lim_{\epsilon \to 0^+} \int_0^\infty \mathe^{\frac{\mathi \sigma}{2 \tau}} \frac{\mathe^{- \mathi M_C^2 \tau} - \mathe^{- \mathi M_A^2 \tau}}{m^2} \tau^{-1 - \frac{n}{2}} \mathe^{- \epsilon \tau} \total \tau \\
\end{splitequation*}
\begin{splitequation*}
&\qquad+ \frac{\sqrt{\Delta}}{(4 \pi)^\frac{n}{2}} \mathe^{- \mathi \frac{n}{4} \pi} \sum_{k=0}^\infty \mathi^k \bigg[ \bigg( 4 \nabla_{(\rho'} \nabla_{|(\mu} + 2 \nabla_{\rho'} \ln \Delta \nabla_{|(\mu} + 2 \nabla_{(\mu} \ln \Delta \nabla_{|(\rho'} + 2 \nabla_{\rho'} \nabla_{|(\mu} \ln \Delta \\
&\qquad\qquad+ \nabla_{(\rho'} \ln \Delta \nabla_{|(\mu} \ln \Delta \bigg) A_{k|\nu)|\sigma')}(x,x') - \bigg( 2 \nabla_{(\rho'} \sigma \nabla_{(\mu} + 2 \nabla_{(\mu} \sigma \nabla_{|(\rho'} + 2 \nabla_{(\rho'} \nabla_{|(\mu} \sigma \\
&\qquad\qquad+ \nabla_{(\rho'} \ln \Delta \nabla_{|(\mu} \sigma + \nabla_{(\rho'} \sigma \nabla_{|(\mu} \ln \Delta \bigg) A_{k+1|\nu)|\sigma')}(x,x') + \nabla_{(\rho'} \sigma \nabla_{|(\mu} \sigma A_{k+2|\nu)|\sigma')}(x,x') \bigg] \\
&\qquad\quad\times \lim_{\epsilon \to 0^+} \int_0^\infty \mathe^{\frac{\mathi \sigma}{2 \tau}} \frac{\mathe^{- \mathi M_C^2 \tau} - \mathe^{- \mathi M_A^2 \tau}}{m^2} \tau^{k - \frac{n}{2}} \mathe^{- \epsilon \tau} \total \tau \\
&\qquad+ \frac{1}{2 (n-1) - (n-2) \lambda} \frac{\sqrt{\Delta}}{(4 \pi)^\frac{n}{2}} \mathe^{- \mathi \frac{n}{4} \pi} \left( g_{\rho'\sigma'} \nabla_\nu \sigma \nabla_\mu \sigma + g_{\mu\nu} \nabla_{\rho'} \sigma \nabla_{\sigma'} \sigma \right) A_0(x,x') \\
&\qquad\quad\times \lim_{\epsilon \to 0^+} \int_0^\infty \frac{1 - \mathe^{- \mathi M_C^2 \tau}}{m^2} \mathe^{\frac{\mathi \sigma}{2 \tau}} \tau^{-2-\frac{n}{2}} \mathe^{- \epsilon \tau} \total \tau \\
&\qquad+ \frac{1}{2 (n-1) - (n-2) \lambda} \frac{\sqrt{\Delta}}{(4 \pi)^\frac{n}{2}} \mathe^{- \mathi \frac{n+2}{4} \pi} \bigg[ - \left( g_{\rho'\sigma'} \nabla_\nu \sigma \nabla_\mu \sigma + g_{\mu\nu} \nabla_{\rho'} \sigma \nabla_{\sigma'} \sigma \right) A_1(x,x') \\
&\qquad\qquad+ 2 g_{\rho'\sigma'} \left( 2 \nabla_{(\mu} \sigma \nabla_{\nu)} + \nabla_\mu \nabla_\nu \sigma + \nabla_{(\mu} \ln \Delta \nabla_{\nu)} \sigma \right) A_0(x,x') \\
&\qquad\qquad+ 2 g_{\mu\nu} \left( 2 \nabla_{(\rho'} \sigma \nabla_{\sigma')} + \nabla_{\rho'} \nabla_{\sigma'} \sigma + \nabla_{(\rho'} \ln \Delta \nabla_{\sigma')} \sigma \right) A_0(x,x') \bigg] \\
&\qquad\quad\times \lim_{\epsilon \to 0^+} \int_0^\infty \frac{1 - \mathe^{- \mathi M_C^2 \tau}}{m^2} \mathe^{\frac{\mathi \sigma}{2 \tau}} \tau^{-1-\frac{n}{2}} \mathe^{- \epsilon \tau} \total \tau \\
&\qquad+ \frac{1}{2 (n-1) - (n-2) \lambda} \frac{\sqrt{\Delta}}{(4 \pi)^\frac{n}{2}} \mathe^{- \mathi \frac{n}{4} \pi} \sum_{k=0}^\infty \mathi^k \bigg[ - \left( g_{\rho'\sigma'} \nabla_\nu \sigma \nabla_\mu \sigma + g_{\mu\nu} \nabla_{\rho'} \sigma \nabla_{\sigma'} \sigma \right) A_{k+2}(x,x') \\
&\qquad\qquad+ 2 g_{\rho'\sigma'} \left( 2 \nabla_{(\mu} \sigma \nabla_{\nu)} + \nabla_\mu \nabla_\nu \sigma + \nabla_{(\mu} \ln \Delta \nabla_{\nu)} \sigma \right) A_{k+1}(x,x') \\
&\qquad\qquad+ 2 g_{\mu\nu} \left( 2 \nabla_{(\rho'} \sigma \nabla_{\sigma')} + \nabla_{\rho'} \nabla_{\sigma'} \sigma + \nabla_{(\rho'} \ln \Delta \nabla_{\sigma')} \sigma \right) A_{k+1}(x,x') \\
&\qquad\qquad- g_{\rho'\sigma'} \left( 4 \nabla_\mu \nabla_\nu + 4 \nabla_{(\mu} \ln \Delta \nabla_{\nu)} + 2 \nabla_\mu \nabla_\nu \ln \Delta + \nabla_\mu \ln \Delta \nabla_\nu \ln \Delta \right) A_k(x,x') \\
&\qquad\qquad- g_{\mu\nu} \left( 4 \nabla_{\rho'} \nabla_{\sigma'} + 4 \nabla_{(\rho'} \ln \Delta \nabla_{\sigma')} + 2 \nabla_{\rho'} \nabla_{\sigma'} \ln \Delta + \nabla_{\rho'} \ln \Delta \nabla_{\sigma'} \right) A_k(x,x') \bigg] \\
&\qquad\quad\times \lim_{\epsilon \to 0^+} \int_0^\infty \frac{1 - \mathe^{- \mathi M_C^2 \tau}}{m^2} \mathe^{\frac{\mathi \sigma}{2 \tau}} \tau^{k-\frac{n}{2}} \mathe^{- \epsilon \tau} \total \tau \\
&\qquad+ \mathe^{- \mathi \frac{n}{4} \pi} \frac{\sqrt{\Delta}}{(4 \pi)^\frac{n}{2}} \frac{4}{(n-2) [ 2 (n-1) - (n-2) \lambda ]} g_{\mu\nu} g_{\rho'\sigma'} \sum_{k=0}^\infty \mathi^k A_k(x,x') \\
&\qquad\quad\times \lim_{\epsilon \to 0^+} \int_0^\infty \mathe^{\frac{\mathi \sigma}{2 \tau}} \mathe^{- \mathi M_C^2 \tau} \tau^{k - \frac{n}{2}} \mathe^{- \epsilon \tau} \total \tau \\
&\qquad+ \mathe^{- \mathi \frac{n}{4} \pi} \frac{\sqrt{\Delta}}{(4 \pi)^\frac{n}{2}} \frac{1}{4} \nabla_\mu \sigma \nabla_\nu \sigma \nabla_{\rho'} \sigma \nabla_{\sigma'} \sigma A_0(x,x') \lim_{\epsilon \to 0^+} \int_0^\infty F(\tau) \, \mathe^{\frac{\mathi \sigma}{2 \tau}} \tau^{-4-\frac{n}{2}} \mathe^{- \epsilon \tau} \total \tau \\
&\qquad+ \mathe^{- \mathi \frac{n+2}{4} \pi} \frac{\sqrt{\Delta}}{(4 \pi)^\frac{n}{2}} \bigg[ \nabla_{(\mu} \sigma \nabla_\nu \sigma \Big( 2 \nabla_{\rho'} \sigma \nabla_{\sigma')} + 3 \nabla_{\rho'} \nabla_{\sigma')} \sigma + \nabla_{\rho'} \sigma \nabla_{\sigma')} \ln \Delta \Big) A_0(x,x') \\
&\qquad\qquad- \frac{1}{4} \nabla_{\rho'} \sigma \nabla_{\sigma'} \sigma \nabla_\mu \sigma \nabla_\nu \sigma A_1(x,x') \bigg] \lim_{\epsilon \to 0^+} \int_0^\infty F(\tau) \, \mathe^{\frac{\mathi \sigma}{2 \tau}} \tau^{-3-\frac{n}{2}} \mathe^{- \epsilon \tau} \total \tau \\
&\qquad+ \mathe^{- \mathi \frac{n-4}{4} \pi} \frac{\sqrt{\Delta}}{(4 \pi)^\frac{n}{2}} \bigg[ \frac{3}{2} \bigg( 4 \nabla_{(\mu} \sigma \nabla_\nu \sigma \nabla_{\rho'} \nabla_{\sigma')} + 4 \nabla_{(\mu} \sigma \nabla_\nu \sigma \nabla_{\rho'} \ln \Delta \nabla_{\sigma')} + 6 \nabla_{(\mu} \sigma \nabla_\nu \nabla_{\rho')} \sigma \nabla_{\sigma')} \\
&\qquad\qquad\quad+ \nabla_{(\mu} \sigma \nabla_\nu \sigma \nabla_{\rho'} \ln \Delta \nabla_{\sigma')} \ln \Delta + 4 \nabla_{(\mu} \sigma \nabla_\nu \nabla_{\rho'} \sigma \nabla_{\sigma')} \ln \Delta + 2 \nabla_{(\mu} \sigma \nabla_\nu \sigma \nabla_{\sigma'} \nabla_{\rho')} \ln \Delta \\
&\qquad\qquad\quad+ 2 \nabla_{(\mu} \nabla_\nu \sigma \nabla_{\rho'} \nabla_{\sigma')} \sigma + \frac{8}{3} \nabla_{(\mu} \sigma \nabla_\nu \nabla_{\rho'} \nabla_{\sigma')} \sigma \bigg) A_0(x,x') \\
&\qquad\qquad- \left( 2 \nabla_{(\mu} \sigma \nabla_\nu \sigma \nabla_{\rho'} \sigma \nabla_{\sigma')} + 3 \nabla_{\sigma'} \sigma \nabla_{(\rho'} \sigma \nabla_\mu \nabla_{\nu)} \sigma + \nabla_{(\mu} \ln \Delta \nabla_\nu \sigma \nabla_{\rho'} \sigma \nabla_{\sigma')} \sigma \right) A_1(x,x') \\
&\qquad\qquad+ \frac{1}{4} \nabla_\mu \sigma \nabla_\nu \sigma \nabla_{\rho'} \sigma \nabla_{\sigma'} \sigma A_2(x,x') \bigg] \lim_{\epsilon \to 0^+} \int_0^\infty F(\tau) \, \mathe^{\frac{\mathi \sigma}{2 \tau}} \tau^{-2-\frac{n}{2}} \mathe^{- \epsilon \tau} \total \tau \\
\end{splitequation*}
\begin{splitequation}
\label{eq:field_propagator_munurhosigma_heatkernel}
&\qquad+ \mathe^{- \mathi \frac{n-2}{4} \pi} \frac{\sqrt{\Delta}}{(4 \pi)^\frac{n}{2}} \bigg[ \bigg( 8 \nabla_{(\mu} \sigma \nabla_\nu \nabla_{\rho'} \nabla_{\sigma')} + 12 \nabla_{(\mu} \nabla_\nu \sigma \nabla_{\rho'} \nabla_{\sigma')} + 12 \nabla_{(\mu} \sigma \nabla_\nu \ln \Delta \nabla_{\rho'} \nabla_{\sigma')} \\
&\qquad\qquad\quad+ 8 \nabla_{(\mu} \nabla_\nu \nabla_{\rho'} \sigma \nabla_{\sigma')} + 12 \nabla_{(\mu} \ln \Delta \nabla_\nu \nabla_{\rho'} \sigma \nabla_{\sigma')} + 12 \nabla_{(\mu} \sigma \nabla_{\nu} \nabla_{\rho'} \ln \Delta \nabla_{\sigma')} \\
&\qquad\qquad\quad+ 6 \nabla_{(\mu} \sigma \nabla_\nu \ln \Delta \nabla_{\rho'} \ln \Delta \nabla_{\sigma')} + 2 \nabla_\mu \nabla_\nu \nabla_{\rho'} \nabla_{\sigma'} \sigma + 4 \nabla_{(\mu} \ln \Delta \nabla_\nu \nabla_{\rho'} \nabla_{\sigma')} \sigma \\
&\qquad\qquad\quad+ 6 \nabla_{(\mu} \nabla_\nu \sigma \nabla_{\rho'} \nabla_{\sigma')} \ln \Delta + 4 \nabla_{(\mu} \sigma \nabla_\nu \nabla_{\rho'} \nabla_{\sigma')} \ln \Delta + 3 \nabla_{(\mu} \nabla_\nu \sigma \nabla_{\rho'} \ln \Delta \nabla_{\sigma')} \ln \Delta \\
&\qquad\qquad\quad+ 6 \nabla_{(\mu} \sigma \nabla_\nu \nabla_{\rho'} \ln \Delta \nabla_{\sigma')} \ln \Delta + \nabla_{(\mu} \sigma \nabla_\nu \ln \Delta \nabla_{\rho'} \ln \Delta \nabla_{\sigma')} \ln \Delta \bigg) A_0(x,x') \\
&\qquad\qquad- \bigg( 6 \nabla_{(\mu} \sigma \nabla_\nu \sigma \nabla_{\rho'} \nabla_{\sigma'} + 6 \nabla_{(\mu} \sigma \nabla_\nu \sigma \nabla_{\rho'} \ln \Delta \nabla_{\sigma')} + 12 \nabla_{(\mu} \sigma \nabla_\nu \nabla_{\rho'} \sigma \nabla_{\sigma')} \\
&\qquad\qquad\quad+ 3 \nabla_{(\mu} \nabla_\nu \sigma \nabla_{\rho'} \nabla_{\sigma')} \sigma + 4 \nabla_{(\mu} \sigma \nabla_\nu \nabla_{\rho'} \nabla_{\sigma')} \sigma + 6 \nabla_{(\mu} \sigma \nabla_\nu \nabla_{\rho'} \sigma \nabla_{\sigma')} \ln \Delta \\
&\qquad\qquad\quad+ 3 \nabla_{(\mu} \sigma \nabla_\nu \sigma \nabla_{\rho'} \nabla_{\sigma')} \ln \Delta + \frac{3}{2} \nabla_{(\mu} \sigma \nabla_\nu \sigma \nabla_{\rho'} \ln \Delta \nabla_{\sigma')} \ln \Delta \bigg) A_1(x,x') \\
&\qquad\qquad\quad+ \nabla_{(\mu} \sigma \nabla_\nu \sigma \left( 2 \nabla_{\rho'} \sigma \nabla_{\sigma')} + 3 \nabla_{\rho'} \nabla_{\sigma')} \sigma + \nabla_{\rho'} \sigma \nabla_{\sigma')} \ln \Delta \right) A_2(x,x') \\
&\qquad\qquad- \frac{1}{4} \nabla_{\sigma'} \sigma \nabla_{\rho'} \sigma \nabla_\mu \sigma \nabla_\nu \sigma A_3(x,x') \bigg] \lim_{\epsilon \to 0^+} \int_0^\infty F(\tau) \, \mathe^{\frac{\mathi \sigma}{2 \tau}} \tau^{-1-\frac{n}{2}} \mathe^{- \epsilon \tau} \total \tau \\
&\qquad+ \mathe^{- \mathi \frac{n-2}{4} \pi} \frac{\sqrt{\Delta}}{(4 \pi)^\frac{n}{2}} \sum_{k=0}^\infty \mathi^k \bigg[ \bigg( 4 \nabla_\mu \nabla_\nu \nabla_{\rho'} \nabla_{\sigma'} + 8 \nabla_{(\mu} \ln \Delta \nabla_\nu \nabla_{\rho'} \nabla_{\sigma')} + 12 \nabla_{(\mu} \nabla_\nu \ln \Delta \nabla_{\rho'} \nabla_{\sigma')} \\
&\qquad\qquad\quad+ 6 \nabla_{(\mu} \ln \Delta \nabla_\nu \ln \Delta \nabla_{\rho'} \nabla_{\sigma')} + 8 \nabla_{(\mu} \nabla_\nu \nabla_{\rho'} \ln \Delta \nabla_{\sigma')} + 12 \nabla_{(\mu} \nabla_\nu \ln \Delta \nabla_{\rho'} \ln \Delta \nabla_{\sigma')} \\
&\qquad\qquad\quad+ 2 \nabla_{(\mu} \ln \Delta \nabla_\nu \ln \Delta \nabla_{\rho'} \ln \Delta \nabla_{\sigma')} + 2 \nabla_\mu \nabla_\nu \nabla_{\rho'} \nabla_{\sigma'} \ln \Delta + 4 \nabla_{(\mu} \ln \Delta \nabla_\nu \nabla_{\rho'} \nabla_{\sigma')} \ln \Delta \\
&\qquad\qquad\quad+ 3 \nabla_{(\mu} \nabla_\nu \ln \Delta \nabla_{\rho'} \nabla_{\sigma')} \ln \Delta + 3 \nabla_{(\mu} \ln \Delta \nabla_\nu \ln \Delta \nabla_{\rho'} \nabla_{\sigma')} \ln \Delta \\
&\qquad\qquad\quad+ \frac{1}{4} \nabla_\mu \ln \Delta \nabla_\nu \ln \Delta \nabla_{\rho'} \ln \Delta \nabla_{\sigma'} \ln \Delta \bigg) A_k(x,x') \\
&\qquad\qquad- \bigg( 8 \nabla_{(\mu} \sigma \nabla_\nu \nabla_{\rho'} \nabla_{\sigma')} + 12 \nabla_{(\mu} \nabla_\nu \sigma \nabla_{\rho'} \nabla_{\sigma')} + 12 \nabla_{(\mu} \sigma \nabla_\nu \ln \Delta \nabla_{\rho'} \nabla_{\sigma')} \\
&\qquad\qquad\quad+ 8 \nabla_{(\mu} \nabla_\nu \nabla_{\rho'} \sigma \nabla_{\sigma')} + 12 \nabla_{(\mu} \sigma \nabla_\nu \nabla_{\rho')} \ln \Delta \nabla_{\sigma')} + 12 \nabla_{(\mu} \nabla_\nu \sigma \nabla_{\rho'} \ln \Delta \nabla_{\sigma')} \\
&\qquad\qquad\quad+ 6 \nabla_{(\mu} \sigma \nabla_\nu \ln \Delta \nabla_{\rho'} \ln \Delta \nabla_{\sigma')} + 2 \nabla_\mu \nabla_\nu \nabla_{\rho'} \nabla_{\sigma'} \sigma + 4 \nabla_{(\mu} \sigma \nabla_\nu \nabla_{\rho'} \nabla_{\sigma')} \ln \Delta \\
&\qquad\qquad\quad+ 4 \nabla_{(\mu} \nabla_\nu \nabla_{\rho'} \sigma \nabla_{\sigma')} \ln \Delta + 6 \nabla_{(\mu} \nabla_\nu \sigma \nabla_{\rho'} \nabla_{\sigma')} \ln \Delta + 6 \nabla_{(\mu} \sigma \nabla_\nu \ln \Delta \nabla_{\rho'} \nabla_{\sigma')} \ln \Delta \\
&\qquad\qquad\quad+ 3 \nabla_{(\mu} \nabla_\nu \sigma \nabla_{\rho'} \ln \Delta \nabla_{\sigma')} \ln \Delta + \nabla_{(\mu} \sigma \nabla_\nu \ln \Delta \nabla_{\rho'} \ln \Delta \nabla_{\sigma'} \ln \Delta \bigg) A_{k+1}(x,x') \\
&\qquad\qquad+ \bigg( 6 \nabla_{(\mu} \sigma \nabla_\nu \sigma \nabla_{\rho'} \nabla_{\sigma')} + 12 \nabla_{(\mu} \sigma \nabla_\nu \nabla_{\rho'} \sigma \nabla_{\sigma')} + 6 \nabla_{(\mu} \sigma \nabla_\nu \sigma \nabla_{\rho'} \ln \Delta \nabla_{\sigma')} \\
&\qquad\qquad\quad+ 4 \nabla_{(\mu} \sigma \nabla_\nu \nabla_{\rho'} \nabla_{\sigma')} \sigma + 3 \nabla_{(\mu} \sigma \nabla_\nu \sigma \nabla_{\rho'} \nabla_{\sigma')} \ln \Delta + 3 \nabla_{(\mu} \sigma \nabla_\nu \nabla_{\rho'} \sigma \nabla_{\sigma')} \ln \Delta \\
&\qquad\qquad\quad+ \frac{3}{2} \nabla_{(\mu} \sigma \nabla_\nu \sigma \nabla_{\rho'} \ln \Delta \nabla_{\sigma')} \ln \Delta + 3 \nabla_{(\mu} \nabla_\nu \sigma \nabla_{\rho'} \nabla_{\sigma')} \sigma + 3 \nabla_{(\mu} \nabla_\nu \sigma \nabla_{\rho'} \sigma \nabla_{\sigma')} \ln \Delta \bigg) A_{k+2}(x,x') \\
&\qquad\qquad- \nabla_{(\mu} \sigma \nabla_\nu \sigma \left( 2 \nabla_{\rho'} \sigma \nabla_{\sigma')} + 3 \nabla_{\rho'} \nabla_{\sigma')} \sigma + \nabla_{\rho'} \sigma \nabla_{\sigma')} \ln \Delta \right) A_{k+3}(x,x') \\
&\qquad\qquad+ \frac{1}{4} \nabla_\mu \sigma \nabla_\nu \sigma \nabla_{\rho'} \sigma \nabla_{\sigma'} \sigma A_{k+4}(x,x') \bigg] \lim_{\epsilon \to 0^+} \int_0^\infty F(\tau) \, \mathe^{\frac{\mathi \sigma}{2 \tau}} \tau^{k-\frac{n}{2}} \mathe^{- \epsilon \tau} \total \tau \eqend{,}
\end{splitequation}
where we defined
\begin{splitequation}
\label{eq:field_propagator_munurhosigma_heatkernel_fdef}
F(\tau) &\equiv \frac{1}{m^2} \bigg[ \frac{2 \mu}{m^2} \left( 1 - \mathe^{- \mathi M_B^2 \tau} \right) - \frac{1}{M_A^2} \left( \mathe^{- \mathi M_A^2 \tau} - \mathe^{- \mathi M_B^2 \tau} \right) \\
&\qquad\qquad+ \frac{n-2}{2 (n-1) - (n-2) \lambda} \frac{1}{m^2} \left( \mathe^{- \mathi M_C^2 \tau} - 1 \right) - \mathi \tau \left( \rho + M_B^2 \frac{\mu}{m^2} \mathe^{- \mathi M_B^2 \tau} \right) \bigg] \\
&= \frac{1}{2} \left[ (\xi-\lambda) \frac{(1-\zeta) (1-3\zeta)}{(2\zeta-1)^2} - \frac{(n-2) (1-\lambda)^2}{2 (n-1) - (n-2) \lambda} \right] \tau^2 + \bigo{m^2} \eqend{,}
\end{splitequation}
\begin{splitequation}
\label{eq:field_propagator_munurho0_heatkernel}
&G_{\mu\nu\rho'\circ}(x,x') = - \frac{2 \mathi \mu}{m^3} \lim_{\epsilon \to 0^+} \int_0^\infty \nabla_\mu \nabla_\nu \nabla_{\rho'} K(x,x',\tau) \left[ 1 - \left( 1 + \mathi M_B^2 \tau \right) \mathe^{- \mathi M_B^2 \tau} \right] \mathe^{- \epsilon \tau} \total \tau \\
&\quad\sim \mathe^{- \mathi \frac{n+2}{4} \pi} \frac{\sqrt{\Delta}}{(4 \pi)^\frac{n}{2}} \frac{\mu}{4 m^3} \nabla_\mu \sigma \nabla_\nu \sigma \nabla_{\rho'} \sigma A_0(x,x') \\
&\qquad\quad\times \lim_{\epsilon \to 0^+} \int_0^\infty \left[ 1 - \left( 1 + \mathi M_B^2 \tau \right) \mathe^{- \mathi M_B^2 \tau} \right] \mathe^{\frac{\mathi \sigma}{2 \tau}} \tau^{-3-\frac{n}{2}} \mathe^{- \epsilon \tau} \total \tau \\
&\qquad+ \mathe^{- \mathi \frac{n+4}{4} \pi} \frac{\sqrt{\Delta}}{(4 \pi)^\frac{n}{2}} \frac{\mu}{4 m^3} \bigg[ 6 \nabla_{(\mu} \sigma \left( \nabla_\nu \sigma \nabla_{\rho')} + \nabla_\nu \nabla_{\rho')} \sigma + \frac{1}{2} \nabla_\nu \sigma \nabla_{\rho')} \ln \Delta \right) A_0(x,x') \\
&\qquad\qquad- \nabla_{\rho'} \sigma \nabla_\mu \sigma \nabla_\nu \sigma A_1(x,x') \bigg] \lim_{\epsilon \to 0^+} \int_0^\infty \left[ 1 - \left( 1 + \mathi M_B^2 \tau \right) \mathe^{- \mathi M_B^2 \tau} \right] \mathe^{\frac{\mathi \sigma}{2 \tau}} \tau^{-2-\frac{n}{2}} \mathe^{- \epsilon \tau} \total \tau \\
&\qquad+ \mathe^{- \mathi \frac{n-2}{4} \pi} \frac{\sqrt{\Delta}}{(4 \pi)^\frac{n}{2}} \frac{\mu}{m^3} \bigg[ \bigg( 3 \nabla_{(\mu} \sigma \nabla_\nu \nabla_{\rho')} + 3 \nabla_{(\rho'} \nabla_\mu \sigma \nabla_{\nu)} + 3 \nabla_{(\rho'} \ln \Delta \nabla_\mu \sigma \nabla_{\nu)} \\
&\qquad\qquad\qquad+ \nabla_{\rho'} \nabla_\mu \nabla_\nu \sigma + \frac{3}{2} \nabla_{(\mu} \ln \Delta \nabla_\nu \nabla_{\rho')} \sigma + \frac{3}{2} \nabla_{(\rho'} \sigma \nabla_\mu \nabla_{\nu)} \ln \Delta \\
&\qquad\qquad\qquad+ \frac{3}{4} \nabla_{(\rho'} \sigma \nabla_\mu \ln \Delta \nabla_{\nu)} \ln \Delta \bigg) A_0(x,x') + \frac{1}{4} \nabla_{\rho'} \sigma \nabla_\mu \sigma \nabla_\nu \sigma A_2(x,x') \\
&\qquad\qquad- \frac{3}{2} \left( \nabla_{(\rho'} \sigma \nabla_\mu \sigma \nabla_{\nu)} + \nabla_{(\rho'} \sigma \nabla_\mu \nabla_{\nu)} \sigma + \frac{1}{2} \nabla_{(\rho'} \ln \Delta \nabla_\mu \sigma \nabla_{\nu)} \sigma \right) A_1(x,x') \bigg] \\
&\qquad\qquad\times \lim_{\epsilon \to 0^+} \int_0^\infty \left[ 1 - \left( 1 + \mathi M_B^2 \tau \right) \mathe^{- \mathi M_B^2 \tau} \right] \mathe^{\frac{\mathi \sigma}{2 \tau}} \tau^{-1-\frac{n}{2}} \mathe^{- \epsilon \tau} \total \tau \\
&\qquad+ \mathe^{- \mathi \frac{n-2}{4} \pi} \frac{\sqrt{\Delta}}{(4 \pi)^\frac{n}{2}} \frac{\mu}{m^3} \sum_{k=0}^\infty \mathi^k \bigg[ \bigg( 2 \nabla_\mu \nabla_\nu \nabla_{\rho'} + 3 \nabla_{(\mu} \ln \Delta \nabla_\nu \nabla_{\rho')} + 3 \nabla_{(\mu} \nabla_\nu \ln \Delta \nabla_{\rho')} \\
&\qquad\qquad\qquad+ \nabla_{\rho'} \nabla_\mu \nabla_\nu \ln \Delta + \frac{3}{2} \nabla_{(\mu} \ln \Delta \nabla_\nu \ln \Delta \nabla_{\rho')} + \frac{3}{2} \nabla_{(\rho'} \nabla_\mu \ln \Delta \nabla_{\nu)} \ln \Delta \\
&\qquad\qquad\qquad+ \frac{1}{4} \nabla_{\rho'} \ln \Delta \nabla_\mu \ln \Delta \nabla_\nu \ln \Delta \bigg) A_k(x,x') - \frac{1}{4} \nabla_{\rho'} \sigma \nabla_\mu \sigma \nabla_\nu \sigma A_{k+3}(x,x') \\
&\qquad\qquad- 3 \bigg( \nabla_{(\mu} \sigma \nabla_\nu \nabla_{\rho')} + \nabla_{(\mu} \nabla_\nu \sigma \nabla_{\rho')} + \nabla_{(\mu} \ln \Delta \nabla_\nu \sigma \nabla_{\rho')} + \frac{1}{4} \nabla_{(\rho'} \sigma \nabla_\mu \ln \Delta \nabla_{\nu)} \ln \Delta \\
&\qquad\qquad\qquad+ \frac{1}{2} \nabla_{(\rho'} \sigma \nabla_\mu \nabla_{\nu)} \ln \Delta + \frac{1}{2} \nabla_{(\rho'} \ln \Delta \nabla_\mu \nabla_{\nu)} \sigma + \nabla_{\rho'} \nabla_\mu \nabla_\nu \sigma \bigg) A_{k+1}(x,x') \\
&\qquad\qquad+ \frac{3}{2} \left( \nabla_{(\mu} \sigma \nabla_\nu \sigma \nabla_{\rho')} + \nabla_{(\rho'} \sigma \nabla_\mu \nabla_{\nu)} \sigma + \frac{1}{2} \nabla_{(\rho'} \ln \Delta \nabla_\mu \sigma \nabla_{\nu)} \sigma \right) A_{k+2}(x,x') \bigg] \\
&\qquad\qquad\times \lim_{\epsilon \to 0^+} \int_0^\infty \left[ 1 - \left( 1 + \mathi M_B^2 \tau \right) \mathe^{- \mathi M_B^2 \tau} \right] \mathe^{\frac{\mathi \sigma}{2 \tau}} \tau^{k-\frac{n}{2}} \mathe^{- \epsilon \tau} \total \tau \eqend{,}
\end{splitequation}
\begin{splitequation}
\label{eq:field_propagator_munu00_heatkernel}
&G_{\mu\nu\circ\circ}(x,x') = 2 \lim_{\epsilon \to 0^+} \int_0^\infty \bigg[ \left[ \mathi \frac{\mu}{m^2} \left( 1 - \mathe^{- \mathi M_B^2 \tau} \right) + \rho \tau \right] \nabla_\mu \nabla_\nu K(x,x',\tau) \\
&\hspace{14em}- \frac{\mathi}{2 (n-1) - (n-2) \lambda} g_{\mu\nu} K(x,x',\tau) \bigg] \mathe^{- \epsilon \tau} \total \tau \\
&\quad\sim \mathe^{- \mathi \frac{n}{4} \pi} \frac{\sqrt{\Delta}}{(4 \pi)^\frac{n}{2}} \frac{\mu}{2 m^2} \nabla_\mu \sigma \nabla_\nu \sigma A_0(x,x') \lim_{\epsilon \to 0^+} \int_0^\infty \left( 1 - \mathe^{- \mathi M_B^2 \tau} \right) \mathe^{\frac{\mathi \sigma}{2 \tau}} \tau^{-2-\frac{n}{2}} \mathe^{- \epsilon \tau} \total \tau \\
&\qquad+ \mathe^{- \mathi \frac{n+2}{4} \pi} \frac{\sqrt{\Delta}}{(4 \pi)^\frac{n}{2}} \frac{\mu}{m^2} \bigg[ \left( 2 \nabla_{(\mu} \sigma \nabla_{\nu)} + \nabla_\mu \nabla_\nu \sigma + \nabla_{(\mu} \ln \Delta \nabla_{\nu)} \sigma \right) A_0(x,x') \\
&\qquad\qquad- \frac{1}{2} \nabla_\mu \sigma \nabla_\nu \sigma A_1(x,x') \bigg] \lim_{\epsilon \to 0^+} \int_0^\infty \left( 1 - \mathe^{- \mathi M_B^2 \tau} \right) \mathe^{\frac{\mathi \sigma}{2 \tau}} \tau^{-1-\frac{n}{2}} \mathe^{- \epsilon \tau} \total \tau \\
&\qquad+ \mathe^{- \mathi \frac{n+2}{4} \pi} \frac{\sqrt{\Delta}}{(4 \pi)^\frac{n}{2}} \frac{\rho}{2} \nabla_\mu \sigma \nabla_\nu \sigma A_0(x,x') \lim_{\epsilon \to 0^+} \int_0^\infty \mathe^{\frac{\mathi \sigma}{2 \tau}} \tau^{-1-\frac{n}{2}} \mathe^{- \epsilon \tau} \total \tau \\
&\qquad+ \mathe^{- \mathi \frac{n-4}{4} \pi} \frac{\sqrt{\Delta}}{(4 \pi)^\frac{n}{2}} \rho \bigg[ \left( 2 \nabla_{(\mu} \sigma \nabla_{\nu)} + \nabla_{(\mu} \ln \Delta \nabla_{\nu)} \sigma + \nabla_\mu \nabla_\nu \sigma \right) A_0(x,x') \\
&\qquad\qquad- \frac{1}{2} \nabla_\mu \sigma \nabla_\nu \sigma A_1(x,x') \bigg] \lim_{\epsilon \to 0^+} \int_0^\infty \mathe^{\frac{\mathi \sigma}{2 \tau}} \tau^{-\frac{n}{2}} \mathe^{- \epsilon \tau} \total \tau \\
&\qquad+ 2 \mathe^{- \mathi \frac{n+2}{4} \pi} \frac{\sqrt{\Delta}}{(4 \pi)^\frac{n}{2}} \frac{\mu}{m^2} \sum_{k=0}^\infty \mathi^k \bigg[ - \left( \nabla_{(\mu} \sigma \nabla_{\nu)} + \frac{1}{2} \nabla_{(\mu} \ln \Delta \nabla_{\nu)} \sigma + \frac{1}{2} \nabla_\mu \nabla_\nu \sigma \right) A_{k+1}(x,x') \\
&\qquad\qquad+ \left( \nabla_\mu \nabla_\nu + \nabla_{(\mu} \ln \Delta \nabla_{\nu)} + \frac{1}{2} \nabla_\mu \nabla_\nu \ln \Delta + \frac{1}{4} \nabla_\mu \ln \Delta \nabla_\nu \ln \Delta \right) A_k(x,x') \\
&\qquad\qquad+ \frac{1}{4} \nabla_\mu \sigma \nabla_\nu \sigma A_{k+2}(x,x') \bigg] \lim_{\epsilon \to 0^+} \int_0^\infty \left( 1 - \mathe^{- \mathi M_B^2 \tau} \right) \mathe^{\frac{\mathi \sigma}{2 \tau}} \tau^{k-\frac{n}{2}} \mathe^{- \epsilon \tau} \total \tau \\
&\qquad+ 2 \mathe^{- \mathi \frac{n-2}{4} \pi} \frac{\sqrt{\Delta}}{(4 \pi)^\frac{n}{2}} \rho \sum_{k=0}^\infty \mathi^k \bigg[ - \left( \nabla_{(\mu} \sigma \nabla_{\nu)} + \frac{1}{2} \nabla_{(\mu} \ln \Delta \nabla_{\nu)} \sigma + \frac{1}{2} \nabla_\mu \nabla_\nu \sigma \right) A_{k+1}(x,x') \\
&\qquad\qquad+ \left( \nabla_\mu \nabla_\nu + \nabla_{(\mu} \ln \Delta \nabla_{\nu)} + \frac{1}{2} \nabla_\mu \nabla_\nu \ln \Delta + \frac{1}{4} \nabla_\mu \ln \Delta \nabla_\nu \ln \Delta \right) A_k(x,x') \\
&\qquad\qquad+ \frac{1}{4} \nabla_\mu \sigma \nabla_\nu \sigma A_{k+2}(x,x') \bigg] \lim_{\epsilon \to 0^+} \int_0^\infty \mathe^{\frac{\mathi \sigma}{2 \tau}} \tau^{k+1-\frac{n}{2}} \mathe^{- \epsilon \tau} \total \tau \\
&\qquad+ 2 \mathe^{- \mathi \frac{n-2}{4} \pi} \frac{\sqrt{\Delta}}{(4 \pi)^\frac{n}{2}} \frac{g_{\mu\nu}}{2 (n-1) - (n-2) \lambda} \sum_{k=0}^\infty \mathi^k A_k(x,x') \lim_{\epsilon \to 0^+} \int_0^\infty \mathe^{\frac{\mathi \sigma}{2 \tau}} \tau^{k-\frac{n}{2}} \mathe^{- \epsilon \tau} \total \tau \eqend{,}
\end{splitequation}
\begin{splitequation}
\label{eq:field_propagator_mu0rho0_heatkernel}
&G_{\mu\circ\rho'\circ}(x,x') = - \mathi \lim_{\epsilon \to 0^+} \int_0^\infty K_{\mu\rho'}(x,x',\tau) \, \mathe^{- \mathi ( M_A^2 - \mathi \epsilon ) \tau} \total \tau \\
&\qquad+ \mathi \lim_{\epsilon \to 0^+} \int_0^\infty \nabla_\mu \nabla_{\rho'} K(x,x',\tau) \left[ \frac{1 - \mathe^{\mathi (M_B^2-M_A^2) \tau}}{M_A^2} + \mathi \tau \frac{\mu}{m^2} M_B^2 \right] \mathe^{- \mathi ( M_B^2 - \mathi \epsilon ) \tau} \total \tau \\
&\quad\sim \mathe^{- \mathi \frac{n}{4} \pi} \frac{\sqrt{\Delta}}{(4 \pi)^\frac{n}{2}} \nabla_\mu \sigma \nabla_{\rho'} \sigma A_0(x,x') \lim_{\epsilon \to 0^+} \int_0^\infty \frac{\mathe^{- \mathi M_B^2 \tau} - \mathe^{- \mathi M_A^2 \tau}}{4 M_A^2} \mathe^{\frac{\mathi \sigma}{2 \tau}} \tau^{-2-\frac{n}{2}} \mathe^{- \epsilon \tau} \total \tau \\
&\qquad+ \mathe^{- \mathi \frac{n-2}{4} \pi} \frac{\sqrt{\Delta}}{(4 \pi)^\frac{n}{2}} \frac{\mu}{4 m^2} M_B^2 \nabla_\mu \sigma \nabla_{\rho'} \sigma A_0(x,x') \lim_{\epsilon \to 0^+} \int_0^\infty \mathe^{- \mathi M_B^2 \tau} \mathe^{\frac{\mathi \sigma}{2 \tau}} \tau^{-1-\frac{n}{2}} \mathe^{- \epsilon \tau} \total \tau \\
&\qquad+ \mathe^{- \mathi \frac{n+2}{4} \pi} \frac{\sqrt{\Delta}}{(4 \pi)^\frac{n}{2}} \bigg[ \left( 4 \nabla_{(\mu} \sigma \nabla_{\rho')} + 2 \nabla_\mu \nabla_{\rho'} \sigma + 2 \nabla_{(\mu} \ln \Delta \nabla_{\rho')} \sigma \right) A_0(x,x') \\
&\qquad\qquad- \nabla_{\rho'} \sigma \nabla_\mu \sigma A_1(x,x') \bigg] \lim_{\epsilon \to 0^+} \int_0^\infty \frac{\mathe^{- \mathi M_B^2 \tau} - \mathe^{- \mathi M_A^2 \tau}}{4 M_A^2} \mathe^{\frac{\mathi \sigma}{2 \tau}} \tau^{-1-\frac{n}{2}} \mathe^{- \epsilon \tau} \total \tau \\
&\qquad+ \mathe^{- \mathi \frac{n}{4} \pi} \frac{\sqrt{\Delta}}{(4 \pi)^\frac{n}{2}} \bigg[ \frac{\mu}{4 m^2} M_B^2 \left( 4 \nabla_{(\mu} \sigma \nabla_{\rho')} + 2 \nabla_\mu \nabla_{\rho'} \sigma + 2 \nabla_{(\mu} \ln \Delta \nabla_{\rho')} \sigma \right) A_0(x,x') \\
&\qquad\qquad- \frac{1}{4 M_A^2} \left( 4 \nabla_\mu \nabla_{\rho'} + 4 \nabla_{(\mu} \ln \Delta \nabla_{\rho')} + 2 \nabla_\mu \nabla_{\rho'} \ln \Delta + \nabla_\mu \ln \Delta \nabla_{\rho'} \ln \Delta \right) A_0(x,x') \\
&\qquad\qquad+ \frac{1}{2 M_A^2} \left( 2 \nabla_{(\mu} \sigma \nabla_{\rho')} + \nabla_{(\mu} \ln \Delta \nabla_{\rho')} \sigma + \nabla_\mu \nabla_{\rho'} \sigma \right) A_1(x,x') \\
&\qquad\qquad- \frac{\mu}{4 m^2} M_B^2 \nabla_{\rho'} \sigma \nabla_\mu \sigma A_1(x,x') - \frac{1}{4 M_A^2} \nabla_\mu \sigma \nabla_{\rho'} \sigma A_2(x,x') \bigg] \\
&\qquad\quad\times \lim_{\epsilon \to 0^+} \int_0^\infty \mathe^{- \mathi M_B^2 \tau} \mathe^{\frac{\mathi \sigma}{2 \tau}} \tau^{-\frac{n}{2}} \mathe^{- \epsilon \tau} \total \tau \\
&\qquad+ \mathe^{- \mathi \frac{n}{4} \pi} \frac{\sqrt{\Delta}}{(4 \pi)^\frac{n}{2}} \sum_{k=0}^\infty \mathi^k \bigg[ A_{k\mu\rho'}(x,x') + \frac{1}{4 M_A^2} \nabla_\mu \sigma \nabla_{\rho'} \sigma A_{k+2}(x,x') \\
&\qquad\qquad- \frac{1}{2 M_A^2} \left( 2 \nabla_{(\mu} \sigma \nabla_{\rho')} + \nabla_{(\mu} \ln \Delta \nabla_{\rho')} \sigma + \nabla_\mu \nabla_{\rho'} \sigma \right) A_{k+1}(x,x') \\
&\qquad\qquad+ \frac{1}{4 M_A^2} \left( 4 \nabla_\mu \nabla_{\rho'} + 4 \nabla_{(\mu} \ln \Delta \nabla_{\rho')} + 2 \nabla_\mu \nabla_{\rho'} \ln \Delta + \nabla_\mu \ln \Delta \nabla_{\rho'} \ln \Delta \right) A_k(x,x') \bigg] \\
&\qquad\quad\times \lim_{\epsilon \to 0^+} \int_0^\infty \mathe^{- \mathi M_A^2 \tau} \mathe^{\frac{\mathi \sigma}{2 \tau}} \tau^{k-\frac{n}{2}} \mathe^{- \epsilon \tau} \total \tau \\
&\qquad+ \mathe^{- \mathi \frac{n+2}{4} \pi} \frac{\sqrt{\Delta}}{(4 \pi)^\frac{n}{2}} \sum_{k=0}^\infty \mathi^k \bigg[ - \frac{1}{2 M_A^2} \left( 2 \nabla_{(\mu} \sigma \nabla_{\rho')} + \nabla_{(\mu} \ln \Delta \nabla_{\rho')} \sigma + \nabla_\mu \nabla_{\rho'} \sigma \right) A_{k+2}(x,x') \\
&\qquad\qquad+ \frac{\mu}{4 m^2} M_B^2 \left( 4 \nabla_\mu \nabla_{\rho'} + 4 \nabla_{(\mu} \ln \Delta \nabla_{\rho')} + 2 \nabla_\mu \nabla_{\rho'} \ln \Delta + \nabla_\mu \ln \Delta \nabla_{\rho'} \ln \Delta \right) A_k(x,x') \\
&\qquad\qquad+ \frac{1}{4 M_A^2} \left( 4 \nabla_\mu \nabla_{\rho'} + 4 \nabla_{(\mu} \ln \Delta \nabla_{\rho')} + 2 \nabla_\mu \nabla_{\rho'} \ln \Delta + \nabla_\mu \ln \Delta \nabla_{\rho'} \ln \Delta \right) A_{k+1}(x,x') \\
&\qquad\qquad- \frac{\mu}{2 m^2} M_B^2 \left( 2 \nabla_{(\mu} \sigma \nabla_{\rho')} + \nabla_{(\mu} \ln \Delta \nabla_{\rho')} \sigma + \nabla_\mu \nabla_{\rho'} \sigma \right) A_{k+1}(x,x') \\
&\qquad\qquad+ \frac{\mu}{4 m^2} M_B^2 \nabla_\mu \sigma \nabla_{\rho'} \sigma A_{k+2}(x,x') + \frac{1}{4 M_A^2} \nabla_\mu \sigma \nabla_{\rho'} \sigma A_{k+3}(x,x') \bigg] \\
&\qquad\quad\times \lim_{\epsilon \to 0^+} \int_0^\infty \mathe^{- \mathi M_B^2 \tau} \mathe^{\frac{\mathi \sigma}{2 \tau}} \tau^{k+1-\frac{n}{2}} \mathe^{- \epsilon \tau} \total \tau \eqend{,}
\end{splitequation}
\begin{splitequation}
\label{eq:field_propagator_mu000_heatkernel}
&G_{\mu\circ\circ\circ}(x,x') = \mathi \frac{\mu}{m} \lim_{\epsilon \to 0^+} \int_0^\infty \nabla_\mu K(x,x',\tau) \, \left[ 1 - \mathe^{- \mathi M_B^2 \tau} \right] \mathe^{- \epsilon \tau} \total \tau \\
&\quad\sim \mathe^{- \mathi \frac{n+2}{4} \pi} \frac{\sqrt{\Delta}}{(4 \pi)^\frac{n}{2}} \frac{\mu}{2 m} \nabla_\mu \sigma A_0(x,x') \lim_{\epsilon \to 0^+} \int_0^\infty \left[ 1 - \mathe^{- \mathi M_B^2 \tau} \right] \mathe^{\frac{\mathi \sigma}{2 \tau}} \tau^{-1-\frac{n}{2}} \mathe^{- \epsilon \tau} \total \tau \\
&\qquad+ \mathe^{- \mathi \frac{n-4}{4} \pi} \frac{\sqrt{\Delta}}{(4 \pi)^\frac{n}{2}} \frac{\mu}{m} \lim_{\epsilon \to 0^+} \sum_{k=0}^\infty \mathi^k \left[ \left( \nabla_\mu + \frac{1}{2} \nabla_\mu \ln \Delta \right) A_k(x,x') - \frac{1}{2} \nabla_\mu \sigma A_{k+1}(x,x') \right] \\
&\qquad\quad\times \int_0^\infty \left[ 1 - \mathe^{- \mathi M_B^2 \tau} \right] \mathe^{\frac{\mathi \sigma}{2 \tau}} \tau^{k-\frac{n}{2}} \mathe^{- \epsilon \tau} \total \tau \eqend{,}
\end{splitequation}
and
\begin{splitequation}
\label{eq:field_propagator_0000_heatkernel}
&G_{\circ\circ\circ\circ}(x,x') = - \lim_{\epsilon \to 0^+} \int_0^\infty \left[ \frac{(n-2) \mathi}{2 (n-1) - (n-2) \lambda} + \rho m^2 \tau \right] K(x,x',\tau) \, \mathe^{- \epsilon \tau} \total \tau \\
&\quad\sim \mathe^{- \mathi \frac{n}{4} \pi} \frac{\sqrt{\Delta}}{(4 \pi)^\frac{n}{2}} \frac{(n-2)}{2 (n-1) - (n-2) \lambda} A_0(x,x') \lim_{\epsilon \to 0^+} \int_0^\infty \mathe^{\frac{\mathi \sigma}{2 \tau}} \tau^{-\frac{n}{2}} \mathe^{- \epsilon \tau} \total \tau \\
&\qquad+ \mathe^{- \mathi \frac{n-2}{4} \pi} \frac{\sqrt{\Delta}}{(4 \pi)^\frac{n}{2}} \lim_{\epsilon \to 0^+} \sum_{k=0}^\infty \mathi^k \left[ \frac{(n-2)}{2 (n-1) - (n-2) \lambda} A_{k+1}(x,x') - \rho m^2 A_k(x,x') \right] \\
&\qquad\quad\times \int_0^\infty \mathe^{\frac{\mathi \sigma}{2 \tau}} \tau^{k+1-\frac{n}{2}} \mathe^{- \epsilon \tau} \total \tau
\end{splitequation}
with the masses~\eqref{eq:field_masses} and the abbreviations~\eqref{eq:field_abbreviations}
\begin{equation}
\rho \equiv \frac{n}{2 (n-1) - (n-2) \lambda} - \alpha \eqend{,} \quad \mu \equiv \frac{m^2}{M_A^2} - \alpha \frac{m^2}{M_B^2} = \frac{\zeta - \alpha (2\zeta-1)}{\zeta (\xi-\lambda)} \eqend{.}
\end{equation}
The expansions of the remaining propagators $G_{\mu\circ\rho'\sigma'}(x,x')$, $G_{\circ\circ\rho'\sigma'}$ and $G_{\circ\circ\rho'\circ}(x,x')$ can be obtained by renaming indices of the propagators $G_{\mu\nu\rho'\circ}(x,x')$, $G_{\mu\nu\circ\circ}$ and $G_{\mu\circ\circ\circ}(x,x')$ appropriately, and we thus do not list them separately.

It is again straightforward to take the massless limit, and as for the ghost sector we obtain a sensible limit $m \to 0$ while keeping $\lambda = \frac{4}{n-2} \frac{\Lambda}{m^2}$ fixed. Since the massless limit of the propagators $G_{\mu\nu\rho'\circ}(x,x')$~\eqref{eq:field_propagator_massless_munurho0}, $G_{\mu\circ\rho'\circ}(x,x')$~\eqref{eq:field_propagator_massless_mu0rho0} and $G_{\mu\circ\circ\circ}(x,x')$~\eqref{eq:field_propagator_massless_mu000} vanishes, we don't list them separately. Moreover, since the expression for the $G_{\mu\nu\rho'\sigma'}$ propagator~\eqref{eq:field_propagator_munurhosigma_heatkernel} is already very long and does not simplify in the massless limit, we do not list this limit separately; it is obtained by performing the replacements
\begin{equation}
\frac{1 - \mathe^{- \mathi M_C^2 \tau}}{m^2} \to \mathi (1-\lambda) \tau \eqend{,} \quad \frac{\mathe^{- \mathi M_C^2 \tau} - \mathe^{- \mathi M_A^2 \tau}}{m^2} \to \mathi (\xi-1) \tau \eqend{,}
\end{equation}
in the result~\eqref{eq:field_propagator_munurhosigma_heatkernel}, and taking also the massless limit of $F(\tau)$~\eqref{eq:field_propagator_munurhosigma_heatkernel_fdef}. For the other propagators, the massless limit results in
\begin{splitequation}
\label{eq:field_propagator_massless_munu00_heatkernel}
&\lim_{m \to 0} G_{\mu\nu\circ\circ}(x,x') \\
&\quad\sim \frac{1}{2} \mathe^{- \mathi \frac{n-2}{4} \pi} \frac{\sqrt{\Delta}}{(4 \pi)^\frac{n}{2}} \left[ \frac{\zeta}{2\zeta-1} - \frac{n}{2 (n-1) - (n-2) \lambda} \right] \nabla_\mu \sigma \nabla_\nu \sigma A_0(x,x') \\
&\qquad\qquad\times \lim_{\epsilon \to 0^+} \int_0^\infty \mathe^{\frac{\mathi \sigma}{2 \tau}} \tau^{-1-\frac{n}{2}} \mathe^{- \epsilon \tau} \total \tau \\
&\qquad+ \mathe^{- \mathi \frac{n}{4} \pi} \frac{\sqrt{\Delta}}{(4 \pi)^\frac{n}{2}} \left[ \frac{\zeta}{2\zeta-1} - \frac{n}{2 (n-1) - (n-2) \lambda} \right] \bigg[ - \frac{1}{2} \nabla_\mu \sigma \nabla_\nu \sigma A_1(x,x') \\
&\qquad\qquad+ \left( 2 \nabla_{(\mu} \sigma \nabla_{\nu)} + \nabla_\mu \nabla_\nu \sigma + \nabla_{(\mu} \ln \Delta \nabla_{\nu)} \sigma \right) A_0(x,x') \bigg] \lim_{\epsilon \to 0^+} \int_0^\infty \mathe^{\frac{\mathi \sigma}{2 \tau}} \tau^{-\frac{n}{2}} \mathe^{- \epsilon \tau} \total \tau \\
&\qquad+ 2 \mathe^{- \mathi \frac{n}{4} \pi} \frac{\sqrt{\Delta}}{(4 \pi)^\frac{n}{2}} \left[ \frac{\zeta}{2\zeta-1} - \frac{n}{2 (n-1) - (n-2) \lambda} \right] \sum_{k=0}^\infty \mathi^k \bigg[ \\
&\qquad\qquad- \left( \nabla_{(\mu} \sigma \nabla_{\nu)} + \frac{1}{2} \nabla_{(\mu} \ln \Delta \nabla_{\nu)} \sigma + \frac{1}{2} \nabla_\mu \nabla_\nu \sigma \right) A_{k+1}(x,x') \\
&\qquad\qquad+ \left( \nabla_\mu \nabla_\nu + \nabla_{(\mu} \ln \Delta \nabla_{\nu)} + \frac{1}{2} \nabla_\mu \nabla_\nu \ln \Delta + \frac{1}{4} \nabla_\mu \ln \Delta \nabla_\nu \ln \Delta \right) A_k(x,x') \\
&\qquad\qquad+ \frac{1}{4} \nabla_\mu \sigma \nabla_\nu \sigma A_{k+2}(x,x') \bigg] \lim_{\epsilon \to 0^+} \int_0^\infty \mathe^{\frac{\mathi \sigma}{2 \tau}} \tau^{k+1-\frac{n}{2}} \mathe^{- \epsilon \tau} \total \tau \\
&\qquad+ 2 \mathe^{- \mathi \frac{n-2}{4} \pi} \frac{\sqrt{\Delta}}{(4 \pi)^\frac{n}{2}} \frac{g_{\mu\nu}}{2 (n-1) - (n-2) \lambda} \sum_{k=0}^\infty \mathi^k A_k(x,x') \lim_{\epsilon \to 0^+} \int_0^\infty \mathe^{\frac{\mathi \sigma}{2 \tau}} \tau^{k-\frac{n}{2}} \mathe^{- \epsilon \tau} \total \tau \eqend{,}
\end{splitequation}
where we used the limit
\begin{equation}
\lim_{m \to 0} \frac{1 - \mathe^{- \mathi M_B^2 \tau}}{m^2} = \mathi \frac{\zeta}{2\zeta-1} (\xi-\lambda) \tau \eqend{,}
\end{equation}
and
\begin{splitequation}
\label{eq:field_propagator_massless_0000_heatkernel}
\lim_{m \to 0} G_{\circ\circ\circ\circ}(x,x') &\sim \mathe^{- \mathi \frac{n}{4} \pi} \frac{\sqrt{\Delta}}{(4 \pi)^\frac{n}{2}} \frac{(n-2)}{2 (n-1) - (n-2) \lambda} \lim_{\epsilon \to 0^+} \sum_{k=0}^\infty \mathi^k A_k(x,x') \\
&\qquad\times \int_0^\infty \mathe^{\frac{\mathi \sigma}{2 \tau}} \tau^{k-\frac{n}{2}} \mathe^{- \epsilon \tau} \total \tau \eqend{.}
\end{splitequation}

Since we wanted to compute the heat kernel expansion of the various propagators in full generality, the resulting expressions are quite lengthy. For practical computations, it is clearly advisable to use a computer algebra system, such as the \textsc{xAct} tensor package~\cite{xact,martingarcia2008,brizuelaetal2009,nutma2014} and the \textsc{FieldsX} extension~\cite{fieldsx}, and we include a Mathematica notebook as supplementary material. However, we note that the expansion simplifies substantially in the coincidence limit, which is all that is needed to determine the counterterms needed for renormalization, or anomalies.

\section{Discussion}
\label{sec:discussion}

In this paper we have presented for the first time the propagators for linearized massive quantum gravity in a family of general linear covariant gauges. We have considered the usual Einstein--Hilbert action for linearized gravity with a cosmological constant and a Fierz--Pauli mass term, and assumed that the background spacetime is solution to the vacuum Einstein equation. To preserve the diffeomorphism invariance of the massless theory and make that limit smooth, we have used Stueckelberg's trick~\cite{Hinterbichler:2011tt,derham2014} of introducing compensating vector and scalar fields to the action. The resulting theory is invariant under linearized diffeomorphisms and has an additional symmetry under the transformation~\eqref{eq:amuphi_gauge} of the Stueckelberg fields, although only the Stueckelberg vector field changes under both transformations. Classicaly, the tensor, vector and scalars sectors of the theory decouple for $\Lambda = 0$ in the massless limit, and we have rescaled the vector and scalar fields using the mass parameter $m$ to make this decoupling evident.

In order to quantize this theory, we have employed the familiar BRST formalism, where we introduced ghost, antighost and auxiliary fields and imposed a general linear covariant gauge condition on the gauge fields. To construct the gauge-fixed propagators of the various fields, we have first considered scalar, vector and rank-2 tensor propagators in Feynman-type gauges. In Feynman-type gauges the field equations are normally hyperbolic, i.e., the second derivative only enters in the form of the d'Alembert operator $\nabla^2$. Normally hyperbolic equations have a well-posed Cauchy problem on globally hyperbolic spacetimes and therefore have a unique solution. The differential operator appearing in the field equations in more general gauges, as in our case, have second derivatives other than $\nabla^2$ and are only Green hyperbolic, for which uniqueness of the solution is guaranteed but not their existence~\cite{baerginouxpfaeffle2007,baerginoux2012,baer2015}.

To determine the propagators in general linear covariant gauges, we have derived identities that relate the scalar, vector and rank-2 tensor propagators and their derivatives in Feynman-type gauges. Using these identities, which might also be of independent interest themselves, we have then expressed the propagators in general linear covariant gauges as linear combinations of the propagators in the Feynman-type gauges, of different masses and including their derivatives.

We have then analyzed the massless limit $m \to 0$ and the limit of vanishing cosmological constant $\Lambda \to 0$ of our results for the propagators in general linear covariant gauges. The limits of the propagators are well defined when both $m$ and $\Lambda$ vanish, but their ratio $\lambda = 4/(n-2) \Lambda m^{-2}$ is held fixed. In this limit, the vector field $A_\mu$ decouples from the other fields, such that the mixed propagators involving $A_\mu$ vanish. However, the scalar field $\phi$ remains coupled to $h_{\mu\nu}$, a consequence of the so-called vDVZ discontinuity~\cite{vandamveltman1970,zakharov1970}. The limit where $\Lambda$ is kept fixed but $m$ vanishes is not well defined as some propagators diverge for small masses, ultimately a result of our rescaling of the Stueckelberg scalar and vector fields. However, the diverging terms in this case are pure gauge, i.e., they have the form of a (operator-valued) gauge transformation, and thus do not contribute to the correlators of invariant observables. The limit where we first send $\Lambda \to 0$ and then take $m \to 0$, on the other hand, produces well defined propagators. Here again the vDVZ discontinuity arises, where $A_\mu$ decouples from other fields, but $\phi$ and $h_{\mu\nu}$ remain coupled. Note that the limits $\Lambda \to 0$ and $m \to 0$ do not commute, which is another manifestation of the vDVZ discontinuity.

We also considered the heat kernel expansion of the massive gravity propagators in a general linear covariant gauge, which are given in Eqs.~\eqref{eq:field_propagator_munurhosigma_heatkernel}--\eqref{eq:field_propagator_0000_heatkernel} for the field sector and Eqs.~\eqref{eq:ghost_propagator_munu_heatkernel}--\eqref{eq:ghost_propagator_00_heatkernel} for the ghost sector. The asymptotic expansion of the heat kernel for small values of the proper time parameter $\tau$ then gives information about the short distance behavior of the propagators. For general linear covariant gauges, the heat kernel expansion is hard to obtain as the field equations are only Green hyperbolic. However, since in these gauges the propagators can be expressed as a combination of (derivatives of) propagators in Feynman-type gauges, we can focus on the heat kernel expansion of the latter. The heat kernel expansion of the propagators in the Feynman-type gauges is much easier to work out as they satisfy normally hyperbolic field equations. Hence, we have first written down the transport equation satisfied by the coefficients of the heat kernel expansion of the propagator of tensor fields on a general bundle (in particular of general spin) satisfying a normally hyperbolic equation and non-minimally coupled to the background geometry. As usual, the transport equation can be solved recursively, starting from the initial condition on the most singular term as $\tau \to 0$, where the propagator becomes proportional to a (spacetime) Dirac $\delta$ distribution. The transport equation is simple to solve exactly for the zeroth order coefficient, which essentially is given by the parallel propagator on the bundle. The solution for the higher-order coefficients were then obtained recursively, using Riemann normal coordinates.

For a range of applications, it is also interesting to have the explicit formulas for the coincidence limit of the heat kernel coefficients and some of their derivatives written in terms of the geometric background curvature tensors. Using the transport equation, we have computed the formulas for the first- and second-order heat kernel coefficients and some of their derivatives for a tensor field on a generic bundle, and then specialized the formulas for the second-order coefficient in the case of scalar, vector and rank-2 tensor fields. In the vector and rank-2 tensor cases, the heat kernel coefficients presented here have been computed in full generality for the first time. We remark that no decomposition of the tensor fields into transverse and longitudinal components nor any subtraction of traces is needed in our approach, which certainly simplifies the use of our results. Moreover, they can also be used for the standard massless Einstein--Hilbert gravity, simply by taking the massless limit of the coefficients, or in general the heat kernel expansion, which are given in Eqs.~\eqref{eq:field_propagator_massless_munu00_heatkernel}--\eqref{eq:field_propagator_massless_0000_heatkernel} for the field sector and Eq.~\eqref{eq:ghost_propagator_massless_munu_heatkernel} for the ghost sector.

Some more technical results of this work include novel relations between the heat kernel coefficients of different fields, which stem from the identities for traces and divergences of the propagators in Feynman-type gauges. These identities relate (the coincidence limit of) derivatives of the coefficients to coefficients of lower spin, as well as the trace of the rank-2 tensor heat kernel coefficients to scalar heat kernel coefficients. Summarizing, our computations are aimed to furnish a toolbox of technical instruments towards the evaluation of renormalized physical quantities in massive (and massless) gravity. To facilitate this work, we include a Mathematica notebook as supplementary material. As checks on our results, we have verified that they have a finite massless limit (if this limit is taken in a suitable way), that we recover the known vDVZ discontinuity, and that our results reduce to known expressions for certain choices of the gauge parameters which had been studied before. In particular, in the analogue~\eqref{eq:gauge_feynman} of Feynman gauge it is straightforward to verify from the results~\eqref{eq:field_propagator_massless_munu00_heatkernel}--\eqref{eq:field_propagator_massless_0000_heatkernel} and~\eqref{eq:ghost_propagator_massless_munu_heatkernel} that all derivatives of heat kernel coefficients which appear for non-minimal operators drop out and only the coefficients of the minimal operators remain, which is an important check.

As a future application, we plan to apply our results in order to compute the trace of the energy-momentum tensor in massive gravity and its anomaly. While there are many proposals for an energy-momentum tensor of gravity itself, the proposed quantities are either not a tensor, not conserved or trivial on-shell. Ultimately, this failure can be traced back to the equivalence principle: locally, it is always possible to pass to an inertial system where gravity is absent, and thus a local energy-momentum tensor for the gravitational field cannot exist. Even if one restricts to small perturbations around a background spacetime (as we did in this work), where one can obtain a quantity quadratic in the perturbations that integrates to the total energy and momentum of these perturbations, this quantity is not gauge-invariant. Therefore, also its trace anomaly will be gauge-dependent. To resolve this problem, one has to employ relational observables, which are gauge-invariant but necessarily non-local objects. Apart from the quantity itself that one wants to consider, relational observables also depends on a choice of physical coordinate system, which can for example be furnished by matter fields, or constructed from metric perturbations. Various examples of such systems have been worked out in detail, see for example Refs.~\onlinecite{brownmarolf1996,dittrich2006,nakamura2007,ponsetal2009,gieseletal2010,gasperinietal2011,tambornino2012,donnellygiddings2016,brunettietal2016,gieseletal2018,froeblima2018,fanizzaetal2021,mitsouetal2021,froeblima2022,baldazzietal2022,froeblima2023,goelleretal2022} and references therein. By constructing a relational observable from the energy-momentum tensor for small perturbations, we will be able to obtain a gauge-invariant relational trace anomaly.

\section{Supplementary material}

We include a Mathematica notebook as supplementary material, which contains all the computations and results of this work.

\begin{acknowledgments}
R.~F. thanks Martin Reuter for helpful discussions. M.~B.~F. thanks Martin Reuter, R.~F. and the THEP group at the University of Mainz for their hospitality.

This work is supported by the Deutsche Forschungsgemeinschaft (DFG, German Research Foundation) --- project no. 396692871 within the Emmy Noether grant CA1850/1-1.
\end{acknowledgments}

\bibliography{references}

\begin{thebibliography}{179}%
\makeatletter
\providecommand \@ifxundefined [1]{%
 \@ifx{#1\undefined}
}%
\providecommand \@ifnum [1]{%
 \ifnum #1\expandafter \@firstoftwo
 \else \expandafter \@secondoftwo
 \fi
}%
\providecommand \@ifx [1]{%
 \ifx #1\expandafter \@firstoftwo
 \else \expandafter \@secondoftwo
 \fi
}%
\providecommand \natexlab [1]{#1}%
\providecommand \enquote  [1]{``#1''}%
\providecommand \bibnamefont  [1]{#1}%
\providecommand \bibfnamefont [1]{#1}%
\providecommand \citenamefont [1]{#1}%
\providecommand \href@noop [0]{\@secondoftwo}%
\providecommand \href [0]{\begingroup \@sanitize@url \@href}%
\providecommand \@href[1]{\@@startlink{#1}\@@href}%
\providecommand \@@href[1]{\endgroup#1\@@endlink}%
\providecommand \@sanitize@url [0]{\catcode `\\12\catcode `\$12\catcode
  `\&12\catcode `\#12\catcode `\^12\catcode `\_12\catcode `\%12\relax}%
\providecommand \@@startlink[1]{}%
\providecommand \@@endlink[0]{}%
\providecommand \url  [0]{\begingroup\@sanitize@url \@url }%
\providecommand \@url [1]{\endgroup\@href {#1}{\urlprefix }}%
\providecommand \urlprefix  [0]{URL }%
\providecommand \Eprint [0]{\href }%
\providecommand \doibase [0]{https://doi.org/}%
\providecommand \selectlanguage [0]{\@gobble}%
\providecommand \bibinfo  [0]{\@secondoftwo}%
\providecommand \bibfield  [0]{\@secondoftwo}%
\providecommand \translation [1]{[#1]}%
\providecommand \BibitemOpen [0]{}%
\providecommand \bibitemStop [0]{}%
\providecommand \bibitemNoStop [0]{.\EOS\space}%
\providecommand \EOS [0]{\spacefactor3000\relax}%
\providecommand \BibitemShut  [1]{\csname bibitem#1\endcsname}%
\let\auto@bib@innerbib\@empty
\bibitem [{\citenamefont {Avramidi}(2000)}]{Avramidi:2000bm}%
  \BibitemOpen
  \bibfield  {author} {\bibinfo {author} {\bibfnamefont {I.~G.}\ \bibnamefont
  {Avramidi}},\ }\href {https://doi.org/10.1007/3-540-46523-5} {\emph {\bibinfo
  {title} {{Heat kernel and quantum gravity}}}},\ Vol.~\bibinfo {volume} {64}\
  (\bibinfo  {publisher} {Springer},\ \bibinfo {address} {New York, USA},\
  \bibinfo {year} {2000})\BibitemShut {NoStop}%
\bibitem [{\citenamefont {Barvinsky}\ and\ \citenamefont
  {Vilkovisky}(1985)}]{Barvinsky:1985an}%
  \BibitemOpen
  \bibfield  {author} {\bibinfo {author} {\bibfnamefont {A.~O.}\ \bibnamefont
  {Barvinsky}}\ and\ \bibinfo {author} {\bibfnamefont {G.~A.}\ \bibnamefont
  {Vilkovisky}},\ }\bibfield  {title} {\enquote {\bibinfo {title} {{The
  generalized Schwinger-Dewitt technique in gauge theories and quantum
  gravity}},}\ }\href {https://doi.org/10.1016/0370-1573(85)90148-6} {\bibfield
   {journal} {\bibinfo  {journal} {Phys. Rept.}\ }\textbf {\bibinfo {volume}
  {119}},\ \bibinfo {pages} {1--74} (\bibinfo {year} {1985})}\BibitemShut
  {NoStop}%
\bibitem [{\citenamefont {Barvinsky}\ and\ \citenamefont
  {Vilkovisky}(1990)}]{Barvinsky:1990up}%
  \BibitemOpen
  \bibfield  {author} {\bibinfo {author} {\bibfnamefont {A.~O.}\ \bibnamefont
  {Barvinsky}}\ and\ \bibinfo {author} {\bibfnamefont {G.~A.}\ \bibnamefont
  {Vilkovisky}},\ }\bibfield  {title} {\enquote {\bibinfo {title} {{Covariant
  perturbation theory (II). Second order in the curvature. General
  algorithms}},}\ }\href {https://doi.org/10.1016/0550-3213(90)90047-H}
  {\bibfield  {journal} {\bibinfo  {journal} {Nucl. Phys. B}\ }\textbf
  {\bibinfo {volume} {333}},\ \bibinfo {pages} {471--511} (\bibinfo {year}
  {1990})}\BibitemShut {NoStop}%
\bibitem [{\citenamefont {Vilkovisky}(1992)}]{Vilkovisky:1992za}%
  \BibitemOpen
  \bibfield  {author} {\bibinfo {author} {\bibfnamefont {G.~A.}\ \bibnamefont
  {Vilkovisky}},\ }\bibfield  {title} {\enquote {\bibinfo {title} {{Heat
  kernel: Rencontre entre physiciens et math{\'e}maticiens}},}\ }in\ \href
  {http://www.numdam.org/item/RCP25_1992__43__203_0/} {\emph {\bibinfo
  {booktitle} {Les rencontres physiciens-math\'ematiciens de Strasbourg -
  RCP25}}},\ Vol.~\bibinfo {volume} {43}\ (\bibinfo {year} {1992})\ pp.\
  \bibinfo {pages} {203--224}\BibitemShut {NoStop}%
\bibitem [{\citenamefont {Vassilevich}(2003)}]{Vassilevich:2003xt}%
  \BibitemOpen
  \bibfield  {author} {\bibinfo {author} {\bibfnamefont {D.~V.}\ \bibnamefont
  {Vassilevich}},\ }\bibfield  {title} {\enquote {\bibinfo {title} {{Heat
  kernel expansion: User's manual}},}\ }\href
  {https://doi.org/10.1016/j.physrep.2003.09.002} {\bibfield  {journal}
  {\bibinfo  {journal} {Phys. Rept.}\ }\textbf {\bibinfo {volume} {388}},\
  \bibinfo {pages} {279--360} (\bibinfo {year} {2003})},\ \Eprint
  {https://arxiv.org/abs/hep-th/0306138} {arXiv:hep-th/0306138} \BibitemShut
  {NoStop}%
\bibitem [{\citenamefont {Codello}\ and\ \citenamefont
  {Zanusso}(2013)}]{Codello:2012kq}%
  \BibitemOpen
  \bibfield  {author} {\bibinfo {author} {\bibfnamefont {A.}~\bibnamefont
  {Codello}}\ and\ \bibinfo {author} {\bibfnamefont {O.}~\bibnamefont
  {Zanusso}},\ }\bibfield  {title} {\enquote {\bibinfo {title} {{On the
  non-local heat kernel expansion}},}\ }\href
  {https://doi.org/10.1063/1.4776234} {\bibfield  {journal} {\bibinfo
  {journal} {J. Math. Phys.}\ }\textbf {\bibinfo {volume} {54}},\ \bibinfo
  {pages} {013513} (\bibinfo {year} {2013})},\ \Eprint
  {https://arxiv.org/abs/1203.2034} {arXiv:1203.2034 [math-ph]} \BibitemShut
  {NoStop}%
\bibitem [{\citenamefont {Reuter}\ and\ \citenamefont
  {Saueressig}(2019)}]{Reuter:2019byg}%
  \BibitemOpen
  \bibfield  {author} {\bibinfo {author} {\bibfnamefont {M.}~\bibnamefont
  {Reuter}}\ and\ \bibinfo {author} {\bibfnamefont {F.}~\bibnamefont
  {Saueressig}},\ }\href@noop {} {\emph {\bibinfo {title} {{Quantum Gravity and
  the Functional Renormalization Group}: {The Road towards Asymptotic
  Safety}}}}\ (\bibinfo  {publisher} {Cambridge University Press},\ \bibinfo
  {year} {2019})\BibitemShut {NoStop}%
\bibitem [{\citenamefont {Percacci}(2017)}]{Percacci:2017fkn}%
  \BibitemOpen
  \bibfield  {author} {\bibinfo {author} {\bibfnamefont {R.}~\bibnamefont
  {Percacci}},\ }\href {https://doi.org/10.1142/10369} {\emph {\bibinfo {title}
  {{An Introduction to Covariant Quantum Gravity and Asymptotic Safety}}}},\
  \bibinfo {series} {100 Years of General Relativity}, Vol.~\bibinfo {volume}
  {3}\ (\bibinfo  {publisher} {World Scientific},\ \bibinfo {year}
  {2017})\BibitemShut {NoStop}%
\bibitem [{\citenamefont {Reuter}(1998)}]{reuter_prd_1998}%
  \BibitemOpen
  \bibfield  {author} {\bibinfo {author} {\bibfnamefont {M.}~\bibnamefont
  {Reuter}},\ }\bibfield  {title} {\enquote {\bibinfo {title} {{Nonperturbative
  evolution equation for quantum gravity}},}\ }\href
  {https://doi.org/10.1103/PhysRevD.57.971} {\bibfield  {journal} {\bibinfo
  {journal} {Phys. Rev. D}\ }\textbf {\bibinfo {volume} {57}},\ \bibinfo
  {pages} {971--985} (\bibinfo {year} {1998})},\ \Eprint
  {https://arxiv.org/abs/hep-th/9605030} {arXiv:hep-th/9605030} \BibitemShut
  {NoStop}%
\bibitem [{\citenamefont {Lauscher}\ and\ \citenamefont
  {Reuter}(2002)}]{lauscher_reuter_prd_2002}%
  \BibitemOpen
  \bibfield  {author} {\bibinfo {author} {\bibfnamefont {O.}~\bibnamefont
  {Lauscher}}\ and\ \bibinfo {author} {\bibfnamefont {M.}~\bibnamefont
  {Reuter}},\ }\bibfield  {title} {\enquote {\bibinfo {title} {{Flow equation
  of quantum Einstein gravity in a higher derivative truncation}},}\ }\href
  {https://doi.org/10.1103/PhysRevD.66.025026} {\bibfield  {journal} {\bibinfo
  {journal} {Phys. Rev. D}\ }\textbf {\bibinfo {volume} {66}},\ \bibinfo
  {pages} {025026} (\bibinfo {year} {2002})},\ \Eprint
  {https://arxiv.org/abs/hep-th/0205062} {arXiv:hep-th/0205062} \BibitemShut
  {NoStop}%
\bibitem [{\citenamefont {Codello}, \citenamefont {Percacci},\ and\
  \citenamefont {Rahmede}(2009)}]{Codello:2008vh}%
  \BibitemOpen
  \bibfield  {author} {\bibinfo {author} {\bibfnamefont {A.}~\bibnamefont
  {Codello}}, \bibinfo {author} {\bibfnamefont {R.}~\bibnamefont {Percacci}},\
  and\ \bibinfo {author} {\bibfnamefont {C.}~\bibnamefont {Rahmede}},\
  }\bibfield  {title} {\enquote {\bibinfo {title} {{Investigating the
  Ultraviolet Properties of Gravity with a Wilsonian Renormalization Group
  Equation}},}\ }\href {https://doi.org/10.1016/j.aop.2008.08.008} {\bibfield
  {journal} {\bibinfo  {journal} {Annals Phys.}\ }\textbf {\bibinfo {volume}
  {324}},\ \bibinfo {pages} {414--469} (\bibinfo {year} {2009})},\ \Eprint
  {https://arxiv.org/abs/0805.2909} {arXiv:0805.2909 [hep-th]} \BibitemShut
  {NoStop}%
\bibitem [{\citenamefont {Knorr}(2021)}]{Knorr:2021slg}%
  \BibitemOpen
  \bibfield  {author} {\bibinfo {author} {\bibfnamefont {B.}~\bibnamefont
  {Knorr}},\ }\bibfield  {title} {\enquote {\bibinfo {title} {{The derivative
  expansion in asymptotically safe quantum gravity: general setup and quartic
  order}},}\ }\href {https://doi.org/10.21468/SciPostPhysCore.4.3.020}
  {\bibfield  {journal} {\bibinfo  {journal} {SciPost Phys. Core}\ }\textbf
  {\bibinfo {volume} {4}},\ \bibinfo {pages} {020} (\bibinfo {year} {2021})},\
  \Eprint {https://arxiv.org/abs/2104.11336} {arXiv:2104.11336 [hep-th]}
  \BibitemShut {NoStop}%
\bibitem [{\citenamefont {Manrique}, \citenamefont {Rechenberger},\ and\
  \citenamefont {Saueressig}(2011)}]{Manrique:2011jc}%
  \BibitemOpen
  \bibfield  {author} {\bibinfo {author} {\bibfnamefont {E.}~\bibnamefont
  {Manrique}}, \bibinfo {author} {\bibfnamefont {S.}~\bibnamefont
  {Rechenberger}},\ and\ \bibinfo {author} {\bibfnamefont {F.}~\bibnamefont
  {Saueressig}},\ }\bibfield  {title} {\enquote {\bibinfo {title}
  {{Asymptotically Safe Lorentzian Gravity}},}\ }\href
  {https://doi.org/10.1103/PhysRevLett.106.251302} {\bibfield  {journal}
  {\bibinfo  {journal} {Phys. Rev. Lett.}\ }\textbf {\bibinfo {volume} {106}},\
  \bibinfo {pages} {251302} (\bibinfo {year} {2011})},\ \Eprint
  {https://arxiv.org/abs/1102.5012} {arXiv:1102.5012 [hep-th]} \BibitemShut
  {NoStop}%
\bibitem [{\citenamefont {Fehre}\ \emph {et~al.}(2023)\citenamefont {Fehre},
  \citenamefont {Litim}, \citenamefont {Pawlowski},\ and\ \citenamefont
  {Reichert}}]{Fehre:2021eob}%
  \BibitemOpen
  \bibfield  {author} {\bibinfo {author} {\bibfnamefont {J.}~\bibnamefont
  {Fehre}}, \bibinfo {author} {\bibfnamefont {D.~F.}\ \bibnamefont {Litim}},
  \bibinfo {author} {\bibfnamefont {J.~M.}\ \bibnamefont {Pawlowski}},\ and\
  \bibinfo {author} {\bibfnamefont {M.}~\bibnamefont {Reichert}},\ }\bibfield
  {title} {\enquote {\bibinfo {title} {{Lorentzian Quantum Gravity and the
  Graviton Spectral Function}},}\ }\href
  {https://doi.org/10.1103/PhysRevLett.130.081501} {\bibfield  {journal}
  {\bibinfo  {journal} {Phys. Rev. Lett.}\ }\textbf {\bibinfo {volume} {130}},\
  \bibinfo {pages} {081501} (\bibinfo {year} {2023})},\ \Eprint
  {https://arxiv.org/abs/2111.13232} {arXiv:2111.13232 [hep-th]} \BibitemShut
  {NoStop}%
\bibitem [{\citenamefont {Banerjee}\ and\ \citenamefont
  {Niedermaier}(2022)}]{Banerjee:2022xvi}%
  \BibitemOpen
  \bibfield  {author} {\bibinfo {author} {\bibfnamefont {R.}~\bibnamefont
  {Banerjee}}\ and\ \bibinfo {author} {\bibfnamefont {M.}~\bibnamefont
  {Niedermaier}},\ }\bibfield  {title} {\enquote {\bibinfo {title} {{The
  spatial Functional Renormalization Group and Hadamard states on cosmological
  spacetimes}},}\ }\href {https://doi.org/10.1016/j.nuclphysb.2022.115814}
  {\bibfield  {journal} {\bibinfo  {journal} {Nucl. Phys. B}\ }\textbf
  {\bibinfo {volume} {980}},\ \bibinfo {pages} {115814} (\bibinfo {year}
  {2022})},\ \Eprint {https://arxiv.org/abs/2201.02575} {arXiv:2201.02575
  [hep-th]} \BibitemShut {NoStop}%
\bibitem [{\citenamefont {Saueressig}\ and\ \citenamefont
  {Wang}(2023)}]{Saueressig:2023tfy}%
  \BibitemOpen
  \bibfield  {author} {\bibinfo {author} {\bibfnamefont {F.}~\bibnamefont
  {Saueressig}}\ and\ \bibinfo {author} {\bibfnamefont {J.}~\bibnamefont
  {Wang}},\ }\bibfield  {title} {\enquote {\bibinfo {title} {{Foliated
  asymptotically safe gravity in the fluctuation approach}},}\ }\href
  {https://doi.org/10.1007/JHEP09(2023)064} {\bibfield  {journal} {\bibinfo
  {journal} {JHEP}\ }\textbf {\bibinfo {volume} {09}},\ \bibinfo {pages} {064}
  (\bibinfo {year} {2023})},\ \Eprint {https://arxiv.org/abs/2306.10408}
  {arXiv:2306.10408 [hep-th]} \BibitemShut {NoStop}%
\bibitem [{\citenamefont {D'Angelo}\ and\ \citenamefont
  {Pinamonti}(2023)}]{DAngelo:2023ssw}%
  \BibitemOpen
  \bibfield  {author} {\bibinfo {author} {\bibfnamefont {E.}~\bibnamefont
  {D'Angelo}}\ and\ \bibinfo {author} {\bibfnamefont {N.}~\bibnamefont
  {Pinamonti}},\ }\bibfield  {title} {\enquote {\bibinfo {title} {{Local
  solutions of RG flow equations from the Nash-Moser theorem}},}\ }\href@noop
  {} {\  (\bibinfo {year} {2023})},\ \Eprint {https://arxiv.org/abs/2310.20596}
  {arXiv:2310.20596 [math-ph]} \BibitemShut {NoStop}%
\bibitem [{\citenamefont {D'Angelo}(2024)}]{DAngelo:2023wje}%
  \BibitemOpen
  \bibfield  {author} {\bibinfo {author} {\bibfnamefont {E.}~\bibnamefont
  {D'Angelo}},\ }\bibfield  {title} {\enquote {\bibinfo {title} {{Asymptotic
  safety in Lorentzian quantum gravity}},}\ }\href
  {https://doi.org/10.1103/PhysRevD.109.066012} {\bibfield  {journal} {\bibinfo
   {journal} {Phys. Rev. D}\ }\textbf {\bibinfo {volume} {109}},\ \bibinfo
  {pages} {066012} (\bibinfo {year} {2024})},\ \Eprint
  {https://arxiv.org/abs/2310.20603} {arXiv:2310.20603 [hep-th]} \BibitemShut
  {NoStop}%
\bibitem [{\citenamefont {Becchi}, \citenamefont {Rouet},\ and\ \citenamefont
  {Stora}(1976)}]{becchietal1975}%
  \BibitemOpen
  \bibfield  {author} {\bibinfo {author} {\bibfnamefont {C.}~\bibnamefont
  {Becchi}}, \bibinfo {author} {\bibfnamefont {A.}~\bibnamefont {Rouet}},\ and\
  \bibinfo {author} {\bibfnamefont {R.}~\bibnamefont {Stora}},\ }\bibfield
  {title} {\enquote {\bibinfo {title} {{Renormalization of Gauge Theories}},}\
  }\href {https://doi.org/10.1016/0003-4916(76)90156-1} {\bibfield  {journal}
  {\bibinfo  {journal} {Ann.~Phys.}\ }\textbf {\bibinfo {volume} {98}},\
  \bibinfo {pages} {287} (\bibinfo {year} {1976})}\BibitemShut {NoStop}%
\bibitem [{\citenamefont {Weinberg}(2005)}]{weinberg_v2}%
  \BibitemOpen
  \bibfield  {author} {\bibinfo {author} {\bibfnamefont {S.}~\bibnamefont
  {Weinberg}},\ }\href@noop {} {\emph {\bibinfo {title} {{The Quantum Theory of
  Fields, Volume 2: Modern Applications}}}}\ (\bibinfo  {publisher} {Cambridge
  University Press},\ \bibinfo {address} {Cambridge, UK},\ \bibinfo {year}
  {2005})\BibitemShut {NoStop}%
\bibitem [{\citenamefont {Barnich}, \citenamefont {Brandt},\ and\ \citenamefont
  {Henneaux}(2000)}]{barnichetal2000}%
  \BibitemOpen
  \bibfield  {author} {\bibinfo {author} {\bibfnamefont {G.}~\bibnamefont
  {Barnich}}, \bibinfo {author} {\bibfnamefont {F.}~\bibnamefont {Brandt}},\
  and\ \bibinfo {author} {\bibfnamefont {M.}~\bibnamefont {Henneaux}},\
  }\bibfield  {title} {\enquote {\bibinfo {title} {{Local BRST cohomology in
  gauge theories}},}\ }\href {https://doi.org/10.1016/S0370-1573(00)00049-1}
  {\bibfield  {journal} {\bibinfo  {journal} {Phys.~Rept.}\ }\textbf {\bibinfo
  {volume} {338}},\ \bibinfo {pages} {439} (\bibinfo {year} {2000})},\ \Eprint
  {https://arxiv.org/abs/hep-th/0002245} {arXiv:hep-th/0002245 [hep-th]}
  \BibitemShut {NoStop}%
\bibitem [{\citenamefont {Collins}(1986)}]{collinsbook}%
  \BibitemOpen
  \bibfield  {author} {\bibinfo {author} {\bibfnamefont {J.~C.}\ \bibnamefont
  {Collins}},\ }\href {https://doi.org/10.1017/CBO9780511622656} {\emph
  {\bibinfo {title} {{Renormalization}: {An Introduction to Renormalization,
  The Renormalization Group, and the Operator Product Expansion}}}},\ \bibinfo
  {series} {Cambridge Monographs on Mathematical Physics}, Vol.~\bibinfo
  {volume} {26}\ (\bibinfo  {publisher} {Cambridge University Press},\ \bibinfo
  {address} {Cambridge, UK},\ \bibinfo {year} {1986})\BibitemShut {NoStop}%
\bibitem [{\citenamefont {Pagani}\ and\ \citenamefont
  {Reuter}(2017)}]{Pagani:2016dof}%
  \BibitemOpen
  \bibfield  {author} {\bibinfo {author} {\bibfnamefont {C.}~\bibnamefont
  {Pagani}}\ and\ \bibinfo {author} {\bibfnamefont {M.}~\bibnamefont
  {Reuter}},\ }\bibfield  {title} {\enquote {\bibinfo {title} {{Composite
  Operators in Asymptotic Safety}},}\ }\href
  {https://doi.org/10.1103/PhysRevD.95.066002} {\bibfield  {journal} {\bibinfo
  {journal} {Phys. Rev. D}\ }\textbf {\bibinfo {volume} {95}},\ \bibinfo
  {pages} {066002} (\bibinfo {year} {2017})},\ \Eprint
  {https://arxiv.org/abs/1611.06522} {arXiv:1611.06522 [gr-qc]} \BibitemShut
  {NoStop}%
\bibitem [{\citenamefont {Houthoff}, \citenamefont {Kurov},\ and\ \citenamefont
  {Saueressig}(2020)}]{Houthoff:2020zqy}%
  \BibitemOpen
  \bibfield  {author} {\bibinfo {author} {\bibfnamefont {W.}~\bibnamefont
  {Houthoff}}, \bibinfo {author} {\bibfnamefont {A.}~\bibnamefont {Kurov}},\
  and\ \bibinfo {author} {\bibfnamefont {F.}~\bibnamefont {Saueressig}},\
  }\bibfield  {title} {\enquote {\bibinfo {title} {{On the scaling of composite
  operators in asymptotic safety}},}\ }\href
  {https://doi.org/10.1007/JHEP04(2020)099} {\bibfield  {journal} {\bibinfo
  {journal} {JHEP}\ }\textbf {\bibinfo {volume} {04}},\ \bibinfo {pages} {099}
  (\bibinfo {year} {2020})},\ \Eprint {https://arxiv.org/abs/2002.00256}
  {arXiv:2002.00256 [hep-th]} \BibitemShut {NoStop}%
\bibitem [{\citenamefont {Becker}, \citenamefont {Pagani},\ and\ \citenamefont
  {Zanusso}(2020)}]{Becker:2019fhi}%
  \BibitemOpen
  \bibfield  {author} {\bibinfo {author} {\bibfnamefont {M.}~\bibnamefont
  {Becker}}, \bibinfo {author} {\bibfnamefont {C.}~\bibnamefont {Pagani}},\
  and\ \bibinfo {author} {\bibfnamefont {O.}~\bibnamefont {Zanusso}},\
  }\bibfield  {title} {\enquote {\bibinfo {title} {{Fractal Geometry of Higher
  Derivative Gravity}},}\ }\href
  {https://doi.org/10.1103/PhysRevLett.124.151302} {\bibfield  {journal}
  {\bibinfo  {journal} {Phys. Rev. Lett.}\ }\textbf {\bibinfo {volume} {124}},\
  \bibinfo {pages} {151302} (\bibinfo {year} {2020})},\ \Eprint
  {https://arxiv.org/abs/1911.02415} {arXiv:1911.02415 [gr-qc]} \BibitemShut
  {NoStop}%
\bibitem [{\citenamefont {Ambj\o{}rn}, \citenamefont {Durhuus},\ and\
  \citenamefont {Jonsson}(2005)}]{Ambjorn:1997di}%
  \BibitemOpen
  \bibfield  {author} {\bibinfo {author} {\bibfnamefont {J.}~\bibnamefont
  {Ambj\o{}rn}}, \bibinfo {author} {\bibfnamefont {B.}~\bibnamefont
  {Durhuus}},\ and\ \bibinfo {author} {\bibfnamefont {T.}~\bibnamefont
  {Jonsson}},\ }\href {https://doi.org/10.1017/CBO9780511524417} {\emph
  {\bibinfo {title} {{Quantum Geometry}: {A Statistical Field Theory
  Approach}}}},\ Cambridge Monographs on Mathematical Physics\ (\bibinfo
  {publisher} {Cambridge Univ. Press},\ \bibinfo {address} {Cambridge, UK},\
  \bibinfo {year} {2005})\BibitemShut {NoStop}%
\bibitem [{\citenamefont {Hamber}(2009)}]{Hamber:2009zz}%
  \BibitemOpen
  \bibfield  {author} {\bibinfo {author} {\bibfnamefont {H.~W.}\ \bibnamefont
  {Hamber}},\ }\href {https://doi.org/10.1007/978-3-540-85293-3} {\emph
  {\bibinfo {title} {{Quantum gravitation: The Feynman path integral
  approach}}}}\ (\bibinfo  {publisher} {Springer},\ \bibinfo {address}
  {Berlin},\ \bibinfo {year} {2009})\BibitemShut {NoStop}%
\bibitem [{\citenamefont {DeWitt}(1964)}]{DeWitt:1964mxt}%
  \BibitemOpen
  \bibfield  {author} {\bibinfo {author} {\bibfnamefont {B.~S.}\ \bibnamefont
  {DeWitt}},\ }\bibfield  {title} {\enquote {\bibinfo {title} {{Dynamical
  theory of groups and fields}},}\ }\href@noop {} {\bibfield  {journal}
  {\bibinfo  {journal} {Conf. Proc. C}\ }\textbf {\bibinfo {volume} {630701}},\
  \bibinfo {pages} {585--820} (\bibinfo {year} {1964})}\BibitemShut {NoStop}%
\bibitem [{\citenamefont {DeWitt}(1975)}]{DeWitt:1975ys}%
  \BibitemOpen
  \bibfield  {author} {\bibinfo {author} {\bibfnamefont {B.~S.}\ \bibnamefont
  {DeWitt}},\ }\bibfield  {title} {\enquote {\bibinfo {title} {{Quantum Field
  Theory in Curved Space-Time}},}\ }\href
  {https://doi.org/10.1016/0370-1573(75)90051-4} {\bibfield  {journal}
  {\bibinfo  {journal} {Phys. Rept.}\ }\textbf {\bibinfo {volume} {19}},\
  \bibinfo {pages} {295--357} (\bibinfo {year} {1975})}\BibitemShut {NoStop}%
\bibitem [{\citenamefont {Christensen}(1976)}]{Christensen:1976vb}%
  \BibitemOpen
  \bibfield  {author} {\bibinfo {author} {\bibfnamefont {S.~M.}\ \bibnamefont
  {Christensen}},\ }\bibfield  {title} {\enquote {\bibinfo {title} {{Vacuum
  Expectation Value of the Stress Tensor in an Arbitrary Curved Background: The
  Covariant Point Separation Method}},}\ }\href
  {https://doi.org/10.1103/PhysRevD.14.2490} {\bibfield  {journal} {\bibinfo
  {journal} {Phys. Rev. D}\ }\textbf {\bibinfo {volume} {14}},\ \bibinfo
  {pages} {2490--2501} (\bibinfo {year} {1976})}\BibitemShut {NoStop}%
\bibitem [{\citenamefont {Mandelstam}(1962)}]{mandelstam1962}%
  \BibitemOpen
  \bibfield  {author} {\bibinfo {author} {\bibfnamefont {S.}~\bibnamefont
  {Mandelstam}},\ }\bibfield  {title} {\enquote {\bibinfo {title}
  {{Quantization of the gravitational field}},}\ }\href
  {https://doi.org/10.1016/0003-4916(62)90233-6} {\bibfield  {journal}
  {\bibinfo  {journal} {Annals Phys.}\ }\textbf {\bibinfo {volume} {19}},\
  \bibinfo {pages} {25--66} (\bibinfo {year} {1962})}\BibitemShut {NoStop}%
\bibitem [{\citenamefont {Mandelstam}(1968)}]{mandelstam1968}%
  \BibitemOpen
  \bibfield  {author} {\bibinfo {author} {\bibfnamefont {S.}~\bibnamefont
  {Mandelstam}},\ }\bibfield  {title} {\enquote {\bibinfo {title} {{Feynman
  Rules for the Gravitational Field from the Coordinate-Independent
  Field-Theoretic Formalism}},}\ }\href
  {https://doi.org/10.1103/PhysRev.175.1604} {\bibfield  {journal} {\bibinfo
  {journal} {Phys. Rev.}\ }\textbf {\bibinfo {volume} {175}},\ \bibinfo {pages}
  {1604--1623} (\bibinfo {year} {1968})}\BibitemShut {NoStop}%
\bibitem [{\citenamefont {Tsamis}\ and\ \citenamefont
  {Woodard}(1992)}]{tsamiswoodard1992}%
  \BibitemOpen
  \bibfield  {author} {\bibinfo {author} {\bibfnamefont {N.~C.}\ \bibnamefont
  {Tsamis}}\ and\ \bibinfo {author} {\bibfnamefont {R.~P.}\ \bibnamefont
  {Woodard}},\ }\bibfield  {title} {\enquote {\bibinfo {title} {{Physical
  Green's Functions in Quantum Gravity}},}\ }\href
  {https://doi.org/10.1016/0003-4916(92)90301-2} {\bibfield  {journal}
  {\bibinfo  {journal} {Annals Phys.}\ }\textbf {\bibinfo {volume} {215}},\
  \bibinfo {pages} {96--155} (\bibinfo {year} {1992})}\BibitemShut {NoStop}%
\bibitem [{\citenamefont {Teitelboim}(1993)}]{teitelboim1993}%
  \BibitemOpen
  \bibfield  {author} {\bibinfo {author} {\bibfnamefont {C.}~\bibnamefont
  {Teitelboim}},\ }\bibfield  {title} {\enquote {\bibinfo {title} {{Gravitation
  theory in path space}},}\ }\href
  {https://doi.org/10.1016/0550-3213(93)90268-T} {\bibfield  {journal}
  {\bibinfo  {journal} {Nucl. Phys. B}\ }\textbf {\bibinfo {volume} {396}},\
  \bibinfo {pages} {303--325} (\bibinfo {year} {1993})}\BibitemShut {NoStop}%
\bibitem [{\citenamefont {Hamber}(1994)}]{hamber1994}%
  \BibitemOpen
  \bibfield  {author} {\bibinfo {author} {\bibfnamefont {H.~W.}\ \bibnamefont
  {Hamber}},\ }\bibfield  {title} {\enquote {\bibinfo {title} {{Invariant
  correlations in simplicial gravity}},}\ }\href
  {https://doi.org/10.1103/PhysRevD.50.3932} {\bibfield  {journal} {\bibinfo
  {journal} {Phys. Rev. D}\ }\textbf {\bibinfo {volume} {50}},\ \bibinfo
  {pages} {3932--3941} (\bibinfo {year} {1994})},\ \Eprint
  {https://arxiv.org/abs/hep-th/9311024} {arXiv:hep-th/9311024} \BibitemShut
  {NoStop}%
\bibitem [{\citenamefont {Modanese}(1994)}]{modanese1994}%
  \BibitemOpen
  \bibfield  {author} {\bibinfo {author} {\bibfnamefont {G.}~\bibnamefont
  {Modanese}},\ }\bibfield  {title} {\enquote {\bibinfo {title} {{Vacuum
  correlations at geodesic distance in quantum gravity}},}\ }\href
  {https://doi.org/10.1007/BF02724514} {\bibfield  {journal} {\bibinfo
  {journal} {Riv. Nuovo Cim.}\ }\textbf {\bibinfo {volume} {17N8}},\ \bibinfo
  {pages} {1--62} (\bibinfo {year} {1994})},\ \Eprint
  {https://arxiv.org/abs/hep-th/9410086} {arXiv:hep-th/9410086} \BibitemShut
  {NoStop}%
\bibitem [{\citenamefont {Ambj{\o}rn}\ and\ \citenamefont
  {Anagnostopoulos}(1997)}]{ambjorn1997}%
  \BibitemOpen
  \bibfield  {author} {\bibinfo {author} {\bibfnamefont {J.}~\bibnamefont
  {Ambj{\o}rn}}\ and\ \bibinfo {author} {\bibfnamefont {K.~N.}\ \bibnamefont
  {Anagnostopoulos}},\ }\bibfield  {title} {\enquote {\bibinfo {title}
  {{Quantum geometry of 2-D gravity coupled to unitary matter}},}\ }\href
  {https://doi.org/10.1016/S0550-3213(97)00259-9} {\bibfield  {journal}
  {\bibinfo  {journal} {Nucl. Phys. B}\ }\textbf {\bibinfo {volume} {497}},\
  \bibinfo {pages} {445--478} (\bibinfo {year} {1997})},\ \Eprint
  {https://arxiv.org/abs/hep-lat/9701006} {arXiv:hep-lat/9701006} \BibitemShut
  {NoStop}%
\bibitem [{\citenamefont {Khavkine}(2012)}]{khavkine2012}%
  \BibitemOpen
  \bibfield  {author} {\bibinfo {author} {\bibfnamefont {I.}~\bibnamefont
  {Khavkine}},\ }\bibfield  {title} {\enquote {\bibinfo {title} {{Quantum
  astrometric observables I: time delay in classical and quantum gravity}},}\
  }\href {https://doi.org/10.1103/PhysRevD.85.124014} {\bibfield  {journal}
  {\bibinfo  {journal} {Phys. Rev. D}\ }\textbf {\bibinfo {volume} {85}},\
  \bibinfo {pages} {124014} (\bibinfo {year} {2012})},\ \Eprint
  {https://arxiv.org/abs/1111.7127} {arXiv:1111.7127 [gr-qc]} \BibitemShut
  {NoStop}%
\bibitem [{\citenamefont {Bonga}\ and\ \citenamefont
  {Khavkine}(2014)}]{bongakhavkine2014}%
  \BibitemOpen
  \bibfield  {author} {\bibinfo {author} {\bibfnamefont {B.}~\bibnamefont
  {Bonga}}\ and\ \bibinfo {author} {\bibfnamefont {I.}~\bibnamefont
  {Khavkine}},\ }\bibfield  {title} {\enquote {\bibinfo {title} {{Quantum
  astrometric observables II: time delay in linearized quantum gravity}},}\
  }\href {https://doi.org/10.1103/PhysRevD.89.024039} {\bibfield  {journal}
  {\bibinfo  {journal} {Phys. Rev. D}\ }\textbf {\bibinfo {volume} {89}},\
  \bibinfo {pages} {024039} (\bibinfo {year} {2014})},\ \Eprint
  {https://arxiv.org/abs/1307.0256} {arXiv:1307.0256 [gr-qc]} \BibitemShut
  {NoStop}%
\bibitem [{\citenamefont {Fr{\"o}b}(2018)}]{froeb2018}%
  \BibitemOpen
  \bibfield  {author} {\bibinfo {author} {\bibfnamefont {M.~B.}\ \bibnamefont
  {Fr{\"o}b}},\ }\bibfield  {title} {\enquote {\bibinfo {title} {{One-loop
  quantum gravitational corrections to the scalar two-point function at fixed
  geodesic distance}},}\ }\href {https://doi.org/10.1088/1361-6382/aa9ad1}
  {\bibfield  {journal} {\bibinfo  {journal} {Class. Quant. Grav.}\ }\textbf
  {\bibinfo {volume} {35}},\ \bibinfo {pages} {035005} (\bibinfo {year}
  {2018})},\ \Eprint {https://arxiv.org/abs/1706.01891} {arXiv:1706.01891
  [hep-th]} \BibitemShut {NoStop}%
\bibitem [{\citenamefont {Klitgaard}\ and\ \citenamefont
  {Loll}(2018)}]{klitgaardloll2018}%
  \BibitemOpen
  \bibfield  {author} {\bibinfo {author} {\bibfnamefont {N.}~\bibnamefont
  {Klitgaard}}\ and\ \bibinfo {author} {\bibfnamefont {R.}~\bibnamefont
  {Loll}},\ }\bibfield  {title} {\enquote {\bibinfo {title} {{Introducing
  Quantum Ricci Curvature}},}\ }\href
  {https://doi.org/10.1103/PhysRevD.97.046008} {\bibfield  {journal} {\bibinfo
  {journal} {Phys. Rev. D}\ }\textbf {\bibinfo {volume} {97}},\ \bibinfo
  {pages} {046008} (\bibinfo {year} {2018})},\ \Eprint
  {https://arxiv.org/abs/1712.08847} {arXiv:1712.08847 [hep-th]} \BibitemShut
  {NoStop}%
\bibitem [{\citenamefont {Becker}\ and\ \citenamefont
  {Pagani}(2019)}]{beckerpagani2019}%
  \BibitemOpen
  \bibfield  {author} {\bibinfo {author} {\bibfnamefont {M.}~\bibnamefont
  {Becker}}\ and\ \bibinfo {author} {\bibfnamefont {C.}~\bibnamefont
  {Pagani}},\ }\bibfield  {title} {\enquote {\bibinfo {title} {{Geometric
  operators in the asymptotic safety scenario for quantum gravity}},}\ }\href
  {https://doi.org/10.1103/PhysRevD.99.066002} {\bibfield  {journal} {\bibinfo
  {journal} {Phys. Rev. D}\ }\textbf {\bibinfo {volume} {99}},\ \bibinfo
  {pages} {066002} (\bibinfo {year} {2019})},\ \Eprint
  {https://arxiv.org/abs/1810.11816} {arXiv:1810.11816 [gr-qc]} \BibitemShut
  {NoStop}%
\bibitem [{\citenamefont {Fr{\"o}b}\ and\ \citenamefont
  {Taslimi~Tehrani}(2018)}]{froebtaslimitehrani2018}%
  \BibitemOpen
  \bibfield  {author} {\bibinfo {author} {\bibfnamefont {M.~B.}\ \bibnamefont
  {Fr{\"o}b}}\ and\ \bibinfo {author} {\bibfnamefont {M.}~\bibnamefont
  {Taslimi~Tehrani}},\ }\bibfield  {title} {\enquote {\bibinfo {title}
  {{Green's functions and Hadamard parametrices for vector and tensor fields in
  general linear covariant gauges}},}\ }\href
  {https://doi.org/10.1103/PhysRevD.97.025022} {\bibfield  {journal} {\bibinfo
  {journal} {Phys. Rev. D}\ }\textbf {\bibinfo {volume} {97}},\ \bibinfo
  {pages} {025022} (\bibinfo {year} {2018})},\ \Eprint
  {https://arxiv.org/abs/1708.00444} {arXiv:1708.00444 [gr-qc]} \BibitemShut
  {NoStop}%
\bibitem [{\citenamefont {D'Amico}\ \emph {et~al.}(2011)\citenamefont
  {D'Amico}, \citenamefont {de~Rham}, \citenamefont {Dubovsky}, \citenamefont
  {Gabadadze}, \citenamefont {Pirtskhalava},\ and\ \citenamefont
  {Tolley}}]{damicoetal2011}%
  \BibitemOpen
  \bibfield  {author} {\bibinfo {author} {\bibfnamefont {G.}~\bibnamefont
  {D'Amico}}, \bibinfo {author} {\bibfnamefont {C.}~\bibnamefont {de~Rham}},
  \bibinfo {author} {\bibfnamefont {S.}~\bibnamefont {Dubovsky}}, \bibinfo
  {author} {\bibfnamefont {G.}~\bibnamefont {Gabadadze}}, \bibinfo {author}
  {\bibfnamefont {D.}~\bibnamefont {Pirtskhalava}},\ and\ \bibinfo {author}
  {\bibfnamefont {A.~J.}\ \bibnamefont {Tolley}},\ }\bibfield  {title}
  {\enquote {\bibinfo {title} {{Massive cosmologies}},}\ }\href
  {https://doi.org/10.1103/PhysRevD.84.124046} {\bibfield  {journal} {\bibinfo
  {journal} {Phys. Rev. D}\ }\textbf {\bibinfo {volume} {84}},\ \bibinfo
  {pages} {124046} (\bibinfo {year} {2011})},\ \Eprint
  {https://arxiv.org/abs/1108.5231} {arXiv:1108.5231 [hep-th]} \BibitemShut
  {NoStop}%
\bibitem [{\citenamefont {de~Rham}(2014)}]{derham2014}%
  \BibitemOpen
  \bibfield  {author} {\bibinfo {author} {\bibfnamefont {C.}~\bibnamefont
  {de~Rham}},\ }\bibfield  {title} {\enquote {\bibinfo {title} {{Massive
  Gravity}},}\ }\href {https://doi.org/10.12942/lrr-2014-7} {\bibfield
  {journal} {\bibinfo  {journal} {Living Rev. Rel.}\ }\textbf {\bibinfo
  {volume} {17}},\ \bibinfo {pages} {7} (\bibinfo {year} {2014})},\ \Eprint
  {https://arxiv.org/abs/1401.4173} {arXiv:1401.4173 [hep-th]} \BibitemShut
  {NoStop}%
\bibitem [{\citenamefont {Koyama}(2016)}]{koyama2015}%
  \BibitemOpen
  \bibfield  {author} {\bibinfo {author} {\bibfnamefont {K.}~\bibnamefont
  {Koyama}},\ }\bibfield  {title} {\enquote {\bibinfo {title} {{Cosmological
  tests of modified gravity}},}\ }\href
  {https://doi.org/10.1088/0034-4885/79/4/046902} {\bibfield  {journal}
  {\bibinfo  {journal} {Rept. Prog. Phys.}\ }\textbf {\bibinfo {volume} {79}},\
  \bibinfo {pages} {046902} (\bibinfo {year} {2016})},\ \Eprint
  {https://arxiv.org/abs/1504.04623} {arXiv:1504.04623 [astro-ph.CO]}
  \BibitemShut {NoStop}%
\bibitem [{\citenamefont {Abbott}\ \emph {et~al.}(2017)\citenamefont {Abbott}
  \emph {et~al.}}]{gravitonmass1}%
  \BibitemOpen
  \bibfield  {author} {\bibinfo {author} {\bibfnamefont {B.~P.}\ \bibnamefont
  {Abbott}} \emph {et~al.} (\bibinfo {collaboration} {LIGO Scientific,
  VIRGO}),\ }\bibfield  {title} {\enquote {\bibinfo {title} {{GW170104:
  Observation of a 50-Solar-Mass Binary Black Hole Coalescence at Redshift
  0.2}},}\ }\href {https://doi.org/10.1103/PhysRevLett.118.221101} {\bibfield
  {journal} {\bibinfo  {journal} {Phys. Rev. Lett.}\ }\textbf {\bibinfo
  {volume} {118}},\ \bibinfo {pages} {221101} (\bibinfo {year} {2017})},\
  \bibinfo {note} {[Erratum: \href{10.1103/PhysRevLett.121.129901}{Phys. Rev.
  Lett. \textbf{121}, 129901 (2018)}]},\ \Eprint
  {https://arxiv.org/abs/1706.01812} {arXiv:1706.01812 [gr-qc]} \BibitemShut
  {NoStop}%
\bibitem [{\citenamefont {Bernus}\ \emph {et~al.}(2019)\citenamefont {Bernus},
  \citenamefont {Minazzoli}, \citenamefont {Fienga}, \citenamefont {Gastineau},
  \citenamefont {Laskar},\ and\ \citenamefont {Deram}}]{gravitonmass2}%
  \BibitemOpen
  \bibfield  {author} {\bibinfo {author} {\bibfnamefont {L.}~\bibnamefont
  {Bernus}}, \bibinfo {author} {\bibfnamefont {O.}~\bibnamefont {Minazzoli}},
  \bibinfo {author} {\bibfnamefont {A.}~\bibnamefont {Fienga}}, \bibinfo
  {author} {\bibfnamefont {M.}~\bibnamefont {Gastineau}}, \bibinfo {author}
  {\bibfnamefont {J.}~\bibnamefont {Laskar}},\ and\ \bibinfo {author}
  {\bibfnamefont {P.}~\bibnamefont {Deram}},\ }\bibfield  {title} {\enquote
  {\bibinfo {title} {{Constraining the Mass of the Graviton with the Planetary
  Ephemeris INPOP}},}\ }\href {https://doi.org/10.1103/PhysRevLett.123.161103}
  {\bibfield  {journal} {\bibinfo  {journal} {Phys. Rev. Lett.}\ }\textbf
  {\bibinfo {volume} {123}},\ \bibinfo {pages} {161103} (\bibinfo {year}
  {2019})},\ \Eprint {https://arxiv.org/abs/1901.04307} {arXiv:1901.04307
  [gr-qc]} \BibitemShut {NoStop}%
\bibitem [{\citenamefont {van Dam}\ and\ \citenamefont
  {Veltman}(1970)}]{vandamveltman1970}%
  \BibitemOpen
  \bibfield  {author} {\bibinfo {author} {\bibfnamefont {H.}~\bibnamefont {van
  Dam}}\ and\ \bibinfo {author} {\bibfnamefont {M.}~\bibnamefont {Veltman}},\
  }\bibfield  {title} {\enquote {\bibinfo {title} {{Massive and mass-less
  Yang-Mills and gravitational fields}},}\ }\href
  {https://doi.org/10.1016/0550-3213(70)90416-5} {\bibfield  {journal}
  {\bibinfo  {journal} {Nucl. Phys. B}\ }\textbf {\bibinfo {volume} {22}},\
  \bibinfo {pages} {397--411} (\bibinfo {year} {1970})}\BibitemShut {NoStop}%
\bibitem [{\citenamefont {Zakharov}(1970)}]{zakharov1970}%
  \BibitemOpen
  \bibfield  {author} {\bibinfo {author} {\bibfnamefont {V.~I.}\ \bibnamefont
  {Zakharov}},\ }\bibfield  {title} {\enquote {\bibinfo {title} {{Linearized
  gravitation theory and the graviton mass}},}\ }\href
  {http://jetpletters.ru/ps/1734/article_26353.shtml} {\bibfield  {journal}
  {\bibinfo  {journal} {JETP Lett.}\ }\textbf {\bibinfo {volume} {12}},\
  \bibinfo {pages} {312} (\bibinfo {year} {1970})}\BibitemShut {NoStop}%
\bibitem [{\citenamefont {Fierz}(1939)}]{fierz1939}%
  \BibitemOpen
  \bibfield  {author} {\bibinfo {author} {\bibfnamefont {M.}~\bibnamefont
  {Fierz}},\ }\bibfield  {title} {\enquote {\bibinfo {title} {{{\"U}ber die
  relativistische Theorie kr{\"a}ftefreier Teilchen mit beliebigem Spin}},}\
  }\href {https://doi.org/10.5169/seals-110930} {\bibfield  {journal} {\bibinfo
   {journal} {Helv. Phys. Acta}\ }\textbf {\bibinfo {volume} {12}},\ \bibinfo
  {pages} {3--37} (\bibinfo {year} {1939})}\BibitemShut {NoStop}%
\bibitem [{\citenamefont {Pauli}\ and\ \citenamefont
  {Fierz}(1939)}]{paulifierz1939}%
  \BibitemOpen
  \bibfield  {author} {\bibinfo {author} {\bibfnamefont {W.}~\bibnamefont
  {Pauli}}\ and\ \bibinfo {author} {\bibfnamefont {M.}~\bibnamefont {Fierz}},\
  }\bibfield  {title} {\enquote {\bibinfo {title} {{{\"U}ber relativistische
  Feldgleichungen von Teilchen mit beliebigem Spin im elektromagnetischen
  Feld}},}\ }\href
  {https://www.e-periodica.ch/digbib/view?pid=hpa-001%3A1939%3A12%3A%3A87#307}
  {\bibfield  {journal} {\bibinfo  {journal} {Helv. Phys. Acta}\ }\textbf
  {\bibinfo {volume} {12}},\ \bibinfo {pages} {297--300} (\bibinfo {year}
  {1939})}\BibitemShut {NoStop}%
\bibitem [{\citenamefont {Fierz}\ and\ \citenamefont
  {Pauli}(1939)}]{fierzpauli1939}%
  \BibitemOpen
  \bibfield  {author} {\bibinfo {author} {\bibfnamefont {M.}~\bibnamefont
  {Fierz}}\ and\ \bibinfo {author} {\bibfnamefont {W.}~\bibnamefont {Pauli}},\
  }\bibfield  {title} {\enquote {\bibinfo {title} {{On relativistic wave
  equations for particles of arbitrary spin in an electromagnetic field}},}\
  }\href {https://doi.org/10.1098/rspa.1939.0140} {\bibfield  {journal}
  {\bibinfo  {journal} {Proc. Roy. Soc. Lond. A}\ }\textbf {\bibinfo {volume}
  {173}},\ \bibinfo {pages} {211--232} (\bibinfo {year} {1939})}\BibitemShut
  {NoStop}%
\bibitem [{\citenamefont
  {Stueckelberg}(1938{\natexlab{a}})}]{stueckelberg1938a}%
  \BibitemOpen
  \bibfield  {author} {\bibinfo {author} {\bibfnamefont {E.~C.~G.}\
  \bibnamefont {Stueckelberg}},\ }\bibfield  {title} {\enquote {\bibinfo
  {title} {{Die Wechselwirkungskr{\"a}fte in der Elektrodynamik und in der
  Feldtheorie der Kernkr{\"a}fte. (Teil I)}},}\ }\href
  {https://doi.org/10.5169/seals-110852} {\bibfield  {journal} {\bibinfo
  {journal} {Helv.~Phys.~Acta}\ }\textbf {\bibinfo {volume} {11}},\ \bibinfo
  {pages} {225} (\bibinfo {year} {1938}{\natexlab{a}})}\BibitemShut {NoStop}%
\bibitem [{\citenamefont
  {Stueckelberg}(1938{\natexlab{b}})}]{stueckelberg1938b}%
  \BibitemOpen
  \bibfield  {author} {\bibinfo {author} {\bibfnamefont {E.~C.~G.}\
  \bibnamefont {Stueckelberg}},\ }\bibfield  {title} {\enquote {\bibinfo
  {title} {{Die Wechselwirkungskr{\"a}fte in der Elektrodynamik und in der
  Feldtheorie der Kernkr{\"a}fte. (Teil II und III)}},}\ }\href
  {https://doi.org/10.5169/seals-110855} {\bibfield  {journal} {\bibinfo
  {journal} {Helv.~Phys.~Acta}\ }\textbf {\bibinfo {volume} {11}},\ \bibinfo
  {pages} {299} (\bibinfo {year} {1938}{\natexlab{b}})}\BibitemShut {NoStop}%
\bibitem [{\citenamefont {Parker}\ and\ \citenamefont
  {Toms}(2009)}]{Parker:2009uva}%
  \BibitemOpen
  \bibfield  {author} {\bibinfo {author} {\bibfnamefont {L.~E.}\ \bibnamefont
  {Parker}}\ and\ \bibinfo {author} {\bibfnamefont {D.}~\bibnamefont {Toms}},\
  }\href {https://doi.org/10.1017/CBO9780511813924} {\emph {\bibinfo {title}
  {{Quantum Field Theory in Curved Spacetime}: {Quantized Field and
  Gravity}}}},\ Cambridge Monographs on Mathematical Physics\ (\bibinfo
  {publisher} {Cambridge University Press},\ \bibinfo {year}
  {2009})\BibitemShut {NoStop}%
\bibitem [{\citenamefont {Vilkovisky}(1984)}]{VILKOVISKY1984125}%
  \BibitemOpen
  \bibfield  {author} {\bibinfo {author} {\bibfnamefont {G.~A.}\ \bibnamefont
  {Vilkovisky}},\ }\bibfield  {title} {\enquote {\bibinfo {title} {{The unique
  effective action in quantum field theory}},}\ }\href
  {https://doi.org/10.1016/0550-3213(84)90228-1} {\bibfield  {journal}
  {\bibinfo  {journal} {Nucl. Phys. B}\ }\textbf {\bibinfo {volume} {234}},\
  \bibinfo {pages} {125--137} (\bibinfo {year} {1984})}\BibitemShut {NoStop}%
\bibitem [{\citenamefont {Will}(2014)}]{will_lrr_2014}%
  \BibitemOpen
  \bibfield  {author} {\bibinfo {author} {\bibfnamefont {C.~M.}\ \bibnamefont
  {Will}},\ }\bibfield  {title} {\enquote {\bibinfo {title} {{The Confrontation
  between General Relativity and Experiment}},}\ }\href
  {https://doi.org/10.12942/lrr-2014-4} {\bibfield  {journal} {\bibinfo
  {journal} {Living Rev. Rel.}\ }\textbf {\bibinfo {volume} {17}},\ \bibinfo
  {pages} {4} (\bibinfo {year} {2014})},\ \Eprint
  {https://arxiv.org/abs/1403.7377} {arXiv:1403.7377 [gr-qc]} \BibitemShut
  {NoStop}%
\bibitem [{\citenamefont {Corbelli}\ and\ \citenamefont
  {Salucci}(2000)}]{corbelli_salucci_mnras_1999}%
  \BibitemOpen
  \bibfield  {author} {\bibinfo {author} {\bibfnamefont {E.}~\bibnamefont
  {Corbelli}}\ and\ \bibinfo {author} {\bibfnamefont {P.}~\bibnamefont
  {Salucci}},\ }\bibfield  {title} {\enquote {\bibinfo {title} {{The Extended
  Rotation Curve and the Dark Matter Halo of M33}},}\ }\href
  {https://doi.org/10.1046/j.1365-8711.2000.03075.x} {\bibfield  {journal}
  {\bibinfo  {journal} {Mon. Not. Roy. Astron. Soc.}\ }\textbf {\bibinfo
  {volume} {311}},\ \bibinfo {pages} {441--447} (\bibinfo {year} {2000})},\
  \Eprint {https://arxiv.org/abs/astro-ph/9909252} {arXiv:astro-ph/9909252}
  \BibitemShut {NoStop}%
\bibitem [{\citenamefont {{de Blok}}(2010)}]{deblok_aa_2010}%
  \BibitemOpen
  \bibfield  {author} {\bibinfo {author} {\bibfnamefont {W.~J.~G.}\
  \bibnamefont {{de Blok}}},\ }\bibfield  {title} {\enquote {\bibinfo {title}
  {{The Core-Cusp Problem}},}\ }\href {https://doi.org/10.1155/2010/789293}
  {\bibfield  {journal} {\bibinfo  {journal} {Adv. Astron.}\ }\textbf {\bibinfo
  {volume} {2010}},\ \bibinfo {pages} {789293} (\bibinfo {year} {2010})},\
  \Eprint {https://arxiv.org/abs/0910.3538} {arXiv:0910.3538 [astro-ph.CO]}
  \BibitemShut {NoStop}%
\bibitem [{\citenamefont {Guth}(1981)}]{guth_prd_1981}%
  \BibitemOpen
  \bibfield  {author} {\bibinfo {author} {\bibfnamefont {A.~H.}\ \bibnamefont
  {Guth}},\ }\bibfield  {title} {\enquote {\bibinfo {title} {{Inflationary
  universe: A possible solution to the horizon and flatness problems}},}\
  }\href {https://doi.org/10.1103/PhysRevD.23.347} {\bibfield  {journal}
  {\bibinfo  {journal} {Phys. Rev. D}\ }\textbf {\bibinfo {volume} {23}},\
  \bibinfo {pages} {347--356} (\bibinfo {year} {1981})}\BibitemShut {NoStop}%
\bibitem [{\citenamefont {Linde}(1982)}]{linde_plb_1982}%
  \BibitemOpen
  \bibfield  {author} {\bibinfo {author} {\bibfnamefont {A.~D.}\ \bibnamefont
  {Linde}},\ }\bibfield  {title} {\enquote {\bibinfo {title} {{A new
  inflationary universe scenario: A possible solution of the horizon, flatness,
  homogeneity, isotropy and primordial monopole problems}},}\ }\href
  {https://doi.org/10.1016/0370-2693(82)91219-9} {\bibfield  {journal}
  {\bibinfo  {journal} {Phys. Lett. B}\ }\textbf {\bibinfo {volume} {108}},\
  \bibinfo {pages} {389--393} (\bibinfo {year} {1982})}\BibitemShut {NoStop}%
\bibitem [{\citenamefont {Aghanim}\ \emph {et~al.}(2020)\citenamefont {Aghanim}
  \emph {et~al.}}]{planck_vi_2020}%
  \BibitemOpen
  \bibfield  {author} {\bibinfo {author} {\bibfnamefont {N.}~\bibnamefont
  {Aghanim}} \emph {et~al.} (\bibinfo {collaboration} {Planck}),\ }\bibfield
  {title} {\enquote {\bibinfo {title} {{Planck 2018 results. VI. Cosmological
  parameters}},}\ }\href {https://doi.org/10.1051/0004-6361/201833910}
  {\bibfield  {journal} {\bibinfo  {journal} {Astron. Astrophys.}\ }\textbf
  {\bibinfo {volume} {641}},\ \bibinfo {pages} {A6} (\bibinfo {year} {2020})},\
  \bibinfo {note} {[Erratum:
  \href{https://doi.org/10.1051/0004-6361/201833910e}{Astron. Astrophys.
  \textbf{652} C4 (2021)}]},\ \Eprint {https://arxiv.org/abs/1807.06209}
  {arXiv:1807.06209 [astro-ph.CO]} \BibitemShut {NoStop}%
\bibitem [{\citenamefont {Akrami}\ \emph {et~al.}(2020)\citenamefont {Akrami}
  \emph {et~al.}}]{planck_x_2020}%
  \BibitemOpen
  \bibfield  {author} {\bibinfo {author} {\bibfnamefont {Y.}~\bibnamefont
  {Akrami}} \emph {et~al.} (\bibinfo {collaboration} {Planck}),\ }\bibfield
  {title} {\enquote {\bibinfo {title} {{Planck 2018 results. X. Constraints on
  inflation}},}\ }\href {https://doi.org/10.1051/0004-6361/201833887}
  {\bibfield  {journal} {\bibinfo  {journal} {Astron. Astrophys.}\ }\textbf
  {\bibinfo {volume} {641}},\ \bibinfo {pages} {A10} (\bibinfo {year}
  {2020})},\ \Eprint {https://arxiv.org/abs/1807.06211} {arXiv:1807.06211
  [astro-ph.CO]} \BibitemShut {NoStop}%
\bibitem [{\citenamefont {Castello}, \citenamefont {Ili{\'c}},\ and\
  \citenamefont {Kunz}(2021)}]{castelloilickunz2021}%
  \BibitemOpen
  \bibfield  {author} {\bibinfo {author} {\bibfnamefont {S.}~\bibnamefont
  {Castello}}, \bibinfo {author} {\bibfnamefont {S.}~\bibnamefont {Ili{\'c}}},\
  and\ \bibinfo {author} {\bibfnamefont {M.}~\bibnamefont {Kunz}},\ }\bibfield
  {title} {\enquote {\bibinfo {title} {{Updated dark energy view of
  inflation}},}\ }\href {https://doi.org/10.1103/PhysRevD.104.023522}
  {\bibfield  {journal} {\bibinfo  {journal} {Phys. Rev. D}\ }\textbf {\bibinfo
  {volume} {104}},\ \bibinfo {pages} {023522} (\bibinfo {year} {2021})},\
  \Eprint {https://arxiv.org/abs/2104.15091} {arXiv:2104.15091 [astro-ph.CO]}
  \BibitemShut {NoStop}%
\bibitem [{\citenamefont {Escamilla}\ \emph {et~al.}(2024)\citenamefont
  {Escamilla}, \citenamefont {Giar\`e}, \citenamefont {Di~Valentino},
  \citenamefont {Nunes},\ and\ \citenamefont
  {Vagnozzi}}]{escamillaetal_arxiv_2023}%
  \BibitemOpen
  \bibfield  {author} {\bibinfo {author} {\bibfnamefont {L.~A.}\ \bibnamefont
  {Escamilla}}, \bibinfo {author} {\bibfnamefont {W.}~\bibnamefont {Giar\`e}},
  \bibinfo {author} {\bibfnamefont {E.}~\bibnamefont {Di~Valentino}}, \bibinfo
  {author} {\bibfnamefont {R.~C.}\ \bibnamefont {Nunes}},\ and\ \bibinfo
  {author} {\bibfnamefont {S.}~\bibnamefont {Vagnozzi}},\ }\bibfield  {title}
  {\enquote {\bibinfo {title} {{The state of the dark energy equation of state
  circa 2023}},}\ }\href {https://doi.org/10.1088/1475-7516/2024/05/091}
  {\bibfield  {journal} {\bibinfo  {journal} {JCAP}\ }\textbf {\bibinfo
  {volume} {05}},\ \bibinfo {pages} {091} (\bibinfo {year} {2024})},\ \Eprint
  {https://arxiv.org/abs/2307.14802} {arXiv:2307.14802 [astro-ph.CO]}
  \BibitemShut {NoStop}%
\bibitem [{\citenamefont {Shankaranarayanan}\ and\ \citenamefont
  {Johnson}(2022)}]{shankaranarayanan_johnson_grg_2022}%
  \BibitemOpen
  \bibfield  {author} {\bibinfo {author} {\bibfnamefont {S.}~\bibnamefont
  {Shankaranarayanan}}\ and\ \bibinfo {author} {\bibfnamefont {J.~P.}\
  \bibnamefont {Johnson}},\ }\bibfield  {title} {\enquote {\bibinfo {title}
  {{Modified theories of gravity: Why, how and what?}}}\ }\href
  {https://doi.org/10.1007/s10714-022-02927-2} {\bibfield  {journal} {\bibinfo
  {journal} {Gen. Rel. Grav.}\ }\textbf {\bibinfo {volume} {54}},\ \bibinfo
  {pages} {44} (\bibinfo {year} {2022})},\ \Eprint
  {https://arxiv.org/abs/2204.06533} {arXiv:2204.06533 [gr-qc]} \BibitemShut
  {NoStop}%
\bibitem [{\citenamefont {Hinterbichler}(2012)}]{Hinterbichler:2011tt}%
  \BibitemOpen
  \bibfield  {author} {\bibinfo {author} {\bibfnamefont {K.}~\bibnamefont
  {Hinterbichler}},\ }\bibfield  {title} {\enquote {\bibinfo {title}
  {{Theoretical Aspects of Massive Gravity}},}\ }\href
  {https://doi.org/10.1103/RevModPhys.84.671} {\bibfield  {journal} {\bibinfo
  {journal} {Rev. Mod. Phys.}\ }\textbf {\bibinfo {volume} {84}},\ \bibinfo
  {pages} {671--710} (\bibinfo {year} {2012})},\ \Eprint
  {https://arxiv.org/abs/1105.3735} {arXiv:1105.3735 [hep-th]} \BibitemShut
  {NoStop}%
\bibitem [{\citenamefont {Tolley}(2015)}]{Tolley:2015oxa}%
  \BibitemOpen
  \bibfield  {author} {\bibinfo {author} {\bibfnamefont {A.~J.}\ \bibnamefont
  {Tolley}},\ }\bibfield  {title} {\enquote {\bibinfo {title} {{Cosmological
  Applications of Massive Gravity}},}\ }\href
  {https://doi.org/10.1007/978-3-319-10070-8_8} {\bibfield  {journal} {\bibinfo
   {journal} {Lect. Notes Phys.}\ }\textbf {\bibinfo {volume} {892}},\ \bibinfo
  {pages} {203--224} (\bibinfo {year} {2015})}\BibitemShut {NoStop}%
\bibitem [{\citenamefont {de~Rham}, \citenamefont {Heisenberg},\ and\
  \citenamefont {Ribeiro}(2015)}]{deRham:2014naa}%
  \BibitemOpen
  \bibfield  {author} {\bibinfo {author} {\bibfnamefont {C.}~\bibnamefont
  {de~Rham}}, \bibinfo {author} {\bibfnamefont {L.}~\bibnamefont
  {Heisenberg}},\ and\ \bibinfo {author} {\bibfnamefont {R.~H.}\ \bibnamefont
  {Ribeiro}},\ }\bibfield  {title} {\enquote {\bibinfo {title} {{On couplings
  to matter in massive (bi-)gravity}},}\ }\href
  {https://doi.org/10.1088/0264-9381/32/3/035022} {\bibfield  {journal}
  {\bibinfo  {journal} {Class. Quant. Grav.}\ }\textbf {\bibinfo {volume}
  {32}},\ \bibinfo {pages} {035022} (\bibinfo {year} {2015})},\ \Eprint
  {https://arxiv.org/abs/1408.1678} {arXiv:1408.1678 [hep-th]} \BibitemShut
  {NoStop}%
\bibitem [{\citenamefont {de~Rham}\ \emph {et~al.}(2023)\citenamefont
  {de~Rham}, \citenamefont {Ko\.zuszek}, \citenamefont {Tolley},\ and\
  \citenamefont {Wiseman}}]{deRham:2023ngf}%
  \BibitemOpen
  \bibfield  {author} {\bibinfo {author} {\bibfnamefont {C.}~\bibnamefont
  {de~Rham}}, \bibinfo {author} {\bibfnamefont {J.}~\bibnamefont {Ko\.zuszek}},
  \bibinfo {author} {\bibfnamefont {A.~J.}\ \bibnamefont {Tolley}},\ and\
  \bibinfo {author} {\bibfnamefont {T.}~\bibnamefont {Wiseman}},\ }\bibfield
  {title} {\enquote {\bibinfo {title} {{Dynamical formulation of ghost-free
  massive gravity}},}\ }\href {https://doi.org/10.1103/PhysRevD.108.084052}
  {\bibfield  {journal} {\bibinfo  {journal} {Phys. Rev. D}\ }\textbf {\bibinfo
  {volume} {108}},\ \bibinfo {pages} {084052} (\bibinfo {year} {2023})},\
  \Eprint {https://arxiv.org/abs/2302.04876} {arXiv:2302.04876 [hep-th]}
  \BibitemShut {NoStop}%
\bibitem [{\citenamefont {Deser}(1970)}]{deser_grg_1970}%
  \BibitemOpen
  \bibfield  {author} {\bibinfo {author} {\bibfnamefont {S.}~\bibnamefont
  {Deser}},\ }\bibfield  {title} {\enquote {\bibinfo {title} {{Selfinteraction
  and gauge invariance}},}\ }\href {https://doi.org/10.1007/BF00759198}
  {\bibfield  {journal} {\bibinfo  {journal} {Gen. Rel. Grav.}\ }\textbf
  {\bibinfo {volume} {1}},\ \bibinfo {pages} {9--18} (\bibinfo {year}
  {1970})},\ \Eprint {https://arxiv.org/abs/gr-qc/0411023}
  {arXiv:gr-qc/0411023} \BibitemShut {NoStop}%
\bibitem [{\citenamefont {Wald}(1986)}]{wald_prd_1986}%
  \BibitemOpen
  \bibfield  {author} {\bibinfo {author} {\bibfnamefont {R.~M.}\ \bibnamefont
  {Wald}},\ }\bibfield  {title} {\enquote {\bibinfo {title} {{Spin-2 Fields and
  General Covariance}},}\ }\href {https://doi.org/10.1103/PhysRevD.33.3613}
  {\bibfield  {journal} {\bibinfo  {journal} {Phys. Rev. D}\ }\textbf {\bibinfo
  {volume} {33}},\ \bibinfo {pages} {3613} (\bibinfo {year}
  {1986})}\BibitemShut {NoStop}%
\bibitem [{\citenamefont {Deser}(1987)}]{deser_cqg_1987}%
  \BibitemOpen
  \bibfield  {author} {\bibinfo {author} {\bibfnamefont {S.}~\bibnamefont
  {Deser}},\ }\bibfield  {title} {\enquote {\bibinfo {title} {{Gravity From
  Selfinteraction in a Curved Background}},}\ }\href
  {https://doi.org/10.1088/0264-9381/4/4/006} {\bibfield  {journal} {\bibinfo
  {journal} {Class. Quant. Grav.}\ }\textbf {\bibinfo {volume} {4}},\ \bibinfo
  {pages} {L99} (\bibinfo {year} {1987})}\BibitemShut {NoStop}%
\bibitem [{\citenamefont {Randall}\ and\ \citenamefont
  {Sundrum}(1999)}]{randall_sundrum_prl_1999}%
  \BibitemOpen
  \bibfield  {author} {\bibinfo {author} {\bibfnamefont {L.}~\bibnamefont
  {Randall}}\ and\ \bibinfo {author} {\bibfnamefont {R.}~\bibnamefont
  {Sundrum}},\ }\bibfield  {title} {\enquote {\bibinfo {title} {{An Alternative
  to Compactification}},}\ }\href {https://doi.org/10.1103/PhysRevLett.83.4690}
  {\bibfield  {journal} {\bibinfo  {journal} {Phys. Rev. Lett.}\ }\textbf
  {\bibinfo {volume} {83}},\ \bibinfo {pages} {4690--4693} (\bibinfo {year}
  {1999})},\ \Eprint {https://arxiv.org/abs/hep-th/9906064}
  {arXiv:hep-th/9906064} \BibitemShut {NoStop}%
\bibitem [{\citenamefont {Gregory}, \citenamefont {Rubakov},\ and\
  \citenamefont {Sibiryakov}(2000)}]{gregory_rubakov_sibiryakov_prl_2000}%
  \BibitemOpen
  \bibfield  {author} {\bibinfo {author} {\bibfnamefont {R.}~\bibnamefont
  {Gregory}}, \bibinfo {author} {\bibfnamefont {V.~A.}\ \bibnamefont
  {Rubakov}},\ and\ \bibinfo {author} {\bibfnamefont {S.~M.}\ \bibnamefont
  {Sibiryakov}},\ }\bibfield  {title} {\enquote {\bibinfo {title} {{Opening up
  Extra Dimensions at Ultralarge Scales}},}\ }\href
  {https://doi.org/10.1103/PhysRevLett.84.5928} {\bibfield  {journal} {\bibinfo
   {journal} {Phys. Rev. Lett.}\ }\textbf {\bibinfo {volume} {84}},\ \bibinfo
  {pages} {5928--5931} (\bibinfo {year} {2000})}\BibitemShut {NoStop}%
\bibitem [{\citenamefont {Hassan}\ and\ \citenamefont
  {Rosen}(2012)}]{Hassan:2011zd}%
  \BibitemOpen
  \bibfield  {author} {\bibinfo {author} {\bibfnamefont {S.~F.}\ \bibnamefont
  {Hassan}}\ and\ \bibinfo {author} {\bibfnamefont {R.~A.}\ \bibnamefont
  {Rosen}},\ }\bibfield  {title} {\enquote {\bibinfo {title} {{Bimetric Gravity
  from Ghost-free Massive Gravity}},}\ }\href
  {https://doi.org/10.1007/JHEP02(2012)126} {\bibfield  {journal} {\bibinfo
  {journal} {JHEP}\ }\textbf {\bibinfo {volume} {02}},\ \bibinfo {pages} {126}
  (\bibinfo {year} {2012})},\ \Eprint {https://arxiv.org/abs/1109.3515}
  {arXiv:1109.3515 [hep-th]} \BibitemShut {NoStop}%
\bibitem [{\citenamefont {Caravano}, \citenamefont {L{\"u}ben},\ and\
  \citenamefont {Weller}(2021)}]{Caravano:2021aum}%
  \BibitemOpen
  \bibfield  {author} {\bibinfo {author} {\bibfnamefont {A.}~\bibnamefont
  {Caravano}}, \bibinfo {author} {\bibfnamefont {M.}~\bibnamefont
  {L{\"u}ben}},\ and\ \bibinfo {author} {\bibfnamefont {J.}~\bibnamefont
  {Weller}},\ }\bibfield  {title} {\enquote {\bibinfo {title} {{Combining
  cosmological and local bounds on bimetric theory}},}\ }\href
  {https://doi.org/10.1088/1475-7516/2021/09/035} {\bibfield  {journal}
  {\bibinfo  {journal} {JCAP}\ }\textbf {\bibinfo {volume} {09}},\ \bibinfo
  {pages} {035} (\bibinfo {year} {2021})},\ \Eprint
  {https://arxiv.org/abs/2101.08791} {arXiv:2101.08791 [gr-qc]} \BibitemShut
  {NoStop}%
\bibitem [{\citenamefont {H{\"o}g{\r a}s}\ and\ \citenamefont
  {M{\"o}rtsell}(2021{\natexlab{a}})}]{Hogas:2021fmr}%
  \BibitemOpen
  \bibfield  {author} {\bibinfo {author} {\bibfnamefont {M.}~\bibnamefont
  {H{\"o}g{\r a}s}}\ and\ \bibinfo {author} {\bibfnamefont {E.}~\bibnamefont
  {M{\"o}rtsell}},\ }\bibfield  {title} {\enquote {\bibinfo {title}
  {{Constraints on bimetric gravity. Part I. Analytical constraints}},}\ }\href
  {https://doi.org/10.1088/1475-7516/2021/05/001} {\bibfield  {journal}
  {\bibinfo  {journal} {JCAP}\ }\textbf {\bibinfo {volume} {05}},\ \bibinfo
  {pages} {001} (\bibinfo {year} {2021}{\natexlab{a}})},\ \Eprint
  {https://arxiv.org/abs/2101.08794} {arXiv:2101.08794 [gr-qc]} \BibitemShut
  {NoStop}%
\bibitem [{\citenamefont {H{\"o}g{\r a}s}\ and\ \citenamefont
  {M{\"o}rtsell}(2021{\natexlab{b}})}]{Hogas:2021lns}%
  \BibitemOpen
  \bibfield  {author} {\bibinfo {author} {\bibfnamefont {M.}~\bibnamefont
  {H{\"o}g{\r a}s}}\ and\ \bibinfo {author} {\bibfnamefont {E.}~\bibnamefont
  {M{\"o}rtsell}},\ }\bibfield  {title} {\enquote {\bibinfo {title}
  {{Constraints on bimetric gravity. Part II. Observational constraints}},}\
  }\href {https://doi.org/10.1088/1475-7516/2021/05/002} {\bibfield  {journal}
  {\bibinfo  {journal} {JCAP}\ }\textbf {\bibinfo {volume} {05}},\ \bibinfo
  {pages} {002} (\bibinfo {year} {2021}{\natexlab{b}})},\ \Eprint
  {https://arxiv.org/abs/2101.08795} {arXiv:2101.08795 [gr-qc]} \BibitemShut
  {NoStop}%
\bibitem [{\citenamefont {H{\"o}g{\r a}s}\ and\ \citenamefont
  {M{\"o}rtsell}(2021{\natexlab{c}})}]{Hogas:2021saw}%
  \BibitemOpen
  \bibfield  {author} {\bibinfo {author} {\bibfnamefont {M.}~\bibnamefont
  {H{\"o}g{\r a}s}}\ and\ \bibinfo {author} {\bibfnamefont {E.}~\bibnamefont
  {M{\"o}rtsell}},\ }\bibfield  {title} {\enquote {\bibinfo {title}
  {{Constraints on bimetric gravity from Big Bang nucleosynthesis}},}\ }\href
  {https://doi.org/10.1088/1475-7516/2021/11/001} {\bibfield  {journal}
  {\bibinfo  {journal} {JCAP}\ }\textbf {\bibinfo {volume} {11}},\ \bibinfo
  {pages} {001} (\bibinfo {year} {2021}{\natexlab{c}})},\ \Eprint
  {https://arxiv.org/abs/2106.09030} {arXiv:2106.09030 [astro-ph.CO]}
  \BibitemShut {NoStop}%
\bibitem [{\citenamefont {Deser}, \citenamefont {Jackiw},\ and\ \citenamefont
  {Templeton}(1982{\natexlab{a}})}]{deser_jackiw_templeton_prl_1982}%
  \BibitemOpen
  \bibfield  {author} {\bibinfo {author} {\bibfnamefont {S.}~\bibnamefont
  {Deser}}, \bibinfo {author} {\bibfnamefont {R.}~\bibnamefont {Jackiw}},\ and\
  \bibinfo {author} {\bibfnamefont {S.}~\bibnamefont {Templeton}},\ }\bibfield
  {title} {\enquote {\bibinfo {title} {{Three-Dimensional Massive Gauge
  Theories}},}\ }\href {https://doi.org/10.1103/PhysRevLett.48.975} {\bibfield
  {journal} {\bibinfo  {journal} {Phys. Rev. Lett.}\ }\textbf {\bibinfo
  {volume} {48}},\ \bibinfo {pages} {975--978} (\bibinfo {year}
  {1982}{\natexlab{a}})}\BibitemShut {NoStop}%
\bibitem [{\citenamefont {Deser}, \citenamefont {Jackiw},\ and\ \citenamefont
  {Templeton}(1982{\natexlab{b}})}]{deser_jackiw_templeton_annals_1982}%
  \BibitemOpen
  \bibfield  {author} {\bibinfo {author} {\bibfnamefont {S.}~\bibnamefont
  {Deser}}, \bibinfo {author} {\bibfnamefont {R.}~\bibnamefont {Jackiw}},\ and\
  \bibinfo {author} {\bibfnamefont {S.}~\bibnamefont {Templeton}},\ }\bibfield
  {title} {\enquote {\bibinfo {title} {{Topologically Massive Gauge
  Theories}},}\ }\href {https://doi.org/10.1016/0003-4916(82)90164-6}
  {\bibfield  {journal} {\bibinfo  {journal} {Annals Phys.}\ }\textbf {\bibinfo
  {volume} {140}},\ \bibinfo {pages} {372--411} (\bibinfo {year}
  {1982}{\natexlab{b}})},\ \bibinfo {note} {[Erratum:
  \href{https://doi.org/10.1016/0003-4916(88)90053-X}{Annals Phys. \textbf{185}
  406 (1988)}]}\BibitemShut {NoStop}%
\bibitem [{\citenamefont {Bergshoeff}, \citenamefont {Hohm},\ and\
  \citenamefont {Townsend}(2009)}]{bergshoeff_hohm_townsend_prl_2009}%
  \BibitemOpen
  \bibfield  {author} {\bibinfo {author} {\bibfnamefont {E.~A.}\ \bibnamefont
  {Bergshoeff}}, \bibinfo {author} {\bibfnamefont {O.}~\bibnamefont {Hohm}},\
  and\ \bibinfo {author} {\bibfnamefont {P.~K.}\ \bibnamefont {Townsend}},\
  }\bibfield  {title} {\enquote {\bibinfo {title} {{Massive Gravity in Three
  Dimensions}},}\ }\href {https://doi.org/10.1103/PhysRevLett.102.201301}
  {\bibfield  {journal} {\bibinfo  {journal} {Phys. Rev. Lett.}\ }\textbf
  {\bibinfo {volume} {102}},\ \bibinfo {pages} {201301} (\bibinfo {year}
  {2009})},\ \Eprint {https://arxiv.org/abs/0901.1766} {arXiv:0901.1766
  [hep-th]} \BibitemShut {NoStop}%
\bibitem [{\citenamefont {Ohta}\ and\ \citenamefont
  {Percacci}(2014)}]{ohta_percacci_cqg_2014}%
  \BibitemOpen
  \bibfield  {author} {\bibinfo {author} {\bibfnamefont {N.}~\bibnamefont
  {Ohta}}\ and\ \bibinfo {author} {\bibfnamefont {R.}~\bibnamefont
  {Percacci}},\ }\bibfield  {title} {\enquote {\bibinfo {title} {{Higher
  Derivative Gravity and Asymptotic Safety in Diverse Dimensions}},}\ }\href
  {https://doi.org/10.1088/0264-9381/31/1/015024} {\bibfield  {journal}
  {\bibinfo  {journal} {Class. Quant. Grav.}\ }\textbf {\bibinfo {volume}
  {31}},\ \bibinfo {pages} {015024} (\bibinfo {year} {2014})},\ \Eprint
  {https://arxiv.org/abs/1308.3398} {arXiv:1308.3398 [hep-th]} \BibitemShut
  {NoStop}%
\bibitem [{\citenamefont {Blas}\ \emph {et~al.}(2009)\citenamefont {Blas},
  \citenamefont {Comelli}, \citenamefont {Nesti},\ and\ \citenamefont
  {Pilo}}]{Blas:2009my}%
  \BibitemOpen
  \bibfield  {author} {\bibinfo {author} {\bibfnamefont {D.}~\bibnamefont
  {Blas}}, \bibinfo {author} {\bibfnamefont {D.}~\bibnamefont {Comelli}},
  \bibinfo {author} {\bibfnamefont {F.}~\bibnamefont {Nesti}},\ and\ \bibinfo
  {author} {\bibfnamefont {L.}~\bibnamefont {Pilo}},\ }\bibfield  {title}
  {\enquote {\bibinfo {title} {{Lorentz Breaking Massive Gravity in Curved
  Space}},}\ }\href {https://doi.org/10.1103/PhysRevD.80.044025} {\bibfield
  {journal} {\bibinfo  {journal} {Phys. Rev. D}\ }\textbf {\bibinfo {volume}
  {80}},\ \bibinfo {pages} {044025} (\bibinfo {year} {2009})},\ \Eprint
  {https://arxiv.org/abs/0905.1699} {arXiv:0905.1699 [hep-th]} \BibitemShut
  {NoStop}%
\bibitem [{\citenamefont {Rubakov}(2004)}]{Rubakov:2004eb}%
  \BibitemOpen
  \bibfield  {author} {\bibinfo {author} {\bibfnamefont {V.~A.}\ \bibnamefont
  {Rubakov}},\ }\bibfield  {title} {\enquote {\bibinfo {title}
  {{Lorentz-violating graviton masses: Getting around ghosts, low strong
  coupling scale and VDVZ discontinuity}},}\ }\href@noop {} {\  (\bibinfo
  {year} {2004})},\ \Eprint {https://arxiv.org/abs/hep-th/0407104}
  {arXiv:hep-th/0407104} \BibitemShut {NoStop}%
\bibitem [{\citenamefont {Rubakov}\ and\ \citenamefont
  {Tinyakov}(2008)}]{Rubakov:2008nh}%
  \BibitemOpen
  \bibfield  {author} {\bibinfo {author} {\bibfnamefont {V.~A.}\ \bibnamefont
  {Rubakov}}\ and\ \bibinfo {author} {\bibfnamefont {P.~G.}\ \bibnamefont
  {Tinyakov}},\ }\bibfield  {title} {\enquote {\bibinfo {title}
  {{Infrared-modified gravities and massive gravitons}},}\ }\href
  {https://doi.org/10.1070/PU2008v051n08ABEH006600} {\bibfield  {journal}
  {\bibinfo  {journal} {Phys. Usp.}\ }\textbf {\bibinfo {volume} {51}},\
  \bibinfo {pages} {759--792} (\bibinfo {year} {2008})},\ \Eprint
  {https://arxiv.org/abs/0802.4379} {arXiv:0802.4379 [hep-th]} \BibitemShut
  {NoStop}%
\bibitem [{\citenamefont {Dubovsky}, \citenamefont {Tinyakov},\ and\
  \citenamefont {Tkachev}(2005)}]{Dubovsky:2004ud}%
  \BibitemOpen
  \bibfield  {author} {\bibinfo {author} {\bibfnamefont {S.~L.}\ \bibnamefont
  {Dubovsky}}, \bibinfo {author} {\bibfnamefont {P.~G.}\ \bibnamefont
  {Tinyakov}},\ and\ \bibinfo {author} {\bibfnamefont {I.~I.}\ \bibnamefont
  {Tkachev}},\ }\bibfield  {title} {\enquote {\bibinfo {title} {{Massive
  graviton as a testable cold dark matter candidate}},}\ }\href
  {https://doi.org/10.1103/PhysRevLett.94.181102} {\bibfield  {journal}
  {\bibinfo  {journal} {Phys. Rev. Lett.}\ }\textbf {\bibinfo {volume} {94}},\
  \bibinfo {pages} {181102} (\bibinfo {year} {2005})},\ \Eprint
  {https://arxiv.org/abs/hep-th/0411158} {arXiv:hep-th/0411158} \BibitemShut
  {NoStop}%
\bibitem [{\citenamefont {Arkani-Hamed}\ \emph {et~al.}(2004)\citenamefont
  {Arkani-Hamed}, \citenamefont {Cheng}, \citenamefont {Luty},\ and\
  \citenamefont {Mukohyama}}]{Arkani-Hamed:2003pdi}%
  \BibitemOpen
  \bibfield  {author} {\bibinfo {author} {\bibfnamefont {N.}~\bibnamefont
  {Arkani-Hamed}}, \bibinfo {author} {\bibfnamefont {H.-C.}\ \bibnamefont
  {Cheng}}, \bibinfo {author} {\bibfnamefont {M.~A.}\ \bibnamefont {Luty}},\
  and\ \bibinfo {author} {\bibfnamefont {S.}~\bibnamefont {Mukohyama}},\
  }\bibfield  {title} {\enquote {\bibinfo {title} {{Ghost condensation and a
  consistent infrared modification of gravity}},}\ }\href
  {https://doi.org/10.1088/1126-6708/2004/05/074} {\bibfield  {journal}
  {\bibinfo  {journal} {JHEP}\ }\textbf {\bibinfo {volume} {05}},\ \bibinfo
  {pages} {074} (\bibinfo {year} {2004})},\ \Eprint
  {https://arxiv.org/abs/hep-th/0312099} {arXiv:hep-th/0312099} \BibitemShut
  {NoStop}%
\bibitem [{\citenamefont {Hell}(2022)}]{hell2022}%
  \BibitemOpen
  \bibfield  {author} {\bibinfo {author} {\bibfnamefont {A.}~\bibnamefont
  {Hell}},\ }\bibfield  {title} {\enquote {\bibinfo {title} {{The strong
  couplings of massive Yang-Mills theory}},}\ }\href
  {https://doi.org/10.1007/JHEP03(2022)167} {\bibfield  {journal} {\bibinfo
  {journal} {JHEP}\ }\textbf {\bibinfo {volume} {03}},\ \bibinfo {pages} {167}
  (\bibinfo {year} {2022})},\ \Eprint {https://arxiv.org/abs/2111.00017}
  {arXiv:2111.00017 [hep-th]} \BibitemShut {NoStop}%
\bibitem [{\citenamefont {Kogan}, \citenamefont {Mouslopoulos},\ and\
  \citenamefont {Papazoglou}(2001)}]{Kogan:2000uy}%
  \BibitemOpen
  \bibfield  {author} {\bibinfo {author} {\bibfnamefont {I.~I.}\ \bibnamefont
  {Kogan}}, \bibinfo {author} {\bibfnamefont {S.}~\bibnamefont
  {Mouslopoulos}},\ and\ \bibinfo {author} {\bibfnamefont {A.}~\bibnamefont
  {Papazoglou}},\ }\bibfield  {title} {\enquote {\bibinfo {title} {{The $m \to
  0$ limit for massive graviton in dS${}_4$ and AdS${}_4$: How to circumvent
  the van Dam--Veltman--Zakharov discontinuity}},}\ }\href
  {https://doi.org/10.1016/S0370-2693(01)00209-X} {\bibfield  {journal}
  {\bibinfo  {journal} {Phys. Lett. B}\ }\textbf {\bibinfo {volume} {503}},\
  \bibinfo {pages} {173--180} (\bibinfo {year} {2001})},\ \Eprint
  {https://arxiv.org/abs/hep-th/0011138} {arXiv:hep-th/0011138} \BibitemShut
  {NoStop}%
\bibitem [{\citenamefont {Porrati}(2001)}]{Porrati:2000cp}%
  \BibitemOpen
  \bibfield  {author} {\bibinfo {author} {\bibfnamefont {M.}~\bibnamefont
  {Porrati}},\ }\bibfield  {title} {\enquote {\bibinfo {title} {{No van
  Dam--Veltman--Zakharov discontinuity in AdS space}},}\ }\href
  {https://doi.org/10.1016/S0370-2693(00)01380-0} {\bibfield  {journal}
  {\bibinfo  {journal} {Phys. Lett. B}\ }\textbf {\bibinfo {volume} {498}},\
  \bibinfo {pages} {92--96} (\bibinfo {year} {2001})},\ \Eprint
  {https://arxiv.org/abs/hep-th/0011152} {arXiv:hep-th/0011152} \BibitemShut
  {NoStop}%
\bibitem [{\citenamefont {Higuchi}(1987)}]{Higuchi:1986py}%
  \BibitemOpen
  \bibfield  {author} {\bibinfo {author} {\bibfnamefont {A.}~\bibnamefont
  {Higuchi}},\ }\bibfield  {title} {\enquote {\bibinfo {title} {{Forbidden Mass
  Range for Spin-2 Field Theory in De Sitter Space-time}},}\ }\href
  {https://doi.org/10.1016/0550-3213(87)90691-2} {\bibfield  {journal}
  {\bibinfo  {journal} {Nucl. Phys. B}\ }\textbf {\bibinfo {volume} {282}},\
  \bibinfo {pages} {397--436} (\bibinfo {year} {1987})}\BibitemShut {NoStop}%
\bibitem [{\citenamefont {Arkani-Hamed}, \citenamefont {Georgi},\ and\
  \citenamefont {Schwartz}(2003)}]{Arkani-Hamed:2002bjr}%
  \BibitemOpen
  \bibfield  {author} {\bibinfo {author} {\bibfnamefont {N.}~\bibnamefont
  {Arkani-Hamed}}, \bibinfo {author} {\bibfnamefont {H.}~\bibnamefont
  {Georgi}},\ and\ \bibinfo {author} {\bibfnamefont {M.~D.}\ \bibnamefont
  {Schwartz}},\ }\bibfield  {title} {\enquote {\bibinfo {title} {{Effective
  field theory for massive gravitons and gravity in theory space}},}\ }\href
  {https://doi.org/10.1016/S0003-4916(03)00068-X} {\bibfield  {journal}
  {\bibinfo  {journal} {Annals Phys.}\ }\textbf {\bibinfo {volume} {305}},\
  \bibinfo {pages} {96--118} (\bibinfo {year} {2003})},\ \Eprint
  {https://arxiv.org/abs/hep-th/0210184} {arXiv:hep-th/0210184} \BibitemShut
  {NoStop}%
\bibitem [{\citenamefont {Dubovsky}\ and\ \citenamefont
  {Rubakov}(2003)}]{Dubovsky:2002jm}%
  \BibitemOpen
  \bibfield  {author} {\bibinfo {author} {\bibfnamefont {S.~L.}\ \bibnamefont
  {Dubovsky}}\ and\ \bibinfo {author} {\bibfnamefont {V.~A.}\ \bibnamefont
  {Rubakov}},\ }\bibfield  {title} {\enquote {\bibinfo {title} {{Brane induced
  gravity in more than one extra dimensions: Violation of equivalence principle
  and ghost}},}\ }\href {https://doi.org/10.1103/PhysRevD.67.104014} {\bibfield
   {journal} {\bibinfo  {journal} {Phys. Rev. D}\ }\textbf {\bibinfo {volume}
  {67}},\ \bibinfo {pages} {104014} (\bibinfo {year} {2003})},\ \Eprint
  {https://arxiv.org/abs/hep-th/0212222} {arXiv:hep-th/0212222} \BibitemShut
  {NoStop}%
\bibitem [{\citenamefont {Buoninfante}(2023)}]{Buoninfante:2023ryt}%
  \BibitemOpen
  \bibfield  {author} {\bibinfo {author} {\bibfnamefont {L.}~\bibnamefont
  {Buoninfante}},\ }\bibfield  {title} {\enquote {\bibinfo {title} {{Massless
  and partially massless limits in Quadratic Gravity}},}\ }\href
  {https://doi.org/10.1007/JHEP12(2023)111} {\bibfield  {journal} {\bibinfo
  {journal} {JHEP}\ }\textbf {\bibinfo {volume} {12}},\ \bibinfo {pages} {111}
  (\bibinfo {year} {2023})},\ \Eprint {https://arxiv.org/abs/2308.11324}
  {arXiv:2308.11324 [hep-th]} \BibitemShut {NoStop}%
\bibitem [{\citenamefont {Kol\'a\v{r}}\ and\ \citenamefont
  {M\'alek}(2023)}]{Kolar:2023mkw}%
  \BibitemOpen
  \bibfield  {author} {\bibinfo {author} {\bibfnamefont {I.}~\bibnamefont
  {Kol\'a\v{r}}}\ and\ \bibinfo {author} {\bibfnamefont {T.}~\bibnamefont
  {M\'alek}},\ }\bibfield  {title} {\enquote {\bibinfo {title} {{Propagators in
  AdS for higher-derivative and nonlocal gravity: Heat kernel approach}},}\
  }\href@noop {} {\  (\bibinfo {year} {2023})},\ \Eprint
  {https://arxiv.org/abs/2307.13056} {arXiv:2307.13056 [gr-qc]} \BibitemShut
  {NoStop}%
\bibitem [{\citenamefont {Belokogne}\ and\ \citenamefont
  {Folacci}(2016)}]{Belokogne:2015etf}%
  \BibitemOpen
  \bibfield  {author} {\bibinfo {author} {\bibfnamefont {A.}~\bibnamefont
  {Belokogne}}\ and\ \bibinfo {author} {\bibfnamefont {A.}~\bibnamefont
  {Folacci}},\ }\bibfield  {title} {\enquote {\bibinfo {title} {{Stueckelberg
  massive electromagnetism in curved spacetime: Hadamard renormalization of the
  stress-energy tensor and the Casimir effect}},}\ }\href
  {https://doi.org/10.1103/PhysRevD.93.044063} {\bibfield  {journal} {\bibinfo
  {journal} {Phys. Rev. D}\ }\textbf {\bibinfo {volume} {93}},\ \bibinfo
  {pages} {044063} (\bibinfo {year} {2016})},\ \Eprint
  {https://arxiv.org/abs/1512.06326} {arXiv:1512.06326 [gr-qc]} \BibitemShut
  {NoStop}%
\bibitem [{\citenamefont {Belokogne}, \citenamefont {Folacci},\ and\
  \citenamefont {Queva}(2016)}]{Belokogne:2016dvd}%
  \BibitemOpen
  \bibfield  {author} {\bibinfo {author} {\bibfnamefont {A.}~\bibnamefont
  {Belokogne}}, \bibinfo {author} {\bibfnamefont {A.}~\bibnamefont {Folacci}},\
  and\ \bibinfo {author} {\bibfnamefont {J.}~\bibnamefont {Queva}},\ }\bibfield
   {title} {\enquote {\bibinfo {title} {{Stueckelberg massive electromagnetism
  in de Sitter and anti\textendash{}de Sitter spacetimes: Two-point functions
  and renormalized stress-energy tensors}},}\ }\href
  {https://doi.org/10.1103/PhysRevD.94.105028} {\bibfield  {journal} {\bibinfo
  {journal} {Phys. Rev. D}\ }\textbf {\bibinfo {volume} {94}},\ \bibinfo
  {pages} {105028} (\bibinfo {year} {2016})},\ \Eprint
  {https://arxiv.org/abs/1610.00244} {arXiv:1610.00244 [gr-qc]} \BibitemShut
  {NoStop}%
\bibitem [{\citenamefont {Buchbinder}, \citenamefont {de~Paula~Netto},\ and\
  \citenamefont {Shapiro}(2017)}]{Buchbinder:2017zaa}%
  \BibitemOpen
  \bibfield  {author} {\bibinfo {author} {\bibfnamefont {I.~L.}\ \bibnamefont
  {Buchbinder}}, \bibinfo {author} {\bibfnamefont {T.}~\bibnamefont
  {de~Paula~Netto}},\ and\ \bibinfo {author} {\bibfnamefont {I.~L.}\
  \bibnamefont {Shapiro}},\ }\bibfield  {title} {\enquote {\bibinfo {title}
  {{Massive vector field on curved background: Nonminimal coupling,
  quantization, and divergences}},}\ }\href
  {https://doi.org/10.1103/PhysRevD.95.085009} {\bibfield  {journal} {\bibinfo
  {journal} {Phys. Rev. D}\ }\textbf {\bibinfo {volume} {95}},\ \bibinfo
  {pages} {085009} (\bibinfo {year} {2017})},\ \Eprint
  {https://arxiv.org/abs/1703.00526} {arXiv:1703.00526 [hep-th]} \BibitemShut
  {NoStop}%
\bibitem [{\citenamefont {B{\"a}r}, \citenamefont {Ginoux},\ and\ \citenamefont
  {Pf{\"a}ffle}(2007)}]{baerginouxpfaeffle2007}%
  \BibitemOpen
  \bibfield  {author} {\bibinfo {author} {\bibfnamefont {C.}~\bibnamefont
  {B{\"a}r}}, \bibinfo {author} {\bibfnamefont {N.}~\bibnamefont {Ginoux}},\
  and\ \bibinfo {author} {\bibfnamefont {F.}~\bibnamefont {Pf{\"a}ffle}},\
  }\href@noop {} {\emph {\bibinfo {title} {{Wave Equations on Lorentzian
  Manifolds and Quantization}}}}\ (\bibinfo  {publisher} {{European
  Mathematical Society Publishing House}},\ \bibinfo {address} {Z{\"u}rich,
  Switzerland},\ \bibinfo {year} {2007})\BibitemShut {NoStop}%
\bibitem [{\citenamefont {B{\"a}r}\ and\ \citenamefont
  {Ginoux}(2012)}]{baerginoux2012}%
  \BibitemOpen
  \bibfield  {author} {\bibinfo {author} {\bibfnamefont {C.}~\bibnamefont
  {B{\"a}r}}\ and\ \bibinfo {author} {\bibfnamefont {N.}~\bibnamefont
  {Ginoux}},\ }\enquote {\bibinfo {title} {{Classical and Quantum Fields on
  Lorentzian Manifolds}},}\ in\ \href
  {https://doi.org/10.1007/978-3-642-22842-1_12} {\emph {\bibinfo {booktitle}
  {{Global Differential Geometry}}}},\ \bibinfo {editor} {edited by\ \bibinfo
  {editor} {\bibfnamefont {C.}~\bibnamefont {B{\"a}r}}, \bibinfo {editor}
  {\bibfnamefont {N.}~\bibnamefont {Ginoux}},\ and\ \bibinfo {editor}
  {\bibfnamefont {M.}~\bibnamefont {Schwarz}}}\ (\bibinfo  {publisher}
  {Springer-Verlag},\ \bibinfo {address} {Berlin/Heidelberg, Germany},\
  \bibinfo {year} {2012})\ p.\ \bibinfo {pages} {359}\BibitemShut {NoStop}%
\bibitem [{\citenamefont {B{\"a}r}(2015)}]{baer2015}%
  \BibitemOpen
  \bibfield  {author} {\bibinfo {author} {\bibfnamefont {C.}~\bibnamefont
  {B{\"a}r}},\ }\bibfield  {title} {\enquote {\bibinfo {title}
  {{Green-Hyperbolic Operators on Globally Hyperbolic Spacetimes}},}\ }\href
  {https://doi.org/10.1007/s00220-014-2097-7} {\bibfield  {journal} {\bibinfo
  {journal} {Commun.~Math.~Phys.}\ }\textbf {\bibinfo {volume} {333}},\
  \bibinfo {pages} {1585} (\bibinfo {year} {2015})},\ \Eprint
  {https://arxiv.org/abs/1310.0738} {arXiv:1310.0738 [math-ph]} \BibitemShut
  {NoStop}%
\bibitem [{\citenamefont {Synge}(1960)}]{synge1960}%
  \BibitemOpen
  \bibfield  {author} {\bibinfo {author} {\bibfnamefont {J.~L.}\ \bibnamefont
  {Synge}},\ }\href@noop {} {\emph {\bibinfo {title} {{Relativity: The General
  Theory}}}}\ (\bibinfo  {publisher} {North-Holland},\ \bibinfo {address}
  {Amsterdam, The Netherlands},\ \bibinfo {year} {1960})\BibitemShut {NoStop}%
\bibitem [{\citenamefont {DeWitt}\ and\ \citenamefont
  {Brehme}(1960)}]{dewittbrehme1960}%
  \BibitemOpen
  \bibfield  {author} {\bibinfo {author} {\bibfnamefont {B.~S.}\ \bibnamefont
  {DeWitt}}\ and\ \bibinfo {author} {\bibfnamefont {R.~W.}\ \bibnamefont
  {Brehme}},\ }\bibfield  {title} {\enquote {\bibinfo {title} {{Radiation
  damping in a gravitational field}},}\ }\href
  {https://doi.org/10.1016/0003-4916(60)90030-0} {\bibfield  {journal}
  {\bibinfo  {journal} {Ann.~Phys.}\ }\textbf {\bibinfo {volume} {9}},\
  \bibinfo {pages} {220} (\bibinfo {year} {1960})}\BibitemShut {NoStop}%
\bibitem [{\citenamefont {Poisson}, \citenamefont {Pound},\ and\ \citenamefont
  {Vega}(2011)}]{poissonpoundvega2011}%
  \BibitemOpen
  \bibfield  {author} {\bibinfo {author} {\bibfnamefont {E.}~\bibnamefont
  {Poisson}}, \bibinfo {author} {\bibfnamefont {A.}~\bibnamefont {Pound}},\
  and\ \bibinfo {author} {\bibfnamefont {I.}~\bibnamefont {Vega}},\ }\bibfield
  {title} {\enquote {\bibinfo {title} {{The Motion of Point Particles in Curved
  Spacetime}},}\ }\href {https://doi.org/10.12942/lrr-2011-7} {\bibfield
  {journal} {\bibinfo  {journal} {Liv. Rev. Rel.}\ }\textbf {\bibinfo {volume}
  {14}},\ \bibinfo {pages} {7} (\bibinfo {year} {2011})},\ \Eprint
  {https://arxiv.org/abs/1102.0529} {arXiv:1102.0529 [gr-qc]} \BibitemShut
  {NoStop}%
\bibitem [{\citenamefont {Lichnerowicz}(1961)}]{lichnerowicz1961}%
  \BibitemOpen
  \bibfield  {author} {\bibinfo {author} {\bibfnamefont {A.}~\bibnamefont
  {Lichnerowicz}},\ }\bibfield  {title} {\enquote {\bibinfo {title}
  {{Propagateurs et commutateurs en Relativit{\'e} G{\'e}n{\'e}rale}},}\ }\href
  {http://www.numdam.org/item?id=PMIHES_1961__10__5_0} {\bibfield  {journal}
  {\bibinfo  {journal} {Publications Math{\'e}matiques de l'IH{\'E}S}\ }\textbf
  {\bibinfo {volume} {10}},\ \bibinfo {pages} {5} (\bibinfo {year}
  {1961})}\BibitemShut {NoStop}%
\bibitem [{\citenamefont {Mart{\'\i}n-Garc{\'\i}a}\ \emph
  {et~al.}(2022)\citenamefont {Mart{\'\i}n-Garc{\'\i}a} \emph {et~al.}}]{xact}%
  \BibitemOpen
  \bibfield  {author} {\bibinfo {author} {\bibfnamefont {J.~M.}\ \bibnamefont
  {Mart{\'\i}n-Garc{\'\i}a}} \emph {et~al.},\ }\href {http://www.xact.es}
  {\enquote {\bibinfo {title} {{xAct: Efficient tensor computer algebra for the
  Wolfram Language }},}\ }\bibinfo {howpublished}
  {\href{http://www.xact.es}{http://www.xact.es}} (\bibinfo {year}
  {2022})\BibitemShut {NoStop}%
\bibitem [{\citenamefont {Mart{\'\i}n-Garc{\'\i}a}(2008)}]{martingarcia2008}%
  \BibitemOpen
  \bibfield  {author} {\bibinfo {author} {\bibfnamefont {J.~M.}\ \bibnamefont
  {Mart{\'\i}n-Garc{\'\i}a}},\ }\bibfield  {title} {\enquote {\bibinfo {title}
  {{xPerm: fast index canonicalization for tensor computer algebra}},}\ }\href
  {https://doi.org/10.1016/j.cpc.2008.05.009} {\bibfield  {journal} {\bibinfo
  {journal} {Comput. Phys. Commun.}\ }\textbf {\bibinfo {volume} {179}},\
  \bibinfo {pages} {597--603} (\bibinfo {year} {2008})}\BibitemShut {NoStop}%
\bibitem [{\citenamefont {Brizuela}, \citenamefont {Martin-Garcia},\ and\
  \citenamefont {Mena~Marugan}(2009)}]{brizuelaetal2009}%
  \BibitemOpen
  \bibfield  {author} {\bibinfo {author} {\bibfnamefont {D.}~\bibnamefont
  {Brizuela}}, \bibinfo {author} {\bibfnamefont {J.~M.}\ \bibnamefont
  {Martin-Garcia}},\ and\ \bibinfo {author} {\bibfnamefont {G.~A.}\
  \bibnamefont {Mena~Marugan}},\ }\bibfield  {title} {\enquote {\bibinfo
  {title} {{\textit{xPert}: computer algebra for metric perturbation
  theory}},}\ }\href {https://doi.org/10.1007/s10714-009-0773-2} {\bibfield
  {journal} {\bibinfo  {journal} {Gen. Rel. Grav.}\ }\textbf {\bibinfo {volume}
  {41}},\ \bibinfo {pages} {2415--2431} (\bibinfo {year} {2009})},\ \Eprint
  {https://arxiv.org/abs/0807.0824} {arXiv:0807.0824 [gr-qc]} \BibitemShut
  {NoStop}%
\bibitem [{\citenamefont {Nutma}(2014)}]{nutma2014}%
  \BibitemOpen
  \bibfield  {author} {\bibinfo {author} {\bibfnamefont {T.}~\bibnamefont
  {Nutma}},\ }\bibfield  {title} {\enquote {\bibinfo {title} {{\textit{xTras}:
  A field-theory inspired xAct package for Mathematica}},}\ }\href
  {https://doi.org/10.1016/j.cpc.2014.02.006} {\bibfield  {journal} {\bibinfo
  {journal} {Comput. Phys. Commun.}\ }\textbf {\bibinfo {volume} {185}},\
  \bibinfo {pages} {1719--1738} (\bibinfo {year} {2014})},\ \Eprint
  {https://arxiv.org/abs/1308.3493} {arXiv:1308.3493 [cs.SC]} \BibitemShut
  {NoStop}%
\bibitem [{\citenamefont {Ward}(1950)}]{ward1950}%
  \BibitemOpen
  \bibfield  {author} {\bibinfo {author} {\bibfnamefont {J.~C.}\ \bibnamefont
  {Ward}},\ }\bibfield  {title} {\enquote {\bibinfo {title} {{An Identity in
  Quantum Electrodynamics}},}\ }\href {https://doi.org/10.1103/PhysRev.78.182}
  {\bibfield  {journal} {\bibinfo  {journal} {Phys.~Rev.}\ }\textbf {\bibinfo
  {volume} {78}},\ \bibinfo {pages} {182} (\bibinfo {year} {1950})}\BibitemShut
  {NoStop}%
\bibitem [{\citenamefont {Takahashi}(1957)}]{takahashi1957}%
  \BibitemOpen
  \bibfield  {author} {\bibinfo {author} {\bibfnamefont {Y.}~\bibnamefont
  {Takahashi}},\ }\bibfield  {title} {\enquote {\bibinfo {title} {{On the
  generalized Ward identity}},}\ }\href {https://doi.org/10.1007/BF02832514}
  {\bibfield  {journal} {\bibinfo  {journal} {Nuovo~Cim.}\ }\textbf {\bibinfo
  {volume} {6}},\ \bibinfo {pages} {371} (\bibinfo {year} {1957})}\BibitemShut
  {NoStop}%
\bibitem [{\citenamefont {Taylor}(1971)}]{taylor1971}%
  \BibitemOpen
  \bibfield  {author} {\bibinfo {author} {\bibfnamefont {J.~C.}\ \bibnamefont
  {Taylor}},\ }\bibfield  {title} {\enquote {\bibinfo {title} {{Ward Identities
  and Charge Renormalization of the Yang-Mills Field}},}\ }\href
  {https://doi.org/10.1016/0550-3213(71)90297-5} {\bibfield  {journal}
  {\bibinfo  {journal} {Nucl.~Phys.~B}\ }\textbf {\bibinfo {volume} {33}},\
  \bibinfo {pages} {436} (\bibinfo {year} {1971})}\BibitemShut {NoStop}%
\bibitem [{\citenamefont {Slavnov}(1972)}]{slavnov1972}%
  \BibitemOpen
  \bibfield  {author} {\bibinfo {author} {\bibfnamefont {A.~A.}\ \bibnamefont
  {Slavnov}},\ }\bibfield  {title} {\enquote {\bibinfo {title} {{Ward
  Identities in Gauge Theories}},}\ }\href {https://doi.org/10.1007/BF01090719}
  {\bibfield  {journal} {\bibinfo  {journal} {Theor.~Math.~Phys.}\ }\textbf
  {\bibinfo {volume} {10}},\ \bibinfo {pages} {99} (\bibinfo {year}
  {1972})}\BibitemShut {NoStop}%
\bibitem [{\citenamefont {Choquet-Bruhat}, \citenamefont {DeWitt-Morette},\
  and\ \citenamefont {Dillard-Bleick}(1982)}]{choquet1982analysis}%
  \BibitemOpen
  \bibfield  {author} {\bibinfo {author} {\bibfnamefont {Y.}~\bibnamefont
  {Choquet-Bruhat}}, \bibinfo {author} {\bibfnamefont {C.}~\bibnamefont
  {DeWitt-Morette}},\ and\ \bibinfo {author} {\bibfnamefont {M.}~\bibnamefont
  {Dillard-Bleick}},\ }\href@noop {} {\emph {\bibinfo {title} {Analysis,
  Manifolds and Physics}}},\ \bibinfo {edition} {revised}\ ed.,\ Vol.~\bibinfo
  {volume} {I}\ (\bibinfo  {publisher} {North-Holland},\ \bibinfo {address}
  {Amsterdam, The Netherlands},\ \bibinfo {year} {1982})\BibitemShut {NoStop}%
\bibitem [{\citenamefont {Fulling}(1989)}]{Fulling:1989nb}%
  \BibitemOpen
  \bibfield  {author} {\bibinfo {author} {\bibfnamefont {S.~A.}\ \bibnamefont
  {Fulling}},\ }\href@noop {} {\emph {\bibinfo {title} {{Aspects of Quantum
  Field Theory in Curved Space-time}}}},\ \bibinfo {series} {London
  Mathematical Society Student Texts}, Vol.~\bibinfo {volume} {17}\ (\bibinfo
  {publisher} {London Mathematical Society},\ \bibinfo {address} {London, UK},\
  \bibinfo {year} {1989})\BibitemShut {NoStop}%
\bibitem [{\citenamefont {Camporesi}(1990)}]{camporesi1990harmonic}%
  \BibitemOpen
  \bibfield  {author} {\bibinfo {author} {\bibfnamefont {R.}~\bibnamefont
  {Camporesi}},\ }\href {https://doi.org/10.1016/0370-1573(90)90120-Q} {\emph
  {\bibinfo {title} {{Harmonic analysis and propagators on homogeneous
  spaces}}}},\ Vol.\ \bibinfo {volume} {196}\ (\bibinfo {year} {1990})\ pp.\
  \bibinfo {pages} {1--134}\BibitemShut {NoStop}%
\bibitem [{\citenamefont {Berline}, \citenamefont {Getzler},\ and\
  \citenamefont {Vergne}(1992)}]{berline2003heat}%
  \BibitemOpen
  \bibfield  {author} {\bibinfo {author} {\bibfnamefont {N.}~\bibnamefont
  {Berline}}, \bibinfo {author} {\bibfnamefont {E.}~\bibnamefont {Getzler}},\
  and\ \bibinfo {author} {\bibfnamefont {M.}~\bibnamefont {Vergne}},\
  }\href@noop {} {\emph {\bibinfo {title} {Heat kernels and Dirac
  operators}}},\ \bibinfo {edition} {corr. 2nd}\ ed.,\ \bibinfo {series}
  {Grundlehren der mathematischen Wissenschaften}, Vol.\ \bibinfo {volume}
  {298}\ (\bibinfo  {publisher} {Springer-Verlag},\ \bibinfo {address} {Berlin,
  Heidelberg, Germany},\ \bibinfo {year} {1992})\BibitemShut {NoStop}%
\bibitem [{\citenamefont {Kawakami}(2010)}]{kawakami2010global}%
  \BibitemOpen
  \bibfield  {author} {\bibinfo {author} {\bibfnamefont {T.}~\bibnamefont
  {Kawakami}},\ }\bibfield  {title} {\enquote {\bibinfo {title} {{Global
  existence of solutions for the heat equation with a nonlinear boundary
  condition}},}\ }\href@noop {} {\bibfield  {journal} {\bibinfo  {journal} {J.
  Math. Anal. Appl.}\ }\textbf {\bibinfo {volume} {368}},\ \bibinfo {pages}
  {320--329} (\bibinfo {year} {2010})}\BibitemShut {NoStop}%
\bibitem [{\citenamefont {Synge}(1931)}]{Synge:1931zz}%
  \BibitemOpen
  \bibfield  {author} {\bibinfo {author} {\bibfnamefont {J.~L.}\ \bibnamefont
  {Synge}},\ }\bibfield  {title} {\enquote {\bibinfo {title} {{A characteristic
  function in Riemannian space and its applications to the solution of geodesic
  triangles}},}\ }\href {https://doi.org/10.1112/plms/s2-32.1.241} {\bibfield
  {journal} {\bibinfo  {journal} {Proc. Lond. Math. Soc.}\ }\textbf {\bibinfo
  {volume} {32}},\ \bibinfo {pages} {241} (\bibinfo {year} {1931})}\BibitemShut
  {NoStop}%
\bibitem [{\citenamefont {Van~Vleck}(1928)}]{VanVleck:1928zz}%
  \BibitemOpen
  \bibfield  {author} {\bibinfo {author} {\bibfnamefont {J.~H.}\ \bibnamefont
  {Van~Vleck}},\ }\bibfield  {title} {\enquote {\bibinfo {title} {{The
  Correspondence Principle in the Statistical Interpretation of Quantum
  Mechanics}},}\ }\href {https://doi.org/10.1073/pnas.14.2.178} {\bibfield
  {journal} {\bibinfo  {journal} {Proc. Nat. Acad. Sci.}\ }\textbf {\bibinfo
  {volume} {14}},\ \bibinfo {pages} {178--188} (\bibinfo {year}
  {1928})}\BibitemShut {NoStop}%
\bibitem [{\citenamefont {Barvinsky}\ and\ \citenamefont
  {Vilkovisky}(1987)}]{Barvinsky:1987uw}%
  \BibitemOpen
  \bibfield  {author} {\bibinfo {author} {\bibfnamefont {A.~O.}\ \bibnamefont
  {Barvinsky}}\ and\ \bibinfo {author} {\bibfnamefont {G.~A.}\ \bibnamefont
  {Vilkovisky}},\ }\bibfield  {title} {\enquote {\bibinfo {title} {{Beyond the
  Schwinger-DeWitt technique: Converting loops into trees and in-in
  currents}},}\ }\href {https://doi.org/10.1016/0550-3213(87)90681-X}
  {\bibfield  {journal} {\bibinfo  {journal} {Nucl. Phys. B}\ }\textbf
  {\bibinfo {volume} {282}},\ \bibinfo {pages} {163--188} (\bibinfo {year}
  {1987})}\BibitemShut {NoStop}%
\bibitem [{\citenamefont {Groh}, \citenamefont {Saueressig},\ and\
  \citenamefont {Zanusso}(2011)}]{Groh:2011dw}%
  \BibitemOpen
  \bibfield  {author} {\bibinfo {author} {\bibfnamefont {K.}~\bibnamefont
  {Groh}}, \bibinfo {author} {\bibfnamefont {F.}~\bibnamefont {Saueressig}},\
  and\ \bibinfo {author} {\bibfnamefont {O.}~\bibnamefont {Zanusso}},\
  }\bibfield  {title} {\enquote {\bibinfo {title} {{Off-diagonal heat-kernel
  expansion and its application to fields with differential constraints}},}\
  }\href@noop {} {\  (\bibinfo {year} {2011})},\ \Eprint
  {https://arxiv.org/abs/1112.4856} {arXiv:1112.4856 [math-ph]} \BibitemShut
  {NoStop}%
\bibitem [{\citenamefont {Decanini}\ and\ \citenamefont
  {Folacci}(2006)}]{Decanini:2005gt}%
  \BibitemOpen
  \bibfield  {author} {\bibinfo {author} {\bibfnamefont {Y.}~\bibnamefont
  {Decanini}}\ and\ \bibinfo {author} {\bibfnamefont {A.}~\bibnamefont
  {Folacci}},\ }\bibfield  {title} {\enquote {\bibinfo {title} {{Off-diagonal
  coefficients of the Dewitt-Schwinger and Hadamard representations of the
  Feynman propagator}},}\ }\href {https://doi.org/10.1103/PhysRevD.73.044027}
  {\bibfield  {journal} {\bibinfo  {journal} {Phys. Rev. D}\ }\textbf {\bibinfo
  {volume} {73}},\ \bibinfo {pages} {044027} (\bibinfo {year} {2006})},\
  \Eprint {https://arxiv.org/abs/gr-qc/0511115} {arXiv:gr-qc/0511115}
  \BibitemShut {NoStop}%
\bibitem [{\citenamefont {Anselmi}\ and\ \citenamefont
  {Benini}(2007)}]{Anselmi:2007eq}%
  \BibitemOpen
  \bibfield  {author} {\bibinfo {author} {\bibfnamefont {D.}~\bibnamefont
  {Anselmi}}\ and\ \bibinfo {author} {\bibfnamefont {A.}~\bibnamefont
  {Benini}},\ }\bibfield  {title} {\enquote {\bibinfo {title} {{Improved
  Schwinger-DeWitt techniques for higher-derivative corrections to operator
  determinants}},}\ }\href {https://doi.org/10.1088/1126-6708/2007/10/099}
  {\bibfield  {journal} {\bibinfo  {journal} {JHEP}\ }\textbf {\bibinfo
  {volume} {10}},\ \bibinfo {pages} {099} (\bibinfo {year} {2007})},\ \Eprint
  {https://arxiv.org/abs/0704.2840} {arXiv:0704.2840 [hep-th]} \BibitemShut
  {NoStop}%
\bibitem [{\citenamefont {Benedetti}\ \emph {et~al.}(2011)\citenamefont
  {Benedetti}, \citenamefont {Groh}, \citenamefont {Machado},\ and\
  \citenamefont {Saueressig}}]{Benedetti:2010nr}%
  \BibitemOpen
  \bibfield  {author} {\bibinfo {author} {\bibfnamefont {D.}~\bibnamefont
  {Benedetti}}, \bibinfo {author} {\bibfnamefont {K.}~\bibnamefont {Groh}},
  \bibinfo {author} {\bibfnamefont {P.~F.}\ \bibnamefont {Machado}},\ and\
  \bibinfo {author} {\bibfnamefont {F.}~\bibnamefont {Saueressig}},\ }\bibfield
   {title} {\enquote {\bibinfo {title} {{The Universal RG Machine}},}\ }\href
  {https://doi.org/10.1007/JHEP06(2011)079} {\bibfield  {journal} {\bibinfo
  {journal} {JHEP}\ }\textbf {\bibinfo {volume} {06}},\ \bibinfo {pages} {079}
  (\bibinfo {year} {2011})},\ \Eprint {https://arxiv.org/abs/1012.3081}
  {arXiv:1012.3081 [hep-th]} \BibitemShut {NoStop}%
\bibitem [{\citenamefont {Groh}\ \emph {et~al.}(2011)\citenamefont {Groh},
  \citenamefont {Rechenberger}, \citenamefont {Saueressig},\ and\ \citenamefont
  {Zanusso}}]{Groh:2011vn}%
  \BibitemOpen
  \bibfield  {author} {\bibinfo {author} {\bibfnamefont {K.}~\bibnamefont
  {Groh}}, \bibinfo {author} {\bibfnamefont {S.}~\bibnamefont {Rechenberger}},
  \bibinfo {author} {\bibfnamefont {F.}~\bibnamefont {Saueressig}},\ and\
  \bibinfo {author} {\bibfnamefont {O.}~\bibnamefont {Zanusso}},\ }\bibfield
  {title} {\enquote {\bibinfo {title} {{Higher Derivative Gravity from the
  Universal Renormalization Group Machine}},}\ }\href
  {https://doi.org/10.22323/1.134.0124} {\bibfield  {journal} {\bibinfo
  {journal} {PoS}\ }\textbf {\bibinfo {volume} {134 (EPS-HEP2011)}},\ \bibinfo
  {pages} {124} (\bibinfo {year} {2011})},\ \Eprint
  {https://arxiv.org/abs/1111.1743} {arXiv:1111.1743 [hep-th]} \BibitemShut
  {NoStop}%
\bibitem [{\citenamefont {Hollands}\ and\ \citenamefont
  {Wald}(2001)}]{Hollands:2001nf}%
  \BibitemOpen
  \bibfield  {author} {\bibinfo {author} {\bibfnamefont {S.}~\bibnamefont
  {Hollands}}\ and\ \bibinfo {author} {\bibfnamefont {R.~M.}\ \bibnamefont
  {Wald}},\ }\bibfield  {title} {\enquote {\bibinfo {title} {{Local Wick
  polynomials and time ordered products of quantum fields in curved
  space-time}},}\ }\href {https://doi.org/10.1007/s002200100540} {\bibfield
  {journal} {\bibinfo  {journal} {Commun. Math. Phys.}\ }\textbf {\bibinfo
  {volume} {223}},\ \bibinfo {pages} {289--326} (\bibinfo {year} {2001})},\
  \Eprint {https://arxiv.org/abs/gr-qc/0103074} {arXiv:gr-qc/0103074}
  \BibitemShut {NoStop}%
\bibitem [{\citenamefont {Sakai}(1971)}]{Sakai}%
  \BibitemOpen
  \bibfield  {author} {\bibinfo {author} {\bibfnamefont {T.}~\bibnamefont
  {Sakai}},\ }\bibfield  {title} {\enquote {\bibinfo {title} {{On eigen-values
  of Laplacian and curvature of Riemannian manifold}},}\ }\href
  {https://doi.org/10.2748/tmj/1178242547} {\bibfield  {journal} {\bibinfo
  {journal} {Tohoku Math. J.}\ }\textbf {\bibinfo {volume} {23}},\ \bibinfo
  {pages} {589 -- 603} (\bibinfo {year} {1971})}\BibitemShut {NoStop}%
\bibitem [{\citenamefont {Berger}, \citenamefont {Gauduchon},\ and\
  \citenamefont {Mazet}(1971)}]{Berger1971LeSD}%
  \BibitemOpen
  \bibfield  {author} {\bibinfo {author} {\bibfnamefont {M.}~\bibnamefont
  {Berger}}, \bibinfo {author} {\bibfnamefont {P.}~\bibnamefont {Gauduchon}},\
  and\ \bibinfo {author} {\bibfnamefont {E.}~\bibnamefont {Mazet}},\
  }\href@noop {} {\emph {\bibinfo {title} {Le spectre d'une variété
  Riemannienne}}},\ \bibinfo {series} {Lect. Notes Math.}, Vol.\ \bibinfo
  {volume} {194}\ (\bibinfo  {publisher} {Springer-Verlag},\ \bibinfo {address}
  {Berlin, Heidelberg, Germany},\ \bibinfo {year} {1971})\ pp.\ \bibinfo
  {pages} {141--241}\BibitemShut {NoStop}%
\bibitem [{\citenamefont {Chavel}, \citenamefont {Randol},\ and\ \citenamefont
  {Dodziuk}(1984)}]{chavel1984eigenvalues}%
  \BibitemOpen
  \bibfield  {author} {\bibinfo {author} {\bibfnamefont {I.}~\bibnamefont
  {Chavel}}, \bibinfo {author} {\bibfnamefont {B.}~\bibnamefont {Randol}},\
  and\ \bibinfo {author} {\bibfnamefont {J.}~\bibnamefont {Dodziuk}},\
  }\href@noop {} {\emph {\bibinfo {title} {Eigenvalues in Riemannian
  Geometry}}},\ \bibinfo {series} {Pure and Applied Mathematics}, Vol.\
  \bibinfo {volume} {115}\ (\bibinfo  {publisher} {Academic Press},\ \bibinfo
  {address} {Cambridge, Mass., USA},\ \bibinfo {year} {1984})\BibitemShut
  {NoStop}%
\bibitem [{\citenamefont {Avramidi}\ and\ \citenamefont
  {Schimming}(1996)}]{Avramidi:1995qe}%
  \BibitemOpen
  \bibfield  {author} {\bibinfo {author} {\bibfnamefont {I.~G.}\ \bibnamefont
  {Avramidi}}\ and\ \bibinfo {author} {\bibfnamefont {R.}~\bibnamefont
  {Schimming}},\ }\bibfield  {title} {\enquote {\bibinfo {title} {{Algorithms
  for the calculation of the heat kernel coefficients}},}\ }in\ \href@noop {}
  {\emph {\bibinfo {booktitle} {{Quantum Field Theory under the Influence of
  External Conditions}}}},\ \bibinfo {series} {Teubner-Texte zur Physik},
  Vol.~\bibinfo {volume} {30},\ \bibinfo {editor} {edited by\ \bibinfo {editor}
  {\bibfnamefont {M.}~\bibnamefont {Bordag}}}\ (\bibinfo  {publisher} {B. G.
  Teubner Verlagsgesellschaft},\ \bibinfo {address} {Stuttgart, Leipzig,
  Germany},\ \bibinfo {year} {1996})\ pp.\ \bibinfo {pages} {150--162},\
  \Eprint {https://arxiv.org/abs/hep-th/9510206} {arXiv:hep-th/9510206}
  \BibitemShut {NoStop}%
\bibitem [{\citenamefont {Gilkey}(1979)}]{Gilkey1979}%
  \BibitemOpen
  \bibfield  {author} {\bibinfo {author} {\bibfnamefont {P.~B.}\ \bibnamefont
  {Gilkey}},\ }\bibfield  {title} {\enquote {\bibinfo {title} {Recursion
  relations and the asymptotic behavior of the eigenvalues of the laplacian},}\
  }\href {http://eudml.org/doc/89401} {\bibfield  {journal} {\bibinfo
  {journal} {Compos. Math.}\ }\textbf {\bibinfo {volume} {38}},\ \bibinfo
  {pages} {201--240} (\bibinfo {year} {1979})}\BibitemShut {NoStop}%
\bibitem [{\citenamefont {Kirsten}(1999)}]{Kirsten:1999qjn}%
  \BibitemOpen
  \bibfield  {author} {\bibinfo {author} {\bibfnamefont {K.}~\bibnamefont
  {Kirsten}},\ }\bibfield  {title} {\enquote {\bibinfo {title} {{Spectral
  functions in mathematics and physics}},}\ }\href
  {https://doi.org/10.1063/1.59656} {\bibfield  {journal} {\bibinfo  {journal}
  {AIP Conf. Proc.}\ }\textbf {\bibinfo {volume} {484}},\ \bibinfo {pages}
  {106--146} (\bibinfo {year} {1999})},\ \Eprint
  {https://arxiv.org/abs/hep-th/0005133} {arXiv:hep-th/0005133} \BibitemShut
  {NoStop}%
\bibitem [{\citenamefont {{van de Ven}}(1985)}]{vandeven1985}%
  \BibitemOpen
  \bibfield  {author} {\bibinfo {author} {\bibfnamefont {A.~E.~M.}\
  \bibnamefont {{van de Ven}}},\ }\bibfield  {title} {\enquote {\bibinfo
  {title} {Explicit counteraction algorithms in higher dimensions},}\ }\href
  {https://doi.org/10.1016/0550-3213(85)90496-1} {\bibfield  {journal}
  {\bibinfo  {journal} {Nucl. Phys. B}\ }\textbf {\bibinfo {volume} {250}},\
  \bibinfo {pages} {593--617} (\bibinfo {year} {1985})}\BibitemShut {NoStop}%
\bibitem [{\citenamefont {Avramidi}(1986)}]{Avramidi:1986mj}%
  \BibitemOpen
  \bibfield  {author} {\bibinfo {author} {\bibfnamefont {I.~G.}\ \bibnamefont
  {Avramidi}},\ }\emph {\bibinfo {title} {{Covariant methods for the
  calculation of the effective action in quantum field theory and investigation
  of higher derivative quantum gravity}}},\ \href@noop {} {\bibinfo {type}
  {Other thesis}} (\bibinfo {year} {1986}),\ \Eprint
  {https://arxiv.org/abs/hep-th/9510140} {arXiv:hep-th/9510140} \BibitemShut
  {NoStop}%
\bibitem [{\citenamefont {Avramidi}(1989)}]{Avramidi:1989ik}%
  \BibitemOpen
  \bibfield  {author} {\bibinfo {author} {\bibfnamefont {I.~G.}\ \bibnamefont
  {Avramidi}},\ }\bibfield  {title} {\enquote {\bibinfo {title} {{Background
  field calculations in quantum field theory (vacuum polarization)}},}\ }\href
  {https://doi.org/10.1007/BF01016530} {\bibfield  {journal} {\bibinfo
  {journal} {Theor. Math. Phys.}\ }\textbf {\bibinfo {volume} {79}},\ \bibinfo
  {pages} {494--502} (\bibinfo {year} {1989})}\BibitemShut {NoStop}%
\bibitem [{\citenamefont {van~de Ven}(1998)}]{vandeVen:1997pf}%
  \BibitemOpen
  \bibfield  {author} {\bibinfo {author} {\bibfnamefont {A.~E.~M.}\
  \bibnamefont {van~de Ven}},\ }\bibfield  {title} {\enquote {\bibinfo {title}
  {{Index free heat kernel coefficients}},}\ }\href
  {https://doi.org/10.1088/0264-9381/15/8/014} {\bibfield  {journal} {\bibinfo
  {journal} {Class. Quant. Grav.}\ }\textbf {\bibinfo {volume} {15}},\ \bibinfo
  {pages} {2311--2344} (\bibinfo {year} {1998})},\ \Eprint
  {https://arxiv.org/abs/hep-th/9708152} {arXiv:hep-th/9708152} \BibitemShut
  {NoStop}%
\bibitem [{\citenamefont {Franchino-Vi\~nas}\ \emph {et~al.}(2024)\citenamefont
  {Franchino-Vi\~nas}, \citenamefont {Garc\'\i{}a-P\'erez}, \citenamefont
  {Mazzitelli}, \citenamefont {Vitagliano},\ and\ \citenamefont
  {Haimovichi}}]{FranchinoVinas2023}%
  \BibitemOpen
  \bibfield  {author} {\bibinfo {author} {\bibfnamefont {S.~A.}\ \bibnamefont
  {Franchino-Vi\~nas}}, \bibinfo {author} {\bibfnamefont {C.}~\bibnamefont
  {Garc\'\i{}a-P\'erez}}, \bibinfo {author} {\bibfnamefont {F.~D.}\
  \bibnamefont {Mazzitelli}}, \bibinfo {author} {\bibfnamefont
  {V.}~\bibnamefont {Vitagliano}},\ and\ \bibinfo {author} {\bibfnamefont
  {U.~W.}\ \bibnamefont {Haimovichi}},\ }\bibfield  {title} {\enquote {\bibinfo
  {title} {{Resummed heat kernel and effective action for Yukawa and QED}},}\
  }\href {https://doi.org/10.1016/j.physletb.2024.138684} {\bibfield  {journal}
  {\bibinfo  {journal} {Phys. Lett. B}\ }\textbf {\bibinfo {volume} {854}},\
  \bibinfo {pages} {138684} (\bibinfo {year} {2024})},\ \Eprint
  {https://arxiv.org/abs/2312.16303} {arXiv:2312.16303 [hep-th]} \BibitemShut
  {NoStop}%
\bibitem [{\citenamefont {Strohmaier}\ and\ \citenamefont
  {Zelditch}(2023)}]{Strohmaier:2023wsv}%
  \BibitemOpen
  \bibfield  {author} {\bibinfo {author} {\bibfnamefont {A.}~\bibnamefont
  {Strohmaier}}\ and\ \bibinfo {author} {\bibfnamefont {S.}~\bibnamefont
  {Zelditch}},\ }\bibfield  {title} {\enquote {\bibinfo {title} {{Heat and Wave
  kernel expansions for stationary spacetimes}},}\ }\href@noop {} {\  (\bibinfo
  {year} {2023})},\ \Eprint {https://arxiv.org/abs/2308.12148}
  {arXiv:2308.12148 [math.SP]} \BibitemShut {NoStop}%
\bibitem [{\citenamefont {Estrada}\ and\ \citenamefont
  {Fulling}(1999)}]{Estrada:1997qt}%
  \BibitemOpen
  \bibfield  {author} {\bibinfo {author} {\bibfnamefont {R.}~\bibnamefont
  {Estrada}}\ and\ \bibinfo {author} {\bibfnamefont {S.~A.}\ \bibnamefont
  {Fulling}},\ }\bibfield  {title} {\enquote {\bibinfo {title} {{Distributional
  asymptotic expansions of spectral functions and of the associated Green
  kernels}},}\ }\href {https://ejde.math.unt.edu/Volumes/1999/07/estrada.pdf}
  {\bibfield  {journal} {\bibinfo  {journal} {Electron. J. Diff. Eqns.}\
  }\textbf {\bibinfo {volume} {1999}},\ \bibinfo {pages} {1--37} (\bibinfo
  {year} {1999})},\ \Eprint {https://arxiv.org/abs/funct-an/9710003}
  {arXiv:funct-an/9710003} \BibitemShut {NoStop}%
\bibitem [{\citenamefont {Moretti}(1999)}]{Moretti:1999ez}%
  \BibitemOpen
  \bibfield  {author} {\bibinfo {author} {\bibfnamefont {V.}~\bibnamefont
  {Moretti}},\ }\bibfield  {title} {\enquote {\bibinfo {title} {{Proof of the
  Symmetry of the Off-Diagonal Heat-Kernel and Hadamard's Expansion
  Coefficients in General $C^\infty$ Riemannian Manifolds}},}\ }\href
  {https://doi.org/10.1007/s002200050759} {\bibfield  {journal} {\bibinfo
  {journal} {Commun. Math. Phys.}\ }\textbf {\bibinfo {volume} {208}},\
  \bibinfo {pages} {283--309} (\bibinfo {year} {1999})},\ \Eprint
  {https://arxiv.org/abs/gr-qc/9902034} {arXiv:gr-qc/9902034} \BibitemShut
  {NoStop}%
\bibitem [{\citenamefont {Piguet}, \citenamefont {Sibold},\ and\ \citenamefont
  {Schweda}(1980)}]{Piguet:1980fa}%
  \BibitemOpen
  \bibfield  {author} {\bibinfo {author} {\bibfnamefont {O.}~\bibnamefont
  {Piguet}}, \bibinfo {author} {\bibfnamefont {K.}~\bibnamefont {Sibold}},\
  and\ \bibinfo {author} {\bibfnamefont {M.}~\bibnamefont {Schweda}},\
  }\bibfield  {title} {\enquote {\bibinfo {title} {{General solution of the
  supersymmetry consistency conditions}},}\ }\href
  {https://doi.org/10.1016/0550-3213(80)90197-2} {\bibfield  {journal}
  {\bibinfo  {journal} {Nucl. Phys. B}\ }\textbf {\bibinfo {volume} {174}},\
  \bibinfo {pages} {183--188} (\bibinfo {year} {1980})}\BibitemShut {NoStop}%
\bibitem [{\citenamefont {Itoyama}, \citenamefont {Nair},\ and\ \citenamefont
  {Ren}(1985)}]{Itoyama:1985qi}%
  \BibitemOpen
  \bibfield  {author} {\bibinfo {author} {\bibfnamefont {H.}~\bibnamefont
  {Itoyama}}, \bibinfo {author} {\bibfnamefont {V.~P.}\ \bibnamefont {Nair}},\
  and\ \bibinfo {author} {\bibfnamefont {H.-c.}\ \bibnamefont {Ren}},\
  }\bibfield  {title} {\enquote {\bibinfo {title} {{Supersymmetry Anomalies and
  Some Aspects of Renormalization}},}\ }\href
  {https://doi.org/10.1016/0550-3213(85)90289-5} {\bibfield  {journal}
  {\bibinfo  {journal} {Nucl. Phys. B}\ }\textbf {\bibinfo {volume} {262}},\
  \bibinfo {pages} {317--330} (\bibinfo {year} {1985})}\BibitemShut {NoStop}%
\bibitem [{\citenamefont {Bardeen}\ and\ \citenamefont
  {Zumino}(1984)}]{Bardeen:1984pm}%
  \BibitemOpen
  \bibfield  {author} {\bibinfo {author} {\bibfnamefont {W.~A.}\ \bibnamefont
  {Bardeen}}\ and\ \bibinfo {author} {\bibfnamefont {B.}~\bibnamefont
  {Zumino}},\ }\bibfield  {title} {\enquote {\bibinfo {title} {{Consistent and
  Covariant Anomalies in Gauge and Gravitational Theories}},}\ }\href
  {https://doi.org/10.1016/0550-3213(84)90322-5} {\bibfield  {journal}
  {\bibinfo  {journal} {Nucl. Phys. B}\ }\textbf {\bibinfo {volume} {244}},\
  \bibinfo {pages} {421--453} (\bibinfo {year} {1984})}\BibitemShut {NoStop}%
\bibitem [{\citenamefont {Morette}(1951)}]{Morette:1951zz}%
  \BibitemOpen
  \bibfield  {author} {\bibinfo {author} {\bibfnamefont {C.}~\bibnamefont
  {Morette}},\ }\bibfield  {title} {\enquote {\bibinfo {title} {{On the
  definition and approximation of Feynman's path integrals}},}\ }\href
  {https://doi.org/10.1103/PhysRev.81.848} {\bibfield  {journal} {\bibinfo
  {journal} {Phys. Rev.}\ }\textbf {\bibinfo {volume} {81}},\ \bibinfo {pages}
  {848--852} (\bibinfo {year} {1951})}\BibitemShut {NoStop}%
\bibitem [{\citenamefont {Avramidi}\ and\ \citenamefont
  {Buckman}(2016)}]{Avramidi_2016}%
  \BibitemOpen
  \bibfield  {author} {\bibinfo {author} {\bibfnamefont {I.~G.}\ \bibnamefont
  {Avramidi}}\ and\ \bibinfo {author} {\bibfnamefont {B.~J.}\ \bibnamefont
  {Buckman}},\ }\bibfield  {title} {\enquote {\bibinfo {title} {{Heat
  determinant on manifolds}},}\ }\href
  {https://doi.org/10.1016/j.geomphys.2016.02.004} {\bibfield  {journal}
  {\bibinfo  {journal} {J. Geom. Phys.}\ }\textbf {\bibinfo {volume} {104}},\
  \bibinfo {pages} {64--88} (\bibinfo {year} {2016})}\BibitemShut {NoStop}%
\bibitem [{\citenamefont {Ottewill}\ and\ \citenamefont
  {Wardell}(2011)}]{Ottewill:2009uj}%
  \BibitemOpen
  \bibfield  {author} {\bibinfo {author} {\bibfnamefont {A.~C.}\ \bibnamefont
  {Ottewill}}\ and\ \bibinfo {author} {\bibfnamefont {B.}~\bibnamefont
  {Wardell}},\ }\bibfield  {title} {\enquote {\bibinfo {title} {{A Transport
  Equation Approach to Calculations of Hadamard Green functions and
  non-coincident DeWitt coefficients}},}\ }\href
  {https://doi.org/10.1103/PhysRevD.84.104039} {\bibfield  {journal} {\bibinfo
  {journal} {Phys. Rev. D}\ }\textbf {\bibinfo {volume} {84}},\ \bibinfo
  {pages} {104039} (\bibinfo {year} {2011})},\ \bibinfo {note} {[Erratum:
  \href{https://doi.org/10.1103/PhysRevD.101.029901}{Phys. Rev. D \textbf{101},
  029901 (2020)}]},\ \Eprint {https://arxiv.org/abs/0906.0005} {arXiv:0906.0005
  [gr-qc]} \BibitemShut {NoStop}%
\bibitem [{\citenamefont {Allen}\ and\ \citenamefont
  {L{\"u}tken}(1986)}]{Allen:1986qj}%
  \BibitemOpen
  \bibfield  {author} {\bibinfo {author} {\bibfnamefont {B.}~\bibnamefont
  {Allen}}\ and\ \bibinfo {author} {\bibfnamefont {C.~A.}\ \bibnamefont
  {L{\"u}tken}},\ }\bibfield  {title} {\enquote {\bibinfo {title} {{Spinor
  two-point functions in maximally symmetric spaces}},}\ }\href
  {https://doi.org/10.1007/BF01454972} {\bibfield  {journal} {\bibinfo
  {journal} {Commun. Math. Phys.}\ }\textbf {\bibinfo {volume} {106}},\
  \bibinfo {pages} {201} (\bibinfo {year} {1986})}\BibitemShut {NoStop}%
\bibitem [{\citenamefont {Kazdan}(1981)}]{kazdan1981}%
  \BibitemOpen
  \bibfield  {author} {\bibinfo {author} {\bibfnamefont {J.~L.}\ \bibnamefont
  {Kazdan}},\ }\bibfield  {title} {\enquote {\bibinfo {title} {{Another proof
  of Bianchi's identity in Riemannian geometry}},}\ }\href
  {https://doi.org/10.1090/S0002-9939-1981-0593487-3} {\bibfield  {journal}
  {\bibinfo  {journal} {Proc. Amer. Math. Soc.}\ }\textbf {\bibinfo {volume}
  {81}},\ \bibinfo {pages} {341--342} (\bibinfo {year} {1981})}\BibitemShut
  {NoStop}%
\bibitem [{\citenamefont {Schlicht}(1995)}]{schlicht1995}%
  \BibitemOpen
  \bibfield  {author} {\bibinfo {author} {\bibfnamefont {M.}~\bibnamefont
  {Schlicht}},\ }\bibfield  {title} {\enquote {\bibinfo {title} {{Another proof
  of Bianchi's identity in arbitrary bundles.}}}\ }\href
  {https://doi.org/10.1007/BF00774563} {\bibfield  {journal} {\bibinfo
  {journal} {Ann. Glob. Anal. Geom.}\ }\textbf {\bibinfo {volume} {13}},\
  \bibinfo {pages} {19--22} (\bibinfo {year} {1995})}\BibitemShut {NoStop}%
\bibitem [{\citenamefont {Fr{\"o}b}(2020)}]{fieldsx}%
  \BibitemOpen
  \bibfield  {author} {\bibinfo {author} {\bibfnamefont {M.~B.}\ \bibnamefont
  {Fr{\"o}b}},\ }\bibfield  {title} {\enquote {\bibinfo {title} {{FieldsX -- An
  extension package for the xAct tensor computer algebra suite to include
  fermions, gauge fields and BRST cohomology}},}\ }\href@noop {} {\  (\bibinfo
  {year} {2020})},\ \Eprint {https://arxiv.org/abs/2008.12422}
  {arXiv:2008.12422 [hep-th]} \BibitemShut {NoStop}%
\bibitem [{\citenamefont {Moretti}(2000)}]{Moretti:1999fb}%
  \BibitemOpen
  \bibfield  {author} {\bibinfo {author} {\bibfnamefont {V.}~\bibnamefont
  {Moretti}},\ }\bibfield  {title} {\enquote {\bibinfo {title} {{Proof of the
  Symmetry of the Off-Diagonal Hadamard/Seeley--deWitt's Coefficients in
  $C^\infty$ Lorentzian Manifolds by a ``Local Wick Rotation''}},}\ }\href
  {https://doi.org/10.1007/s002200000202} {\bibfield  {journal} {\bibinfo
  {journal} {Commun. Math. Phys.}\ }\textbf {\bibinfo {volume} {212}},\
  \bibinfo {pages} {165--189} (\bibinfo {year} {2000})},\ \Eprint
  {https://arxiv.org/abs/gr-qc/9908068} {arXiv:gr-qc/9908068} \BibitemShut
  {NoStop}%
\bibitem [{\citenamefont {Kami{\'n}ski}(2019)}]{Kaminski:2019adk}%
  \BibitemOpen
  \bibfield  {author} {\bibinfo {author} {\bibfnamefont {W.}~\bibnamefont
  {Kami{\'n}ski}},\ }\bibfield  {title} {\enquote {\bibinfo {title}
  {{Elementary proof of symmetry of the off-diagonal Seeley-DeWitt (and related
  Hadamard) coefficients}},}\ }\href@noop {} {\  (\bibinfo {year} {2019})},\
  \Eprint {https://arxiv.org/abs/1904.03708} {arXiv:1904.03708 [math-ph]}
  \BibitemShut {NoStop}%
\bibitem [{\citenamefont {Lovelock}(1970)}]{lovelock1970}%
  \BibitemOpen
  \bibfield  {author} {\bibinfo {author} {\bibfnamefont {D.}~\bibnamefont
  {Lovelock}},\ }\bibfield  {title} {\enquote {\bibinfo {title} {{Dimensionally
  dependent identities}},}\ }\href {https://doi.org/10.1017/S0305004100046144}
  {\bibfield  {journal} {\bibinfo  {journal} {Math. Proc. Cambridge Phil.
  Soc.}\ }\textbf {\bibinfo {volume} {68}},\ \bibinfo {pages} {345–350}
  (\bibinfo {year} {1970})}\BibitemShut {NoStop}%
\bibitem [{\citenamefont {Edgar}\ and\ \citenamefont
  {H{\"o}glund}(2002)}]{edgarhoeglund2002}%
  \BibitemOpen
  \bibfield  {author} {\bibinfo {author} {\bibfnamefont {S.~B.}\ \bibnamefont
  {Edgar}}\ and\ \bibinfo {author} {\bibfnamefont {A.}~\bibnamefont
  {H{\"o}glund}},\ }\bibfield  {title} {\enquote {\bibinfo {title}
  {{Dimensionally dependent tensor identities by double antisymmetrization}},}\
  }\href {https://doi.org/10.1063/1.1425428} {\bibfield  {journal} {\bibinfo
  {journal} {J. Math. Phys.}\ }\textbf {\bibinfo {volume} {43}},\ \bibinfo
  {pages} {659--677} (\bibinfo {year} {2002})},\ \Eprint
  {https://arxiv.org/abs/gr-qc/0105066} {arXiv:gr-qc/0105066} \BibitemShut
  {NoStop}%
\bibitem [{\citenamefont {Gopakumar}, \citenamefont {Gupta},\ and\
  \citenamefont {Lal}(2011)}]{Gopakumar:2011qs}%
  \BibitemOpen
  \bibfield  {author} {\bibinfo {author} {\bibfnamefont {R.}~\bibnamefont
  {Gopakumar}}, \bibinfo {author} {\bibfnamefont {R.~K.}\ \bibnamefont
  {Gupta}},\ and\ \bibinfo {author} {\bibfnamefont {S.}~\bibnamefont {Lal}},\
  }\bibfield  {title} {\enquote {\bibinfo {title} {{The Heat Kernel on
  \emph{AdS}}},}\ }\href {https://doi.org/10.1007/JHEP11(2011)010} {\bibfield
  {journal} {\bibinfo  {journal} {JHEP}\ }\textbf {\bibinfo {volume} {11}},\
  \bibinfo {pages} {010} (\bibinfo {year} {2011})},\ \Eprint
  {https://arxiv.org/abs/1103.3627} {arXiv:1103.3627 [hep-th]} \BibitemShut
  {NoStop}%
\bibitem [{\citenamefont {Bobev}\ \emph {et~al.}(2024)\citenamefont {Bobev},
  \citenamefont {David}, \citenamefont {Hong}, \citenamefont {Reys},\ and\
  \citenamefont {Zhang}}]{Bobev:2023dwx}%
  \BibitemOpen
  \bibfield  {author} {\bibinfo {author} {\bibfnamefont {N.}~\bibnamefont
  {Bobev}}, \bibinfo {author} {\bibfnamefont {M.}~\bibnamefont {David}},
  \bibinfo {author} {\bibfnamefont {J.}~\bibnamefont {Hong}}, \bibinfo {author}
  {\bibfnamefont {V.}~\bibnamefont {Reys}},\ and\ \bibinfo {author}
  {\bibfnamefont {X.}~\bibnamefont {Zhang}},\ }\bibfield  {title} {\enquote
  {\bibinfo {title} {{A compendium of logarithmic corrections in AdS/CFT}},}\
  }\href {https://doi.org/10.1007/JHEP04(2024)020} {\bibfield  {journal}
  {\bibinfo  {journal} {JHEP}\ }\textbf {\bibinfo {volume} {04}},\ \bibinfo
  {pages} {020} (\bibinfo {year} {2024})},\ \Eprint
  {https://arxiv.org/abs/2312.08909} {arXiv:2312.08909 [hep-th]} \BibitemShut
  {NoStop}%
\bibitem [{\citenamefont {Kluth}\ and\ \citenamefont
  {Litim}(2020)}]{Kluth:2019vkg}%
  \BibitemOpen
  \bibfield  {author} {\bibinfo {author} {\bibfnamefont {Y.}~\bibnamefont
  {Kluth}}\ and\ \bibinfo {author} {\bibfnamefont {D.~F.}\ \bibnamefont
  {Litim}},\ }\bibfield  {title} {\enquote {\bibinfo {title} {{Heat kernel
  coefficients on the sphere in any dimension}},}\ }\href
  {https://doi.org/10.1140/epjc/s10052-020-7784-2} {\bibfield  {journal}
  {\bibinfo  {journal} {Eur. Phys. J. C}\ }\textbf {\bibinfo {volume} {80}},\
  \bibinfo {pages} {269} (\bibinfo {year} {2020})},\ \Eprint
  {https://arxiv.org/abs/1910.00543} {arXiv:1910.00543 [hep-th]} \BibitemShut
  {NoStop}%
\bibitem [{dlm()}]{dlmf}%
  \BibitemOpen
  \href {http://dlmf.nist.gov} {\enquote {\bibinfo {title} {{NIST Digital
  Library of Mathematical Functions}},}\ }\bibinfo {howpublished}
  {\href{http://dlmf.nist.gov}{http://dlmf.nist.gov}}\BibitemShut {NoStop}%
\bibitem [{\citenamefont {Brown}\ and\ \citenamefont
  {Marolf}(1996)}]{brownmarolf1996}%
  \BibitemOpen
  \bibfield  {author} {\bibinfo {author} {\bibfnamefont {J.~D.}\ \bibnamefont
  {Brown}}\ and\ \bibinfo {author} {\bibfnamefont {D.}~\bibnamefont {Marolf}},\
  }\bibfield  {title} {\enquote {\bibinfo {title} {{On relativistic material
  reference systems}},}\ }\href {https://doi.org/10.1103/PhysRevD.53.1835}
  {\bibfield  {journal} {\bibinfo  {journal} {Phys. Rev. D}\ }\textbf {\bibinfo
  {volume} {53}},\ \bibinfo {pages} {1835--1844} (\bibinfo {year} {1996})},\
  \Eprint {https://arxiv.org/abs/gr-qc/9509026} {arXiv:gr-qc/9509026}
  \BibitemShut {NoStop}%
\bibitem [{\citenamefont {Dittrich}(2006)}]{dittrich2006}%
  \BibitemOpen
  \bibfield  {author} {\bibinfo {author} {\bibfnamefont {B.}~\bibnamefont
  {Dittrich}},\ }\bibfield  {title} {\enquote {\bibinfo {title} {{Partial and
  complete observables for canonical general relativity}},}\ }\href
  {https://doi.org/10.1088/0264-9381/23/22/006} {\bibfield  {journal} {\bibinfo
   {journal} {Class. Quant. Grav.}\ }\textbf {\bibinfo {volume} {23}},\
  \bibinfo {pages} {6155--6184} (\bibinfo {year} {2006})},\ \Eprint
  {https://arxiv.org/abs/gr-qc/0507106} {arXiv:gr-qc/0507106} \BibitemShut
  {NoStop}%
\bibitem [{\citenamefont {Nakamura}(2007)}]{nakamura2007}%
  \BibitemOpen
  \bibfield  {author} {\bibinfo {author} {\bibfnamefont {K.}~\bibnamefont
  {Nakamura}},\ }\bibfield  {title} {\enquote {\bibinfo {title} {{Second-order
  gauge invariant cosmological perturbation theory: Einstein equations in terms
  of gauge invariant variables}},}\ }\href {https://doi.org/10.1143/PTP.117.17}
  {\bibfield  {journal} {\bibinfo  {journal} {Prog. Theor. Phys.}\ }\textbf
  {\bibinfo {volume} {117}},\ \bibinfo {pages} {17--74} (\bibinfo {year}
  {2007})},\ \Eprint {https://arxiv.org/abs/gr-qc/0605108}
  {arXiv:gr-qc/0605108} \BibitemShut {NoStop}%
\bibitem [{\citenamefont {Pons}, \citenamefont {Salisbury},\ and\ \citenamefont
  {Sundermeyer}(2009)}]{ponsetal2009}%
  \BibitemOpen
  \bibfield  {author} {\bibinfo {author} {\bibfnamefont {J.~M.}\ \bibnamefont
  {Pons}}, \bibinfo {author} {\bibfnamefont {D.~C.}\ \bibnamefont
  {Salisbury}},\ and\ \bibinfo {author} {\bibfnamefont {K.~A.}\ \bibnamefont
  {Sundermeyer}},\ }\bibfield  {title} {\enquote {\bibinfo {title} {{Revisiting
  observables in generally covariant theories in the light of gauge fixing
  methods}},}\ }\href {https://doi.org/10.1103/PhysRevD.80.084015} {\bibfield
  {journal} {\bibinfo  {journal} {Phys. Rev. D}\ }\textbf {\bibinfo {volume}
  {80}},\ \bibinfo {pages} {084015} (\bibinfo {year} {2009})},\ \Eprint
  {https://arxiv.org/abs/0905.4564} {arXiv:0905.4564 [gr-qc]} \BibitemShut
  {NoStop}%
\bibitem [{\citenamefont {Giesel}\ \emph {et~al.}(2010)\citenamefont {Giesel},
  \citenamefont {Hofmann}, \citenamefont {Thiemann},\ and\ \citenamefont
  {Winkler}}]{gieseletal2010}%
  \BibitemOpen
  \bibfield  {author} {\bibinfo {author} {\bibfnamefont {K.}~\bibnamefont
  {Giesel}}, \bibinfo {author} {\bibfnamefont {S.}~\bibnamefont {Hofmann}},
  \bibinfo {author} {\bibfnamefont {T.}~\bibnamefont {Thiemann}},\ and\
  \bibinfo {author} {\bibfnamefont {O.}~\bibnamefont {Winkler}},\ }\bibfield
  {title} {\enquote {\bibinfo {title} {{Manifestly Gauge-invariant general
  relativistic perturbation theory. II. FRW background and first order}},}\
  }\href {https://doi.org/10.1088/0264-9381/27/5/055006} {\bibfield  {journal}
  {\bibinfo  {journal} {Class. Quant. Grav.}\ }\textbf {\bibinfo {volume}
  {27}},\ \bibinfo {pages} {055006} (\bibinfo {year} {2010})},\ \Eprint
  {https://arxiv.org/abs/0711.0117} {arXiv:0711.0117 [gr-qc]} \BibitemShut
  {NoStop}%
\bibitem [{\citenamefont {Gasperini}\ \emph {et~al.}(2011)\citenamefont
  {Gasperini}, \citenamefont {Marozzi}, \citenamefont {Nugier},\ and\
  \citenamefont {Veneziano}}]{gasperinietal2011}%
  \BibitemOpen
  \bibfield  {author} {\bibinfo {author} {\bibfnamefont {M.}~\bibnamefont
  {Gasperini}}, \bibinfo {author} {\bibfnamefont {G.}~\bibnamefont {Marozzi}},
  \bibinfo {author} {\bibfnamefont {F.}~\bibnamefont {Nugier}},\ and\ \bibinfo
  {author} {\bibfnamefont {G.}~\bibnamefont {Veneziano}},\ }\bibfield  {title}
  {\enquote {\bibinfo {title} {{Light-cone averaging in cosmology: Formalism
  and applications}},}\ }\href {https://doi.org/10.1088/1475-7516/2011/07/008}
  {\bibfield  {journal} {\bibinfo  {journal} {JCAP}\ }\textbf {\bibinfo
  {volume} {07}},\ \bibinfo {pages} {008} (\bibinfo {year} {2011})},\ \Eprint
  {https://arxiv.org/abs/1104.1167} {arXiv:1104.1167 [astro-ph.CO]}
  \BibitemShut {NoStop}%
\bibitem [{\citenamefont {Tambornino}(2012)}]{tambornino2012}%
  \BibitemOpen
  \bibfield  {author} {\bibinfo {author} {\bibfnamefont {J.}~\bibnamefont
  {Tambornino}},\ }\bibfield  {title} {\enquote {\bibinfo {title} {{Relational
  Observables in Gravity: a Review}},}\ }\href
  {https://doi.org/10.3842/SIGMA.2012.017} {\bibfield  {journal} {\bibinfo
  {journal} {SIGMA}\ }\textbf {\bibinfo {volume} {8}},\ \bibinfo {pages} {017}
  (\bibinfo {year} {2012})},\ \Eprint {https://arxiv.org/abs/1109.0740}
  {arXiv:1109.0740 [gr-qc]} \BibitemShut {NoStop}%
\bibitem [{\citenamefont {Donnelly}\ and\ \citenamefont
  {Giddings}(2016)}]{donnellygiddings2016}%
  \BibitemOpen
  \bibfield  {author} {\bibinfo {author} {\bibfnamefont {W.}~\bibnamefont
  {Donnelly}}\ and\ \bibinfo {author} {\bibfnamefont {S.~B.}\ \bibnamefont
  {Giddings}},\ }\bibfield  {title} {\enquote {\bibinfo {title}
  {{Diffeomorphism-invariant observables and their nonlocal algebra}},}\ }\href
  {https://doi.org/10.1103/PhysRevD.93.024030} {\bibfield  {journal} {\bibinfo
  {journal} {Phys. Rev. D}\ }\textbf {\bibinfo {volume} {93}},\ \bibinfo
  {pages} {024030} (\bibinfo {year} {2016})},\ \bibinfo {note} {[Erratum:
  \href{https://doi.org/10.1103/PhysRevD.94.029903}{Phys. Rev. D \textbf{94}
  (2016) 029903}]},\ \Eprint {https://arxiv.org/abs/1507.07921}
  {arXiv:1507.07921 [hep-th]} \BibitemShut {NoStop}%
\bibitem [{\citenamefont {Brunetti}\ \emph {et~al.}(2016)\citenamefont
  {Brunetti}, \citenamefont {Fredenhagen}, \citenamefont {Hack}, \citenamefont
  {Pinamonti},\ and\ \citenamefont {Rejzner}}]{brunettietal2016}%
  \BibitemOpen
  \bibfield  {author} {\bibinfo {author} {\bibfnamefont {R.}~\bibnamefont
  {Brunetti}}, \bibinfo {author} {\bibfnamefont {K.}~\bibnamefont
  {Fredenhagen}}, \bibinfo {author} {\bibfnamefont {T.-P.}\ \bibnamefont
  {Hack}}, \bibinfo {author} {\bibfnamefont {N.}~\bibnamefont {Pinamonti}},\
  and\ \bibinfo {author} {\bibfnamefont {K.}~\bibnamefont {Rejzner}},\
  }\bibfield  {title} {\enquote {\bibinfo {title} {{Cosmological perturbation
  theory and quantum gravity}},}\ }\href
  {https://doi.org/10.1007/JHEP08(2016)032} {\bibfield  {journal} {\bibinfo
  {journal} {JHEP}\ }\textbf {\bibinfo {volume} {08}},\ \bibinfo {pages} {032}
  (\bibinfo {year} {2016})},\ \Eprint {https://arxiv.org/abs/1605.02573}
  {arXiv:1605.02573 [gr-qc]} \BibitemShut {NoStop}%
\bibitem [{\citenamefont {Giesel}, \citenamefont {Herzog},\ and\ \citenamefont
  {Singh}(2018)}]{gieseletal2018}%
  \BibitemOpen
  \bibfield  {author} {\bibinfo {author} {\bibfnamefont {K.}~\bibnamefont
  {Giesel}}, \bibinfo {author} {\bibfnamefont {A.}~\bibnamefont {Herzog}},\
  and\ \bibinfo {author} {\bibfnamefont {P.}~\bibnamefont {Singh}},\ }\bibfield
   {title} {\enquote {\bibinfo {title} {{Gauge invariant variables for
  cosmological perturbation theory using geometrical clocks}},}\ }\href
  {https://doi.org/10.1088/1361-6382/aacda2} {\bibfield  {journal} {\bibinfo
  {journal} {Class. Quant. Grav.}\ }\textbf {\bibinfo {volume} {35}},\ \bibinfo
  {pages} {155012} (\bibinfo {year} {2018})},\ \Eprint
  {https://arxiv.org/abs/1801.09630} {arXiv:1801.09630 [gr-qc]} \BibitemShut
  {NoStop}%
\bibitem [{\citenamefont {Fr{\"o}b}\ and\ \citenamefont
  {Lima}(2018)}]{froeblima2018}%
  \BibitemOpen
  \bibfield  {author} {\bibinfo {author} {\bibfnamefont {M.~B.}\ \bibnamefont
  {Fr{\"o}b}}\ and\ \bibinfo {author} {\bibfnamefont {W.~C.~C.}\ \bibnamefont
  {Lima}},\ }\bibfield  {title} {\enquote {\bibinfo {title} {{Propagators for
  gauge-invariant observables in cosmology}},}\ }\href
  {https://doi.org/10.1088/1361-6382/aab427} {\bibfield  {journal} {\bibinfo
  {journal} {Class. Quant. Grav.}\ }\textbf {\bibinfo {volume} {35}},\ \bibinfo
  {pages} {095010} (\bibinfo {year} {2018})},\ \Eprint
  {https://arxiv.org/abs/1711.08470} {arXiv:1711.08470 [gr-qc]} \BibitemShut
  {NoStop}%
\bibitem [{\citenamefont {Fanizza}\ \emph {et~al.}(2021)\citenamefont
  {Fanizza}, \citenamefont {Marozzi}, \citenamefont {Medeiros},\ and\
  \citenamefont {Schiaffino}}]{fanizzaetal2021}%
  \BibitemOpen
  \bibfield  {author} {\bibinfo {author} {\bibfnamefont {G.}~\bibnamefont
  {Fanizza}}, \bibinfo {author} {\bibfnamefont {G.}~\bibnamefont {Marozzi}},
  \bibinfo {author} {\bibfnamefont {M.}~\bibnamefont {Medeiros}},\ and\
  \bibinfo {author} {\bibfnamefont {G.}~\bibnamefont {Schiaffino}},\ }\bibfield
   {title} {\enquote {\bibinfo {title} {{The Cosmological Perturbation Theory
  on the Geodesic Light-Cone background}},}\ }\href
  {https://doi.org/10.1088/1475-7516/2021/02/014} {\bibfield  {journal}
  {\bibinfo  {journal} {JCAP}\ }\textbf {\bibinfo {volume} {02}},\ \bibinfo
  {pages} {014} (\bibinfo {year} {2021})},\ \Eprint
  {https://arxiv.org/abs/2009.14134} {arXiv:2009.14134 [gr-qc]} \BibitemShut
  {NoStop}%
\bibitem [{\citenamefont {Mitsou}\ \emph {et~al.}(2021)\citenamefont {Mitsou},
  \citenamefont {Fanizza}, \citenamefont {Grimm},\ and\ \citenamefont
  {Yoo}}]{mitsouetal2021}%
  \BibitemOpen
  \bibfield  {author} {\bibinfo {author} {\bibfnamefont {E.}~\bibnamefont
  {Mitsou}}, \bibinfo {author} {\bibfnamefont {G.}~\bibnamefont {Fanizza}},
  \bibinfo {author} {\bibfnamefont {N.}~\bibnamefont {Grimm}},\ and\ \bibinfo
  {author} {\bibfnamefont {J.}~\bibnamefont {Yoo}},\ }\bibfield  {title}
  {\enquote {\bibinfo {title} {{Cutting out the cosmological middle man:
  General Relativity in the light-cone coordinates}},}\ }\href
  {https://doi.org/10.1088/1361-6382/abd681} {\bibfield  {journal} {\bibinfo
  {journal} {Class. Quant. Grav.}\ }\textbf {\bibinfo {volume} {38}},\ \bibinfo
  {pages} {055011} (\bibinfo {year} {2021})},\ \Eprint
  {https://arxiv.org/abs/2009.14687} {arXiv:2009.14687 [gr-qc]} \BibitemShut
  {NoStop}%
\bibitem [{\citenamefont {Fr{\"o}b}\ and\ \citenamefont
  {Lima}(2022)}]{froeblima2022}%
  \BibitemOpen
  \bibfield  {author} {\bibinfo {author} {\bibfnamefont {M.~B.}\ \bibnamefont
  {Fr{\"o}b}}\ and\ \bibinfo {author} {\bibfnamefont {W.~C.~C.}\ \bibnamefont
  {Lima}},\ }\bibfield  {title} {\enquote {\bibinfo {title} {{Cosmological
  perturbations and invariant observables in geodesic lightcone
  coordinates}},}\ }\href {https://doi.org/10.1088/1475-7516/2022/01/034}
  {\bibfield  {journal} {\bibinfo  {journal} {JCAP}\ }\textbf {\bibinfo
  {volume} {01}},\ \bibinfo {pages} {034} (\bibinfo {year} {2022})},\ \Eprint
  {https://arxiv.org/abs/2108.11960} {arXiv:2108.11960 [gr-qc]} \BibitemShut
  {NoStop}%
\bibitem [{\citenamefont {Baldazzi}, \citenamefont {Falls},\ and\ \citenamefont
  {Ferrero}(2022)}]{baldazzietal2022}%
  \BibitemOpen
  \bibfield  {author} {\bibinfo {author} {\bibfnamefont {A.}~\bibnamefont
  {Baldazzi}}, \bibinfo {author} {\bibfnamefont {K.}~\bibnamefont {Falls}},\
  and\ \bibinfo {author} {\bibfnamefont {R.}~\bibnamefont {Ferrero}},\
  }\bibfield  {title} {\enquote {\bibinfo {title} {{Relational observables in
  asymptotically safe gravity}},}\ }\href
  {https://doi.org/10.1016/j.aop.2022.168822} {\bibfield  {journal} {\bibinfo
  {journal} {Annals Phys.}\ }\textbf {\bibinfo {volume} {440}},\ \bibinfo
  {pages} {168822} (\bibinfo {year} {2022})},\ \Eprint
  {https://arxiv.org/abs/2112.02118} {arXiv:2112.02118 [hep-th]} \BibitemShut
  {NoStop}%
\bibitem [{\citenamefont {Fr{\"o}b}\ and\ \citenamefont
  {Lima}(2023)}]{froeblima2023}%
  \BibitemOpen
  \bibfield  {author} {\bibinfo {author} {\bibfnamefont {M.~B.}\ \bibnamefont
  {Fr{\"o}b}}\ and\ \bibinfo {author} {\bibfnamefont {W.~C.~C.}\ \bibnamefont
  {Lima}},\ }\bibfield  {title} {\enquote {\bibinfo {title} {{Synchronous
  coordinates and gauge-invariant observables in cosmological spacetimes}},}\
  }\href {https://doi.org/10.1088/1361-6382/acf98a} {\bibfield  {journal}
  {\bibinfo  {journal} {Class. Quant. Grav.}\ }\textbf {\bibinfo {volume}
  {40}},\ \bibinfo {pages} {215006} (\bibinfo {year} {2023})},\ \Eprint
  {https://arxiv.org/abs/2303.16218} {arXiv:2303.16218 [gr-qc]} \BibitemShut
  {NoStop}%
\bibitem [{\citenamefont {Goeller}, \citenamefont {Hoehn},\ and\ \citenamefont
  {Kirklin}(2022)}]{goelleretal2022}%
  \BibitemOpen
  \bibfield  {author} {\bibinfo {author} {\bibfnamefont {C.}~\bibnamefont
  {Goeller}}, \bibinfo {author} {\bibfnamefont {P.~A.}\ \bibnamefont {Hoehn}},\
  and\ \bibinfo {author} {\bibfnamefont {J.}~\bibnamefont {Kirklin}},\
  }\bibfield  {title} {\enquote {\bibinfo {title} {{Diffeomorphism-invariant
  observables and dynamical frames in gravity: reconciling bulk locality with
  general covariance}},}\ }\href@noop {} {\  (\bibinfo {year} {2022})},\
  \Eprint {https://arxiv.org/abs/2206.01193} {arXiv:2206.01193 [hep-th]}
  \BibitemShut {NoStop}%
\end{thebibliography}%

\end{document}